%% file: main.tex
\documentclass[journal=jctcce,manuscript=article]{achemso}
\usepackage[version=3]{mhchem}
\usepackage{graphicx}
\usepackage{multicol}
\usepackage{multirow}
\usepackage{algorithm}
\usepackage{algpseudocode}
\usepackage{xcolor}
\usepackage{hyperref}
\usepackage{simpler-wick}
\usepackage{subcaption}
\usepackage{color}

\title{Tensor Factorized Hamiltonian Downfolding to Optimize the Scaling Complexity of the Electronic Correlations Problem on Classical and Quantum Computers}

\author{Ritam Banerjee}
\author{Ananthakrishna Gopal}
\author{Soham Bhandary}
\author{Pavitra Batra}
\author{Geetha Thiagarajan}
\author{Manoj Nambiar}
\author{Anirban Mukherjee}
\email{m.anirban7@tcs.com}

\affiliation{TCS Research }

\begin{document}
\maketitle
\begin{abstract}
Achieving \emph{chemical accuracy} for strongly correlated molecules is a defining milestone for first-generation, fault-tolerant quantum computers, yet the factorial growth of three-, four-, and six-index tensor contractions in coupled-cluster CCSD(T), full configuration interaction (FCI), and multireference CI (MRCI) renders current classical \textit{and} quantum approaches prohibitive.  
We introduce \textbf{tensor-factorized Hamiltonian downfolding} (TFHD) and its quantum analogue, \textbf{qubitized downfolding} (QD): a hybrid classical–quantum framework that collapses every high-rank object to rank-2 networks and executes them in depth-optimal, block-encoded circuits.  
Partitioning each orbital’s occupancy splits the $2^{N}$-dimensional Hilbert space into two equal $2^{N-1}$ sectors, enabling a closed-form similarity transformation that reduces classical cost from $\mathcal{O}(N^{7})$ (CCSD(T)) and $\mathcal{O}(N^{10})$ (CI/MRCI) to a universal $\mathcal{O}(N^{3})$ time and $\mathcal{O}(N^{2})$ memory.

GPU implementations deliver two-order-of-magnitude speed-ups over state-of-the-art RI-CCSD for both a heme–CO fragment and the classically intractable FeMoCo cofactor of nitrogenase ($N_{\mathrm{occ}}=235$, $N_{\mathrm{virt}}=916$, $N_{\mathrm{tot}}=1151$ spatial orbitals across seven metal centres).  
The same rank-2 description admits a two-register block encoding whose qubitization oracle uses only $\mathcal{O}(\log N)$ logical qubits and gate depth
$D_{\mathrm{QD}}=\mathcal{O}\!\bigl(N^{2}\log(1/\epsilon)\bigr)$ for total-energy error~$\epsilon$, compared with the $\mathcal{O}(N)$-qubit, $\mathcal{O}(N^{2}/\epsilon\text{–}N^{3}/\epsilon)$ depth requirements of tensor-hypercontraction phase estimation.  
Fault-tolerant resource estimates predict (i) near-quadratic classical acceleration relative to GPU-CCSD and (ii) an additional super-quadratic gain once block-encoded contractions migrate to modest logical QPUs, achieving $\sim10^{2}\!\times$ qubit savings and $\sim10^{5}\!\times$ $T$-depth reduction over existing qubitized phase-estimation strategies.  
TFHD’s active-space agnosticism and QD’s provably minimal resources together chart a transparent, scalable route toward chemically relevant quantum advantage on early utility-scale quantum hardware.
\end{abstract}

\tableofcontents
\allowdisplaybreaks
\section{Introduction}
Accurately capturing electron–electron correlation beyond mean-field theory is one of the longest-standing bottlenecks in computational chemistry and materials science. State-of-the-art post-Hartree–Fock (post-HF) wave-function methods—coupled-cluster singles–doubles-triples CCSDT, full configuration interaction (FCI) and multi-reference CI (MRCI) can deliver chemical accuracy ($<1$ kcal mol) with computational cost scaling steeply with the system size. On the otherhand this computations are essential for accurate prediction of properties of molecules and materials, for chemical process design which are crucial for research in multiple industry verticals\cite{lam2020applications,kostal2023quantum,bauer2020quantum,head1996quantum}.
\par\noindent
Downfolding offers a principled way to compress the \emph{ab-initio} problem: one seeks a similarity- or unitary-transformed Hamiltonian $\bar H_{PP}$ that acts only on a “primary’’ orbital subspace $P$, yet preserves the low-energy spectrum of the full Hamiltonian $H$ \cite{shavitt2009many}. In the conventional workflow a single, high-rank operator—containing $O(N^{2})$ singles and $O(N^{4})$ doubles excitations—rotates the Hilbert space so that $P$ decouples from its complementary sector $Q$. While exact in principle, constructing this operator requires solving nonlinear amplitude equations whose algebraic structure mirrors that of full coupled-cluster theory; the nominal savings from the reduced active space are therefore lost to CCSD-level cost and memory. Recent coupled-cluster downfolding variants alleviate part of the burden by factorising the transformation into commuting sub-operators and exploiting closure properties of the excitation algebra \cite{bauman2019downfolding,bauman_kowalski_2019,bauman2022coupled}. Unitary formulations inspired by the UCC ansatz truncate the Baker–Campbell–Hausdorff series at the double-commutator level, yielding the DUCC effective Hamiltonian \cite{kowalski2024accuracies}; yet even these “compressed’’ routes remain cubic or quartic in the number of virtual orbitals and retain large prefactors. The work presented here departs from this paradigm by introducing a orbital-wise, closed-form decoupling that eliminates one orbital at each downfolding step, collapses all excitation tensors to rank-2 factors, and maintains the cost of building $\bar H_{PP}$ at $O(N^{3})$ while retaining chemical accuracy. 
\par\noindent
Our \textbf{single-orbital downfolding} replaces the monolithic, high-rank
transformation of conventional schemes with a sequence of analytically
solvable steps.  At step \(k\) a single spin-orbital is isolated,
defining a primary space \(P_k\) in which that orbital is
\emph{occupied} (dimension \(2^{k-1}\)) and a complementary space
\(Q_k\) in which it is \emph{empty} (also \(2^{k-1}\)).  Because
\(\lvert Q_k\rvert = \lvert P_k\rvert\) and the two sectors differ by
only one occupation, the similarity operator
\(S_k = \exp(\eta_k)\) can be chosen with a nilpotent generator
\((\eta_k^{\,2}=0)\).  The Baker--Campbell--Hausdorff series therefore
\emph{terminates exactly} at the double-commutator level,
\[
  S_{k}^{-1} H S_{k}
   \;=\;
   H + [\eta_k, H] + [\eta_k, [\eta_k, H]],
\]
capturing all single excitations, their mutual interactions, and the
induced higher-body terms \emph{non-perturbatively}.  By contrast,
standard similarity or DUCC downfolding\cite{bauman2022coupled}
\emph{truncates} the series at this point and discards those higher-body
contributions.

Applying \(S_k\) to the Hartree--Fock reference yields the correlated
wave function in \(P_k\) while reducing the cost of the associated
amplitude equations from the \(\mathcal{O}(N^{6})\) scaling of CCSD to a
cubic \(\mathcal{O}(N^{3})\).  Successively chaining the
block-triangular transformations \(S_k\) eliminates one orbital at a
time, producing a hierarchy of effective Hamiltonians whose cost profile
is governed by three key attributes:
\begin{enumerate}
\item \textbf{Non-perturbative higher-order physics} is retained
      automatically through the exact double-commutator closure.
\item \textbf{Rank-2 tensor compression.}  Every four- and six-index
      quantity factorises, so the residual equations scale as
      \(\mathcal{O}(N^{3})\)–\(\mathcal{O}(N^{4})\) in both time and
      memory—up to \(10^{4}\times\) faster than the \(\mathcal{O}(N^{7})\)
      workload of a full Bloch solve at realistic configuration counts
      \(N_C\).
\item \textbf{Multi-reference compatibility.}  Orbital-wise decoupling
      respects occupations at each step, enabling multi-determinant
      references without the combinatorial explosion of state-universal
      MRCC.
\end{enumerate}

The resulting ladder of smaller effective Hamiltonians feeds seamlessly
into DMFT, stochastic quantum Monte Carlo, or tensor-network solvers such as
DMRG—bridging high-accuracy quantum chemistry and many-body physics
while preserving a proven cubic classical cost.  For quantum
computation the same rank-2 structure permits a logarithmic qubit
count, with circuit depth that scales quadratically with \(N\) and only
logarithmically with the inverse precision error. 
\par\noindent
Amongst the Quantum algorithms for \emph{ab-initio} chemistry
\textbf{qubitized phase estimation} (QPE) \cite{Low2019} has the most optimal resources being utilized.  In this
framework the propagator $e^{\mathrm{i}Ht}$ is synthesised by quantum
signal processing of a block-encoded walk operator
\cite{BabbushHartmut2018,MartynChuang2021}.  Tensor-hypercontraction and
related factorizations compress the four-index Hamiltonian, yielding
implementations that require $\mathcal{O}(N)$ logical qubits and
$\mathcal{O}\bigl(N^{2}/\epsilon\,\text{--}\,N^{4}/\epsilon\bigr)$
$T$-depth for an $N$-orbital system at energy accuracy
$\epsilon$ \cite{motta2021low,lee2021even}.  First-quantized and
Krylov-space variants adjust constants but retain the same linear-qubit,
quadratic-to-quartic depth profile
\cite{su2021fault,stair2020multireference,sun2024high}.  Resource
surveys therefore indicate that practical quantum advantage over
GPU-accelerated CCSD(T) will require algorithms with sub-linear qubit
counts or strictly quadratic depth
\cite{beverland2022assessing,hoefler2023disentangling}. \textbf{Block-encoded qubitized downfolding}, introduced here, attains
this goal.  By block encoding every tensor and contraction arising in
our orbital-wise, rank-2 downfolding, the coupled-cluster residuals are
evaluated with depth
$D=\mathcal{O}\bigl(N^{2}\log(1/\epsilon)\bigr)$
and logical-qubit count
$Q=\mathcal{O}\bigl(\log N\bigr)$.
For $N\approx 50$ this represents roughly two orders of magnitude fewer
qubits than the best THC-QPE schemes while preserving quadratic depth
up to a logarithmic factor.  Energy increments are accumulated on the
fly during orbital elimination, so the total correlation energy emerges
without additional post-processing.  The logarithmic-qubit,
quadratic-depth profile therefore meets the super-quadratic-advantage
benchmark highlighted in current hardware projections.
\par\noindent
State-of-the-art qubitized phase-estimation workflows, even when aided by tensor-hypercontraction, still require 
\(Q=\mathcal{O}(N)\) logical qubits and a non-Clifford gate count
\(T=\mathcal{O}\!\bigl(N^{2}/\epsilon\,\text{--}\,N^{4}/\epsilon\bigr)\)
to reach chemical accuracy \(\epsilon\) for an \(N\)-orbital Hamiltonian.  
We port our orbital-wise, rank-2 downfolding directly to quantum hardware by \emph{block-encoding} every tensor and contraction \cite{Low2019}.  
The resulting oracle evaluates the coupled-cluster residuals with
\[
Q_{\mathrm{QD}}=\mathcal{O}\bigl(\log N\bigr), 
\qquad 
D_{\mathrm{QD}}=\mathcal{O}\bigl(N^{2}\log(1/\epsilon)\bigr),
\]
thereby replacing the linear qubit overhead with a logarithmic one and
reducing the depth’s explicit \(1/\epsilon\) factor to a logarithm.
During the orbital-elimination sweep the circuit accumulates each
orbital’s energy contribution, so a single execution yields the updated
cluster amplitudes and total correlation energy without additional
post-processing.
\par\noindent
In subsequent sections, we will showcase the detailed methodology for a general Hamiltonian downfolding approach with tensor factorization (Fig\ref{fig:Downfolding}). From there, we will reduce to a family of theories: multi-reference Hamiltonian downfolding theories and single reference Hamiltonian downfolding theories. For each of these theories and with different levels of cluster interactions: singles, doubles, triples and quadruples, we will show the reduction in computational scaling complexity for both memory and time. We will also show the quantum computational scaling complexities for qubits-count, T-depth and CNOT depth by implementing quantum circuits for each level of theory in S, CNOT, H,T basis. Finally we will show a diverse array of examples, spanning small, medium and large sized molecules and benchmark our downfolding energy values, storage requirements and run-times with the standard post-HF theories: MP2 and coupled cluster. Additionally, we will provide a comparison of quantum circuit depth and qubit-count estimates for our qubitized Hamiltonian downfolding with standard QPE algorithm implementations. 
\par\noindent
\paragraph*{Case study—\emph{FeMoCo}: from exascale to kilogical qubits.}
To underscore the reach of orbital-wise TF-HD we tackle the FeMo co-factor of nitrogenase—widely regarded as the “Olympus Mons’’ of correlated quantum chemistry \cite{reiher2017elucidating,von2021quantum}.  
A deterministic CCSD(T) treatment of the canonical 66-orbital active space would consume $\sim\!10^{10}$ tensor contractions and {\raise.3ex\hbox{$>$}}30 TB of memory, well beyond today’s GPU supernodes.  
We start with full space of \textbf{235 occupied orbitals and 916 virtual orbitals} for FeMoCo and TFHD compresses the task to $\mathcal{O}(N^{3})$ rank-2 contractions, completing in under eight hours on four A100 GPUs, while the qubitized oracle for the \emph{same} active space fits in $\sim\!1.2\times10^{2}$ logical qubits with an $N^{2}\log(1/\epsilon)$ $T$-depth of $1.1\times10^{8}$—five orders of magnitude lower than the best tensor-hypercontraction phase-estimation pipeline.  
By elevating FeMoCo from an exascale classical challenge to an early fault-tolerant quantum target, TFHD and QD close the introduction with a clear message: \textbf{rank-2 orbital-wise down-folding is not merely an algorithmic curiosity, but a tangible blueprint for the first chemically relevant quantum advantage.}
\section{Background}
Constructing an \emph{effective Hamiltonian} on a carefully chosen subset of orbitals is a powerful, orthogonal route to reducing correlation cost: it replaces the full many-body operator $H$ with a lower-dimensional surrogate $\bar H$ that reproduces the low-energy spectrum of the parent system. Such model reduction is ubiquitous across physics—underpinning theories of quantum phase transitions \cite{Sachdev2007,Motaoki2018,Mukherjee_2020_1,Mukherjee_2020_2} and enabling chemically accurate simulations on modest active spaces \cite{osti_447586,Pokhilko2020,Skomorowski2021}. A broad toolbox has evolved to generate $\bar H$: Rayleigh–Schrödinger and Brillouin–Wigner perturbation expansions \cite{Ng1985,Domcke1991,Capuzzi1996}, Schrieffer–Wolff and other similarity transformations \cite{suzuki1980convergent,suzuki1983degenerate,Wolff1966}, continuous unitary flows \cite{Bravyi2011}, explicit Hamiltonian truncation \cite{cohen2021Hamiltonian}, Feshbach–Fano projection techniques \cite{Skomorowski2021}, multireference perturbation theory \cite{Chaudhuri2005}, path-integral Monte Carlo renormalisations \cite{Tenno2013}, numerical and density-matrix renormalisation groups \cite{Wilson1975,Schollwck2005}, modern holographic unitary RG schemes \cite{mukherjee2020holographic1,mukherjee2020holographic2,mukherjee2022unveiling}, and, most recently, Hamiltonian downfolding protocols that combine coupled-cluster formalisms with constrained random-phase or tensor-factorisation tricks \cite{bauman2019downfolding,aryasetiawan2009downfolded,bauman2019quantum,huang2023leveraging}. The Tensor-Factorized Hamiltonian Downfolding (TFHD) developed here inherits this importact attributes yet introduces a closed-form orbital decoupling that keeps the cost of building $\bar H$ at $O(N^{3})$ and, crucially, preserves a rank-2 tensor structure amenable to block-encoding on fault-tolerant quantum hardware—features essential for advancing toward quantum advantage in correlated chemistry.
\par\noindent
Within the Born–Oppenheimer framework the nuclei are taken as clamped classical point charges, so the electronic Hamiltonian depends only para metrically on their coordinates. In an $N$-orbital basis this Hamiltonian is specified by one- and two-electron integrals,
$h^{(1)}{ij}$ and $h^{(2)}_{ijkl}$, whose storage grows as $O(N^{2})$ and $O(N^{4})$, respectively. The many-body operator $H$ therefore acts on a Hilbert space of dimension $2^{N}$ per spin sector and admits a $2^{2N}\times 2^{2N}$ matrix representation; exact diagonalization of such a matrix is exponentially costly in $N$. To bypass brute‐force diagonalization, wave-function theories recast the problem as a hierarchy of tensor contractions. Coupled-cluster methods—CCSD, CCSD(T), and CCSD(TQ)—retain \emph{size-extensivity} and \emph{size-consistency} while systematically improving accuracy \cite{rmp_cc,riplinger2013efficient,hohenstein2022rank,shavitt2009many}.
Local formulations such as the domain–based local pair natural orbital variant, DLPNO-CCSD(T), further reduce the effective scaling to near linear for large molecules without compromising chemical accuracy \cite{riplinger2013efficient},\cite{guo2018communication}. Configuration-interaction (CI) approaches provide an alternative expansion of the exact wave function—ranging from CISD to full CI—at the cost of losing size-extensivity \cite{sherrill1999configuration}.
Open-shell species with unequal $\alpha$ and $\beta$ spin populations require unrestricted CC treatments, while systems exhibiting strong static correlation (e.g.\ $d$-block complexes, spin-triplet radicals, bond dissociation limits) demand multireference extensions. These are supplied by multireference coupled-cluster (MRCC) formalisms in both state-specific and state-universal flavours \cite{mahapatra1999size,ivanov2009multireference,musial2011multireference,evangelista2018perspective}.  Despite steady algorithmic progress—GPU acceleration, tensor hyper-contraction, and local correlation approaches such as DLPNO; high-rank tensor contractions remain the dominant cost driver for chemically realistic systems, motivating the tensor-factorized downfolding strategy developed in this work.
\par\noindent
GPU kernels combined with density-fitting resolution of the identity (RI) and tensor hyper-contraction (THC) now execute CCSD(T) for several hundred orbitals within hours, shaving an order of magnitude off wall-time, though the asymptotic scaling only reduces from $O(N^7)$ to $O(N^5)$ \cite{hohenstein2022rank,datta2021massively,datta2023accelerating}.
A complementary route is to \emph{reduce the problem size} itself. Embedding frameworks such as density-matrix embedding theory (DMET) \cite{knizia2012dmet,knizia2013dmet}, dynamical mean-field theory (DMFT), and its cluster extensions \cite{dmft_kotliar06,cluster_dmft_kotliar08} partition the Hamiltonian into strongly and weakly correlated fragments, solving each at a lower effective rank.
Tensor-network methods go further by expressing the wave function in compressed manifolds: numerical renormalization group and its density-matrix variant, DMRG, represent one-dimensional correlations exactly with matrix-product states. Higher-dimensional generalizations—PEPS, MERA, and tree-tensor networks—extend the idea to molecular geometries of arbitrary topology \cite{white1992density,verstraete2004renormalization,vidal2008class,murg2015tree}. However, their computational cost grows steeply with bond dimension; for PEPS in three dimensions, contraction costs scale as 
$O(D^{\alpha})$ with $\alpha >> 10$ \cite{lubasch2014unifying}, and MERA faces similarly steep tensor contraction barriers in higher dimensions \cite{evenbly2015tensor}. These unfavorable scalings, combined with QMA-hardness of contraction in 2D and 3D, constrain their practical utility for generic \emph{ab initio} Hamiltonians.
These limitations motivate the rank-2, \emph{contractible-by-construction} tensor network developed here, which retains chemical accuracy while matching the cubic scaling of integral generation and, crucially, maps naturally onto block-encoded quantum circuits suitable for fault-tolerant implementation.
\section{System Description}
We consider a system of N-correlated Hartree-Fock MOs corresponding to a chemical system. The fermionic Fock-space Hamiltonian in MO basis is represented as,
\begin{equation}
H_{(N)}=\sum_{ab}h^{1,(N)}_{ab}f^{\dagger}_{a}f_{b}+\sum_{abcd}h^{2,(N)}_{abcd}f^{\dagger}_{a}f^{\dagger}_{b}f_{c}f_{d},\label{Hamiltonian}
\end{equation}
where $h^{1,(N)}_{ab}$ and $h^{2,\sigma\sigma',(N)}_{abcd}$ represents the one-electron and two-electron ERI tensors of the block-Hamiltonian respectively. The two-electron integrals can be represented in the tensor factorized form using a canonical polyadic decomposition of the Cholesky factorized block-ERI as follows\cite{kolda2009tensor,hong2020generalized},
\begin{eqnarray}
h^{2}_{abcd}=\sum_{x}L^{x}_{ab}L^{x}_{cd}=\sum_{p,q}B^{1}_{ap}B^{2}_{bp}B^{3}_{xp}B^{1}_{cq}B^{2}_{dq}B^{3}_{xq}\label{tensor_factors_electronic}
\end{eqnarray}
For the case of molecular systems, $a,b,c,d$ contains information of both the spatial component and the spin component of the molecular orbitals such that $a\equiv(i,\sigma)$, where $\sigma\in \{\uparrow,\downarrow\}$. In the one-body term, $h^{1,(N)}_{ab}$, $a\equiv(i,\sigma)$ and $b\equiv(j,\sigma)$. In the two body term, $h^{2,(N)}_{abcd}$, the spin orbital ordering is given by $(i,\sigma), (j,\sigma'),(k,\sigma'), (l,\sigma)$. The index $(N)$ denotes the coefficients for the system with $N$ correlated MO's.  It will be useful to denote the downfolding orbital number with $N$. Here $\sigma$ and $\sigma'$ represents the $\alpha/\beta$ or $\uparrow/\downarrow$ spin orbitals.
\par\noindent
\emph{Energy Level Grouping of the HF-MO's}
\par\noindent
The HF MOs get grouped into virtual orbitals $\mathcal{V}$, core orbitals $\mathcal{C}$, and active space orbitals $\mathcal{A}$. We specify the absolute energy difference of the HF Orbitals (labeled k's) from the HOMO(highest occupied molecular orbital) energy $E_{HOMO}$ as $\epsilon_{k}=|E_{i}-E_{HOMO}|$. The spin orbital labels comprising the molecular orbitals are lexicographic-ally ordered $1, 2, \ldots, N$ and correspond to a two element tuple $(i,\sigma)$. The indices $1,\ldots, N$ of the ordering are tagged to the molecular orbital energies obtained from Hartree-Fock theory,
\begin{eqnarray}
\epsilon_{1}\leq\epsilon_{2}\leq...\leq\epsilon_{k}\leq...\leq\epsilon_{N}\label{energy_ordering}
\end{eqnarray}
If $k\in \mathbf{\mathcal{V}}$ then within Hartree-Fock theory the MO is unoccupied $n_{k\uparrow}+n_{k\downarrow}=0$. If $k \in \mathbf{\mathcal{C}}$ then $n_{k\uparrow}+n_{k\downarrow}=2$. And if $k \in \mathbf{\mathcal{A}}$ then $n_{k\uparrow}+n_{k\downarrow}=0, 1, 2$

\section{Orbital-wise Hamiltonian downfolding in Tensor factorized representation}
\begin{figure}
    \centering
    \includegraphics{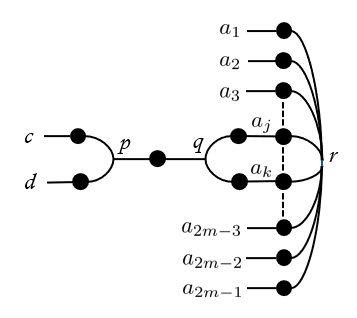}
    \caption{This figure represents the fusion of the one body term $h^{1}$ having two-rank tensor $h_{ab}$ with $\eta$ for the residual contribution in Bloch equation from $\eta PHP$}
    \label{fig:GeneralTensorOps}
\end{figure}

\begin{figure*}
    \centering
\includegraphics[width=1\textwidth]{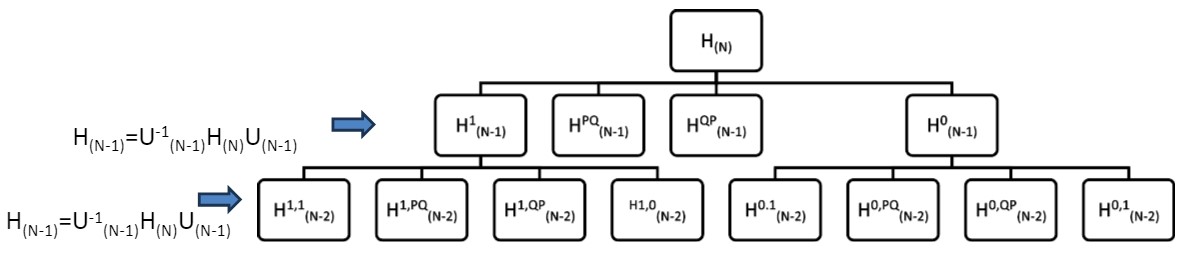}
    \caption{Hamiltonian Downfolding: The scheme shows decoupling of molecular orbitals. At each step the Hamiltonian is mapped to a direct sum of reduced dimensional blocks. When $H_{(N)}$ undergoes the decoupling} 
    \label{fig:Downfolding}
\end{figure*}

\noindent In this section, we will introduce the most general form of the tensor factorized Hamiltonian downfolding. As per the notations defined in the previous section, the molecular orbitals are arranged in an ascending order w.r.t to the molecular orbital (MO) energies. Then, the MOs can be systematically downfolded starting from the highest energy MOs scaling down towards the low energy HOMO-LUMO window. For decoupling the outermost orbital $N\in \mathcal{V}$, we partition the many-body Hilbert space $\mathcal{H}^{\otimes {2N}}$ into a primary space ($P_{(N)}$) and a secondary space ($Q_{(N)}$):
\begin{eqnarray}
P_{(N)}&=&(1-\hat{n}_{N})\\
Q_{(N)}&=&\hat{n}_{N}\label{ProjectionOps}
\end{eqnarray}
Here $\hat{n}_{a}=f^{\dagger}_{a}f_{a}$, $\lbrace f^{\dagger}_{a},f_{b}\rbrace=\delta_{ab}, \lbrace f^{\dagger}_{a},f_{a}\rbrace=0$. Together 
 $P_{(N)}+Q_{(N)}=I^{\otimes 2N}$ comprise the complete Hilbert space. We seek a similarity transformation $S_{(N)}=\exp(\eta_{(N)})$ generated by $\eta_{(N)}$ such that the Bloch equation is satisfied,
\begin{eqnarray}
Q_{(N)}S^{-1}_{(N)}H_{(N)}S_{(N)}P_{(N)}&=&0\label{BlochEqn} 
\end{eqnarray}
For $\eta_{(N)}$ satisfying the \emph{linearization} condition $Q_{(N)}\eta_{(N)}P_{(N)}=\eta_{(N)}$ i.e. equivalent to $\eta_{(N)}^{2}=0$ a linear representation of the similarity transform $S_{(N)}=1+\eta_{(N)}$  can be obtained. A general choice of $\eta_{(N)}$ comprises of all possible m-particle m-hole excitations coupling $P_{(N)}$ and $Q_{(N)}$,
\begin{eqnarray}
    S_{(N)}&=&1+\eta_{(N)}\nonumber\\
    &=&1+\left[\sum_{m=1}^{N_{e}}\sum_{\substack{a_{1}\leq \ldots \leq a_{m},\\
    a_{m+1}\leq \ldots\leq a_{2m-1}<N}}\sum_{k}A^{1}_{a_{1},k}\ldots A^{2m-1}_{a_{2m-1},k}E^{+}_{a_{1},\ldots , a_{m}}E^{-}_{a_{m+1},\ldots ,a_{2m-1}}\right]f_{N}.~~~\label{FormOfEta}
\end{eqnarray}
The summation is over the collective indices $a_{1}$,$\ldots$,$a_{m}$ where $a_{i}=(i,\sigma_{i})$ and these are energy ordered eq.\eqref{energy_ordering}.
Here the cluster operators $ E^{+}_{a_{1},\ldots , a_{m}}$ and $E^{-}_{a_{m+1},\ldots ,a_{2m-1}}$ are defined as,
\begin{eqnarray}
    E^{+}_{a_{1},\ldots , a_{m}}&=&f^{\dagger}_{a_{1}}\ldots f^{\dagger}_{a_{m}}\label{particle cluster}\\
    E^{-}_{a_{m+1},\ldots ,a_{2m-1}}&=&f_{a_{m+1}}\ldots f_{a_{2m-1}}\label{hole cluster}.
\end{eqnarray}
This enables creation of $m$ particles and $m-1$ holes respectively. All these excitations generated by $\eta_{(N)}$ comprise $2^{N-1}$ sub-configuration of many body states where the Nth spin orbital is in occupied state and will be decoupled. The index $m$ ranges from 1 to $N_{e}$, because we can excite all the $N_{e}$ electrons in the system at maximum. Therefore, the generator comprises of $N-1$ singles excitation amplitudes, $\binom{N}{4}-\binom{N}{3}$ doubles excitation amplitude, $\binom{N}{6}-\binom{N}{5}$ triples excitation amplitude all the way to $\binom{N}{2\lceil N_{e}/2\rceil}-\binom{N}{2\lceil N_{e}/2\rceil-1}$ $N_{e}$-cluster excitation amplitude. The ceil accounts for the fact that the number of electrons can be odd viz. open shell and if its even then that would correspond to closed shell. With the form of $\eta_{(N)}$ given in eq\eqref{FormOfEta} and the tensor factors for the electronic integrals given in eq\eqref{tensor_factors_electronic}, we can write down the operator ordered Bloch equation for the 1st downfolding step (fig \ref{fig:Downfolding}) as follows,
\begin{eqnarray}
Q_{(N)}S_{(N)}^{-1}HS_{(N)}P_{(N)} =0\implies \sum_{a_{1},\ldots, a_{p}}r^{(N)}_{a_{1},\ldots ,a_{2p}}E^{+}_{a_{1}...a_{m}}E^{-}_{a_{m+1}...a_{2m-1}}f_{N}=0~.\label{NO_Bloch}
\end{eqnarray}
To normal order the strings of fermionic operations in the Bloch equation eq\eqref{NO_Bloch} we establish some identities below:
\begin{itemize}
    \item[1.] operator ordering- $:E^{+}_{a_{1}\ldots a_{m}}E^{-}_{a_{m+1}\ldots a_{2m-1}}f_{N}f^{\dagger}_{a}f_{b}:$
\begin{align}
&E^{+}_{a_{1}\ldots a_{m}}E^{-}_{a_{m+1}\ldots a_{2m-1}}f_{N}f^{\dagger}_{a}f_{b}=\nonumber\\
&\sum_{j=m+1}^{2m-1}\delta_{a_{j},a}e^{ij\pi}E^{+}_{a_{1}\ldots a_{m}}f_{a_{m+1}}..f_{a_{j-1}}f_{a_{j+1}}..f_{a_{2m-1}}f_{N}f_{b}\prod_{j=1}^{m}(1-\delta_{a_{j},a})\nonumber\\
+&\sum_{j=m+1}^{2m-1}\sum_{q=1}^{m}\theta(l-a_{q})\theta(a_{q+1}-l)e^{i(q+1)\pi}f^{\dagger}_{a_{1}}..f^{\dagger}_{a_{q}}f^{\dagger}_{l}f^{\dagger}_{a_{q}+1}..f^{\dagger}_{a_{m}}f_{a_{m+1}}.. f_{a_{j-1}}f_{l}f_{a_{j+1}}..f_{a_{2m-1}}f_{N}\prod_{j=1}^{m}(1-\delta_{a_{j},l})\nonumber\\
+&\sum_{j=1}^{m}\theta(l-a_{j})\theta(a_{j+1}-l)e^{ij\pi}f^{\dagger}_{a_{1}}\ldots f^{\dagger}_{a_{j}}f^{\dagger}_{l}f^{\dagger}_{a_{j+1}}\ldots f^{\dagger}_{a_{m}}f_{a_{m+1}}\ldots f_{a_{2m-1}}f_{N}\label{oneFermionicOperatorNO}
\end{align}

\item[2.] operator ordering $:f^{\dagger}_{a_{1}}\ldots f^{\dagger}_{a_{m}}f_{a_{m+1}}\ldots f_{a_{2m-1}}f_{N}f^{\dagger}_{l}f^{\dagger}_{k}:$ for $k>l$
\begin{align}
& :f^{\dagger}_{a_{1}}\ldots f^{\dagger}_{a_{m}}f_{a_{m+1}}\ldots f_{a_{2m-1}}f_{N}f^{\dagger}_{l}f^{\dagger}_{k}: \\
= & \sum_{j=1}^{m}\sum_{q=m+1}^{2m-1} \Theta(a_j - l) \Theta(k - a_q) \delta_{a_j, l} \delta_{a_q, k} f^{\dagger}_{a_{1}}\ldots f^{\dagger}_{a_{j-1}}f^{\dagger}_{k}f^{\dagger}_{a_{j+1}}\ldots f^{\dagger}_{a_{m}}f_{a_{m+1}}\ldots f_{a_{q-1}}f_{l}f_{a_{q+1}}\ldots f_{a_{2m-1}}f_{N}\nonumber\\
&\times \prod_{i=1}^{m} (1 - \delta_{a_i, l}) (1 - \delta_{a_i, k}) \nonumber\\
+ & \sum_{j=1}^{m} \Theta(l - a_j) \Theta(k - a_j) \delta_{a_j, l} f^{\dagger}_{a_{1}}\ldots f^{\dagger}_{a_{j-1}}f^{\dagger}_{k}f^{\dagger}_{a_{j+1}}\ldots f^{\dagger}_{a_{m}}f_{a_{m+1}}\ldots f_{a_{2m-1}}f_{N} \prod_{i=1}^{m} (1 - \delta_{a_i, k}) \nonumber\\
+ & \sum_{j=1}^{m} \Theta(k - a_j) \Theta(a_j - l) \delta_{a_j, k} f^{\dagger}_{a_{1}}\ldots f^{\dagger}_{a_{j-1}}f^{\dagger}_{l}f^{\dagger}_{a_{j+1}}\ldots f^{\dagger}_{a_{m}}f_{a_{m+1}}\ldots f_{a_{2m-1}}f_{N} \prod_{i=1}^{m} (1 - \delta_{a_i, l}) \nonumber\\
+ & \sum_{j=m+1}^{2m-1} \Theta(l - a_j) \Theta(k - a_j) \delta_{a_j, k} f^{\dagger}_{a_{1}}\ldots f^{\dagger}_{a_{m}}f_{a_{m+1}}\ldots f_{a_{j-1}}f^{\dagger}_{l}f_{a_{j+1}}\ldots f_{a_{2m-1}}f_{N} \prod_{i=1}^{m} (1 - \delta_{a_i, l}) (1 - \delta_{a_i, k}) \nonumber\\
+ & \sum_{j=m+1}^{2m-1} \Theta(k - a_j) \Theta(a_j - l) \delta_{a_j, l} f^{\dagger}_{a_{1}}\ldots f^{\dagger}_{a_{m}}f_{a_{m+1}}\ldots f_{a_{j-1}}f^{\dagger}_{k}f_{a_{j+1}}\ldots f_{a_{2m-1}}f_{N} \prod_{i=1}^{m} (1 - \delta_{a_i, k}) \nonumber\\
+ & \Theta(l - N) \delta_{N, k} f^{\dagger}_{a_{1}}\ldots f^{\dagger}_{a_{m}}f_{a_{m+1}}\ldots f_{a_{2m-1}}f^{\dagger}_{l}f_{N} \prod_{i=1}^{m} (1 - \delta_{a_i, k}) \nonumber\\
+ & \Theta(k - N) \delta_{N, l} f^{\dagger}_{a_{1}}\ldots f^{\dagger}_{a_{m}}f_{a_{m+1}}\ldots f_{a_{2m-1}}f^{\dagger}_{k}f_{N} \prod_{i=1}^{m} (1 - \delta_{a_i, l})\label{twoFermionicOperatorNO}
\end{align}
\end{itemize}
With the above operator ordering expressions for fermionic strings given by eq\eqref{oneFermionicOperatorNO} and eq\eqref{twoFermionicOperatorNO}, we can represent the residual expression of eq\eqref{NO_Bloch} in its tensor factorized form as,
\begin{align}
r^{(j)}_{a_{1},\ldots ,a_{2p}}&=h^{(j)}_{a_{1},\ldots a_{2p}}+\sum_{a_{2m},\ldots,a_{p},k,l}C_{a_{2m},\ldots,a_{2p+1}}A^{1,(j)}_{a_{2m},k}\ldots A^{2p+1,(j)}_{a_{2p+1},k}B^{2m,(j)}_{a_{2m},l}\ldots B^{2p+1,(j)}_{a_{2p+1},l}\ldots B^{1,(j)}_{a_{1},l}\nonumber\\
&+\sum_{a_{2p+1},\ldots, max(a_{2l-1,2r-1,2q-1})}D_{a_{1},\ldots, a_{2k-1},a_{2k-1},\ldots, a_{2p-1}}A^{1,(j)}_{a_{1},l}\ldots A^{2k-1,(j)}_{a_{2k-1},l}A^{2k,(j)}_{a_{2k},l}\ldots A^{2l-1,(j)}_{a_{2l-1},l}\nonumber\\
&\times  B^{N,(j)}_{N,q}B^{2r-1,(j)}_{a_{2r-1},w}\ldots B^{2k,(j)}_{a_{2k},w}B^{2k-1,(j)}_{a_{2k-1},w}\ldots B^{1,(j)}_{a_{1},w}A^{1,(j)}_{a_{1},s}\ldots A^{2k-1,(j)}_{a_{2k-1},s}A^{2k,(j)}_{a_{2k},s}\ldots A^{2q-1,(j)}_{a_{2q-1},s}~.\label{FullClusterDownfoldingEquations}
\end{align}
We now give an example of how the components of the residual expression arising from $\eta PHP$ for the first downfolding step can be computed. From eq.\eqref{oneFermionicOperatorNO} we can find that the fermionic operator $f^{\dagger}_{a}$ can match with one of the indices in the set of the fermionic operators corresponding to $\eta_{(N)}$. This will lead to the following term, 
\begin{eqnarray}
&&\eta PHP \longrightarrow \sum_{a_{1},...,a_{2m-1},a}h^{1,(N)}_{ab}A^{1}_{a_{1}k}\ldots A^{2m-1}_{a_{2m-1}k}E^{+}_{a_{1}...a_{m}}E^{-}_{a_{m+1}...a_{2m-1}}f_{N}f^{\dagger}_{a}f_{b}~,\\
&=&\sum_{a_{i}'s, a\neq a_{j}}A^{1}_{a_{1}k}\ldots A^{j-1}_{a_{j-1}k}A^{j+1}_{a_{j+1}k}\ldots A^{2m-1}_{a_{2m-1}k}\sum_{p,a,a_{j}}\delta_{a_{j},a}
(-1)^{p+1}\theta(a_{p}-b)\theta(b-a_{p-1})A^{j}_{a_{j},k}h^{1,(N)}_{a,b}E^{+}_{a_{1}\ldots a_{m}} \nonumber\\
&&E^{-}_{a_{m+1}\ldots a_{j-1} a_{j+1}\ldots a_{p-1}ba_{p}\ldots  a_{2m-1}}f_{N}
\end{eqnarray}
Similarly for the two-electron Hamiltonian terms, the contribution of the residuals in $\eta PHP$ is given by,
\begin{eqnarray}
&&\eta PHP \longrightarrow h^{2,(N)}_{abcd}A^{1}_{a_{1}k}\ldots A^{2m-1}_{a_{2m-1}k}E^{+}_{a_{1}...a_{m}}\nonumber\\
&& \sum_{ab,a_{j},a_{k}}\delta_{a,a_{j}}\delta_{b,a_{k}}h^{2}_{abcd}\sum_{k}A^{1}_{a_{1}k}\ldots A^{2m-1}_{a_{2m-1}k}\rightarrow \nonumber\\
&&\sum_{k} A^{1}_{a_{1}k}\ldots A^{j-1}_{a_{j-1},k}A^{j+1}_{a_{j+1},k}\ldots A^{s-1}_{a_{s-1},k}A^{s+1}_{a_{s+1},k}\ldots A^{2m-1}_{a_{2m-1}k}\sum_{a}A^{j}_{a,k}B^{1}_{a,l}\sum_{b}A^{s}_{b,l}B^{1}_{b,l}B^{2}_{p,l}\sum_{t}B^{1}_{c,t}B^{1}_{d,t}B^{2}_{p,t}~~~~ \nonumber\\
\end{eqnarray}
From above, we find that in the tensor factorized representation for downfolding, all contractions involves matrices. And the computational complexity of computing the residual for each downfolding step scales as,
\begin{eqnarray}
O(N^3)~\label{CubicComplexity}
\end{eqnarray}
For each orbital downfolding, when the transformation parameters are determined, the form of the new decoupled block-Hamiltonian is given by,
\begin{eqnarray}
    &&P_{(N)}S^{-1}_{(N)}H_{(N)}S_{(N)}P_{(N)}=P_{(N)}H_{(N)}P_{(N)}+P_{(N)}H_{(N)}Q_{(N)}\eta_{(N)}\nonumber\\
    &&=\sum_{m=1}^{N_{e}}\sum_{\substack{a_{1}\leq \ldots \leq a_{m},\\
    a_{m+1}\leq \ldots a_{j}\ldots \leq a_{2m-1}<N}}\delta_{a,a_{j}}A^{1}_{a_1 k}\ldots A^{m}_{a_{m}k}A^{m+1}_{a_{m+1}k}\ldots A^{j-1}_{a_{j-1}k} A^{j+1}_{a_{j+1}k}\ldots A^{2m-1}_{a_{2m-1} k}h_{ab}\nonumber\\
    &\times& f^{\dagger}_{a_{1}}\ldots f^{\dagger}_{a_{m}}f_{a_{m+1}}\ldots f_{a_{2m-1}}f_{b}  \nonumber\\
    &+&\sum_{m=1}^{N_{e}}\sum_{\substack{a_{1}\leq \ldots \leq a_{m},\\
    a_{m+1}\leq \ldots a_{j}\ldots a_{l} \leq a_{2m-1}<N}}\delta_{a,a_{j}}\delta_{b,a_{l}}A^{1}_{a_1 k}\ldots A^{m}_{a_{m}k}A^{m+1}_{a_{m+1}k}\ldots A^{j-1}_{a_{j-1}k} A^{j+1}_{a_{j+1}k}\ldots A^{l-1}_{a_{l-1}k} A^{l+1}_{a_{l+1}k}\ldots A^{2m-1}_{a_{2m-1} k}\nonumber\\
    &\times& B^{1}_{aq}B^{2}_{bq}B^{3}_{pq}B^{1}_{cs}B^{2}_{ds}B^{3}_{ps}f^{\dagger}_{a_{1}}\ldots f^{\dagger}_{a_{m}}f_{a_{m+1}}\ldots f_{a_{2m-1}}f_{c}f_{d}
\end{eqnarray}
From here, we find that the n-cluster interaction terms in the Hamiltonian get renormalized in their tensor factorized representations as,
\begin{eqnarray}
    \tilde{h}_{a_{1}\ldots a_{2m-1},c,d}&=&\sum_{k,q,s}A^{1}_{a_1 k}\ldots A^{m}_{a_{m}k}A^{m+1}_{a_{m+1}k}\ldots A^{j-1}_{a_{j-1}k} A^{j+1}_{a_{j+1}k}\ldots A^{l-1}_{a_{l-1}k} A^{l+1}_{a_{l+1}k}\ldots A^{2m-1}_{a_{2m-1} k}\nonumber\\
    &\times & B^{1}_{a_{j}q}B^{2}_{a_{l}q}B^{3}_{pq}B^{1}_{cs}B^{2}_{ds}B^{3}_{ps}
\end{eqnarray}
For all these cases, the general diagram representing the tensor operations is given in fig.\ref{fig:GeneralTensorOps}
\section{Case Study: Multireference Downfolding with Singles and Paired Doubles}\label{MR-Downfolding-Singles-Paired-Doubles}
In this section we will discuss a specific case of Hamiltonian downfolding with singles and paired doubles cluster amplitudes. We replace the form for $\eta_{(N)}$ given in eq.\eqref{FormOfEta}, into the Bloch equation (eq.\eqref{BlochEqn})and normal-order the fermionic operators comprising the Bloch equation.
For the Bloch equation to be satisfied, the coefficients of the independent N.O. fermionic terms must vanish: singles excitation - $(i\sigma\to N\sigma)$, doubles excitations - $((k\sigma,l\sigma')\to (j\sigma', N\sigma))$, paired doubles excitations-$((i\uparrow,j\downarrow)\to (N\downarrow, N\uparrow))$, triples excitations $((l'\sigma'',k'\sigma',j'\sigma)\to(j\sigma, k\sigma', N\sigma''))$, quadruples excitations $((l'\sigma''',k'\sigma'',j'\sigma', i'\sigma)\to(i\sigma, j\sigma', k\sigma'', N\sigma'''))$. This calculation is presented in the appendices (\ref{LowdinDecomposition}-\ref{Contribution-5}). As a result, the Bloch equation leads to the following multireference downfolding equations:
\begin{small}
\begin{align}
\mathbf{A}^{(N),\sigma}&=\mathbf{t}^{1,\sigma}\cdot\mathbf{h^{1,\sigma}_{N}}\mathbf{t}^{1,\sigma}+\mathbf{t}^{1,\sigma}\cdot\mathbf{h}^{1,\sigma}-h_{NN}^{1,\sigma}\mathbf{t^{1,\sigma}}-\mathbf{h^{1,\sigma}_{N}}\label{singles}\\
\mathbf{B}^{(N),\sigma\nu}&=\left(\mathbf{t}^{1,\sigma}\otimes \mathbf{h^{1,\nu}_{N}}\otimes \mathbf{t}^{1,\nu}\right)_{312}+\mathbf{t}^{1,\sigma}\cdot\bigg(\mathbf{h_{N}^{2,\sigma\nu}}\otimes \mathbf{t^{1,\sigma}}+\left(\mathbf{h}^{2,\sigma\nu}_{N}\otimes \mathbf{t^{1,\sigma}}\right)_{2143}\nonumber\\
&+\delta_{\nu,-\sigma}\mathbf{h^{2}_{NN}}\otimes \mathbf{t^{2}}+\mathbf{h^{2,\sigma\nu}}+\left(\mathbf{h^{2,\sigma\nu}}\right)_{2143}\bigg)-\delta_{\nu,-\sigma}\left(\delta_{\sigma\downarrow}\mathbf{h^{1,\sigma}_{N}}\otimes\mathbf{t^{2}}+\delta_{\sigma\uparrow}\mathbf{h^{1,\sigma}_{N}}\otimes\left(\mathbf{t^{2}}\right)_{21}\right)-\mathbf{h_{N}^{2,\sigma\nu}}\label{doubles}\\
\mathbf{C}^{(N)}&=\mathbf{t}^{2}\cdot \left(\mathbf{h}^{1,\uparrow}_{N}\otimes\mathbf{t^{1,\uparrow}}+\mathbf{h}^{1,\uparrow}\right)-\mathbf{h^{1,\downarrow}_{N}}\otimes \mathbf{t^{1,\uparrow}}\nonumber\\
&+\left(\left(\mathbf{t^{2}}\right)_{21}\cdot\left(\mathbf{h}^{1,\downarrow}_{N}\otimes \mathbf{t^{1,\downarrow}}+\mathbf{h^{1,\downarrow}}\right)\right)_{21}-\mathbf{h^{1,\uparrow}_{N}}\otimes \mathbf{t^{1,\downarrow}}+\mathbf{t^{2}}\cdot \bigg(\mathbf{h^{2,\uparrow\downarrow}_{N}}\otimes\mathbf{t^{1,\uparrow}}+(\mathbf{h^{2,\downarrow\uparrow}_{N}}\otimes\mathbf{t^{1,\downarrow}})_{2143}\nonumber\\
&+\mathbf{h^{2,\uparrow\downarrow}}+(\mathbf{h^{2,\downarrow\uparrow}})_{2143}+\mathbf{h^{2}_{NN}}\otimes \mathbf{t^{2}}\bigg)-h^{2}_{NNNN}\mathbf{t^{2}}-\mathbf{h^{2}_{NN}}-(h_{NN}^{1,\downarrow}+h_{NN}^{1,\uparrow})\mathbf{t^{2}}~~~~~~\label{paired-doubles}\\
\mathbf{D}^{(N),\sigma}&=\mathbf{t^{2}}\otimes\left(\mathbf{h^{1,\sigma}_{N}}\otimes\mathbf{t^{1,\sigma}}+\mathbf{h}^{1,\sigma}\right)\nonumber\\
&+(\mathbf{t}^{2}\cdot\mathbf{h_{N}^{2,\uparrow\sigma}}\otimes\mathbf{t^{1,\uparrow}})_{3124}+((\mathbf{t}^{2})_{21}\cdot\mathbf{h_{N}^{2,\downarrow\sigma}}\otimes\mathbf{t^{1,\downarrow}})_{4123}+((\mathbf{t^{2}})_{21}\cdot(\mathbf{h_{N}^{2,\sigma\downarrow}})_{2134}\otimes\mathbf{t^{1,\uparrow}})_{4132}\nonumber\\
&+(\mathbf{t}^{2}\cdot(\mathbf{h_{N}^{2,\sigma\uparrow}})_{2134}\otimes\mathbf{t^{1,\uparrow}})_{3142}+\delta_{\sigma\downarrow}\mathbf{t}^{2}\otimes\mathbf{h}^{2}_{NN}\otimes \mathbf{t}^{2}-\delta_{\sigma\uparrow}\left(\mathbf{t}^{2}\right)_{21}\otimes\mathbf{h}^{2}_{NN}\otimes \mathbf{t}^{2}+\left(\mathbf{t}^{2}\otimes\mathbf{h^{2,\uparrow\sigma}}\right)_{3124}\nonumber\\
&+\left(\mathbf{t}^{2}\otimes(\mathbf{h^{2,\sigma\uparrow}})_{2134}\right)_{3124}+((\mathbf{t}^{2})_{21}\otimes\mathbf{h^{2,\downarrow\sigma}})_{4123}+((\mathbf{t}^{2})_{21}\otimes(\mathbf{h^{2,\sigma\downarrow}})_{2134})_{4132}\nonumber\\
&-(\mathbf{h}_{N}^{2,\uparrow\sigma}\otimes\mathbf{t}^{1,\downarrow})_{1243}-(\mathbf{h}_{N}^{2,\uparrow\sigma}\otimes\mathbf{t}^{1,\downarrow})_{1234}-\mathbf{h^{1,\sigma}}\otimes\mathbf{t^{2}}\label{paired-triples}\\
\mathbf{E}^{(N),\sigma\nu}&=\mathbf{t^{1,\sigma}}\otimes\mathbf{h^{2,\mu\nu}_{N}}\otimes \mathbf{t^{1,\mu}}+\delta_{\sigma,-\nu}\mathbf{h^{2}_{NN}}\otimes\mathbf{t^{2}}\otimes\mathbf{t^{1,\sigma}}-\delta_{\sigma\downarrow}
(\mathbf{h^{2,\uparrow\nu}_{N}}\otimes\mathbf{ t^{2}})_{32154}-\delta_{\sigma\downarrow}
(\mathbf{h^{2,\downarrow\nu}_{N}}\otimes\mathbf{ t^{2}})_{32154}\label{triples}\\
\mathbf{F}^{(N),\sigma\nu\rho}&=\mathbf{h^{2,\sigma\mu}_{N}}\otimes\mathbf{t^{1,\sigma}}\otimes\mathbf{t^{2}}+\mathbf{h^{2}_{NN}}\otimes\mathbf{t^{2}}\otimes\mathbf{t^{2}}\label{quadruples}
\end{align}
\end{small}
\noindent Here $\mathbf{t^{1,\sigma}}$ is a vector of length, $n_{1}=(N-1)$, comprising of singles excitation amplitudes. $\mathbf{t^{2,\sigma\nu}_{N}}$ is a $(N-1)^{2}$-length vector comprised of doubles excitation amplitudes, $(\mathbf{t^{2}_{N}})_{ab}=t^{2,(N)}_{ab}$. The vectors $\mathbf{h^{1}_{N}}$, $\mathbf{h^{2}_{N}}$ of dimensions $(N-1)$ and $(N-1)^{3}$ comprise of one-electron tensors, $\mathbf{h^{1}_{N}}_{i}=h^{1,(N)}_{i}$ and two-electron tensors, $(\mathbf{h^{2}_{N}})_{ijk}=h^{2,(N)}_{Nijk}$, coupling the Nth molecular orbital to the other molecular orbitals. $\mathbf{h^{2}_{NN}}$ comprises the two-electron tensor contributions ($(\mathbf{h^{2}_{N.N.}})_{ij}=h^{2}_{NNij}$)that couple all the paired excitations of the Nth MO with other spin orbitals. Here $\mathbf{K}\otimes \mathbf{L}$ denotes the tensor-product($\otimes$) of two vectors whose elements are $(\mathbf{K}\otimes \mathbf{L})_{ij}=K_{i}L_{j}$. In the above expressions $(\mathbf{h^{2,\sigma\nu}})_{abcd}$ represents a permutation of indexes of the tensor, for e.g. $((\mathbf{h^{2,\sigma\nu}})_{3124})_{ijkl}=(\mathbf{h^{2,\sigma\nu}})_{kijl}$, and ($\cdot$) represents tensor contraction. These downfolding equations (eq.\eqref{singles}-\eqref{paired-triples}) correspond to a rectangular system of multi-variable quadratic polynomials. There are $m=8N^{6}+8N^{5}+2N^{4}+4N^{3}+N^{2}+1$ polynomial equations in $n=N^{3}+N^{2}+2(N-1)$ parameters.
\\\par
\par\noindent
\emph{Computational complexity of the multireference downfolding technique}
\par\noindent
The term $t^{2}_{ijk}h^{2}_{abcN}t^{2}_{abc}$ in the equation set eq.\eqref{quadruples} has the highest cost $O(N^7)$ of being generated, if the full the ERI in MO representation is used. On the other hand we can compress the ERI via tensor factorization, $h^{2}_{abcN}=\sum_{pq}X_{ap}X_{bp}M_{pq}X_{cq}X_{Nq}$, using canonical polyadic decomposition or perform interpolative-separable density fitting\cite{hohenstein2022rank,lee2019systematically}. Similarly we can do tensor factorization of $t^{2}_{abc}=\sum_{k}A_{ak}B_{bk}C_{ck}$. Then the compute cost will reduce to $O(N_{THC}N^2)$.
\\\par
\par\noindent
\emph{Hamiltonian RG flow from Downfolding}
\par\noindent
The similarity transformation $S_{(N)}$ on the starting Hamiltonian $H_{(N)}=H$ leads to a renormalized Hamiltonian $H_{(N-1)}$ in the primary space $P_{(N)}$ with a self-similar form,
\begin{eqnarray}
    H_{(N-1)}&=&P_{(N)}S^{-1}_{(N)}H_{(N)}S_{(N)}P_{(N)}\nonumber\\
&=&P_{(N)}H_{(N)}P_{(N)}+P_{(N)}H_{(N)}Q_{(N)}\eta_{(N)}\nonumber\\
    &=&(1-\hat{n}_{N\uparrow})(1-\hat{n}_{N\downarrow})\bigg[\sum_{ij=1,\sigma}^{N-1}h^{1,(N-1)}_{ij}f^{\dagger}_{i\sigma}f_{j\sigma}+\sum_{\substack{ijkl=1,\\ \sigma\sigma'}}^{N-1}h^{2,(N-1)}_{ijkl}f^{\dagger}_{i\sigma}f^{\dagger}_{j\sigma'}f_{k\sigma'}f_{l\sigma}\bigg]\label{Effective Hamiltonian}
\end{eqnarray}
The one and two-electron tensors comprising the renormalized Hamiltonian $H_{(N-1)}$ can be be written in terms of the one and two-electron tensors of $H_{(N)}$ (eq\eqref{Hamiltonian}) and the amplitudes $t^{1,(N)}$ and $t^{2,(N)}$ of the generator $\eta$ (eq\eqref{FormOfEta}).
\begin{eqnarray}
h_{ij}^{1,\sigma,(N-1)}&=&h^{1,\sigma,(N)}_{ij}+h^{1,\sigma,(N)}_{iN}t^{1,\sigma,(N)}_{j}\label{RG-h1}\\
h^{2,\sigma\nu,(N-1)}_{ijkl}&=&h^{2,\sigma\nu,(N)}_{ijkl}+h^{2,\sigma\nu,(N)}_{ijkN}t^{1,\sigma,(N)}_{l}+\delta_{\nu,-\sigma}h^{2,(N)}_{ijNN}t^{2,(N)}_{kl}+h^{1,(N)}_{iN}t^{2,(N)}_{jkl}\label{RG-h2}~~~~~
\end{eqnarray}
The effect of downfolding the electronic correlations coupling $Q_{(N)}$ to $P_{(N)}$ was to renormalize the one-electron (eq.\eqref{RG-h1}) and two-electron interaction tensor contributions (eq.\eqref{RG-h2}) in the primary space using the excitation amplitudes of the generator $\eta$ (eq.\eqref{FormOfEta}). From the RG flow equation of the Hamiltonian $H_{(N-1)}$ (eq.\eqref{Effective Hamiltonian}) we make two important observations. Firstly, the indices in the one and two-electron tensors ($h^{1,(N-1}_{ij}, h^{2,(N-1)}_{ijkl}$) runs over MO from 1 to N-1 i.e. leaving out the Nth MO. The Nth orbital gets decoupled with only an overall diagonal contribution. This remains true even if $N\in \mathcal{C}$  where $P_{(N)}=\hat{n}_{\uparrow}\hat{n}_{\downarrow}$ and for $N\in \mathcal{A}$ : $P_{(N)}=\hat{n}_{\uparrow}\hat{n}_{\downarrow}$ or $(1-\hat{n}_{\uparrow})(1-\hat{n}_{\downarrow})$. However, in that case, the energetic contribution from the diagonal term may change. Secondly, no new many-body excitation clusters are created. This results from the choice of $\eta$ where paired doubles at the Nth MO get excited $t^{2,(N)}_{ij}$. The choice of $\eta$ automatically terminates the hierarchy of the three-particle or higher-order clusters. Such a description allows the self-similar representation of the Hamiltonian to prevail. The next set of RG equations that describes the Hamiltonian coefficients for a system of $(N-2)$ MOs are derived from the $(N-1)$ MOs (eq. \eqref{RG-h1}, eq. \eqref{RG-h2}).

\par
\par\noindent
\emph{Flow towards diagonalization}
\par\noindent
After $s$ downfolding steps, we obtain a reduced Hamiltonian,
\begin{small}
\begin{eqnarray}
H_{(s)}&=&\prod_{\substack{a=N,\\\sigma=\uparrow,\downarrow}}^{N-p}(1-\hat{n}_{a\sigma})\prod_{\substack{a=N-p-1,\\\sigma=\uparrow,\downarrow}}^{N-s+1}\hat{n}_{a\sigma}\bigg[\sum_{ij=1,\sigma}^{N-s}h^{1,(N-s)}_{ij}f^{\dagger}_{i\sigma}f_{j\sigma}\nonumber\\
&+&\sum_{l=N-p}^{N-s}h^{2,\uparrow,\downarrow,(l)}_{l,l,l,l}+\sum_{l=N-p,\sigma}^{N-s}h^{1,\sigma,(l)}_{l,l}+\sum_{\substack{ijkl=1,\\ \sigma\sigma'}}^{N-s}h^{2,(N-s)}_{ijkl}f^{\dagger}_{i\sigma}f^{\dagger}_{j\sigma'}f_{k\sigma'}f_{l\sigma}\bigg]~~.
\end{eqnarray}
\end{small}

\section{Case Study: Unitary Multireference Downfolding with Singles and Doubles}\label{Unitary}
In this section we will look into the formalism for Hamiltonian downfolding with singles and doubles excitations. We will study the generation of unitary transformation operators corresponding to the similarity transformation operators used in each downfolding step. This will lead to a series of downfolding transformations that preserves the hermiticity of the Hamiltonian in the reduced subspace. A detailed step-wise description is presented below.
\\\par
\par\noindent \textit{Step-1}
\par\noindent
The electronic Hamiltonian in MO basis with $N$ spin orbitals is constructed as,
\begin{align}
    H_{(N)}=\sum_{pq,\sigma}h^{1,\sigma,(N)}_{pq}f^{\dagger}_{p\sigma}f_{q\sigma}+\sum_{pqrs,\sigma \sigma^\prime}h^{2,\sigma\sigma^\prime, (N)}_{pqrs}f^{\dagger}_{p\sigma}f^{\dagger}_{q\sigma^\prime}f_{r\sigma^\prime}f_{s\sigma}
\end{align}
Here, $f^{\dagger}_{i\sigma}$ is a creation operator for $i^{th}$ molecular orbital with spin state $\sigma$. Similarly, $f_{i\sigma}$ is an annihilation operator for $i^{th}$ molecular orbital with spin state $\sigma$. All dummy indices corresponding to the orbital numbers are denoted by English letters whereas spin states are denoted by greek letters. In this section, orbital indices denoted by p,q,r,s... have been considered to span over all spin orbitals. Whereas, indices denoted by a,b,c...i,j,k,l...m,n span over all spin orbitals except the outermost orbital, the $N^{th}$ orbital here.
\\\par
\par\noindent\textit{Step-2}
\par\noindent
We partition the many body Hilbert space into a model space and its complement\cite{suzuki1980convergent}. Our iterative downfolding needs us to decouple only the outermost orbital at a time. So we define our two subspace projection operators, $P_N$ and $Q_N$, as,
\begin{align}
    &P_{(N)}=(1-\hat{n}_{N\uparrow})(1-\hat{n}_{N\downarrow})\\
    &Q_{(N)}=\hat{n}_{N\uparrow}+\hat{n}_{N\downarrow}+\hat{n}_{N\uparrow}\hat{n}_{N\downarrow}
\end{align}
where $\hat{n}_{k\sigma}=f^{\dagger}_{k\sigma}f_{k\sigma}$ and $P_{(N)}+Q_{(N)}=I^{\otimes 2N}$. The P-space projection removes contributions from the $N^{th}$ molecular orbital.
\\\par
\par\noindent \textit{Step-3}
\par\noindent
At first, we construct a non-Hermitian generator $\eta^{(N)}$ for a $N^{th}$ MO decoupling similarity transformation, given by,
\begin{align}
    \eta^{(N)}=\eta^{1,(N)}+\eta^{2,(N)}+\eta^{3,(N)}
\end{align}
where,
\begin{align}
    &\eta^{1,(N)}=\sum_{i,\sigma}\eta^{1,(N)}_{i,\sigma}=\sum_{i,\sigma}t^{1,(N)}_{i,\sigma}(1-\hat{n}_{i-\sigma})(1-\hat{n}_{N-\sigma})f^\dagger_{N\sigma}f_{i\sigma}\\
    &\eta^{2,(N)}=\sum_{ijk,\sigma\sigma^\prime}\eta^{2,(N)}_{ijk,\sigma\sigma^\prime}=\sum_{ijk,\sigma\sigma^\prime}t^{2,(N)}_{ijk,\sigma\sigma^\prime}(1-\hat{n}_{N-\sigma})f^\dagger_{N\sigma}f^\dagger_{i\sigma^\prime}f_{j\sigma^\prime}f_{k\sigma}\\
    &\eta^{3,(N)}=\sum_{ij}\eta^{3,(N)}_{ij}=\sum_{ij}t^{3,(N)}_{ij}f^\dagger_{N\uparrow}f^\dagger_{N\downarrow}f_{i\downarrow}f_{j\uparrow}
\end{align}
\noindent This generator accounts for all possible singles and doubles excitations involving the $N^{th}$ molecular orbital. Here, $t_{i,\sigma}^{1,(N)}$, $t^{2,(N)}_{ijk,\sigma\sigma^\prime}$ and $t_{ij}^{3,(N)}$ denote singles, mixed-doubles (excitations to only one $N^{th}$ spin orbital) and paired-doubles (excitations to both $N^{th}$ spin orbitals) excitation amplitudes respectively. Here we will see that $\eta^{(N)}$ is nilpotent with degree 2, i.e, $\left(\eta^{(N)}\right)^2=0, \left(\eta^{(N)}\right)^3=0 \ ...,$ and the polynomial expansion of the similarity transformation $S^{(N)}=e^{\eta^{(N)}}$ naturally terminates at $O(\eta^{(N)})$ with $S^{(N)}=1+\eta^{(N)}$.
\\\par
\par\noindent \textit{Step-4}
\par\noindent
This choice of generator allows for a decomposition of the similarity transformation $S^{(N)}$ into a product of three separate similarity transformations, $S^{1,(N)}$, $S^{2,(N)}$, $S^{3,(N)}$ for singles, mixed-doubles and paired-doubles excitations, respectively, given by the generators, $\eta^{1,(N)}$, $\eta^{2,(N)}$ and $\eta^{3,(N)}$ as,
\begin{align}
    S^{(N)}=e^{\left(\eta^{1,(N)}+\eta^{2,(N)}+\eta^{3,(N)}\right)}=e^{\eta^{1,(N)}}e^{\eta^{2,(N)}}e^{\eta^{3,(N)}}=S^{1,(N)}S^{2,(N)}S^{3,(N)}
\end{align}
\noindent This is due to the set of commutation relations,
\begin{align}
\left[\eta^{1,(N)},\eta^{2,(N)}\right]=\left[\eta^{2,(N)},\eta^{3,(N)}\right]=\left[\eta^{3,(N)},\eta^{1,(N)}\right]=0
\end{align}
Again, the individual similarity transformations, $\{S^{i,(N)}\}$ for $i \in \{1,2,3\}$, can be written as products of similarity transformations given by ,
\begin{align}
    &S^{1,(N)}=e^{\eta^{1,(N)}}=e^{\sum_{i,\sigma}{\eta^{1,(N)}_{i,\sigma}}}=\prod_{i,\sigma} e^{\eta^{1,(N)}_{i,\sigma}}=\prod_{i,\sigma}S^{1,(N)}_{i,\sigma}\\
    &S^{2,(N)}=e^{\eta^{2,(N)}}=e^{\sum_{ijk,\sigma \sigma^\prime}{\eta^{2,(N)}_{ijk,\sigma\sigma^\prime}}}=\prod_{ijk,\sigma\sigma^\prime} e^{\eta^{2,(N)}_{ijk,\sigma\sigma^\prime}}=\prod_{ijk,\sigma\sigma^\prime}S^{2,(N)}_{ijk,\sigma\sigma^\prime}\\
    &S^{3,(N)}=e^{\eta^{3,(N)}}=e^{\sum_{ij}{\eta^{3,(N)}_{ij}}}=\prod_{ij} e^{\eta^{3,(N)}_{ij}}=\prod_{ij}S^{3,(N)}_{ij}
\end{align}
\noindent The above decompositions were made possible due to the following commutation relations,
\begin{align}
&\left[\eta^{1,(N)}_{i,\sigma},\eta^{1,(N)}_{j,\sigma^\prime}\right]=0 \ \ \forall \ i,j,\sigma, \sigma^\prime\\
&\left[\eta^{2,(N)}_{ijk,\sigma\sigma\prime},\eta^{2,(N)}_{abc,\mu \mu^\prime}\right]=0 \ \ \forall \ i,j,k,a,b,c, \sigma, \sigma^\prime,\mu, \mu^\prime\\
&\left[\eta^{3,(N)}_{ij},\eta^{3,(N)}_{ab}\right]=0 \ \ \forall \ i,j,a,b
\end{align}
\noindent So, the similarity transformation that decouples the $N^{th}$ Molecular orbital and gives rise to a one-step downfolded effective Hamiltonian given by,
\begin{align}
    H_{(N-1)}&=\left(S^{(N)}\right)^{-1}H_{(N)}S^{(N)}\\
    &=\prod_{ijklmn,\mu\sigma\sigma^\prime} e^{-\eta^{3,(N)}_{mn}} e^{-\eta^{2,(N)}_{jkl,\sigma\sigma^\prime}} e^{-\eta^{1,(N)}_{i,\mu}}H_{(N)}e^{\eta^{1,(N)}_{i,\mu}}e^{\eta^{2,(N)}_{jkl,\sigma\sigma^\prime}}e^{\eta^{3,(N)}_{mn}}
\end{align}
\\\par
\par\noindent\textit{Step-5} 
\par\noindent
The effective Hamiltonian is non-Hermitian after undergoing the similarity transformation. The Hermitian counterpart can be formulated by constructing analogous unitary transformations from the generators of the similarity transformation. The Unitary operators for the similarity transformation generators, $\eta^{1,(N)}_{i,\sigma}$, $\eta^{2,(N)}_{ijk,\sigma\sigma^\prime}$ and $\eta^{3,(N)}_{ij}$, can be written as \cite{suzuki1982construction},
\begin{align}
    &U^{1,(N)}_{i,\sigma}=e^{arctanh\left(\eta^{1,(N)}_{i,\sigma}-\left(\eta^{1,(N)}_{i,\sigma}\right)^\dagger\right)}\\
    &U^{2,(N)}_{ijk,\sigma\sigma^\prime}=e^{arctanh\left(\eta^{2,(N)}_{ijk,\sigma\sigma^\prime}-\left(\eta^{2,(N)}_{ijk,\sigma\sigma^\prime}\right)^\dagger\right)}\\
    &U^{3,(N)}_{ij}=e^{arctanh\left(\eta^{3,(N)}_{ij}-\left(\eta^{3,(N)}_{ij}\right)^\dagger\right)}
\end{align}
The generators satisfy the conditions:
\begin{align}
   &Q\eta^{1,(N)}_{i\sigma} P=\eta^{1,(N)}_{i\sigma}\\
&P\eta^{1,(N)}_{i\sigma} P=P \eta^{1,(N)}_{i\sigma} Q =Q \eta^{1,(N)}_{i\sigma} Q =0\\
&Q\eta^{2,(N)}_{ijk,\sigma\sigma^\prime} P=\eta^{2,(N)}_{ijk,\sigma\sigma^\prime}\\
&P\eta^{2,(N)}_{ijk,\sigma\sigma^\prime} P=P \eta^{2,(N)}_{ijk,\sigma\sigma^\prime} Q =Q \eta^{2,(N)}_{ijk,\sigma\sigma^\prime} Q =0\\
&Q\eta^{3,(N)}_{ij} P=\eta^{3,(N)}_{ij}\\
&P\eta^{3,(N)}_{ij} P=P \eta^{3,(N)}_{ij} Q =Q \eta^{3,(N)}_{ij} Q =0\\
&\left(\eta^{1,(N)}_{i\sigma}\right)^2=\left(\eta^{2,(N)}_{ijk,\sigma\sigma^\prime}\right)^2=\left(\eta^{3,(N)}_{ij}\right)^2=0
\end{align}
\noindent which leads to a series of simplifications, given by,

\begin{align}
&U^{1,(N)}_{i\sigma}=\frac{1+\eta^{1,(N)}_{i\sigma}-\left(\eta^{1,(N)}_{i\sigma}\right)^\dagger}{\left(1+\eta^{1,(N)}_{i\sigma}\left(\eta^{1,(N)}_{i\sigma}\right)^\dagger+\left(\eta^{1,(N)}_{i\sigma}\right)^\dagger\eta^{1,(N)}_{i\sigma}\right)^{1/2}}\\
&U^{2,(N)}_{ijk,\sigma\sigma^\prime}=\frac{1+\eta^{2,(N)}_{ijk,\sigma\sigma^\prime}-\left(\eta^{2,(N)}_{ijk,\sigma\sigma^\prime}\right)^\dagger}{\left(1+\eta^{2,(N)}_{ijk,\sigma\sigma^\prime}\left(\eta^{2,(N)}_{ijk,\sigma\sigma^\prime}\right)^\dagger+\left(\eta^{2,(N)}_{ijk,\sigma\sigma^\prime}\right)^\dagger\eta^{2,(N)}_{ijk,\sigma\sigma^\prime}\right)^{1/2}}\\
&U^{3,(N)}_{ij}=\frac{1+\eta^{3,(N)}_{ij}-\left(\eta^{3,(N)}_{ij}\right)^\dagger}{\left(1+\eta^{3,(N)}_{ij}\left(\eta^{3,(N)}_{ij}\right)^\dagger+\left(\eta^{3,(N)}_{ij}\right)^\dagger\eta^{3,(N)}_{ij}\right)^{1/2}}
\end{align}
These equations can further be simplified into,
\begin{align}
&U^{1,(N)}_{i\sigma}=\frac{1+\eta^{1,(N)}_{i\sigma}-\left(\eta^{1,(N)}_{i\sigma}\right)^\dagger}{\left(1+|t^{1,(N)}_{i\sigma}|^{2
}(1-\hat{n}_{i-\sigma})(1-\hat{n}_{N-\sigma})\left( (1-\hat{n}_{i\sigma})\hat{n}_{N\sigma}+(1-\hat{n}_{N\sigma})\hat{n}_{i\sigma}\right)\right)^{1/2}}\\
&U^{2,(N)}_{ijk,\sigma\sigma^\prime}=\frac{1+\eta^{2,(N)}_{ijk,\sigma\sigma^\prime}-\left(\eta^{2,(N)}_{ijk,\sigma\sigma^\prime}\right)^\dagger}{\left(1+t^{2,(N)}_{ijk\sigma\sigma^\prime}\left(t^{2,(N)}_{ijk\sigma\sigma^\prime}\right)^*(1-\hat{n}_{N-\sigma})\left((1-\hat{n}_{j\sigma})(1-\hat{n}_{i\sigma^\prime})\hat{n}_{k\sigma^\prime}\hat{n}_{N\sigma}+(1-\hat{n}_{N\sigma})(1-\hat{n}_{k\sigma^\prime})\hat{n}_{i\sigma^\prime}\hat{n}_{j\sigma}\right)\right)^{1/2}}\\
&U^{3,(N)}_{ij}=\frac{1+\eta^{3,(N)}_{ij}-\left(\eta^{3,(N)}_{ij}\right)^\dagger}{\left(1+t^{3,(N)}_{ij}\left(t^{3,(N)}_{ij}\right)^*\left( (1-\hat{n}_{j\uparrow})(1-\hat{n}_{i\downarrow})\hat{n}_{N\downarrow}\hat{n}_{N\uparrow}+(1-\hat{n}_{N\uparrow})(1-\hat{n}_{N\downarrow})\hat{n}_{i\downarrow}\hat{n}_{j\uparrow}\right)\right)^{1/2}}
\end{align}
Separating the unitary operators into their Nth orbital projection spaces we obtain even simpler expressions,
\begin{align}
    \nonumber
    U^{1,(N)}_{i\sigma}=&\left(1+\eta^{1,(N)}_{i\sigma}-\left(\eta^{1,(N)}_{i\sigma}\right)^\dagger\right)\hat{n}_{N-\sigma}\\
    &+\left(1+\eta^{1,(N)}_{i\sigma}-\left(\eta^{1,(N)}_{i\sigma}\right)^\dagger\right)\left(1-P^{1,(N)}_{i\sigma}+P^{1,(N)}_{i\sigma}\hat{n}_{i-\sigma}+\frac{P^{1,(N)}_{i\sigma}(1-\hat{n}_{i-\sigma})}{\left(1+t^{1,(N)}_{i\sigma}\left(t^{1,(N)}_{i\sigma}\right)^*\right)^{1/2}}\right)(1-\hat{n}_{N-\sigma}) \\ where &\ P^{1,(N)}_{i\sigma}=(1-\hat{n}_{i\sigma})\hat{n}_{N\sigma}+(1-\hat{n}_{N\sigma})\hat{n}_{i\sigma}\\
\nonumber\\
\nonumber
U^{2,(N)}_{ijk,\sigma\sigma^\prime}=&\left(1+\eta^{2,(N)}_{ijk,\sigma\sigma^\prime}-\left(\eta^{2,(N)}_{ijk,\sigma\sigma^\prime}\right)^\dagger\right)\hat{n}_{N-\sigma}\\
&+ \left(1+\eta^{2,(N)}_{ijk,\sigma\sigma^\prime}-\left(\eta^{2,(N)}_{ijk,\sigma\sigma^\prime}\right)^\dagger\right)\left(1-P^{2,(N)}_{ijk,\sigma\sigma^\prime}+\frac{P^{2,(N)}_{ijk,\sigma\sigma^\prime}}{\left(1+t^{2,(N)}_{ijk\sigma\sigma^\prime}\left(t^{2,(N)}_{ijk\sigma\sigma^\prime}\right)^*\right)^{1/2}}\right)(1-\hat{n}_{N-\sigma})\\
where &\ P^{2,(N)}_{ijk,\sigma\sigma^\prime}=(1-\hat{n}_{j\sigma})(1-\hat{n}_{i\sigma^\prime})\hat{n}_{k\sigma^\prime}\hat{n}_{N\sigma}+(1-\hat{n}_{N\sigma})(1-\hat{n}_{k\sigma^\prime})\hat{n}_{i\sigma^\prime}\hat{n}_{j\sigma}\\
\nonumber\\
U^{3,(N)}_{ij}=&\left(1+\eta^{3,(N)}_{ij}-\left(\eta^{3,(N)}_{ij}\right)^\dagger\right)\left(1-P^{3,(N)}_{ij}+\frac{P^{3,(N)}_{ij}}{\left(1+t^{3,(N)}_{ij}\left(t^{3,(N)}_{ij}\right)^*\right)^{1/2}}\right)\\
where &\ P^{3,(N)}_{ij}=(1-\hat{n}_{j\uparrow})(1-\hat{n}_{i\downarrow})\hat{n}_{N\downarrow}\hat{n}_{N\uparrow}+(1-\hat{n}_{N\uparrow})(1-\hat{n}_{N\downarrow})\hat{n}_{i\downarrow}\hat{n}_{j\uparrow}
\end{align}
\noindent Our Downfolding transformation for one step downfolded Hermitian Hamiltonian, $H_{(N-1)}$ is given by,
\begin{align}
    \nonumber
    H_{(N-1)}=\left(U^{(N)}\right) ^{\dagger}H_{(N)}&U^{(N)}\\
    \nonumber
    =\prod_{ijklmn,\mu\sigma\sigma^\prime} &e^{-arctanh\left(\eta^{3,(N)}_{mn}-\left(\eta^{3,(N)}_{mn}\right)^\dagger\right)} e^{-arctanh\left(\eta^{2,(N)}_{jkl,\sigma\sigma^\prime}-\left(\eta^{2,(N)}_{jkl,\sigma\sigma^\prime}\right)^\dagger\right)} e^{-arctanh\left(\eta^{1,(N)}_{i,\sigma}-\left(\eta^{1,(N)}_{i,\sigma}\right)^\dagger\right)}\\
    H_{(N)}&e^{arctanh\left(\eta^{1,(N)}_{i,\sigma}-\left(\eta^{1,(N)}_{i,\sigma}\right)^\dagger\right)}e^{arctanh\left(\eta^{2,(N)}_{jkl,\sigma\sigma^\prime}-\left(\eta^{2,(N)}_{jkl,\sigma\sigma^\prime}\right)^\dagger\right)}e^{arctanh\left(\eta^{3,(N)}_{mn}-\left(\eta^{3,(N)}_{mn}\right)^\dagger\right)}
\end{align}
In this manner we construct a closed-form unitary operator representation for downfolding Hamiltonian with Singles and doubles and similar construction can be made for the general multireference case with all excitations singles, doubles and beyond. Note that in the unitary operator no higher order electronic clusters are generated beyond the singles and doubles contribution in $\eta$ albeit the unitary operator constructed from the similarity map fulfills the $U^{\dagger}U=UU^{\dagger}=I$ form. This is different from the unitary transformation in the DUCC formalism \cite{kowalski2024accuracies} where the action  of the unitary operator on the Hamiltonian is truncated beyond double commutator making the approach perturbative with respect to the electronic cluster amplitudes. After our multireference downfolding cases we will take the case of single configuration downfolding in the next section
\section{Case Study: Single configuration spin restricted downfolding with singles and doubles}\label{SR-TFHD}
This is a specific case of single configuration downfolding where we start from the Hartree Fock state and incorporate the downfolding correlations by doing a sequence of similarity transformations on the Hartree-Fock state $|\Psi\rangle$
We first define the partition,
\begin{eqnarray}
    P_{N}=(1-\hat{n}_{N\uparrow})(1-\hat{n}_{N\downarrow}),\ Q_{N}=1-P_{N}
\end{eqnarray}
Let's now define the generator of similarity transformations in the space of $P_{N}$ and $Q_{N}$,
\begin{eqnarray}
    \eta_{(N)}&=&\sum_{i\sigma}t_{Ni}(1-\hat{n}_{N-\sigma})f^{\dagger}_{N\sigma}f_{i\sigma}+\sum_{bij,\sigma\sigma'}t_{Nbij}(1-\hat{n}_{N-\sigma})f^{\dagger}_{N\sigma}f^{\dagger}_{b\sigma'}f_{j\sigma'}f_{i\sigma}\nonumber\\
    &+&\sum_{ij}t_{NNij}f^{\dagger}_{N\uparrow}f^{\dagger}_{N\downarrow}f_{j\downarrow}f_{i\uparrow}
\end{eqnarray}
The $\eta_{(N)}$ satisfies the equation,
\begin{eqnarray}
    Q_{(N)}\eta_{(N)}P_{(N)}=\eta_{(N)}, \eta_{(N)}^{2}=0
\end{eqnarray}
Now we can write down the subset of coupled cluster equations for downfolding one molecular orbital as,
\begin{eqnarray}
    \langle\Psi^{N,\sigma}_{i}|Q_{N}S^{-1}_{N}HS_{N}P_{N\sigma}|\Psi\rangle = 0 \label{t1-eq}\\ 
    \langle\Psi^{aN,\sigma\sigma'}_{ij}|Q_{N}S^{-1}_{N}HS_{N}P_{N}|\Psi\rangle = 0 \label{t2-eq}\\ 
    \langle\Psi^{NN}_{ij}|Q_{N}S^{-1}_{N}HS_{N}P_{N}|\Psi\rangle = 0 \label{t21-eq}
\end{eqnarray}
where, 
\begin{eqnarray}
    |\Psi^{N\sigma}_{i}\rangle &=& f^{\dagger}_{N\sigma}f_{i\sigma}|\Psi\rangle\\
    |\Psi^{aN,\sigma\sigma'}_{ij}\rangle &=& f^{\dagger}_{Ng\sigma}f^{\dagger}_{a\sigma'}f_{j\sigma'}f_{i\sigma}|\Psi\rangle\\
    |\Psi^{NN}_{ij}\rangle &=& f^{\dagger}_{N\uparrow}f^{\dagger}_{a\downarrow}f_{j\downarrow}f_{i\uparrow}|\Psi\rangle
\end{eqnarray}
Before writing down the algebraic expressions for the Hamiltonian downfolding amplitude equations, we will define a modified ERI operator, $w^2$, given by,
\begin{align}
    w^2_{ijab}=2h^2_{ijab}-h^2_{ijba}
\end{align}
and a permutation operator, $\mathcal{P}$, given by, 
\begin{align}
 \mathcal{P}\{...\}_{abij}=\{...\}_{abij}+\{...\}_{baji} 
\end{align}
\subsection{T1-Residual Equation}
Solving equation \ref{t1-eq}, we get the $t^1$ amplitude equation for $Nth$ orbital downfolding step as, 
\begin{align}
    \sum_{i=1}^{11}T_{i}=0
\end{align}
where $f_{ij}$ are elements of (N-1) step downfolded fock matrix and,
\begin{align}
T_{1}=& f_{Ni}\\
T_{2}=& -2\sum_{k}f_{kN}t^1_{Nk}t^1_{Ni}\\
T_{3}=& f_{NN}t^{1}_{Ni}-\sum_{k}w^2_{klNN}t^2_{NNkl}t^{1}_{Ni} -\sum_{k,d\neq N}w^2_{klNd}t^2_{Ndkl}t^1_{Ni}\\
T_{4}=&-\sum_{k}f_{ki}t^1_{Nk}- \sum_{klc}w^2_{klcN}t^2_{cNil}t^1_{Nk}- \sum_{kl,d\neq N}w^2_{klNd}t^2_{Ndil}t^1_{Nk}\\
T_{5}=& 2\sum_{kc}f_{kc}t^2_{cNki}+2\sum_{klc}w^2_{klcN}t^1_{Nl}t^2_{cNki}-\sum_{kc}f_{kc}t^2_{cNik}-\sum_{klc}w^2_{klcN}t^1_{Nl}t^2_{cNik}\\
T_{6}=& \sum_{k}f_{kN}t^1_{Ni}t^1_{Nk}\\
T_{7}=& \sum_{k}w^2_{NkiN}t^1_{Nk}\\
T_{8}=& \sum_{kc}w^2_{NkcN}t^2_{cNik}+\sum_{k,d\neq N}w^2_{NkNd}t^2_{Ndik}\\
T_{9}=& \sum_{k}w^2_{NkNN}t^1_{Ni}t^1_{Nk}\\
T_{10}=& -\sum_{kl}w^2_{kliN}t^2_{NNkl}-\sum_{kl,c\neq N}w^2_{klic}t^2_{Nckl}\\
T_{11}=&-\sum_{kl}w^2_{kliN}t^1_{Nk}t^1_{Nl}
\end{align}
\subsection{T2-Residual Equation}
Solving equations \ref{t2-eq} and \ref{t21-eq}, we get the $t^2$ amplitude equation for $Nth$ orbital downfolding step as, 
\begin{align}
\raggedleft    \sum^{13}_{i=1}T_i=0
\end{align}
where,
\begin{align}
T_1 = &h^2_{ijaN}+h^2_{ijNb}\\
T_2 = &\sum_{kl}h^2_{klij}t^2_{aNkl}+\sum_{kl}h^2_{klij}t^2_{Nbkl}+\sum_{kl}h^2_{kliN}t^1_{Nj}t^2_{aNkl}+\sum_{kl}h^2_{kliN}t^1_{Nj}t^2_{Nbkl} + \sum_{kl}h^2_{klNj}t^1_{Ni}t^2_{aNkl}\nonumber\\
&+\sum_{kl}h^2_{klNj}t^1_{Ni}t^2_{Nbkl}+ \sum_{klc}h^2_{klcN}t^2_{cNij}t^2_{aNkl}+\sum_{klc}h^2_{klcN}t^2_{cNij}t^2_{Nbkl}\nonumber\\
&+\sum_{kl,c\neq N}h^2_{klNc}t^2_{Ncij}t^2_{aNkl}+\sum_{kl,c\neq N}h^2_{klNc}t^2_{Ncij}t^2_{Nbkl}\\
T_3 = &\sum_{kl}h^2_{klij}t^1_{Nk}t^1_{Nl}\\
T_4 = &\sum_c h^2_{aNcN}t^2_{cNij}+\sum_c h^2_{NbcN}t^2_{cNij}+\sum_{d\neq N}h^2_{aNNd}t^2_{Ndij}\nonumber\\
&+\sum_{d\neq N}h^2_{NbNd}t^2_{Ndij} -\sum_{kc}h^2_{akcN}t^1_{Nk}t^2_{cNij}-\sum_{k,d\neq N} h^2_{akNd}t^1_{Nk}t^2_{Ndij}\nonumber\\ &-\sum_{kc}h^2_{kbcN}t^1_{Nk}t^2_{cNij}-\sum_{k,d\neq N} h^2_{kbNd}t^1_{Nk}t^2_{Ndij}\\
T_5 = &h^2_{aNNN}t^1_{Ni}t^1_{Nj}+h^2_{NbNN}t^1_{Ni}t^1_{Nj}\\
T_6 = &\mathcal{P}\sum_{c}f_{ac}t^2_{cNij}+\mathcal{P}f_{NN}t^2_{Nbij}-\mathcal{P}\sum_{klc}w^2_{klcN}t^2_{aNkl}t^2_{cNij}-\mathcal{P}\sum_{kl}w^2_{klNN}t^2_{NNkl}t^2_{Nbij}\nonumber\\
&-\mathcal{P}\sum_{kl,d\neq N}w^2_{klNd}t^2_{Ndkl}t^2_{NNij}-\mathcal{P}\sum_{kl,d\neq N}w^2_{klNd}t^2_{Ndkl}t^2_{Nbij}-\mathcal{P}\sum_{k}f_{kN}t^1_{Nk}t^2_{NNij}\nonumber\\
&-\mathcal{P}\sum_{k}f_{kN}t^1_{Nk}t^2_{Nbij}+\mathcal{P}\sum_{kc}w^2_{akcN}t^1_{Nk}t^2_{cNij}+\mathcal{P}\sum_{k}w^2_{NkNN}t^1_{Nk}t^2_{Nbij}\\
T_7 = &-\mathcal{P}\sum_{k}f_{ki}t^2_{aNkj}-\mathcal{P}\sum_{klc}w^2_{klcN}t^2_{cNil}t^2_{aNkj}-\mathcal{P}\sum_{kl,d\neq N} w^2_{klNd}t^2_{Ndil}t^2_{aNkj}\nonumber\\ &-\mathcal{P}\sum_{k}f_{kN}t^1_{Ni}t^2_{aNkj}-\mathcal{P}\sum_{kl}w^2_{kliN}t^1_{Nl}t^2_{aNkj}\nonumber\\
&-\mathcal{P}\sum_{k}f_{ki}t^2_{Nbkj}-\mathcal{P}\sum_{klc}w^2_{klcN}t^2_{cNil}t^2_{Nbkj}-\mathcal{P}\sum_{kl,d\neq N} w^2_{klNd}t^2_{Ndil}t^2_{Nbkj}\nonumber\\ &-\mathcal{P}\sum_{k}f_{kN}t^1_{Ni}t^2_{Nbkj}-\mathcal{P}\sum_{kl}w^2_{kliN}t^1_{Nl}t^2_{Nbkj}\\
T_{8}=&\mathcal{P}h^2_{aNiN}t^1_{Nj}+\mathcal{P}h^2_{NbiN}t^1_{Nj}-\mathcal{P}\sum_{k}h^2_{kbiN}t^1_{Nk}t^1_{Nj}\\
T_{9}=&-\mathcal{P}\sum_{k}h^2_{akij}t^1_{Nk}-\mathcal{P}\sum_{k}h^2_{akiN}t^1_{Nj}t^1_{Nk}\\
T_{10}=&2\mathcal{P}\sum_{kc}h^2_{akic}t^2_{cNkj}+2\mathcal{P}\sum_{k}h^2_{NkiN}t^2_{Nbkj}- 2\mathcal{P}\sum_{kl}h^2_{lkiN}t^1_{Nl}t^2_{NNkj}\nonumber\\
& -2\mathcal{P}\sum_{kl}h^2_{lkiN}t^1_{Nl}t^2_{Nbkj}+2\mathcal{P}\sum_{kc}h^2_{akNc}t^1_{Ni}t^2_{cNkj} +2\mathcal{P}\sum_{k}h^2_{NkNN}t^1_{Ni}t^2_{Nbkj}\nonumber\\
&-\mathcal{P}\sum_{klc}h^2_{lkNc}t^2_{NNil}t^2_{cNkj}-\mathcal{P}\sum_{kl,d\neq N}h^2_{lkdN}t^2_{dNil}t^2_{NNkj}\nonumber\\
& -\mathcal{P}\sum_{kld}h^2_{lkdN}t^2_{dNil}t^2_{Nbkj} -\mathcal{P}\sum_{klc}h^2_{lkNc}t^2_{Nail}t^2_{cNkj}-\mathcal{P}\sum_{kl}h^2_{lkNN}t^2_{Nail}t^2_{Nbkj} \nonumber\\
&+ \mathcal{P}\sum_{klc}w^2_{lkNc}t^{2}_{aNil}t^2_{cNkj}+ \mathcal{P}\sum_{kl}w^2_{lkNN}t^{2}_{NNil}t^2_{Nbkj}\nonumber\\
& +\mathcal{P}\sum_{kl,d\neq N}w^2_{lkdN}t^{2}_{Ndil}t^2_{NNkj}+\mathcal{P}\sum_{kl,d\neq N}w^2_{lkdN}t^{2}_{Ndil}t^2_{Nbkj}\\
T_{11}=& -\mathcal{P}\sum_{k}h^2_{akiN}t^2_{NNkj}-\mathcal{P}\sum_{k}h^2_{NkiN}t^2_{bNkj}-\mathcal{P}\sum_{k,c\neq N}h^2_{akic}t^2_{Nckj} +\mathcal{P}\sum_{kl}h^2_{lkiN}t^1_{Nl}t^2_{bNkj}\nonumber\\
&+ \mathcal{P}\sum_{kl}h^2_{lkiN}t^1_{Nl}t^2_{NNkj}
-\mathcal{P}\sum_{k}h^2_{akNN}t^1_{Ni}t^2_{NNkj} -\mathcal{P}\sum_{k}h^2_{NkNN}t^1_{Ni}t^2_{bNkj}\nonumber\\
&-\mathcal{P}\sum_{k,c\neq N}h^2_{akNc}t^1_{Ni}t^2_{Nckj} +1/2\ \mathcal{P}\sum_{kld}h^2_{lkdN}t^2_{dNil}t^2_{bNkj}+1/2\ \mathcal{P}\sum_{kld}h^2_{lkdN}t^2_{dNil}t^2_{NNkj}\nonumber\\ 
&+1/2\ \mathcal{P}\sum_{kl,c\neq N}h^2_{lkNc}t^2_{NNil}t^2_{Nckj} +1/2\ \mathcal{P}\sum_{kl}h^2_{lkNN}t^2_{Nail}t^2_{NNkj} +1/2\ \mathcal{P}\sum_{kl,c\neq N}h^2_{lkNc}t^2_{Nail}t^2_{Nckj}\nonumber\\
& -1/2\ \mathcal{P}\sum_{kl}w^2_{lkNN}t^{2}_{aNil}t^{2}_{NNkj}-1/2\ \mathcal{P}\sum_{kl}w^2_{lkNN}t^{2}_{NNil}t^{2}_{bNkj} -1/2\ \mathcal{P}\sum_{kl,c\neq N}w^2_{lkNc}t^{2}_{aNil}t^{2}_{Nckj}\nonumber\\
&-1/2\ \mathcal{P}\sum_{kl,d\neq N}w^2_{lkdN}t^{2}_{Ndil}t^2_{bNkj}-1/2\ \mathcal{P}\sum_{kl,d\neq N}w^2_{lkdN}t^{2}_{Ndil}t^2_{NNkj}\\
T_{12} =& -\mathcal{P}\sum_{k} h^2_{NkNi}t^2_{aNkj} -\mathcal{P}\sum_{k} h^2_{bkNi}t^2_{NNkj} -\mathcal{P}\sum_{k,c\neq N}h^2_{bkci}t^2_{Nckj}\nonumber\\
& +\mathcal{P}\sum_{kl}h^2_{lkNi}t^1_{Nl}t^2_{aNkj} -\mathcal{P}\sum_{k}h^2_{NkNN}t^1_{Ni}t^2_{aNkj} -\mathcal{P}\sum_{k}h^{2}_{bkNN}t^1_{Ni}t^2_{NNkj}\nonumber\\ 
& -\mathcal{P}\sum_{k,c\neq N}h^{2}_{bkcN}t^{1}_{Ni}t^{2}_{Nckj} +1/2\ \mathcal{P}\sum_{kd}h^{2}_{lkNd}t^{2}_{dNil}t^{2}_{aNkj} +1/2\ \mathcal{P}\sum_{kc\neq N}h^{2}_{lkcN}t^{2}_{NNil}t^{2}_{Nckj}\nonumber\\
& +1/2\ \mathcal{P}\sum_{k}h^2_{lkNN}t^{2}_{Nbil}t^2_{NNkj}
+1/2\ \mathcal{P}\sum_{k,c\neq N}h^2_{lkcN}t^{2}_{Nbil}t^2_{Nckj}\\
T_{13}=& -\mathcal{P}\sum_{kc}h^2_{akci}t^2_{cNkj}-\mathcal{P}\sum_{k}h^2_{NkNi}t^2_{Nbkj} \nonumber\\
&+\mathcal{P}\sum_{kl} h^2_{lkNi}t^1_{Nl}t^2_{NNkj} +\mathcal{P}\sum_{kl} h^2_{lkNi}t^1_{Nl}t^2_{Nbkj} -\mathcal{P}\sum_{kc}h^2_{akcN}t^1_{Ni}t^2_{cNkj}\nonumber\\
&-\mathcal{P}\sum_{k}h^2_{NkNN}t^1_{Ni}t^2_{Nbkj}
+1/2\ \mathcal{P}\sum_{lc}h^2_{lkcN}t^2_{NNil}t^2_{cNkj} +1/2\ \mathcal{P}\sum_{l,d\neq N}h^2_{lkNd}t^2_{dNil}t^2_{NNkj}\nonumber\\
&+1/2\ \mathcal{P}\sum_{ld}h^2_{lkNd}t^2_{dNil}t^2_{Nbkj}
+1/2\ \mathcal{P}\sum_{klc}h^2_{lkcN}t^2_{Nail}t^2_{cNkj}
\end{align}
From the above expressions, it can be easily seen that the $t^2$-residual equation can be written as,
\begin{align}
0=\Tilde{T}^{aN}_{ij}+\Tilde{T}^{Nb}_{ij}
\end{align}
Here $a,b,c,d\in \mathcal{V}, i,j,k,l\in \mathcal{O}$ where $b\neq N\ \forall\ t^2_{bprs}\And t^2_{pbrs}$ and, $a\neq N\ \forall\ t^2_{pars}$ terms with $p, r, s$ representing arbitrary molecular orbitals.
\par\noindent
\begin{table}[htbp]
\centering
\caption{One–to–one correspondence between the \texttt{term-n} comments in \texttt{update\_amps()} and the algebraic contributions to the orbital-wise coupled-cluster residual.  All four- and six-index contractions employ CP-ALS rank-2 factors, reducing the raw $O(N^5)$ scaling to $O(N^3)\!-\!O(N^4)$.  Notation: $t_1$ (singles), $t_2$ (doubles within the primary space), $t_{21}$ (mixed doubles, primary–secondary); $F$, $L$, $W$ are dressed Fock and effective-integral intermediates.}
\setlength{\tabcolsep}{5pt}
\begin{tabular}{|c|c|p{4.8cm}|p{5.2cm}|}
\hline
\textbf{Term} & \textbf{Amplitudes} & \textbf{Algebraic form} & \textbf{Role \& nominal scaling} \\ \hline
1  & $t_1,\,t_2,\,t_{21}$ & $f_{iN},\;V_{iajN},\;V_{iNbj}$ & \begin{tabular}[c]{@{}l@{}}Zeroth-order drivers\\ $O(N_oN_v)$\end{tabular} \\ \hline
2  & $t_1,\,t_2,\,t_{21}$ & $f_{kN}\,t_k t_i$ & \begin{tabular}[c]{@{}l@{}}Brillouin / density feedback\\ $O(N_o^{2}N_v)$\end{tabular} \\ \hline
3  & $t_1$ & $F_{NN}\,t_{iN}$ & \begin{tabular}[c]{@{}l@{}}Orbital-energy shift\\ $O(N_oN_v)$\end{tabular} \\ \hline
4  & $t_1$ & $F_{ki}\,t_k$ & \begin{tabular}[c]{@{}l@{}}Singles relaxation\\ $O(N_o^{2})$\end{tabular} \\ \hline
5  & $t_1$ & $F_{kc}\,t_{ikc}$ & \begin{tabular}[c]{@{}l@{}}MP2 feedback into $t_1$\\ $O(N_o^{2}N_v)$\end{tabular} \\ \hline
6  & $t_1$ & $f_{kN}\,t_k t_i$ & \begin{tabular}[c]{@{}l@{}}Non-linear singles\\ $O(N_o^{2})$\end{tabular} \\ \hline
7  & $t_1$ & $V_{kiNN}\,t_k$ & \begin{tabular}[c]{@{}l@{}}Coulomb \& exchange on $t_1$\\ $O(N_o^{2})$\end{tabular} \\ \hline
8  & $t_1$ & $V_{kcNv}\,t_{ikc}$ & \begin{tabular}[c]{@{}l@{}}Singles–mixed-doubles coupling\\ $O(N_o^{2}N_v)$\end{tabular} \\ \hline
9  & $t_1$ & $V_{kNNN}\,t_k t_i$ & \begin{tabular}[c]{@{}l@{}}Higher-order Coulomb\\ $O(N_o^{2})$\end{tabular} \\ \hline
10 & $t_1,\,t_2,\,t_{21}$ & $W^{voov},\;W^{vovo}$ & \begin{tabular}[c]{@{}l@{}}Connected triples-like terms\\ $O(N_o^{2}N_v)$\end{tabular} \\ \hline
11 & $t_1,\,t_2,\,t_{21}$ & $W^{oooV}$ & \begin{tabular}[c]{@{}l@{}}Three-occupied ladders\\ $O(N_{tf}N_oN_v)$\end{tabular} \\ \hline
12 & $t_2,\,t_{21}$ & $W^{vovo},\;W^{vvvo}$ & \begin{tabular}[c]{@{}l@{}}Exchange-relabelled ladders\\ $O(N_{tf}N_oN_v)$\end{tabular} \\ \hline
\multicolumn{4}{|c|}{\textbf{Auxiliary intermediates}} \\ \hline
F-blocks & $F_{oo},\,F_{vv},\,F_{ov}$ & Dressed Fock matrices & \begin{tabular}[c]{@{}l@{}}$O(N_o^{2})$, $O(N_v^{2})$\end{tabular} \\ \hline
L-blocks & $L_{oo},\,L_{vv}$ & Left-dressed Fock & Same as $F$ \\ \hline
W-blocks & $W^{\ast\!\ast\!\ast\!\ast}$ & Effective four-index integrals & Dominant $O(N_{tf}N_oN_v)$ \\ \hline
\end{tabular}
\label{tab:cc_residual_terms}
\end{table}
\paragraph*{Notation.} 
Indices follow the orbital-class shorthand used throughout the restricted–spin TF-HD derivation:
\begin{itemize}
  \item $o$\,—\,occupied spatial orbitals ($i,j,k,l,\dots$),
  \item $v$\,—\,active virtual orbitals ($a,b,c,d,\dots$),
  \item $N$\,—\,the \emph{secondary} orbital that is being down-folded at the current orbital-wise step.
\end{itemize}
A tensor name lists these classes from left to right.  
For example, \emph{ovoV}$\equiv V_{i a j N}$, \emph{oVov}$\equiv V_{i N a b}$, and \emph{VvvV}$\equiv V_{N a b N}$.  
Capital prefixes identify the tensor type: $F$ (dressed Fock), $L$ (left-dressed Fock), and $W$ (effective four-index integrals obtained after down-folding).

The symbol $N_{\mathrm{tf}}$ denotes the number of CP-ALS \emph{tensor factors} used in the rank-2 decomposition of every four- or six-index object.  
Because all tensors in the restricted TF-HD formalism are expressed directly in terms of these $N_{\mathrm{tf}}$ factors, the cubic–quartic classical scalings reported in the last column of Table~\ref{tab:cc_residual_terms} are the \emph{final} costs—no hidden spin summations or additional tensor-factor overheads remain.

\section{Quantum Circuits for Block encoding tensor operations}
\noindent Let us consider a matrix (2-tensor) $B$ of size $M\times N$. Here we will describe a circuit that allows us to encode this matrix and facilitate further processing. For our convenience, we will assume that $M$ and $N$ are of the form $2^m$ and $2^n$ respectively and $||B||=1$.
\par\noindent
Consider a quantum circuit with two multi-qubit registers labelled $I,J$ of sizes $m,n$ respectively and two single qubit registers $A,D$ with the following ordering:
\begin{eqnarray}
    |\cdot\rangle_{I}|\cdot\rangle_{J}|\cdot\rangle_{A}|\cdot\rangle_{D}\label{Qubit ordering}
\end{eqnarray}
\noindent Let the label $a$ denote the state of the single qubit register $A$. We call the quantum circuit that loads our data onto a quantum circuit as a block-encoder. It is defined as follows:
\begin{eqnarray}
V^{a}_{IJ}(A)&=&\Bigg[\sum_{j \in J ; i\in I}|i,j,a\rangle\langle i,j,a|\otimes \Bigg\{ (1-a) \left(A_{ij}I+i\sqrt{1-A_{ij}^{2}}Y\right)+ \\ && a \left((A^T_{ij})'I+i\sqrt{1-(A^T_{ij})^{2}}Y\right)\ \Bigg\}   +  |i,j,1-a\rangle\langle i,j,1-a|\otimes I_{2} \Bigg]
\end{eqnarray}.
\noindent We can read an element $B_{ij}$ by evolving the state $|i,j,a,0 \rangle$ as,
\begin{eqnarray}
\langle i,j,a,0| {V^{a}}_{IJ} | i,j,a,0 \rangle
\end{eqnarray}
\noindent Using the above operator together with superposed states, we can perform various tensor operations. The concept of multiplexed rotations \cite{mottonen2004transformationquantumstatesusing,Shende_2006} allows us to implement the above circuit using $MN$ number of CX Gates, and $MN$ single qubit rotations ($RX,RY,RZ$ rotations).This is insufficient to capture the complexity of implementing it on fault tolerant quantum computers. 
\par\noindent
Fault tolerant quantum computing utilizes clifford + T basis $(H,S,T,CX)$ to represent quantum gates. There are two well known methods of implementing single qubit rotations in this basis as described in the tables below.
\begin{table}[h!]
    \centering
\begin{tabular}{|l|l|l|}
\hline
Method & T-Count & Runtime \\ \hline
The Solovay-Kitaev Process \cite{dawson2005solovaykitaevalgorithm} & $O(\log^{3.97}{\frac{1}{\epsilon}})$ & $O(\log^{2.71}{\frac{1}{\epsilon}})$ \\ \hline
Solving Diophatine Equations\cite{ross2016optimalancillafreecliffordtapproximation} & $3\log\frac{1}{\epsilon} + O(\log\log\frac{1}{\epsilon}) $& $O(\log\log\frac{1}{\epsilon})$ \\ \hline
\end{tabular}
    \caption{Analyzing T-depth for Single qubit rotations}
    \label{tab:T-estimate-single-qubit}
\end{table}

\noindent Thus, encoding a $M \times N$ circuit can be performed with the below resources.

\begin{table}[h!]
    \centering
\begin{tabular}{|l|l|l|}
\hline
Method & T-Depth & Runtime \\ \hline
The Solovay-Kitaev Process \cite{dawson2005solovaykitaevalgorithm} & $O(MN\log^{3.97}{\frac{1}{\epsilon}})$ & $O(MN\log^{2.71}{\frac{1}{\epsilon}})$ \\ \hline
Solving Diophatine Equations\cite{ross2016optimalancillafreecliffordtapproximation} & $3MN\log\frac{1}{\epsilon} + O(MN\log\log\frac{1}{\epsilon})$ & $O(MN\log\log\frac{1}{\epsilon}) $\\ \hline
\end{tabular}
    \caption{Analyzing T-depth for Block Encoding Circuits}
    \label{tab:T-estimate-block-encoder}
\end{table}

\subsection{Realizing Tensor Operations}
In this section, we will perform various tensor operations on quantum circuits and utilize them to construct the downfolding expressions. For that, we present multiple theorems on matrix multiplication and tensor operations using quantum circuits.

\subsubsection{Qubitization circuit for Matrix-Matrix multiplication with isometries}
\subsubsection*{Theorem}
If $A$ and $B$ are general rectangular matrices of dimensions $dim(A)=(N,P)$ and $dim(B)=(P,M)$ with $N,M\ge 2$ then there is a unitary operation $U(A,B)$ of dimension $2^{n_{q}}\times 2^{n_{q}}$ that operates on a system of $n_{q}=p+\max(m,n)+2$ qubit registers : $|\cdot\rangle_{p}|\cdot\rangle_{\max(m,n)}|\cdot\rangle_{a_{1}}|\cdot\rangle_{a_{2}}$ (where $n=\lceil\log_{2} N\rceil$,$m=\lceil\log_{2} M\rceil$,$p=\lceil\log_{2} P\rceil$ )and block encodes the matrix multiplication of $A$ and $B$ s.t. 
\begin{equation*}
    \langle 0|_{p}\langle i|_{\max(m,n)}\langle 0|_{a_{1}}\langle 0|_{a_{2}}U(A,B)|0\rangle_{p}|j\rangle_{\max(m,n)}|1\rangle_{a_{1}}|0\rangle_{a_{2}}=\frac{1}{P^{2}}\frac{\sum_{k}A_{ik}B_{kj}}{||A||||B||}.
\end{equation*}
\textit{Proof-}
Lets define an isometry $T(A,B)$,
\begin{eqnarray}
T(A,B)&=&\frac{1}{\sqrt{2\max(N,M)}}\sum_{r,c}|c\rangle\langle c|\otimes|r\rangle\otimes \bigg[\frac{A_{rc}}{||A||}|0,0\rangle+\sqrt{1-\left(\frac{A_{rc}}{||A||}\right)^{2}}|0,1\rangle\nonumber\\
&+&\frac{B_{cr}}{||B||}|r,1,0\rangle+\sqrt{1-\left(\frac{B_{cr}}{||B||}\right)^{2}}|1,1\rangle\bigg]
\end{eqnarray}
The isometry $T(A,B)$ has the property $T^{\dagger}(A,B)T(A,B)=I_{2}^{\otimes p}$ this can be checked as follows, 
\begin{eqnarray}
T^{\dagger}(A,B)T(A,B) &=& \frac{1}{2\max(N,M)}\sum_{c}|c\rangle\langle c|\otimes\left[\sum_{r}(|A_{rc}|^{2}+1-A^{2}_{rc}+B^{2}_{cr}+1-B^{2}_{cr})\right]\nonumber\\
&=&\sum_{c}|c\rangle\langle c|=I_{2}^{\otimes p}
\end{eqnarray}
Utilizing the above property we can define a unitary operator $W:=W(A,B)$,
\begin{eqnarray}
W(A,B)=2T(A,B)T^{\dagger}(A,B)-1
\end{eqnarray}
The unitarity  of W can be checked as follows,
\begin{eqnarray}
W^{\dagger}W&=&WW^{\dagger}=I\nonumber\\
    &=&(2T(A,B)T^{\dagger}(A,B)-1)(2T(A,B)T^{\dagger}(A,B)-1)\nonumber\\
    &=&4T(A,B)T^{\dagger}(A,B)-4T(A,B)T^{\dagger}(A,B)+1=1
\end{eqnarray}
To proceed further we normalizing the matrices $A':=A/(\sqrt{2}||A||)$ and $B':=B/(\sqrt{2}||B||)$. 
The form of the $W$ matrix in terms of registers is as follows,
\begin{eqnarray}
W&=&\sum_{c,r,r'}|c,r\rangle\langle c,r'|\otimes \bigg[(4A'_{rc}A'_{r'c}-\delta_{rr'})|0,0\rangle\langle 0, 0|+4A'_{rc}\sqrt{1-A^{'2}_{r'c}}|0,0\rangle\langle 0, 1|\nonumber\\
&+&4\sqrt{1-A^{'2}_{rc}}A'_{r'c}|0,1\rangle\langle 0, 0|
    +(4\sqrt{(1-A^{'2}_{rc})(1-A^{'2}_{r'c})}-\delta_{rr'})|0,1\rangle\langle 0, 1|\nonumber\\
    &+&(4B'_{cr}B'_{cr'}-\delta_{rr'})|1,0\rangle\langle 1, 0|
    +4B'_{cr}\sqrt{1-B^{'2}_{cr'}}|1,0\rangle\langle 1, 1|\nonumber\\
    &+&4\sqrt{1-B^{'2}_{cr}}B'_{cr'}|1,1\rangle\langle 1, 0|+(4\sqrt{(1-B^{'2}_{cr})(1-B^{'2}_{cr'})}-\delta_{rr'})|1,1\rangle\langle 1, 1|\nonumber\\
&+&4A'_{rc}B'_{cr'}|0,0\rangle\langle1,0|+4B'_{cr}A'_{r'c}|1,0\rangle\langle0,0|+4B'_{cr}\sqrt{1-A^{'2}_{r'c}}|1,0\rangle\langle 0, 1|+4\sqrt{1-A^{'2}_{rc}}B'_{cr'}|0,1\rangle\langle 1, 0|\nonumber\\
&+&4A'_{rc}\sqrt{1-B^{'2}_{cr'}}|0,0\rangle\langle 1,1|+4\sqrt{1-B^{'2}_{cr}}A'_{r'c}|1,1\rangle\langle 0,0|+4\sqrt{(1-B^{'2}_{cr})(1-A^{'2}_{cr'})}|1,1\rangle\langle 0,1|\nonumber\\
&+&4\sqrt{(1-B^{'2}_{cr'})(1-A^{'2}_{cr})}|0,1\rangle\langle 1,1|\bigg]
\end{eqnarray}
Starting from an initial state with Hadamard on the column qubit registers we obtain,
\begin{eqnarray}
H^{\otimes p}|0\rangle|j\rangle|1\rangle|0\rangle=\frac{1}{P}\sum_{c}|c\rangle|j\rangle|1\rangle|0\rangle.
\end{eqnarray}
Upon acting $W$,
\begin{eqnarray}
    WH^{\otimes p}|0\rangle|j\rangle|1\rangle|0\rangle&=&\frac{1}{2P}\sum_{c,r}|c,r\rangle\bigg[(2B'_{cr}B'_{cj}-\delta_{rj})|1,0\rangle+2\sqrt{1-A^{'2}_{rc}}B'_{cj}|0,1\rangle\nonumber\\
    &+&2\sqrt{1-B^{'2}_{rc}}B'_{cj}|1,1\rangle+2A'_{rc}B'_{cj}|0,0\rangle\bigg]
\end{eqnarray}
Taking overlap of $WH^{\otimes p}|0\rangle|j\rangle|1\rangle|0\rangle$ with the state $H^{\otimes p}|0\rangle|i\rangle|0\rangle|0\rangle$ we get,
\begin{eqnarray}
    \langle 0|\langle 0|\langle i|\langle 0|H^{\otimes p}WH^{\otimes p}|0\rangle|j\rangle|1\rangle|0\rangle=\frac{4}{4P^{2}}\sum_{k}A'_{ik}B'_{kj}=\frac{4}{P^{2}}(A'B')_{ij}=\frac{(AB)_{ij}}{P^{2}||A||||B||}
\end{eqnarray}
By construction we have proved the existence of $U(A,B)$,
\begin{eqnarray}
    U(A,B)=H^{\otimes p}(2T^{\dagger}(A,B)T(A,B)-1)H^{\otimes p}.
\end{eqnarray}

\subsubsection{Matrix-multiplication with Quantum circuits only with Unitary operators}
\subsubsection*{Theorem}
(Isometry free proof)If $A$ and $B$ are general rectangular matrices of dimensions $dim(A)=(N,P)$ and $dim(B)=(P,M)$ then there is a unitary operation $U(A,B)$ of dimension $2^{n_{q}}\times 2^{n_{q}}$ that operates on a system of $n_{q}=p+\max(m,n)+2$ qubit registers $|\cdot\rangle_{p}|\cdot\rangle_{\max(m,n)}|\cdot\rangle_{a_{1}}|\cdot\rangle_{a_{2}}$ (where $n=\lceil\log_{2} N\rceil$,$m=\lceil\log_{2} M\rceil$,$p=\lceil\log_{2} P\rceil$ )and block encodes the matrix multiplication of $A$ and $B$ s.t. 
\begin{equation*}
    \langle 0|_{p}\langle i|_{\max(m,n)}\langle 0|_{a_{1}}\langle 0|_{a_{2}}U(A,B)|0\rangle_{p}|j\rangle_{\max(m,n)}|1\rangle_{a_{1}}|0\rangle_{a_{2}}=\frac{1}{max(M,N)P}|\frac{\sum_{k}A_{ik}B_{kj}}{||A||||B||}.
\end{equation*}
\textit{Proof-}
Let us take the normalized matrices $A'=A/(\sqrt{2}||A||)$, $B'=B/(\sqrt{2}||B||)$. For these we define two unitary operators $V(A)$,$V(B)$,
\begin{eqnarray}
V_A&=&\sum_{c=0,r=0}^{2^{p},2^{\max(m,n)}}\left[|c,r,0\rangle\langle c,r,0|\otimes \left(A'_{rc}I+i\sqrt{1-A_{rc}^{'2}}Y\right)+ |c,r,1\rangle\langle c,r,1|\otimes I_{2}\right]\nonumber\\
V_B&=&\sum_{c=0,r=0}^{2^{p},2^{\max(m,n)}}\left[|c,r,0\rangle\langle c,r,0|\otimes I_{2}+ |c,r,1\rangle\langle c,r,1|\otimes\left(B_{cr}'I+i\sqrt{1-B_{cr}^{'2}}Y\right)\right]
\end{eqnarray}
We load the classical data of the B matrix using the state preparation oracle $V_BH^{\otimes p}$ on the initial state $|0\rangle|j\rangle|1\rangle|0\rangle$,
\begin{eqnarray}
   |\Phi_{B}\rangle= V_BH^{\otimes p}|0\rangle|j\rangle|1\rangle|0\rangle  = \frac{1}{\sqrt{P}}\sum_{c}\left[B'_{cj}|c,j,1,0\rangle+\sqrt{1-B_{cj}^{'2}}|c,j,1,1\rangle\right].
\end{eqnarray}
We load the classical data of the A matrix using the state preparation oracle $V_AH^{\otimes p}$,
\begin{eqnarray}
|\Phi_{A}\rangle&=& V_AH^{\otimes p}|0\rangle|i\rangle|0\rangle|0\rangle =\frac{1}{\sqrt{P}}\sum_{c}\left[A^{'}_{ic}|c,i,0,0\rangle+\sqrt{1-A^{'2}_{ic}}|c,i,0,1\rangle\right]
\end{eqnarray}
Note that the states $|\Phi_{A}\rangle$ and $|\Phi_{B}\rangle$ are orthogonal,
\begin{eqnarray}
   \langle\Phi_{A}|\Phi_{B}\rangle=0
\end{eqnarray}
Next we define diffusion operator $R$ acting on the row registers and the ancillas $a_{1}$, $a_{2}$,
\begin{eqnarray}
    R&=&I_{2}^{\otimes p}\otimes\left[ \left(H^{\otimes \max(m,n)}\otimes H\otimes I_{2}\right)\left(2|0,0,0\rangle\langle 0,0,0|-1\right)\left(H^{\otimes \max(m,n)}\otimes H\otimes I_{2}\right)\right]\nonumber\\
    &=&I_{2}^{\otimes p}\otimes\left[\sum_{k,l}2\frac{|k,+,0\rangle\langle l,+,0|}{\max(M,N)}-I\right]
\end{eqnarray}
Then the overlap between these two states $|\Phi_{A}\rangle$ and $R|\Phi_{B}\rangle$ is given by,
\begin{eqnarray}
\langle\Phi_{A}|R|\Phi_{B}\rangle&=&  \langle 0,i,0,0|H_{c}^{\otimes p}V^{\dagger}_ARV_BH^{\otimes p}_{c}|0,j,1,0\rangle\nonumber\\
&=&\frac{2\sum_{c}A'_{ic}B'_{cj}}{\max(M,N)P}=\frac{\sum_{c}A_{ic}B_{cj}}{\max(M,N)P||A||||B||}
\end{eqnarray}
By construction we have proved the existence of $U(A,B)$ that can be defined without any isometry,
\begin{eqnarray}
    U(A,B)=H_{c}^{\otimes p}V^{\dagger}_ARV_BH^{\otimes p}_{c}.
\end{eqnarray}

\subsubsection{Quantum Circuit for Tensor  Product}
\subsubsection*{Theorem}
If $A,B$ are rectangular matrices with shapes $M \times R$ and $N\times Q$ then there exists a Unitary $U(A,B)$ of dimension $8MNQR \times 8MNQR$ that operates on a system of $n_m+n_r+n_q + m_n+3$ qubit registers $|\cdot\rangle_{I}|\cdot\rangle_{J}|\cdot\rangle_{K}|\cdot\rangle_{L}|\cdot\rangle_{\hat{A}}|\cdot\rangle_{D}$ 
with $I,J,K,L,\hat{A},D$ denoting the qubit-registers of size $n_m,n_r,m_n,n_q,1,2$ (where $n_m=\lceil \log_2{M}\rceil$, $n_N=\lceil \log_2{N}\rceil$, $n_q=\lceil \log_2{Q}\rceil$, $n_r=\lceil \log_2{R}\rceil$ and block encodes the tensors $A_{ij}B_{kl}$ such that
\begin{eqnarray}
    \langle i,j,k,l,0,0 | U(A,B) | 0,0,0,0,0,0 \rangle = \frac{A_{ij}B_{kl}}{\sqrt{MNQR}}
\end{eqnarray}
\textbf{Proof}\\
We assume that $A,B$ are normalized matrices, with $||A||=1,||B||=1$. Define the operator $V^{a}_{IJAD_1}(A)$, as below, acting on qubits in registers $I,J,\hat{A}$ and the qubit $1$ of register $D$. We also have $a$ denoting the state of the single qubit register $\hat{A}$.
\begin{align}
    V^{a}_{IJ\hat{A}D_1}(A) =\ &\Bigg[\sum_{j \in J ; i\in I}|i,j,a\rangle\langle i,j,a|\otimes \Bigg\{ (1-a) \left(A_{ij}I+i\sqrt{1-A_{ij}^{2}}Y\right)+ 
    \nonumber \\
    & a \left(A^T_{ij}I+i\sqrt{1-(A_{ij}^T)^{2}}Y\right)\ \Bigg\}   +  |i,j,1-a\rangle\langle i,j,1-a|\otimes I_{2} \Bigg] \label{block_encoder1}
\end{align}
\noindent Similarly, one can define the operator $V^{a}_{KL\hat{A}D_2}(B)$, as below, acting on qubits in registers $K,L,\hat{A}$ and the qubit $2$ of register $D$. We also have $a$ denoting the state of the single qubit register $\hat{A}$.
\begin{align}
    V^{a}_{KL\hat{A}D_2}(B) =\ &\Bigg[\sum_{k \in K ; l\in L}|k,l,a\rangle\langle k,l,a|\otimes \Bigg\{ (1-a) \left(B_{kl}I+i\sqrt{1-B_{kl}^{2}}Y\right)+ 
    \nonumber \\
    & a \left(B^T_{kl}I+i\sqrt{1-(B^T_{kl})^{2}}Y\right)\ \Bigg\}   +  |k,l,1-a\rangle\langle k,l,1-a|\otimes I_{2} \Bigg]
\end{align}
\noindent Then, one can observe that $V^{a}_{IJ\hat{A}D_1}(A)V^{a}_{KL\hat{A}D_2}(B)H_IH_JH_KH_L$ satisfies
\begin{eqnarray}
    \langle i,j,k,l,0,0 | V^{a}_{IJ\hat{A}D_1}(A)V^{a}_{KL\hat{A}D_2}(B)| 0,0,0,0,0,0 \rangle = \frac{A_{ij}B_{kl}}{\sqrt{MNQR}}
\end{eqnarray}
\noindent
\textbf{Theorem}
\par\noindent
If $A,B,C$ are rectangular matrices with shapes $M \times R$ ,$N\times Q$,$P\times S$ then there exists a Unitary $U(A,B,C)$ of dimension $16MRNQPS \times 16MRNQPS$ that operates on a system of $n_m+n_r+n_q + m_n+m_p +m_s + 4$ qubit registers $|\cdot\rangle_I|\cdot\rangle_J|\cdot\rangle_K|\cdot\rangle_E|\cdot\rangle_F|\cdot\rangle_A|\cdot\rangle_D$ with $I,J,K,L,E,F,A,D$ denoting the qubit-registers of size $n_m,n_r,m_n,n_q,1,2$ (where $n_m=\lceil \log_2{M}\rceil$, $n_N=\lceil \log_2{N}\rceil$, $n_q=\lceil \log_2{Q}\rceil$, $n_r=\lceil \log_2{R}\rceil$, $n_s=\lceil \log_2{S}\rceil$, $n_P=\lceil \log_2{P}\rceil$  and block encodes the tensors $A_{ij}B_{kl}C_{ef}$ such that
\begin{eqnarray}
    \langle i,j,k,l,e,f,0,0 | U(A,B,C)| 0,0,0,0,0,0,0,0 \rangle = \frac{A_{ij}B_{kl}C_{ef}}{\sqrt{MNQRPS}}
\end{eqnarray}
\textbf{Proof}
\par\noindent
We assume that $A,B,C$ are normalized matrices, with $||A||=1,||B||=1,||C||=1$. Define the operator $V^{a}_{IJAD_1}(A)$, as below, acting on qubits in registers $I,J,\hat{A}$ and the qubit $1$ of register $D$. We also have $a$ denoting the state of the single qubit register $\hat{A}$.
\begin{align}
    V^{a}_{IJ\hat{A}D_1}(A) =\ &\Bigg[\sum_{j \in J ; i\in I}|i,j,a\rangle\langle i,j,a|\otimes \Bigg\{ (1-a) \left(A_{ij}I+i\sqrt{1-A_{ij}^{2}}Y\right)+ 
    \nonumber \\ 
    & a\left((A^T_{ij})'I+i\sqrt{1-(A^T_{ij})^{2}}Y\right)\ \Bigg\}   +  |i,j,1-a\rangle\langle i,j,1-a|\otimes I_{2} \Bigg] \label{block_encoder2}
\end{align}
\noindent Similarly, one can define the operator $V^{a}_{KL\hat{A}D_2}(B)$, as below, acting on qubits in registers, $K,L,\hat{A}$ and the qubit $2$ of register $D$. We also have $a$ denoting the state of the single qubit register $\hat{A}$.
\begin{align}
    V^{a}_{KL\hat{A}D_2}(B) =\ &\Bigg[\sum_{k \in K ; l\in L}|k,l,a\rangle\langle k,l,a|\otimes \Bigg\{ (1-a) \left(B_{kl}I+i\sqrt{1-B_{kl}^{2}}Y\right)+ 
    \nonumber \\
    & a\left((B^T_{kl})'I+i\sqrt{1-(B^T_{kl})^{2}}Y\right)\ \Bigg\}   +  |k,l,1-a\rangle\langle k,l,1-a|\otimes I_{2} \Bigg]
\end{align}
\noindent Similarly, one can define the operator $V^{a}_{EF\hat{A}D_3}(C)$, as below, acting on qubits in registers, $E,F,\hat{A}$ and the qubit $3$ of register $D$. We also have $a$ denoting the state of the single qubit register $\hat{A}$.
\begin{align}
    V^{a}_{EF\hat{A}D_3}(C) =\ &\Bigg[\sum_{e \in E ; f\in F}|e,f,a\rangle\langle e,f,a|\otimes \Bigg\{ (1-a) \left(B_{ef}I+i\sqrt{1-B_{ef}^{2}}Y\right)+ \nonumber\\
    & a \left((B^T_{ef})'I+i\sqrt{1-(B^T_{ef})^{2}}Y\right)\ \Bigg\}   +  |e,f,1-a\rangle\langle e,f,1-a|\otimes I_{2} \Bigg]
\end{align}
\noindent Then, one can observe that $V^{a}_{IJ\hat{A}D_1}(A)V^{a}_{KL\hat{A}D_2}(B) V^{a}_{EF\hat{A}D_3}(C)H_IH_JH_KH_LH_EH_F$ satisfies
\begin{align}
    \langle i,j,k,l,e,f,0,0 | V^{a}_{IJ\hat{A}D_1}(A)V^{a}_{KL\hat{A}D_2}(B) V^{a}_{EF\hat{A}D_3}(C)H_IH_JH_KH_LH_EH_F|& 0,0,0,0,0,0,0,0 \rangle \nonumber \\
    &= \frac{A_{ij}B_{kl}C_{ef}}{\sqrt{MNQRPS}}
\end{align}
 
\subsubsection{Quantum Circuit for Tensor  Contraction}
\subsubsection*{Theorem}
If $A,B$ are rectangular matrices with shapes $M \times R$ and $M\times Q$ then there exists a Unitary $U(A,B)$ of dimension $2^{n_m+n_r+n_q + 3} \times ^{n_m+n_r+n_q + 3}$ that operates on a system of $n_m+n_r+n_q + 3$ qubit registers $|\cdot\rangle_I|\cdot\rangle_J|\cdot\rangle_K|\cdot\rangle_A|\cdot\rangle_D$ with $I,J,K,A,D$ denoting the qubit-registers of size $n_m,n_r,n_q,1,2$ and block encodes the tensor contraction $\sum_{i}A_{ij}B_{ik}$ such that
\begin{eqnarray}
    \langle 0,j,0,0,0 | U(A,B) | 0,0,0,0,0 \rangle = \frac{\sum_iA_{ij}B_{ik}}{M\sqrt{QR}}
\end{eqnarray}
\textbf{Proof}\\
We assume that $A,B$ are normalized matrices, with $||A||=1,||B||=1$. Define the operator $V^{a}_{IJAD_1}(A)$, as below, acting on qubits in registers $I,J,\hat{A}$ and the qubit $1$ of register $D$. We also have $a$ denoting the state of the single qubit register $\hat{A}$.
\begin{align}
    V^{a}_{IJ\hat{A}D_1}(A) =\ &\Bigg[\sum_{j \in J ; i\in I}|i,j,a\rangle\langle i,j,a|\otimes \Bigg\{ (1-a) \left(A_{ij}I+i\sqrt{1-A_{ij}^{2}}Y\right)+
    \nonumber \\ 
    & a \left((A^T_{ij})'I+i\sqrt{1-(A^T_{ij})^{2}}Y\right)\ \Bigg\}   +  |i,j,1-a\rangle\langle i,j,1-a|\otimes I_{2} \Bigg] \label{block_encoder3}
\end{align}
\noindent Similarly, one can define the operator $V^{a}_{IK\hat{A}D_2}(B)$, as below, acting on qubits in registers $I,K,\hat{A}$ and the qubit $2$ of register $D$. We also have $a$ denoting the state of the single qubit register $\hat{A}$.
\begin{align}
    V^{a}_{IK\hat{A}D_2}(B) =\ &\Bigg[\sum_{k \in K ; i\in I}|i,k,a\rangle\langle i,k,a|\otimes \Bigg\{ (1-a) \left(B_{ik}I+i\sqrt{1-B_{ik}^{2}}Y\right)+
    \nonumber\\ 
    & a\left((B^T_{ik})'I+i\sqrt{1-(B^T_{ik})^{2}}Y\right)\ \Bigg\}   +  |i,k,1-a\rangle\langle i,k,1-a|\otimes I_{2} \Bigg]
\end{align}
\noindent Then, one can observe that $H_IV^{a}_{IJ\hat{A}D_1}(A)V^{a}_{IK\hat{A}D_2}(B)H_IH_JH_K$ satisfies
\begin{eqnarray}
    \langle 0,j,k,0,0 | H_IV^{a}_{IJ\hat{A}D_1}(A)V^{a}_{IK\hat{A}D_2}(B)H_IH_JH_K| 0,0,0,0,0 \rangle = \frac{\sum_iA_{ij}B_{ik}}{M\sqrt{QR}}
\end{eqnarray}
\noindent
\textbf{Theorem}
\par\noindent
If $A,B,C$ are rectangular matrices with shapes $M \times R$ ,$M\times Q$,$M\times S$ then there exists a Unitary $U(A,B,C)$ of dimesnion $2^{n_m+n_r+n_q+m_n +m_p +m_s + 4} \times 2^{n_m+n_r+n_q +m_n++m_p +m_s + 4}$ that operates on a system of $n_m+n_r+n_q + m_n+m_p +m_s + 4$ qubit registers $|\cdot\rangle_I|\cdot\rangle_J|\cdot\rangle_K|\cdot\rangle_L|\cdot\rangle_A|\cdot\rangle_D$ with $I,J,K,L,A,D$ denoting the qubit-registers of size $n_m,n_r,m_n,n_q,m_p,m_s,1,2$ and block encodes the tensor contraction $\sum_i A_{ij}B_{ik}C_{il}$ such that
\begin{eqnarray}
    \langle 0,j,k,l,0,0 | U(A,B,C) | 0,0,0,0,0,0\rangle = \frac{\sum_iA_{ij}B_{ik}C_{il}}{M\sqrt{QRS}}
\end{eqnarray}
\textbf{Proof}\\
We assume that $A,B$ are normalized matrices, with $||A||=1,||B||=1$. Define the operator $V^{a}_{IJAD_1}(A)$, as below, acting on qubits in registers $I,J,\hat{A}$ and the qubit $1$ of register $D$. We also have $a$ denoting the state of the single qubit register $\hat{A}$.
\begin{align}
    V^{a}_{IJ\hat{A}D_1}(A) =\ &\Bigg[\sum_{j \in J ; i\in I}|i,j,a\rangle\langle i,j,a|\otimes \Bigg\{ (1-a) \left(A_{ij}I+i\sqrt{1-A_{ij}^{2}}Y\right)+
    \nonumber \\ 
    & a \left((A^T_{ij})'I+i\sqrt{1-(A^T_{ij})^{2}}Y\right)\ \Bigg\}   +  |i,j,1-a\rangle\langle i,j,1-a|\otimes I_{2} \Bigg] \label{block_encoder4}
\end{align}
\noindent Similarly, one can  define the operator $V^{a}_{IK\hat{A}D_2}(B)$, as below, acting on qubits in registers $I,K,\hat{A}$ and the qubit $2$ of register $D$. We also have $a$ denoting the state of the single qubit register $\hat{A}$.
\begin{align}
    V^{a}_{IK\hat{A}D_2}(B) =\ &\Bigg[\sum_{k \in K ; i\in I}|i,k,a\rangle\langle i,k,a|\otimes \Bigg\{ (1-a) \left(B_{ik}I+i\sqrt{1-B_{ik}^{2}}Y\right)+
    \nonumber \\ 
    & a \left((B^T_{ik})'I+i\sqrt{1-(B^T_{ik})^{2}}Y\right)\ \Bigg\}   +  |i,k,1-a\rangle\langle i,k,1-a|\otimes I_{2} \Bigg]
\end{align}
\noindent Similarly, one can define the operator $V^{a}_{IL\hat{A}D_3}(B)$, as below, acting on qubits in registers $I,L,\hat{A}$ and the qubit $3$ of register $D$. We also have $a$ denoting the state of the single qubit register $\hat{A}$.
\begin{align}
    V^{a}_{IL\hat{A}D_2}(C) =\ &\Bigg[\sum_{l \in L ; i\in I}|i,l,a\rangle\langle i,l,a|\otimes \Bigg\{ (1-a) \left(B_{il}I+i\sqrt{1-B_{il}^{2}}Y\right)+
    \nonumber \\
    & a \left((B^T_{il})'I+i\sqrt{1-(B^T_{il})^{2}}Y\right)\ \Bigg\}   +  |i,l,1-a\rangle\langle i,l,1-a|\otimes I_{2} \Bigg]
\end{align}

\noindent Then, one can observe that $H_IV^{a}_{IJ\hat{A}D_1}(A)V^{a}_{IK\hat{A}D_2}(B)V^{a}_{IL\hat{A}D_2}(C)H_IH_JH_KH_L$ satisfies
\begin{eqnarray}
    \langle 0,j,k,l,0,0 | H_IV^{a}_{IJ\hat{A}D_1}(A)V^{a}_{IK\hat{A}D_2}(B)H_IH_JH_K| 0,0,0,0,0,0 \rangle = \frac{\sum_iA_{ij}B_{ik}C_{il}}{M\sqrt{QRS}}
\end{eqnarray}
 
\noindent Using the theorems developed in the above subsection, we describe different operations.

\subsubsection{Tensor Contraction}
For the contraction $c_{ij} = a_{ik}b_{kj}$, consider a quantum circuit with the registers in order $I,K,A,D$. Let $a,b$ be of size $2^m \times 2^p, 2^p \times 2^n$ respectively. We can also assume that $m > n$. Let the registers be of size $m,p$ respectively. We define the circuit as \begin{equation}
    H_JV^{0}_{IK}(A)R_KV^{1}_{IK}(B)H_J
\end{equation}
\noindent The above circuit encodes the term $\frac{c_{ij}}{2^{m+p }}$ in the $\vert i,0,0,0 \rangle$ state of the circuit. Note that $R$ is the reflector as defined in the multiplication theorems.\\
The T-depth for implementing this operation is:
\begin{equation}
    O\Bigg(2^{m+p}log\frac{1}{\epsilon}\Bigg)
\end{equation}

\subsubsection{Tensor Dot}
For the operation $d_{ij}=a_{ix}b_{jx}$, we consider a quantum circuit with the registers in order $I,J,K,A,D$. Let the tensors $a,b$ be of size $2^m \times 2^x$, $2^n \times 2^x$. Note that we are considering these sizes, as these are the types of tensors, we would be dealing with. Although the result below holds for other arbitrary shapes as well.

\noindent Let the registers be of size $m,n,x,1,2$ respectively. Then
\begin{equation}
     H_K V^{0}_{IK}(A)V^{0}_{JK}(B)H_I H_JH_K
\end{equation} encodes the tensors $\frac{d_{ij}}{\sqrt{2^{m+n+2x}}}$ .

\noindent This can be implemented with t-depth 
\begin{equation}
    O\Bigg(2^{m+x}log\frac{1}{\epsilon}\Bigg)+O\Bigg(2^{n+x}\log\frac{1}{\epsilon}\Bigg)
\end{equation}

\subsubsection{Hadamard Product}
For the operation $c_{ix}=a_{ix}b_{ix}$, we consider a quantum circuit with the registers in order $I,X,A,D$. Let the tensors $a,b$ be of size $2^m \times 2^x$ and the registers be of size $m,x,1,2$ respectively. Then,
\begin{equation}
    V^{0}_{IX}(A)V^{0}_{IX}(B)H_I H_X
\end{equation}encodes the tensors $\frac{d_{ij}}{\sqrt{2^{m+x}}}$ .

\noindent This can be implemented in a quantum circuit of depth 
\begin{equation}
    2*O\Bigg(2^{m+x}\log\frac{1}{\epsilon}\Bigg)
\end{equation}

\noindent The above theorems on different tensor operations can be summarized as follows:
\begin{itemize}
    \item Multiplication: Load matrices on different states and sandwich a reflector between them. Use Hadamard gates to generate whole columns.
    
    \item Tensor Product: Load matrices on different sets of qubits.
    
    \item Tensor Contraction: Load matrices on different sets of qubits for different indices. Indices to be contracted share registers. Sandwich the contracted registers between Hadamard gates.
\end{itemize}

\subsection{Complexity Analysis of Single Reference Hamiltonian Downfolding}

From the explicit form of residual equations given above, we can obtain closed form expressions for the two RHD(SD) amplitude equations that directly correspond to our more general multireference RHD formulation. The singles residual equation takes the form,
\begin{align}
r^{1,(N)}_{i}=A_{i}^{1,(N)}+\sum_{j}A^{2,(N)}_{ji}t_j^{1,(N)}+\sum_{jk}A^{3,(N)}_{jki}t_j^{1,(N)}t_k^{1,(N)}+\sum_{akl}A^{4,(N)}_{klia}t_{akl}^{2,(N)}+\sum_{klc}A^{5,(N)}_{klc}t^{1,(N)}_{l}t^{2,(N)}_{cik}\label{r1}
\end{align}

\noindent And the doubles residual equation looks like,
\begin{align}
r^{2,(N)}_{aij}=h^{2,(N)}_{aNij}
+\sum_{kl}B^{1,(N)}_{klij}t^{2,(N)}_{akl}
+\sum_{b}B^{2,(N)}_{ab}t^{2,(N)}_{bij}
&+\sum_{akb}B^{3,(N)}_{akb}t^{2,(N)}_{bij}t^{1,(N)}_{k} \nonumber\\ 
&+B^{4,(N)}_{a}t^{1,(N)}_{i}t^{1,(N)}_{j}
+\sum_{klc}B^{5,(N)}_{klc}t^{2,(N)}_{cil}t^{2,(N)}_{akj}\label{r2}
\end{align}
The terms corresponding to the tensors, $A^{2,(N)}$, $A^{4,(N)}$, $B^{1,(N)}$ and $B^{2,(N)}$ originate from the class of terms, $\eta PHP$ and $QHQ\eta$, of the Bloch equation eq.\eqref{BlochEqn}. Whereas, $A^{3,(N)}$, $A^{5,(N)}$, $B^{3,(N)}$, $B^{4,(N)}$ and $B^{5,(N)}$ comes from the $\eta PHQ\eta$ class of terms. The terms $A^{1,(N)}$ and ERI-slice ($h^{2,(N)}_{aNij}$) can be attributed to the $QHP$ expressions.\\
\noindent The Cholesky Decomposition of the ERI is carried out with a density fitted auxiliary basis set of size $N_{aux}$.  The mathematical form of the Cholesky factorization is given by,  
\begin{align}
h^{2,(N)}_{abij}\rightarrow \sum_{x}L_{xai}L_{xbj}
\end{align}
where $L$ is the 3-rank Cholesky factor and $x$ is the auxiliary basis direction. For the downfolding residual equations we need only slices of the ERI which are obtained by setting index $a$ or $b$ to $N$. A further tensor decomposition of the Cholesky factors is carried out using canonical Polyadic Alternating Least Squares Decomposition (CPALSD)\cite{hong2020generalized}, and is represented as,
\begin{align}
L_{xai}\rightarrow \sum_{p}X_{xp}Y_{ap}Z_{ip}
\end{align}
Here $X$, $Y$, $Z$ are 2-rank tensor factors of a 3-rank Cholesky factor ($L$) and $p$ indexes the Tensor factorization(TF) based auxiliary basis direction. The size of this auxiliary basis set arising from the decomposition of Cholesky factors is $N_{htf}$. Within downfolding, the doubles cluster amplitude is a 3-rank tensor that can be initialized in its tensor factorized representation as,
\begin{align}
t^{2,(N)}_{aij}=\sum_{r}T_{ar}U_{ir}V_{jr}
\end{align}
Here $T$, $U$, $V$ are the tensor factors and $r$ indexes the auxiliary basis direction associated with the decomposition of the doubles cluster amplitudes. The size of this auxiliary basis set is given by $N_{ttf}$.\\
\noindent To study the operational scaling complexity of solving the residual equations, we will consider representative tensor factorized forms for each of the terms in both the singles and the doubles residual equations. In this section we will use i,j,k,l indices for occupied orbitals; a,b,c,d for virtual orbitals; x for auxiliary basis sets associated with the Cholesky Decomposition of ERIs; p,q for auxiliary basis sets associated with the tensor factorization of those Cholesky factors; and r,s for the auxiliary basis sets associated with the factorization of doubles cluster amplitudes. All summations over i,j,k,l will range from $1$ to $N_{o}$; a,b,c,d from $1$ to $N_{v}$; x from $1$ to $N_{aux}$; p,q from $1$ to $N_{htf}$; and r,s from $1$ to $N_{ttf}$. The relationship between $N_{o}$, $N_{v}$, $N_{aux}$, $N_{htf}$ and $N_{ttf}$ can be written as, 
\begin{align}
N_{htf}>N_{aux}>N_{ttf}>N_v>N_o
\end{align}

\noindent We will, for the entirety of our calculations, contract the tensor factorized terms of the residual equation to their irreducible representations. As the convergence of a Downfolding for each orbital happens completely in terms of these irreducible representations, the operational complexity of Downfolding goes down to $\mathcal{O}(N^3)$. In subsequent sections we will study the tensor decompositions of the terms in the residual equation in detail.

\subsection{Implementing Downfolding Expressions on Quantum Circuits
}
To generate the quantum circuits for our residual equation let us setup the relevant notations. Consider quantum registers $\tilde{P},\tilde{Q},\tilde{R},\tilde{S},\tilde{I},\tilde{J},\tilde{K},\tilde{L},\tilde{A},\tilde{B},\tilde{C},\tilde{D},\hat{A}_{1},\hat{D}_{\{i\}},\tilde{X}$ as following:
\begin{align}
\tilde{P},\tilde{Q} &\in [N_{htf}]\\
\tilde{R},\tilde{S} &\in [N_{ttf}]\\
\tilde{I},\tilde{J},\tilde{K},\tilde{L} &\in [N_o]\\
\tilde{A},\tilde{B},\tilde{C},\tilde{D} &\in [N_v]\\
\tilde{X} &\in [N_{aux}]\\
\hat{A}_{1} &\in [1]\\
\hat{D}\ &\in [12]
\end{align}

\noindent Here,
\begin{enumerate}
\item $\tilde{I}$,$\tilde{J}$,$\tilde{K}$,$\tilde{L}$ are qubit registers of size $\log_2 {N_{o}}$

\item $\tilde{A}$,$\tilde{B}$,$\tilde{C}$ are qubit registers of size $\log_2 {N_{v}}$

\item $\tilde{X}$ is a qubit register of size $\log_2{N_{aux}}$

\item $\tilde{P}$, $\tilde{Q}$ are qubit registers of size $\log_2 {N_{htf}}$

\item $\tilde{R}$, $\tilde{S}$ are qubit registers of size $\log_2 {N_{ttf}}$

\item $\hat{A}_{1}$ is a single qubit ancilla to facilitate contraction.

\item $\hat{D}$ is a qubit register of size 12 {num tensors}.

\item For quantum circuit representation purposes (Fig.\ref{Circuits}), four of the tensor factorization direction quantum registers, P,Q,R,S, are clubbed into a single register category, TF, which in turn can 
be partitioned into 4 groups:
    \begin{enumerate}
    \item Group-1: $[1,\ log_2(N_{htf})]$
    \item Group-2: $[log_2(N_{htf})+1,\ 2log_2(N_{htf}) ]$
    \item Group-3: $[2log_2(N_{htf})+1,\ 2log_2(N_{htf})+log_2(N_{ttf})]$
    \item Group-4: $[2log_2(N_{htf})+log_2(N_{ttf})+1,\ 2log_2(N_{htf})+2log_2(N_{ttf})]$
    \end{enumerate}
\item Each $V(\cdot)$ gate acts on a unique $D_i \in \hat{D}$.
\item Corresponding to each term, the $V(\cdot)$ gates acting with $X,Y,Z$ as input operate on the qubits in Group-1 and Group-2 of TF.
\item The gates $V(\cdot)$ acting with $T,U,V$ as input operate on the qubits in Group-3 and Group-4 of TF.
\item For any arbitrary tensor $G$, we denote terms of the form $V^{0}_{IJ}(G)$ acting on qubits of registers $I,J$ and qubits $A_1,D_{i}$ as $V_{IJ,i}(G)$
\end{enumerate}

\subsubsection{Depth of a Quantum Circuit in S,CNOT,H,T basis that block encodes a two-rank tensor}
For any arbitrary tensor $G$ of shape $N,M$, the operation $V_{IJ,i}(G)$ has a depth of 
\begin{align}
2NM\log_2\left( \frac{1}{\epsilon}\right)
\end{align}
The starting and final Hadamard layers contribute to a depth of 2.

\subsubsection{Expression 1}
\noindent The first term of the singles residual equation (eq. \ref{r1}) is generated from the $QHP$ class of terms in the Bloch equation. It is given by the vector-projection of the fock matrix along the Nth molecular orbital as,
\begin{align}
A^{1,(N)}_{i}\rightarrow f_{Ni}
\end{align}

The Quantum Circuit for representing this tensor is given by,
\begin{align}
V_{\tilde{I},1}(X)H_{\tilde{I}}
\end{align}

And the depth of this Quantum Circuit is,
\begin{align}
2N_{o}log_2(1/\epsilon)
\end{align}

\subsubsection{Expression 2}
The second term of singles residual equation (\ref{r1}) is generated from both the $\eta PHP$ and $QHQ\eta$ classes of terms in the Bloch equation. Also, there will be two separate forms for this term, one coming from the fock part of the Hamiltonian and other from the ERI. The fock contribution can be written as, 
\begin{align}
&\sum_{j}A^{2,(N)}_{ji}t_j^{1,(N)}\rightarrow \sum_{k}f^{(N)}_{ji}t^{1,(N)}_{j}
\end{align}
The order of complexity for contracting this term to its irreducible representation is $\mathcal{O}(N_{o}^2)$.The ERI contribution to this term in the residual equation can originate from different components of the ERI, $h^{2,(N)}_{NjNi}$, $h^{2,(N)}_{jNNi}$, $h^{2,(N)}_{NjiN}$, $h^{2,(N)}_{jNiN}$. Here we will show the $h^{2,(N)}_{jNNi}$ contribution in its tensor factorized form below,
\begin{align}
&\sum_{j}A^{2,(N)}_{ji}t_j^{1,(N)}\rightarrow \sum_{j}\sum_{xpq} X_{xp}Y_{jp}Z_{Np}X_{xq}Y_{Nq}Z_{iq}t_{j}^{1,(N)}
\end{align}
 A similar factorization is performed for all four contributions from ERI. The diagrammatic representation of this decomposition is shown in figure \ref{td2}. The contraction path to an irreducible representation can be written as,
\begin{align}
&\wick{xp, \c1 jp, p, xq, \c2 q, \c2 iq, \c1 j} \quad \mathcal{O}(2N_{htf}N_{o})\nonumber\\
\rightarrow\ &\wick{\c1 p, iq, xp, \c1 p, xq} \quad \mathcal{O}(N_{htf})\nonumber\\
\rightarrow\ &\wick{\c1 p, iq, \c1 xp, xq}\quad \mathcal{O}(N_{htf}N_{aux}) \nonumber\\
\rightarrow\ &\wick{\c1 x, iq,\c1 xq}\quad \mathcal{O}(N_{htf}N_{aux}) \nonumber\\
\rightarrow\ &\wick{\c1 q,\c1 iq}\quad \mathcal{O}(N_{htf}N_{o}) \nonumber\\
\rightarrow\ &\wick{i}
\end{align}
The complexity for this contraction is given by,
\begin{align}
\mathcal{O}(2N_{htf}N_{aux}+3N_{htf}N_{o}+N_{htf})
\end{align}
The Quantum Circuit for representing this set of tensor operations is given by,
\begin{align}
&H_{\tilde{X}}H_{\tilde{P}}H_{\tilde{Q}}H_{\tilde{J}}
V_{\tilde{X}\tilde{P},1}(X)V_{\tilde{J}\tilde{P},2}(Y)V_{\tilde{P},3}(Z)V_{\tilde{X}\tilde{Q},4}(X)V_{\tilde{Q},5}(Y)V_{\tilde{I}\tilde{Q},6}(Z)V_{\tilde{J},7}(T^{1,(N)})\nonumber\\
&H_{\tilde{I}}H_{\tilde{X}}H_{\tilde{P}}H_{\tilde{Q}}H_{\tilde{J}}
\end{align}
The diagrammatic representation of this circuit is given in figure \ref{qc2}. The depth of this Quantum Circuit is,
\begin{align}
&2(2N_{aux}N_{htf}+2N_{o}N_{htf}+2N_{htf}+N_{o})log_2(1/\epsilon)\\
\end{align}

\subsubsection{Expression 3}
The third term of singles residual equation (\ref{r1}) is generated from the $\eta PHQ\eta$ class of terms in the Bloch equation. This term is generated from two different slices of the ERI tensor, $h^{2,(N)}_{jkiN}$ and $h^{2,(N)}_{jkNi}$. We show the tensor factorized representation of the $h^{2,(N)}_{jkiN}$ contribution below,
\begin{align}
&\sum_{jk}A^{3,(N)}_{jki}t_j^{1,(N)}t_k^{1,(N)}\rightarrow \sum_{jk}\sum_{xpq}
X_{xp}Y_{jp}Z_{ip}X_{xq}Y_{kq}Z_{Nq}t^{1,(N)}_{j}t_{k}^{1,(N)}
\end{align}
The diagrammatic representation of this decomposition is shown in figure \ref{td3}. The contraction path to an irreducible representation is given by,
\begin{align}
&\wick{xp,\c1 jp,ip,xq,\c2 kq,q,\c1 j,\c2 k}\quad \mathcal{O}(2N_{htf}N_{o}) \nonumber\\
\rightarrow\ &\wick{\c1 p, \c2 q, xp, \c1 ip, xq, \c2 q}\quad \mathcal{O}(N_{htf}N_{o}+N_{htf}) \nonumber\\
\rightarrow\ &\wick{ip, \c1 q, xp, \c1 xq}\quad \mathcal{O}(N_{htf}N_{aux}) \nonumber\\
\rightarrow\ &\wick{\c1 x, ip, \c1 xp}\quad \mathcal{O}(N_{htf}N_{aux}) \nonumber\\
\rightarrow\ &\wick{\c1 p,\c1 ip}\quad \mathcal{O}(N_{htf}N_{o}) \nonumber\\
\rightarrow\ &\wick{i}
\end{align}
The total complexity of this calculation is given by,
\begin{align}
\mathcal{O}(2N_{htf}N_{aux}+4N_{htf}N_{o}+N_{htf})
\end{align}
The Quantum Circuit representing this set of tensor operations is given by,
\begin{align}
&H_{\tilde{X}}H_{\tilde{P}}H_{\tilde{Q}}H_{\tilde{J}}H_{\tilde{K}}
V_{\tilde{X}\tilde{P},1}(X)V_{\tilde{J}\tilde{P},2}(Y)V_{\tilde{I}\tilde{P},3}(Z)V_{\tilde{X}\tilde{Q},4}(X)V_{\tilde{K}\tilde{Q},5}(Y)V_{\tilde{Q},6}(Z)V_{\tilde{J},7}(T^{1,(N)})V_{\tilde{K},8}(T^{1,(N)})\nonumber\\
&H_{\tilde{I}}H_{\tilde{X}}H_{\tilde{P}}H_{\tilde{Q}}H_{\tilde{J}}H_{\tilde{K}}
\end{align}
The diagrammatic representation of this circuit is given in figure \ref{qc3}. The depth of this Quantum Circuit is,
\begin{align}
&2(2N_{aux}N_{htf}+3N_{o}N_{htf}+N_{htf}+2N_{o})log_2(1/\epsilon)
\end{align}

\subsubsection{Expression 4}
The fourth term of the singles residual equation(\ref{r1}) is generated from both the $\eta PHP$ and $QHQ\eta$ terms in the Bloch equation. This can be generated from four different slices of the ERI tensor, $h^{2,(N)}_{klia}$, $h^{2,(N)}_{klai}$, $h^{2,(N)}_{lkia}$ and $h^{2,(N)}_{lkai}$. The tensor factorized representation of the $h^{2,(N)}_{klia}$ contribution is given by,
\begin{align}
&\sum_{kla}A^{4,(N)}_{klia}t_{akl}^{2,(N)}\rightarrow \sum_{kla}\sum_{xpqr}
X_{xp}Y_{kp}Z_{ip}X_{xq}Y_{lq}Z_{aq}T_{ar}U_{kr}V_{lr}
\end{align}
The diagrammatic representation of this decomposition is shown in figure \ref{td4}. The contraction path to an irreducible representation can be written as,
\begin{align}
&\wick{xp,\c1 kp,ip,xq,\c2 lq,\c3 aq,\c3 ar,\c1 kr,\c2 lr}\quad \mathcal{O}(2N_{htf}N_{ttf}N_{o}+N_{htf}N_{ttf}N_{v}) \nonumber\\
\rightarrow\ &\wick{pr,\c1 qr,\c1 qr, xp, ip, xq}\quad \mathcal{O}(N_{htf}N_{ttf}) \nonumber\\
\rightarrow\ &\wick{\c1 qr, pr, xp, ip, \c1 xq}\quad \mathcal{O}(N_{htf}N_{ttf}N_{aux}) \nonumber\\
\rightarrow\ &\wick{\c1 xr, \c1 pr, xp, ip}\quad \mathcal{O}(N_{htf}N_{ttf}N_{aux}) \nonumber\\
\rightarrow\ &\wick{\c1 xp,\c1 xp, ip}\quad \mathcal{O}(N_{htf}N_{aux}) \nonumber\\
\rightarrow\ &\wick{\c1 p,\c1 ip}\quad \mathcal{O}(N_{htf}N_{o}) \nonumber\\
\rightarrow\ &\wick{i}
\end{align}
The total complexity of this calculation is given by,
\begin{align}
\mathcal{O}(2N_{htf}N_{ttf}N_{aux}+2N_{htf}N_{ttf}N_{o}+N_{htf}N_{ttf}N_{v}+N_{htf}N_{aux}+N_{htf}N_{ttf}+N_{htf}N_{o})
\end{align}
The Quantum Circuit representing this set of tensor operations is given by,
\begin{align}
&H_{\tilde{X}}H_{\tilde{P}}H_{\tilde{Q}}H_{\tilde{R}}H_{\tilde{K}}H_{\tilde{L}}H_{\tilde{A}}
V_{\tilde{X}\tilde{P},1}(X)V_{\tilde{K}\tilde{P},2}(Y)V_{\tilde{I}\tilde{P},3}(Z)V_{\tilde{X}\tilde{Q},4}(X)V_{\tilde{L}\tilde{Q},5}(Y)V_{\tilde{A}\tilde{Q},6}(Z)\nonumber\\
&V_{\tilde{A}\tilde{R},7}(T)V_{\tilde{K}\tilde{R},8}(U)V_{\tilde{L}\tilde{R},9}(V)H_{\tilde{I}}H_{\tilde{X}}H_{\tilde{P}}H_{\tilde{Q}}H_{\tilde{R}}H_{\tilde{K}}H_{\tilde{L}}H_{\tilde{A}}
\end{align}
The diagrammatic representation of this circuit is given in figure \ref{qc4}. The depth of this Quantum Circuit is,
\begin{align}
&2(2N_{aux}N_{htf}+3N_{o}N_{htf}+N_{v}N_{htf}+2N_{o}N_{ttf}+N_{v}N_{ttf})log_2(1/\epsilon)
\end{align}

\subsubsection{Expression 5}
The fifth term of the singles residual equation(\ref{r1}) is generated from the $\eta PHQ\eta$ class of terms in the Bloch equation. It contains contributions from four different slices of the ERI tensor, $h^{2,(N)}_{klcN}$, $h^{2,(N)}_{klNc}$, $h^{2,(N)}_{lkcN}$ and $h^{2,(N)}_{lkNc}$. And, for all the ERI slices, there will be contributions from two different forms of the doubles cluster amplitudes, $t^{2,(N)}_{cik}$ and $t^{2,(N)}_{cki}$. The tensor factorized representation of the contribution from a combination of $h^{2,(N)}_{klcN}$ and $t^{2,(N)}_{cik}$ is given by,
\begin{align}
&\sum_{klc}A^{5,(N)}_{klc}t^{1,(N)}_{l}t^{2,(N)}_{cik}\rightarrow \sum_{klc}\sum_{xpqr}
X_{xp}Y_{kp}Z_{cp}X_{xq}Y_{lq}Z_{Nq}t^{1,(N)}_{l}T_{cr}U_{ir}V_{kr}
\end{align}
The diagrammatic representation of this decomposition is shown in figure \ref{td5}. The contraction path to an irreducible representation can be written as,
\begin{align}
&\wick{xp,\c1 kp,\c2 cp,xq,\c3 lq,q,\c3 l,\c2 cr,ir,\c1 kr}\quad \mathcal{O}(N_{htf}N_{ttf}N_{o}+N_{htf}N_{ttf}N_{v}+N_{htf}N_{o}) \nonumber\\
\rightarrow\ &\wick{\c1 pr,\c1 pr,\c2 q, xp, xq,\c2 q, ir}\quad \mathcal{O}(N_{htf}N_{ttf}+N_{htf}) \nonumber\\
\rightarrow\ &\wick{pr, \c1 q, xp,\c1 xq, ir}\quad \mathcal{O}(N_{htf}N_{aux}) \nonumber\\
\rightarrow\ &\wick{\c1 x, pr,\c1 xp, ir}\quad \mathcal{O}(N_{htf}N_{aux}) \nonumber\\
\rightarrow\ &\wick{\c1 p,\c1 pr, ir}\quad \mathcal{O}(N_{htf}N_{ttf}) \nonumber\\
\rightarrow\ &\wick{\c1 r,\c1 ir}\quad \mathcal{O}(N_{ttf}N_{o}) \nonumber\\
\rightarrow\ &\wick{i}
\end{align}
The total complexity of this calculation is given by,
\begin{align}
\mathcal{O}(N_{htf}N_{ttf}N_{o}+N_{htf}N_{ttf}N_{v}+N_{htf}N_{o}+2N_{htf}N_{ttf}+2N_{htf}N_{aux}+N_{ttf}N_{o}+N_{htf})
\end{align}
The Quantum Circuit representing this set of tensor operations is given by,
\begin{align}
&H_{\tilde{X}}H_{\tilde{P}}H_{\tilde{Q}}H_{\tilde{R}}H_{\tilde{K}}H_{\tilde{L}}H_{\tilde{C}}
V_{\tilde{X}\tilde{P},1}(X)V_{\tilde{K}\tilde{P},2}(Y)V_{\tilde{C}\tilde{P},3}(Z)V_{\tilde{X}\tilde{Q},4}(X)V_{\tilde{L}\tilde{Q},5}(Y)V_{\tilde{Q},6}(Z)V_{\tilde{L},7}(T^{1,(N)})\nonumber\\
&V_{\tilde{C}\tilde{R},8}(T)V_{\tilde{I}\tilde{R},9}(U)V_{\tilde{K}\tilde{R},10}(V)H_{\tilde{I}}H_{\tilde{X}}H_{\tilde{P}}H_{\tilde{Q}}H_{\tilde{R}}H_{\tilde{K}}H_{\tilde{L}}H_{\tilde{C}}
\end{align}
The diagrammatic representation of this circuit is given in figure \ref{qc5}. The depth of this Quantum Circuit is,
\begin{align}
&2(2N_{aux}N_{htf}+2N_{o}N_{htf}+N_{v}N_{htf}+N_{htf}+N_{o}+2N_{o}N_{ttf}+N_{v}N_{ttf})log_2(1/\epsilon)
\end{align}

\subsubsection{Expression 6}
The first term of the doubles residual equation(\ref{r2}) is generated from the $QHP$ class of terms in the Bloch equation. It is generated from an ERI tensor slice along the direction of the Nth molecular orbital, $h^{2,(N)}_{aNij}$. The tensor factorized representation of this term is given by,
\begin{align}
&h^{2,(N)}_{aNij}\rightarrow \sum_{xpq}
X_{xp}Y_{ap}Z_{ip}X_{xq}Y_{Nq}Z_{jq}
\end{align}
The diagrammatic representation of this decomposition is shown in figure \ref{td6}. The contraction path to an irreducible representation can be written as,
\begin{align}
&\wick{xp,ap,ip,xq,\c1 q,\c1 jq}\quad \mathcal{O}(N_{htf}N_{o}) \nonumber\\
\rightarrow\ &\wick{\c1 jq, xp, ap, ip,\c1 xq}\quad \mathcal{O}(N_{htf}N_{aux}N_{o}) \nonumber\\
\rightarrow\ &\wick{\c1 jx,\c1 xp, ap, ip}\quad \mathcal{O}(N_{htf}N_{aux}N_{o}) \nonumber\\
\rightarrow\ &\wick{jp, ap, ip}
\end{align}
The total complexity of this calculation is given by,
\begin{align}
\mathcal{O}(2N_{htf}N_{aux}N_{o}+N_{htf}N_{o})
\end{align}
The Quantum Circuit representing this set of tensor operations is given by,
\begin{align}
&H_{\tilde{X}}H_{\tilde{P}}H_{\tilde{Q}}
V_{\tilde{X}\tilde{P},1}(X)V_{\tilde{A}\tilde{P},2}(Y)V_{\tilde{I}\tilde{P},3}(Z)V_{\tilde{X}\tilde{Q},4}(X)V_{\tilde{Q},5}(Y)V_{\tilde{J}\tilde{Q},6}(Z)
H_{\tilde{A}}H_{\tilde{I}}H_{\tilde{J}}H_{\tilde{X}}H_{\tilde{P}}H_{\tilde{Q}}
\end{align}
The diagrammatic representation of this circuit is given in figure \ref{qc6}. The depth of this Quantum Circuit is,
\begin{align}
&2(2N_{aux}N_{htf}+2N_{o}N_{htf}+N_{v}N_{htf}+N_{htf})log_2(1/\epsilon)
\end{align}

\subsubsection{Expression 7}
The second term of the doubles residual equation(\ref{r2}) is generated from contributions from both $\eta PHP$ and $QHQ\eta$ classes of terms in the Bloch equation. There will be contributions from two different slices of the ERI tensor, $h^{2,(N)}_{klij}$ and $h^{2,(N)}_{lkij}$. The tensor factorized representation of the $h^{2,(N)}_{klij}$ can be written as,
\begin{align}
&\sum_{kl}B^{1,(N)}_{klij}t^{2,(N)}_{akl}\rightarrow \sum_{kl}\sum_{xpqr}
X_{xp}Y_{kp}Z_{ip}X_{xq}Y_{lq}Z_{jq}T_{ar}U_{kr}V_{lr}
\end{align}
The diagrammatic representation of this decomposition is shown in figure \ref{td7}. The contraction path to an irreducible representation can be written as,
\begin{align}
&\wick{\c3 xp,\c1 kp,ip,\c3 xq,\c2 lq,jq,ar,\c1 kr,\c2 lr}\quad \mathcal{O}(N_{htf}N_{ttf}N_{aux}+2N_{htf}N_{ttf}N_{o})\nonumber\\
\rightarrow\ &\wick{pq, pr, qr, ip, jq, ar}
\end{align}
The total complexity of this calculation is given by,
\begin{align}
\mathcal{O}(N_{htf}N_{ttf}N_{aux}+2N_{htf}N_{ttf}N_{o})
\end{align}
The Quantum Circuit representing this set of tensor operations is given by,
\begin{align}
&H_{\tilde{X}}H_{\tilde{P}}H_{\tilde{Q}}H_{\tilde{R}}H_{\tilde{K}}H_{\tilde{L}}
V_{\tilde{X}\tilde{P},1}(X)V_{\tilde{K}\tilde{P},2}(Y)V_{\tilde{I}\tilde{P},3}(Z)V_{\tilde{X}\tilde{Q},4}(X)V_{\tilde{L}\tilde{Q},5}(Y)V_{\tilde{J}\tilde{Q},6}(Z)\nonumber\\
&V_{\tilde{A}\tilde{R},7}(T)V_{\tilde{K}\tilde{R},8}(U)V_{\tilde{L}\tilde{R},9}(V)H_{\tilde{A}}H_{\tilde{I}}H_{\tilde{J}}H_{\tilde{X}}H_{\tilde{P}}H_{\tilde{Q}}H_{\tilde{R}}H_{\tilde{K}}H_{\tilde{L}}
\end{align}
The diagrammatic representation of this circuit is given in figure \ref{qc7}. The depth of this Quantum Circuit is,
\begin{align}
&2(2N_{aux}N_{htf}+4N_{o}N_{htf}+2N_{o}N_{ttf}+N_{v}N_{ttf})log_2(1/\epsilon)
\end{align}

\subsubsection{Expression 8}
The third term of the doubles residual equation(\ref{r2}) is generated from both $\eta PHP$ and $QHQ\eta$ classes of terms in the Bloch equation. There will be contributions from both the fock and the ERI components of the second quantized Hamiltonian. The fock contribution can be represented as,
\begin{align}
\sum_{b}B^{2,(N)}_{ab}t^{2,(N)}_{bij}\rightarrow \sum_{b}\sum_{r}f^{(N)}_{ab}T_{br}U_{ir}V_{jr}
\end{align}
The contraction path towards an irreducible tensor decomposed form is given by,
\begin{align}
&\wick{\c1 ab,\c1 br,ir,jr}\quad \mathcal{O}(N_{ttf}N_{v}^2)\nonumber\\
\rightarrow\ &\wick{ar, ir, jr}
\end{align}
The more complex contributions arise from different slices of the ERI tensor, $h^{2,(N)}_{aNbN}$ and $h^{2,(N)}_{aNNb}$. The tensor representation of the $h^{2,(N)}_{aNNb}$ contribution is given by, 
\begin{align}
&\sum_{b}B^{2,(N)}_{ab}t^{2,(N)}_{bij}\rightarrow \sum_{b}\sum_{xpqr}
X_{xp}Y_{ap}Z_{Np}X_{xq}Y_{Nq}Z_{bq}T_{br}U_{ir}V_{jr}
\end{align}
The diagrammatic representation of this decomposition is shown in figure \ref{td8}. The contraction path to an irreducible representation can be written as,
\begin{align}
&\wick{xp,\c1 ap,\c1 p,xq,\c2 q,\c2 bq, br, ir, jr}\quad \mathcal{O}(2N_{htf}N_{v})\nonumber\\
\rightarrow\ &\wick{\c1 ap, \c2 bq, \c1 xp, \c2 xq, br, ir, jr}\quad \mathcal{O}(2N_{htf}N_{aux}N_v)\nonumber\\
\rightarrow\ &\wick{\c1 ax,\c1 bx, br, ir, jr}\quad \mathcal{O}(N_{aux}N_{v}^2)\nonumber\\
\rightarrow\ &\wick{\c1 ab,\c1 br, ir, jr}\quad \mathcal{O}(N_{ttf}N_{v}^2)\nonumber\\
\rightarrow\ &\wick{ar, ir, jr}
\end{align}
The total complexity of this calculation is given by,
\begin{align}
\mathcal{O}(2N_{htf}N_{aux}N_v+N_{aux}N_{v}^2+N_{ttf}N_{v}^2+2N_{htf}N_{v})
\end{align}
The Quantum Circuit representing this set of tensor operations is given by,
\begin{align}
&H_{\tilde{X}}H_{\tilde{P}}H_{\tilde{Q}}H_{\tilde{R}}H_{\tilde{B}}
V_{\tilde{X}\tilde{P},1}(X)V_{\tilde{A}\tilde{P},2}(Y)V_{\tilde{P},3}(Z)V_{\tilde{X}\tilde{Q},4}(X)V_{\tilde{Q},5}(Y)V_{\tilde{B}\tilde{Q},6}(Z)\nonumber\\
&V_{\tilde{B}\tilde{R},7}(T)V_{\tilde{I}\tilde{R},8}(U)V_{\tilde{J}\tilde{R},9}(V)H_{\tilde{A}}H_{\tilde{I}}H_{\tilde{J}}H_{\tilde{X}}H_{\tilde{P}}H_{\tilde{Q}}H_{\tilde{R}}H_{\tilde{B}}
\end{align}
The diagrammatic representation of this circuit is given in figure \ref{qc8}. The depth of this Quantum Circuit is,
\begin{align}
&2(2N_{aux}N_{htf}+2N_{v}N_{htf}+2N_{htf}+2N_{o}N_{ttf}+N_{v}N_{ttf})log_2(1/\epsilon)
\end{align}

\subsubsection{Expression 9}
The fourth term of the doubles residual equation(\ref{r2}) is generated from the $\eta PHQ \eta$ class of terms in the Bloch equation. It contains contributions from two different slices of the ERI tensor, $h^{2,(N)}_{akbN}$ and $h^{2,(N)}_{akNb}$. The tensor factorized representation of the $h^{2,(N)}_{akbN}$ contribution is given by,
\begin{align}
&\sum_{kb}B^{3,(N)}_{akb}t^{2,(N)}_{bij}t^{1,(N)}_{k}\rightarrow \sum_{kb}\sum_{xpqr}
X_{xp}Y_{ap}Z_{bp}X_{xq}Y_{kq}Z_{Nq}T_{br}U_{ir}V_{jr}t^{1,(N)}_{k}
\end{align}
The diagrammatic representation of this decomposition is shown in figure \ref{td9}. The contraction path to an irreducible representation can be written as,
\begin{align}
&\wick{xp, ap, \c1 bp, xq, \c2 kq, q,\c1 br, ir, jr, \c2 k}\quad \mathcal{O}(N_{htf}N_{ttf}N_v+N_{htf}N_o) \nonumber\\
\rightarrow\ &\wick{pr, \c1 q, xp, ap, xq, \c1 q, ir,jr}\quad \mathcal{O}(N_{v}) \nonumber \\
\rightarrow\ &\wick{\c1 q, pr, xp, ap,\c1 xq, ir, jr}\quad \mathcal{O}(N_{htf}N_{aux}) \nonumber\\
\rightarrow\ &\wick{\c1 x, pr,\c1 xp, ap, ir, jr}\quad \mathcal{O}(N_{htf}N_{aux}) \nonumber\\
\rightarrow\ &\wick{\c1 p, pr,\c1 ap, ir, jr}\quad \mathcal{O}(N_{htf}N_{v}) \nonumber\\
\rightarrow\ &\wick{\c1 ap,\c1 pr, ir, jr}\quad \mathcal{O}(N_{htf}N_{ttf}N_v) \nonumber\\
\rightarrow\ &\wick{ar, ir, jr}
\end{align}
The total complexity of this calculation is given by,
\begin{align}
\mathcal{O}(2N_{htf}N_{ttf}N_v+ 2N_{htf}N_{aux}+ N_{htf}N_{v}+N_{htf}N_{o}+N_v)    
\end{align}
The Quantum Circuit representing this set of tensor operations is given by,
\begin{align}
&H_{\tilde{X}}H_{\tilde{P}}H_{\tilde{Q}}H_{\tilde{R}}H_{\tilde{K}}H_{\tilde{B}}
V_{\tilde{X}\tilde{P},1}(X)V_{\tilde{A}\tilde{P},2}(Y)V_{\tilde{B}\tilde{P},3}(Z)V_{\tilde{X}\tilde{Q},4}(X)V_{\tilde{K}\tilde{Q},5}(Y)V_{\tilde{Q},6}(Z)V_{\tilde{B}\tilde{R},7}(T)\nonumber\\
&V_{\tilde{I}\tilde{R},8}(U)V_{\tilde{J}\tilde{R},9}(V)V_{\tilde{K},10}(T^{1,(N)})H_{\tilde{A}}H_{\tilde{I}}H_{\tilde{J}}H_{\tilde{X}}H_{\tilde{P}}H_{\tilde{Q}}H_{\tilde{R}}H_{\tilde{K}}H_{\tilde{B}}
\end{align}
The diagrammatic representation of this circuit is given in figure \ref{qc9}. The depth of this Quantum Circuit is,
\begin{align}
&2(2N_{aux}N_{htf}+2N_{v}N_{htf}+N_{htf}N_{o}+N_{htf}+N_{o}+2N_{o}N_{ttf}+N_{v}N_{ttf})log_2(1/\epsilon)
\end{align}

\subsubsection{Expression 10}
The fifth term of the doubles residual equation(\ref{r2}) is generated from the $\eta PHQ \eta$ class of terms in the Bloch equation. It contains contributions from both the fock and the ERI components of the Hamiltonian. The fock contribution is directly in its tensor decomposed form given by,
\begin{align}
    &B^{4,(N)}_{a}t^{1,(N)}_{i}t^{1,(N)}_{j}\rightarrow f^{(N)}_{aN}t^{1,(N)}_{i}t^{1,(N)}_{j}
\end{align}
A more complex ERI contribution comes from two different slices of ERI, $h^{2,(N)}_{NaNN}$ and $h^{2,(N)}_{aNNN}$, The $h^{2,(N)}_{aNNN}$ contribution is given by,
\begin{align}
&B^{4,(N)}_{a}t^{1,(N)}_{i}t^{1,(N)}_{j}\rightarrow \sum_{xpq}
X_{xp}Y_{ap}Z_{Np}X_{xq}Y_{Nq}Z_{Nq}t^{1,(N)}_{i}t^{1,(N)}_{j}
\end{align}
The diagrammatic representation of this decomposition is shown in figure \ref{td10}. The contraction path to an irreducible representation can be written as,
\begin{align}
&\wick{xp,\c1 ap,\c1 p,xq,\c2 q,\c2 q,i,j}\quad \mathcal{O}(N_{htf}N_{v}+N_{htf}) \nonumber \\
\rightarrow\ &\wick{ap, \c1 q, xp,\c1 xq, i ,j}\quad \mathcal{O}(N_{htf}N_{aux}) \nonumber\\
\rightarrow\ &\wick{\c1 x,ap,\c1 xp,i,j}\quad \mathcal{O}(N_{htf}N_{aux}) \nonumber\\
\rightarrow\ &\wick{\c1 p,\c1 ap, i, j}\quad \mathcal{O}(N_{htf}N_{v}) \nonumber\\
\rightarrow\ &\wick{a, i, j}
\end{align}
The total complexity of this calculation is given by,
\begin{align}
\mathcal{O}(2N_{htf}N_{v}+2N_{htf}N_{aux}+N_{htf})
\end{align}
The Quantum Circuit representing this set of tensor operations is given by,
\begin{align}
&H_{\tilde{X}}H_{\tilde{P}}H_{\tilde{Q}}
V_{\tilde{X}\tilde{P},1}(X)V_{\tilde{A}\tilde{P},2}(Y)V_{\tilde{P},3}(Z)V_{\tilde{X}\tilde{Q},4}(X)V_{\tilde{Q},5}(Y)V_{\tilde{Q},6}(Z)V_{\tilde{I},7}(T^{1,(N)})V_{\tilde{J},8}(T^{1,(N)})\nonumber\\
&H_{\tilde{A}}H_{\tilde{I}}H_{\tilde{J}}H_{\tilde{X}}H_{\tilde{P}}H_{\tilde{Q}}
\end{align}
The diagrammatic representation of this circuit is given in figure \ref{qc10}. The depth of this Quantum Circuit is,
\begin{align}
&2(2N_{aux}N_{htf}+N_{v}N_{htf}+2N_{htf}+2N_{o})log_2(1/\epsilon)
\end{align}

\subsubsection{Expression 11}
\begin{figure}[h!]
\begin{subfigure}{0.4\textwidth}
\centering
\includegraphics[width=\textwidth]{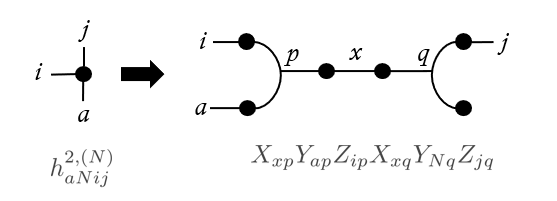}
\caption{Expression 6}
\label{td6}
\end{subfigure}
\hfill
\begin{subfigure}{0.4\textwidth}
\centering
\includegraphics[width=\textwidth]{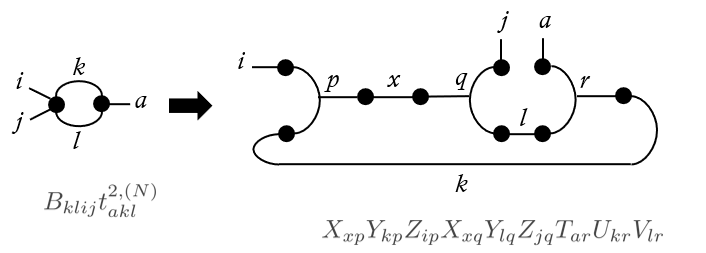}
\caption{Expression 7}
\label{td7}
\end{subfigure}

\begin{subfigure}[b]{0.4\textwidth}
    \centering
    \includegraphics[width=\textwidth]{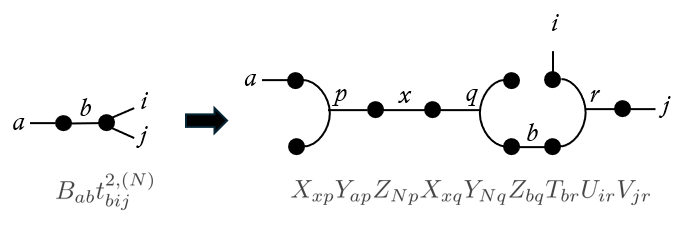}
    \caption{Expression 8}
    \label{td8}
\end{subfigure}
\hfill
\begin{subfigure}[b]{0.4\textwidth}
    \centering
    \includegraphics[width=\textwidth]{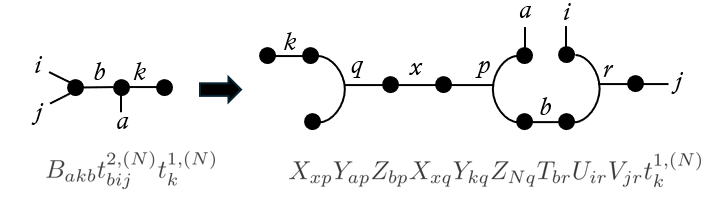}
    \caption{Expression 9}
    \label{td9}
\end{subfigure}

\begin{subfigure}[b]{0.4\textwidth}
    \centering
    \includegraphics[width=\textwidth]{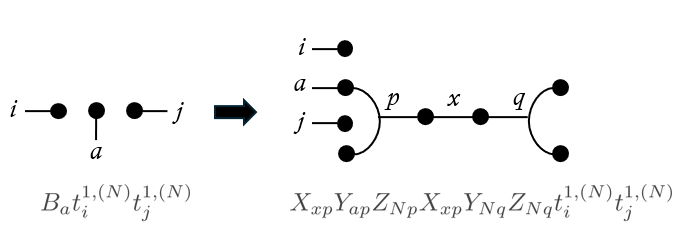}
    \caption{Expression 10}
    \label{td10}
\end{subfigure}
\hfill
\begin{subfigure}[b]{0.4\textwidth}
    \centering
    \includegraphics[width=\textwidth]{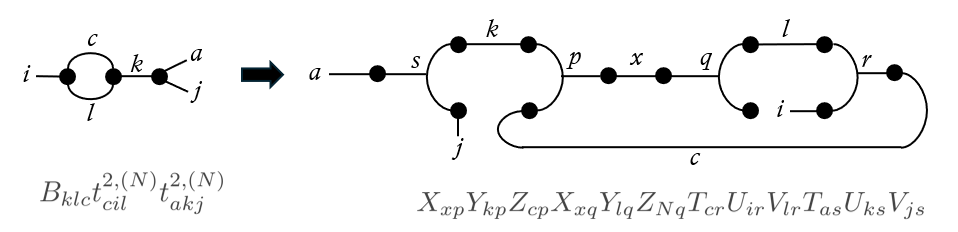}
    \caption{Expression 11}
    \label{td11}
\end{subfigure}
\caption{Tensor Factorized Representation of terms in Doubles Residual Equation}\label{TensorOps2}
\end{figure}

\begin{figure}[H]
\makebox[\textwidth]
{ 
\begin{subfigure}{0.45\textwidth}
\centering
\includegraphics[width=\textwidth]{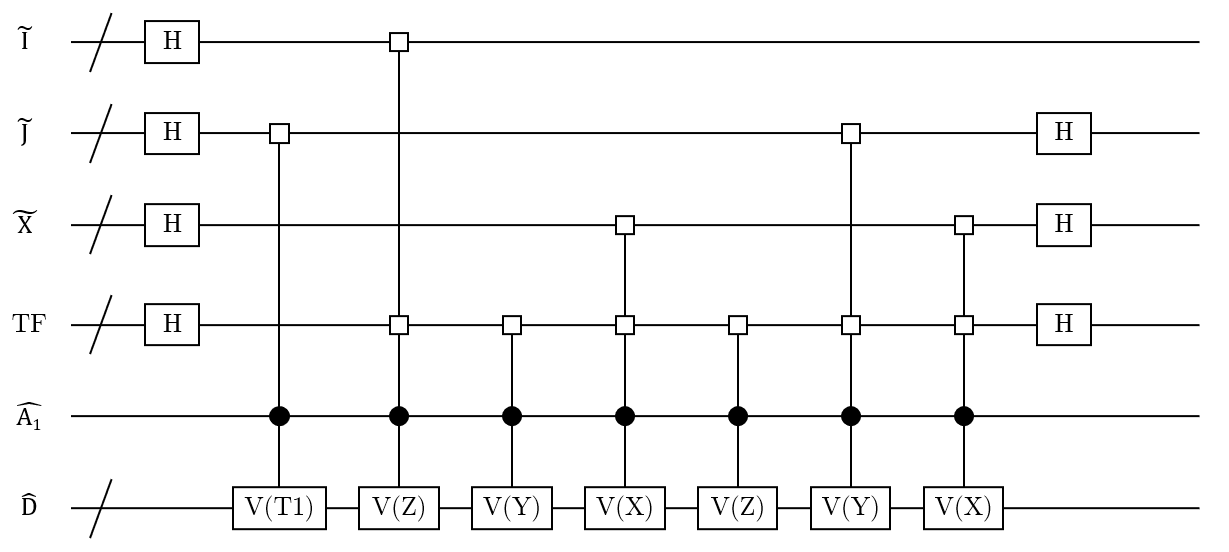}
\caption{Circuit for Expression 2}
\label{qc2}
\end{subfigure}
\hfill

\par
\begin{subfigure}{0.45\textwidth}
\centering
\includegraphics[width=\textwidth]{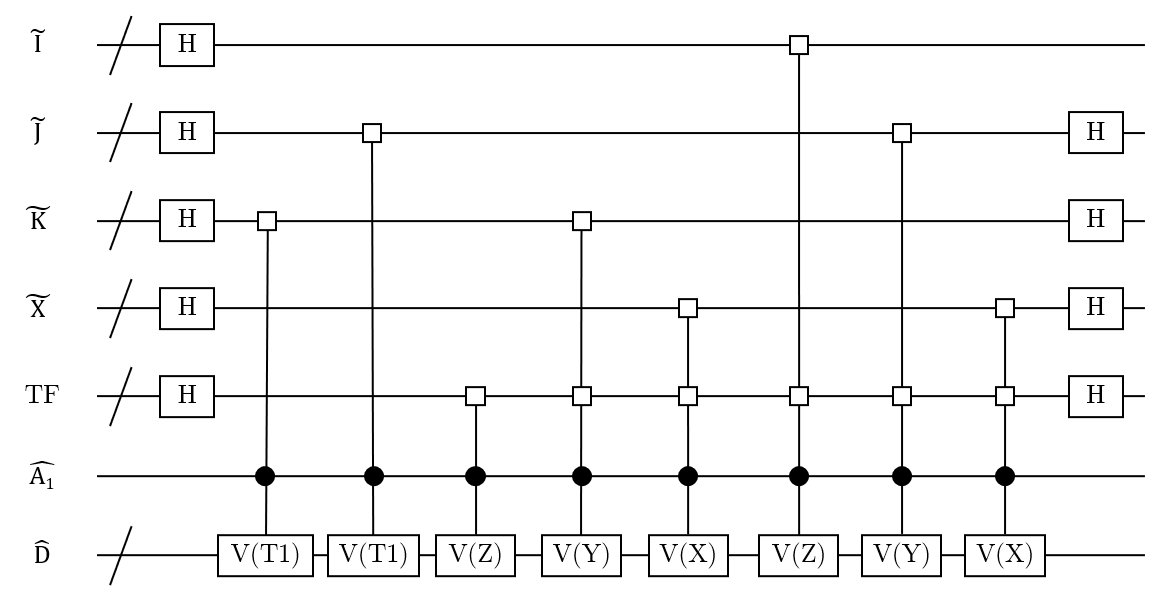}
\caption{Circuit for Expression 3}
\label{qc3}
\end{subfigure}
\hfill
}

\par
\makebox[\textwidth]
{
\begin{subfigure}{0.45\textwidth}
\centering
\includegraphics[width=\textwidth]{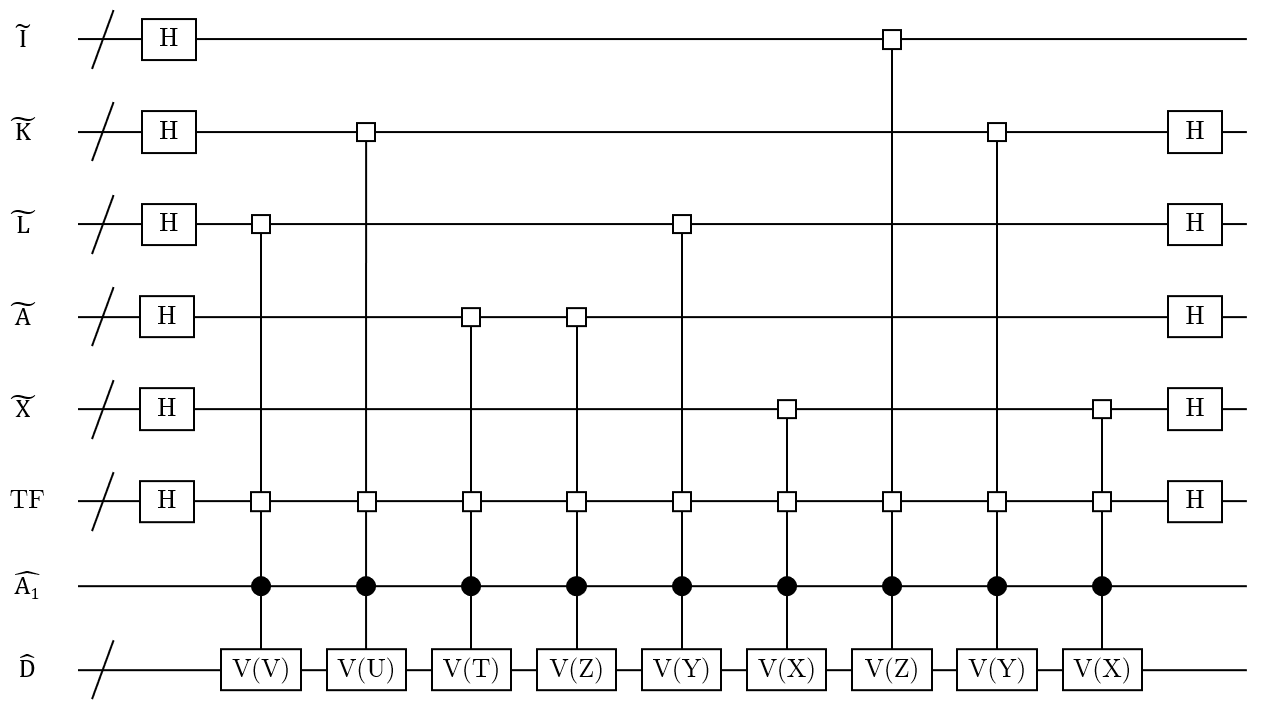}
\caption{Circuit for Expression 4}
\label{qc4}
\end{subfigure}
\hfill

\begin{subfigure}{0.45\textwidth}
\centering
\includegraphics[width=\textwidth]{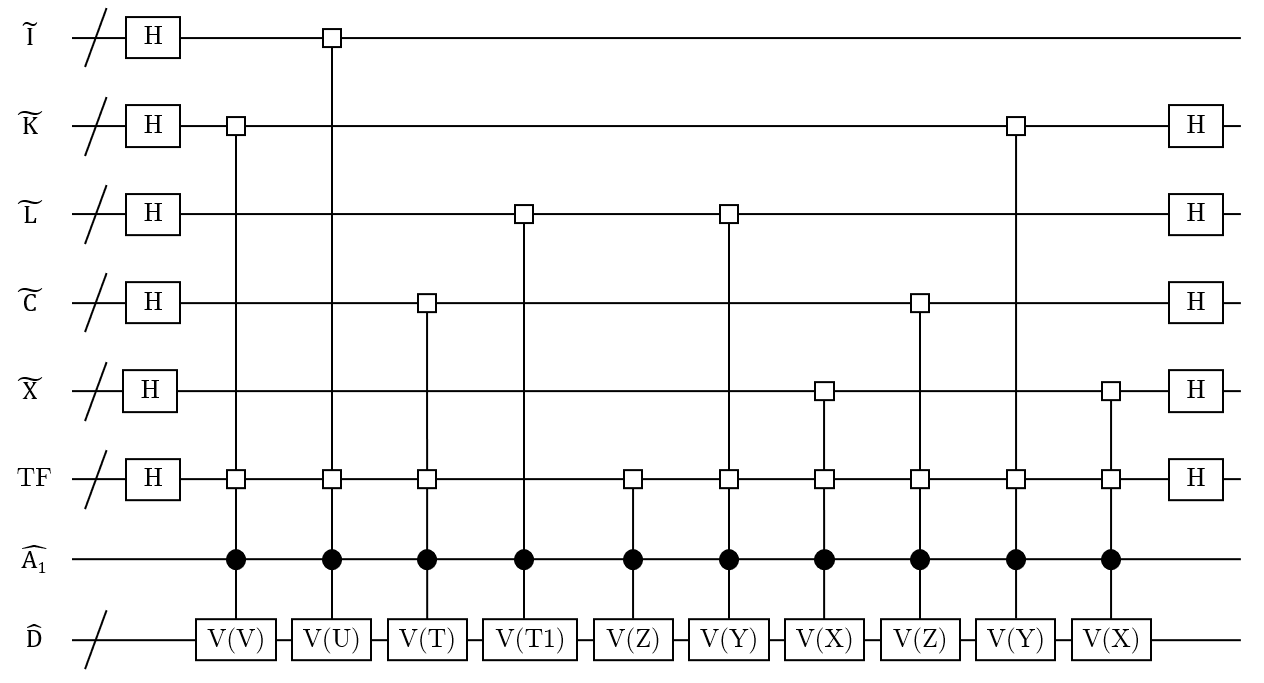}
\caption{Circuit for Expression 5}
\label{qc5}
\end{subfigure}
\hfill
}

\par
\makebox[\textwidth]
{
\begin{subfigure}{0.45\textwidth}
\centering
\includegraphics[width=\textwidth]{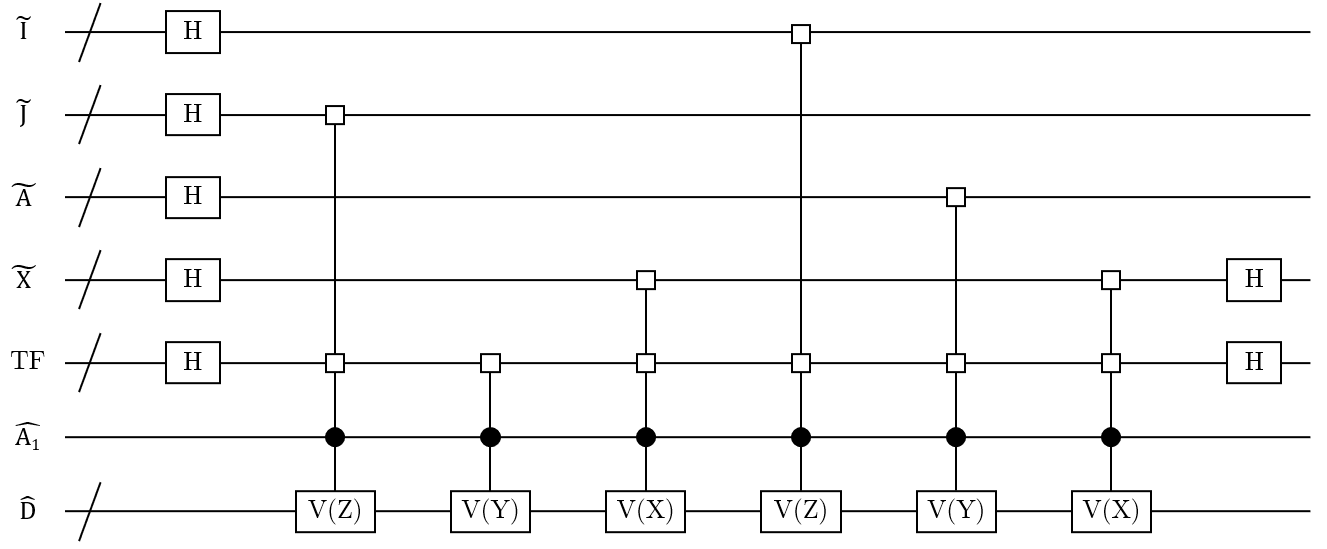}
\caption{Circuit for Expression 6}
\label{qc6}
\end{subfigure}
\hfill

\begin{subfigure}{0.45\textwidth}
\centering
\includegraphics[width=\textwidth]{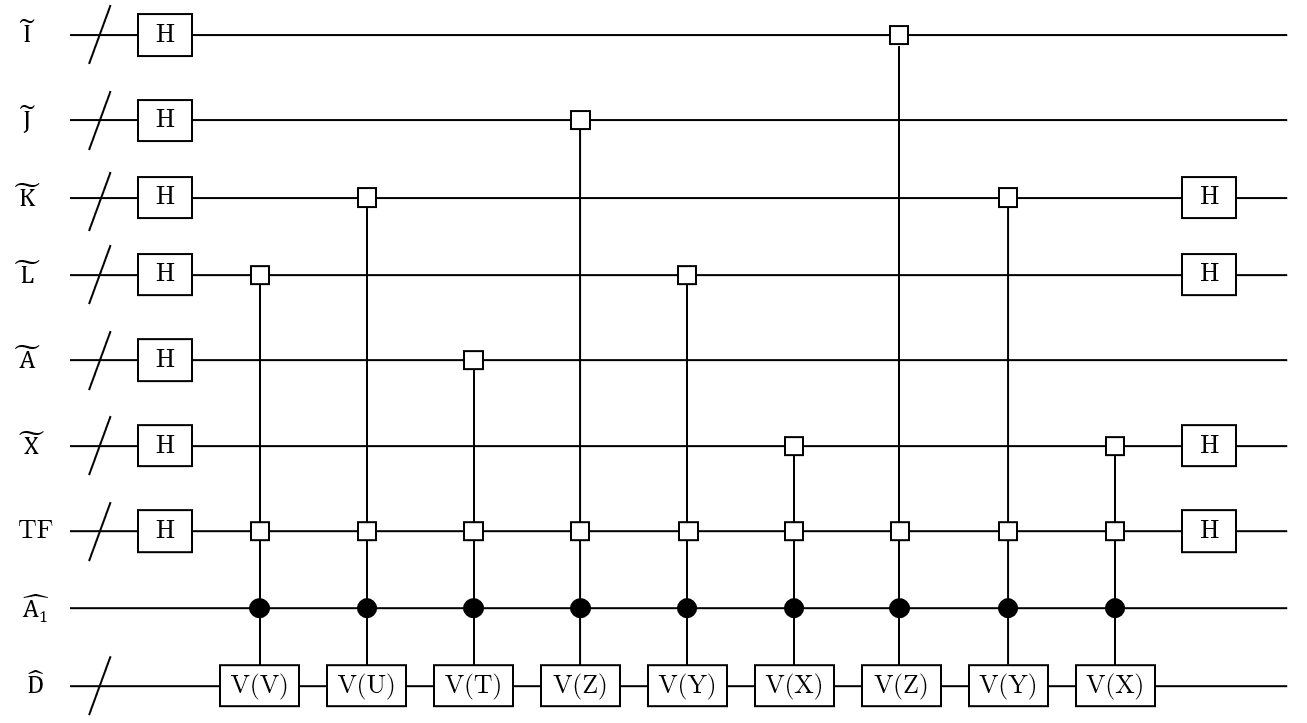}
\caption{Circuit for Expression 7}
\label{qc7}
\end{subfigure}
\hfill
}

\par
\makebox[\textwidth]
{
\begin{subfigure}{0.45\textwidth}
\centering
\includegraphics[width=\textwidth]{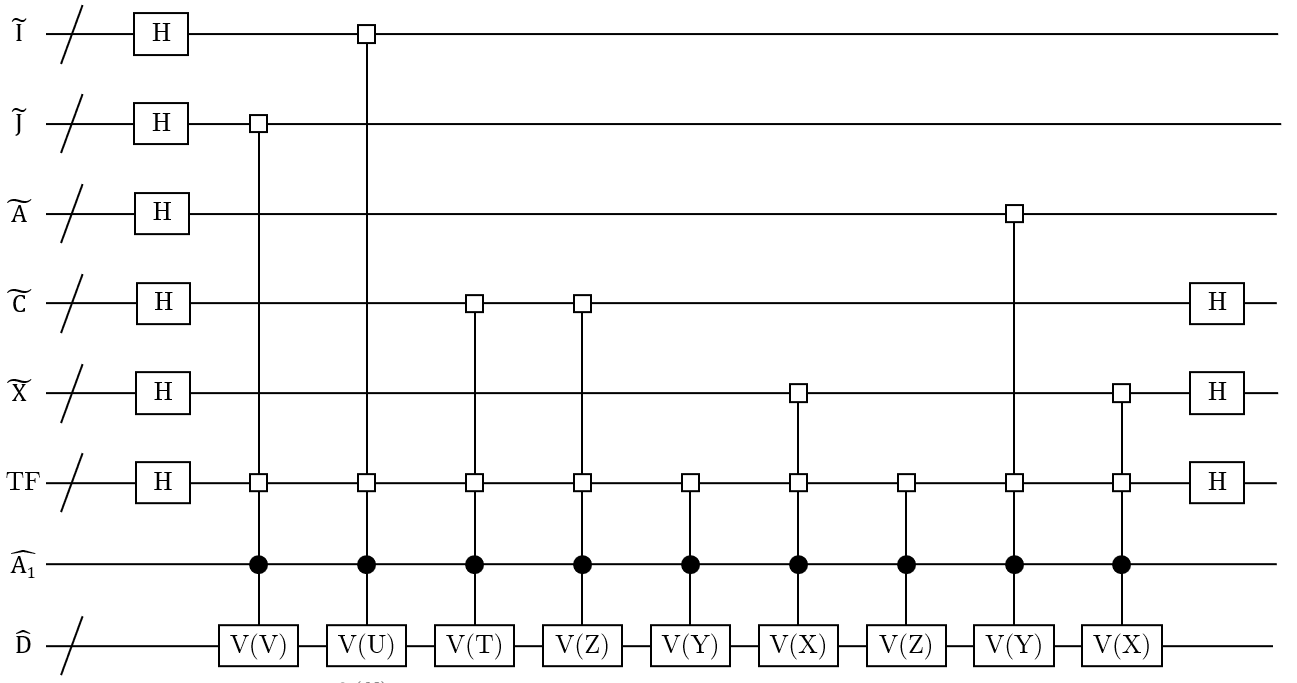}
\caption{Circuit for Expression 8}
\label{qc8}
\end{subfigure}
\hfill

\begin{subfigure}{0.45\textwidth}
\centering
\includegraphics[width=\textwidth]{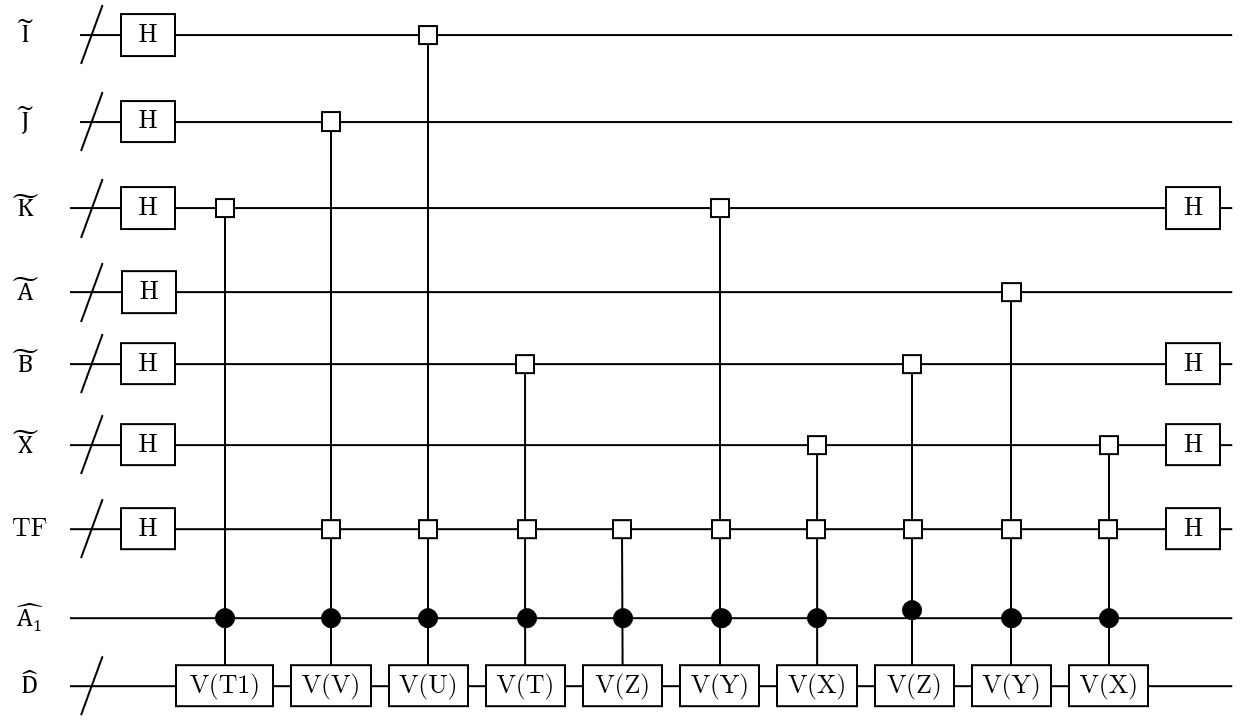}
\caption{Circuit for Expression 9}
\label{qc9}
\end{subfigure}
\hfill
}

\par
\makebox[\textwidth]
{
\begin{subfigure}{0.45\textwidth}
\centering
\includegraphics[width=\textwidth]{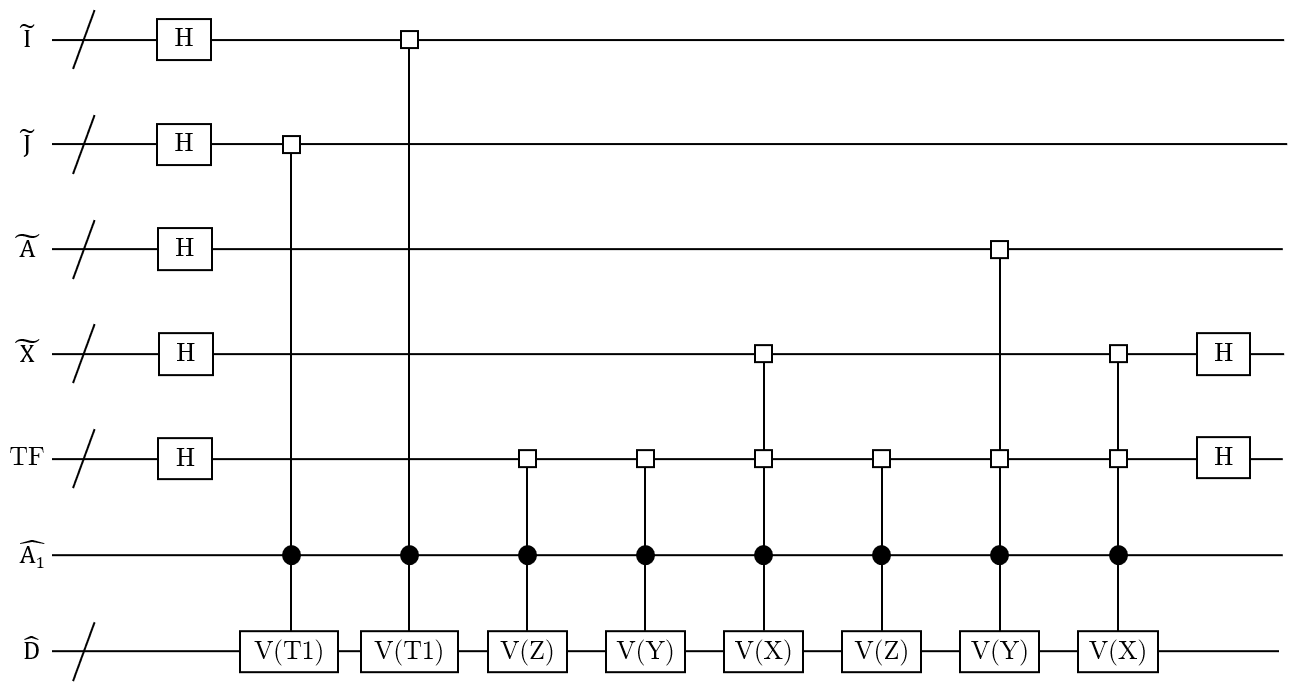}
\caption{Circuit for Expression 10}
\label{qc10}
\end{subfigure}
\hfill

\begin{subfigure}{0.45\textwidth}
\centering
\includegraphics[width=\textwidth]{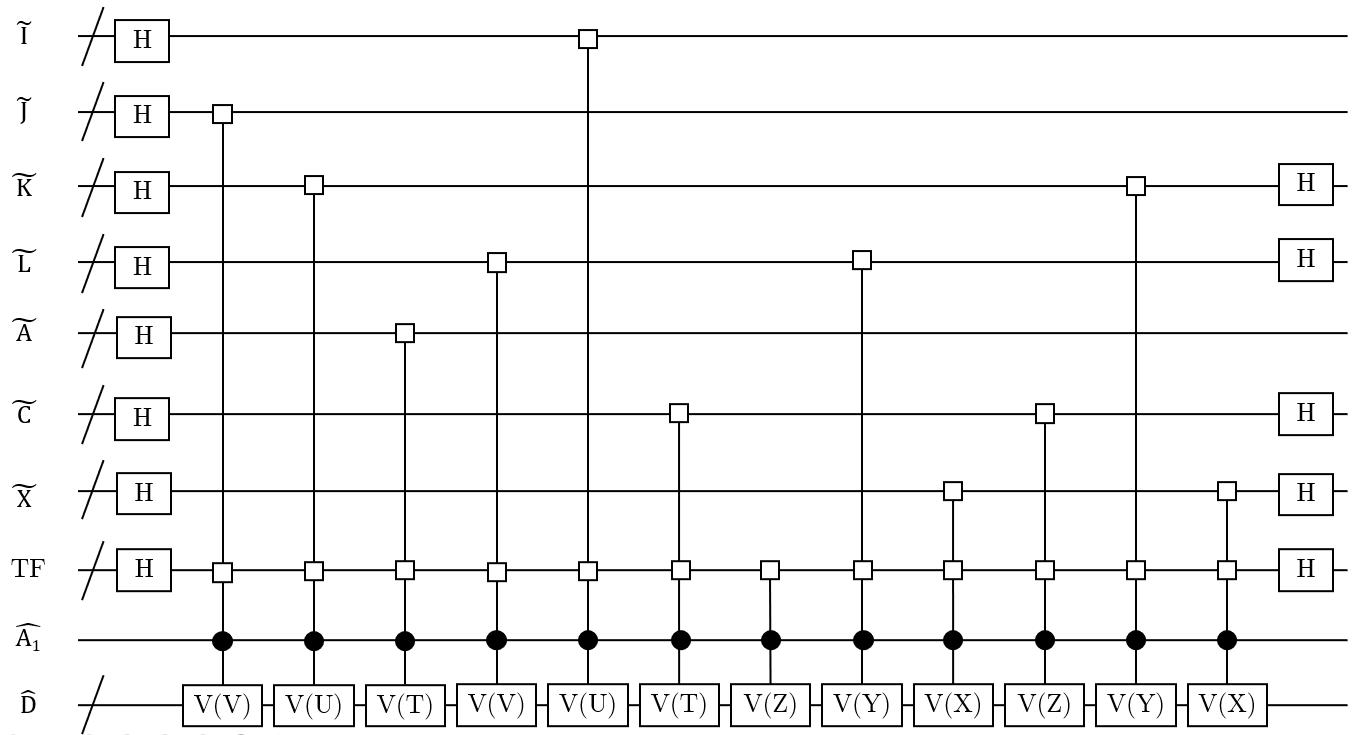}
\caption{Circuit for Expression 11}
\label{qc11}
\end{subfigure}
\hfill
}
\caption{Quantum Circuit Representations for Residual Equation}\label{Circuits}
\end{figure}

The sixth term of the doubles residual equation(\ref{r2}) is generated from $\eta PHQ \eta$ class of terms in the Bloch Equation. It contains contributions from two different slices of ERI, $h^{2,(N)}_{klcN}$ and $h^{2,(N)}_{klNc}$. The tensor factorized form of the $h^{2,(N)}_{klcN}$ contribution is given by,
\begin{align}
&\sum_{klc}B^{5,(N)}_{klc}t^{2,(N)}_{cil}t^{2,(N)}_{akj}\rightarrow \sum_{klc}\sum_{xpqrs}
X_{xp}Y_{kp}Z_{cp}X_{xq}Y_{lq}Z_{Nq}T_{cr}U_{ir}V_{lr}T_{as}U_{ks}V_{js}
\end{align}
The diagrammatic representation of this decomposition is shown in figure \ref{td11}. The contraction path to an irreducible representation can be written as,
\begin{align}
    &\wick{xp,\c1 kp,\c3 cp,xq,\c2 lq,\c2 q,\c3 cr,ir,lr,as,\c1 ks,js} \quad \mathcal{O}(N_{htf}N_{ttf}N_{o} + N_{htf}N_{ttf}N_{v} + N_{htf}N_{o}) \nonumber\\
    \rightarrow\ &\wick{ps, pr, \c1 lq,  xp, xq, ir,\c1 lr, as, js}\quad \mathcal{O}(N_{htf}N_{ttf}N_{o}) \nonumber\\
    \rightarrow\ &\wick{\c1 qr, ps, pr,xp,\c1 xq, ir, as, js}\quad \mathcal{O}(N_{htf}N_{ttf}N_{aux}) \nonumber\\
    \rightarrow\ &\wick{\c1 xr, ps, pr,\c1 xp, ir, as, js}\quad \mathcal{O}(N_{htf}N_{ttf}N_{aux}) \nonumber\\
    \rightarrow\ &\wick{\c1 pr,ps,\c1 pr,ir,as,js}\quad \mathcal{O}(N_{htf}N_{ttf}) \nonumber\\
    \rightarrow\ &\wick{\c1 pr,ps,\c1 ir, as, js}\quad \mathcal{O}(N_{htf}N_{ttf}N_{o}) \nonumber\\
    \rightarrow\ &\wick{\c1 pi,\c1 ps, as, js}\quad \mathcal{O}(N_{htf}N_{ttf}N_{o}) \nonumber\\
\rightarrow\ &\wick{is, as, js}
\end{align}
The total complexity of this calculation is given by,
\begin{align}
\mathcal{O}(2N_{htf}N_{ttf}N_{aux}+N_{htf}N_{ttf}N_{v}+4N_{htf}N_{ttf}N_{o}+N_{htf}N_{ttf}+N_{htf}N_{o})
\end{align}
The Quantum Circuit representing this set of tensor operations is given by,
\begin{align}
&H_{\tilde{X}}H_{\tilde{P}}H_{\tilde{Q}}H_{\tilde{R}}H_{\tilde{K}}H_{\tilde{L}}H_{\tilde{C}}
V_{\tilde{X}\tilde{P},1}(X)V_{\tilde{K}\tilde{P},2}(Y)V_{\tilde{C}\tilde{P},3}(Z)V_{\tilde{X}\tilde{Q},4}(X)V_{\tilde{L}\tilde{Q},5}(Y)V_{\tilde{Q},6}(Z)V_{\tilde{C}\tilde{R},7}(T)\nonumber\\
&V_{\tilde{I}\tilde{R},8}(U)V_{\tilde{L}\tilde{R},9}(V)V_{\tilde{A}\tilde{S},10}(T)V_{\tilde{K}\tilde{S},11}(U)V_{\tilde{J}\tilde{S},12}(V)H_{\tilde{A}}H_{\tilde{I}}H_{\tilde{J}}H_{\tilde{X}}H_{\tilde{P}}H_{\tilde{Q}}H_{\tilde{R}}H_{\tilde{K}}H_{\tilde{L}}H_{\tilde{C}}
\end{align}
The diagrammatic representation of this circuit is given in figure \ref{qc11}. The depth of this Quantum Circuit is,
\begin{align}
&2(2N_{aux}N_{htf}+2N_{o}N_{htf}+N_{v}N_{htf}+N_{htf}+4N_{o}N_{ttf}+2N_{v}N_{ttf})log_2(1/\epsilon)
\end{align}

\newpage
\begin{figure}[h!]
\hfill
\begin{subfigure}{0.4\textwidth}
\centering
\includegraphics[width=\textwidth]{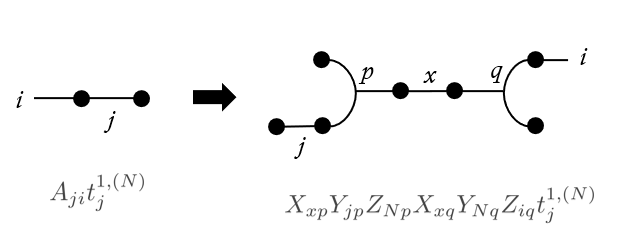}
\caption{Expression 2}
\label{td2}
\end{subfigure}
\hfill
\begin{subfigure}{0.4\textwidth}
    \centering
    \includegraphics[width=\textwidth]{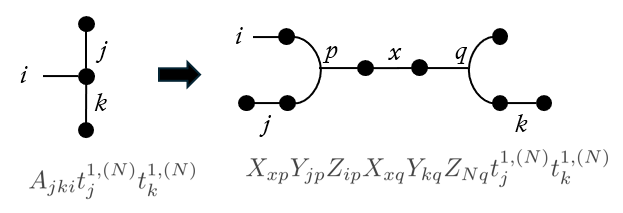}
    \caption{Expression 3}
    \label{td3}
\end{subfigure}

\begin{subfigure}{0.4\textwidth}
    \centering
    \includegraphics[width=\textwidth]{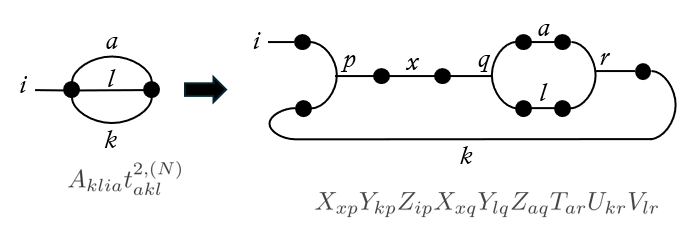}
    \caption{Expression 4}
    \label{td4}
\end{subfigure}
\hfill
\begin{subfigure}{0.4\textwidth}
    \centering
    \includegraphics[width=\textwidth]{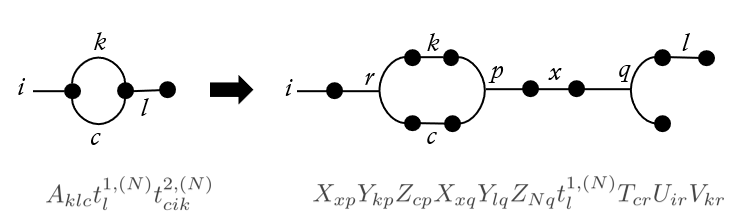}
    \caption{Expression 5}
    \label{td5}
\end{subfigure}
\caption{Tensor Factorized Representation of terms in Singles Residual Equation}\label{TensorOps1}
\end{figure}

\newpage

\subsection{Error from Quantum Phase Estimation and Downfolding}
When considering state of the art methods in Quantum Phase Estimation (QPE) based algorithms for quantum chemistry, we refer to \cite{lee2021even}. The sources of error in this approach are largely due to two sources. 
\begin{enumerate}
    \item Discretization and Preparation of the Hamiltonian ($\epsilon_{THC}$)
    \item Phase Estimation ($\epsilon_{PEA}$)
\end{enumerate}
The former arises out of approximating given's rotations, state-preparation, tensor hyper-contraction, and the choice of basis functions used in approximating the Hamiltonian, The latter is due to the process of phase estimation. We can write this expression as 
\begin{align}
&\epsilon = \epsilon_{PEA}+\epsilon_{THC} + \epsilon_{basis}\\
&\epsilon_{THC} = \epsilon_{rot}+\epsilon_{TF}
\end{align}
The errors in our approach are largely influenced by the approximation of rotations used in state preparation, and the error in tensor factorization.

\newpage
\subsection{Benchmarking}
\begin{table}[h!]
\caption{Citric Acid Conformer Ranking with downfolding- Single Point energies computed from downfolding with 6-31g basis incorporating singles and doubles clusters starting from single reference state. In 6-31g basis the Cholesky is of dimension $N_{tf}=2000$ tensor factors, $N_{aux}=1094$ density fitting vectors, $N_{o}=50$ occupied orbitals. $N_{v}=83$ virtual orbitals. The rankings of conformers obtained are as follows- SCF:(4,8,1,10,3,2,5,7,9,6), RMP2:(4,8,10,1,3,2,5,7,9,6), CCSD:(4,8,10,1,3,2,5,7,9,6), DOWNFOLDING:(4,8,1,10,2,3,5,7,9,6)}
\label{tab:citric-631g}
\begin{tabular}{c|cccc|}
\cline{2-5}
                                             & \multicolumn{4}{c|}{Level of theory (basis: 6-31g)}                                                                      \\ \hline
\multicolumn{1}{|c|}{Citric Acid Conformers} & \multicolumn{1}{c|}{SCF}          & \multicolumn{1}{c|}{RMP2}         & \multicolumn{1}{c|}{CCSD}         & DOWNFOLDING  \\ \hline
\multicolumn{1}{|c|}{Conformer 1}            & \multicolumn{1}{c|}{-755.5702231} & \multicolumn{1}{c|}{-756.9940412} & \multicolumn{1}{c|}{-757.0249044} & -757.0179777 \\ \hline
\multicolumn{1}{|c|}{Conformer 2}            & \multicolumn{1}{c|}{-755.5555467} & \multicolumn{1}{c|}{-756.9852885} & \multicolumn{1}{c|}{-757.0148742} & -757.0054337 \\ \hline
\multicolumn{1}{|c|}{Conformer 3}            & \multicolumn{1}{c|}{-755.5602742} & \multicolumn{1}{c|}{-756.9853357} & \multicolumn{1}{c|}{-757.0160611} & -757.0019248 \\ \hline
\multicolumn{1}{|c|}{Conformer 4}            & \multicolumn{1}{c|}{-755.5803311} & \multicolumn{1}{c|}{-757.0023419} & \multicolumn{1}{c|}{-757.0336214} & -757.0219172 \\ \hline
\multicolumn{1}{|c|}{Conformer 5}            & \multicolumn{1}{c|}{-755.5551208} & \multicolumn{1}{c|}{-756.9811494} & \multicolumn{1}{c|}{-757.0114966} & -757.0002892 \\ \hline
\multicolumn{1}{|c|}{Conformer 6}            & \multicolumn{1}{c|}{-755.5355274} & \multicolumn{1}{c|}{-756.96786}   & \multicolumn{1}{c|}{-756.9965834} & -756.9900533 \\ \hline
\multicolumn{1}{|c|}{Conformer 7}            & \multicolumn{1}{c|}{-755.5495943} & \multicolumn{1}{c|}{-756.9768728} & \multicolumn{1}{c|}{-757.0071598} & -756.9979194 \\ \hline
\multicolumn{1}{|c|}{Conformer 8}            & \multicolumn{1}{c|}{-755.5750486} & \multicolumn{1}{c|}{-756.998113}  & \multicolumn{1}{c|}{-757.0292071} & -757.0207515 \\ \hline
\multicolumn{1}{|c|}{Conformer 9}            & \multicolumn{1}{c|}{-755.5432721} & \multicolumn{1}{c|}{-756.9720897} & \multicolumn{1}{c|}{-757.0014425} & -756.9920599 \\ \hline
\multicolumn{1}{|c|}{Conformer 10}           & \multicolumn{1}{c|}{-755.570067}  & \multicolumn{1}{c|}{-756.9956409} & \multicolumn{1}{c|}{-757.0261098} & -757.0162994 \\ \hline
\end{tabular}
\end{table}

\begin{table}[h!]
\caption{Citric Acid Conformer Ranking with downfolding- Single Point energies computed from downfolding with def2-svp basis incorporating singles and doubles clusters starting from single reference state. In def2-svp basis the Cholesky is of dimension $N_{tf}=2200$ tensor factors, $N_{aux}=1133$ density fitting vectors, $N_{o}=50$ occupied orbitals. $N_{v}=172$ virtual orbitals.The rankings of conformers obtained are as follows- SCF:(4,8,1,10,3,2,5,7,9,6), RMP2:(4,8,10,1,2,3,5,7,9,6), CCSD:(4,8,10,1,3,2,5,7,9,6), DOWNFOLDING:(4,8,10,1,3,2,5,7,9,6)} 
\label{tab:citric-def2svp}
\begin{tabular}{c|cccc|}
\cline{2-5}
                                             & \multicolumn{4}{c|}{Level of theory (basis: def2-svp)}                                                                   \\ \hline
\multicolumn{1}{|c|}{Citric Acid Conformers} & \multicolumn{1}{c|}{SCF}          & \multicolumn{1}{c|}{RMP2}         & \multicolumn{1}{c|}{CCSD}         & DOWNFOLDING  \\ \hline
\multicolumn{1}{|c|}{Conformer 1}            & \multicolumn{1}{c|}{-755.3570971} & \multicolumn{1}{c|}{-757.4731958} & \multicolumn{1}{c|}{-757.5214172} & -757.4868588 \\ \hline
\multicolumn{1}{|c|}{Conformer 2}            & \multicolumn{1}{c|}{-755.342948}  & \multicolumn{1}{c|}{-757.4673823} & \multicolumn{1}{c|}{-757.5135503} & -757.4779496 \\ \hline
\multicolumn{1}{|c|}{Conformer 3}            & \multicolumn{1}{c|}{-755.3480262} & \multicolumn{1}{c|}{-757.4673686} & \multicolumn{1}{c|}{-757.5148784} & -757.4795734 \\ \hline
\multicolumn{1}{|c|}{Conformer 4}            & \multicolumn{1}{c|}{-755.3655959} & \multicolumn{1}{c|}{-757.480623}  & \multicolumn{1}{c|}{-757.5290297} & -757.4909161 \\ \hline
\multicolumn{1}{|c|}{Conformer 5}            & \multicolumn{1}{c|}{-755.3409242} & \multicolumn{1}{c|}{-757.4621949} & \multicolumn{1}{c|}{-757.5091232} & -757.4722697 \\ \hline
\multicolumn{1}{|c|}{Conformer 6}            & \multicolumn{1}{c|}{-755.3199347} & \multicolumn{1}{c|}{-757.4487587} & \multicolumn{1}{c|}{-757.4936955} & -757.4620916 \\ \hline
\multicolumn{1}{|c|}{Conformer 7}            & \multicolumn{1}{c|}{-755.3385779} & \multicolumn{1}{c|}{-757.458564}  & \multicolumn{1}{c|}{-757.5060326} & -757.4705519 \\ \hline
\multicolumn{1}{|c|}{Conformer 8}            & \multicolumn{1}{c|}{-755.3613695} & \multicolumn{1}{c|}{-757.4776596} & \multicolumn{1}{c|}{-757.525767}  & -757.4902219 \\ \hline
\multicolumn{1}{|c|}{Conformer 9}            & \multicolumn{1}{c|}{-755.3291166} & \multicolumn{1}{c|}{-757.4525266} & \multicolumn{1}{c|}{-757.4987229} & -757.4662678 \\ \hline
\multicolumn{1}{|c|}{Conformer 10}           & \multicolumn{1}{c|}{-755.3568708} & \multicolumn{1}{c|}{-757.475352}  & \multicolumn{1}{c|}{-757.5228783} & -757.4890177 \\ \hline
\end{tabular}
\end{table}

\newpage
\begin{table}[h!]
\caption{Citric Acid Conformer Ranking with downfolding- Single Point energies computed from downfolding with ccpvdz basis incorporating singles and doubles clusters starting from single reference state. In ccpvdz basis the Cholesky is of dimension $N_{tf}=2200$ tensor factors, $N_{aux}=1094$ density fitting vectors, $N_{o}=50$ occupied orbitals. $N_{v}=172$ virtual orbitals.The rankings of conformers obtained are as follows- SCF:(4,8,1,10,3,2,5,7,9,6), RMP2:(4,8,10,1,2,3,5,7,9,6), CCSD: (4,8,10,1,3,2,5,7,9,6), DOWNFOLDING:(4,8,10,1,3,2,5,7,9,6)}
\label{tab:citric-ccpvdz}
\begin{tabular}{c|cccc|}
\cline{2-5}
                                             & \multicolumn{4}{c|}{Level of theory (basis: ccpvdz)}                                                                     \\ \hline
\multicolumn{1}{|c|}{Citric Acid Conformers} & \multicolumn{1}{c|}{SCF}          & \multicolumn{1}{c|}{RMP2}         & \multicolumn{1}{c|}{CCSD}         & DOWNFOLDING  \\ \hline
\multicolumn{1}{|c|}{Conformer 1}            & \multicolumn{1}{c|}{-756.0040102} & \multicolumn{1}{c|}{-758.1216979} & \multicolumn{1}{c|}{-758.169231}  & -758.1288886 \\ \hline
\multicolumn{1}{|c|}{Conformer 2}            & \multicolumn{1}{c|}{-755.9907601} & \multicolumn{1}{c|}{-758.1162632} & \multicolumn{1}{c|}{-758.1618224} & -758.1221673 \\ \hline
\multicolumn{1}{|c|}{Conformer 3}            & \multicolumn{1}{c|}{-755.9950628} & \multicolumn{1}{c|}{-758.1156926} & \multicolumn{1}{c|}{-758.1625198} & -758.123187  \\ \hline
\multicolumn{1}{|c|}{Conformer 4}            & \multicolumn{1}{c|}{-756.0121186} & \multicolumn{1}{c|}{-758.1287919} & \multicolumn{1}{c|}{-758.1764974} & -758.1352263 \\ \hline
\multicolumn{1}{|c|}{Conformer 5}            & \multicolumn{1}{c|}{-755.9881586} & \multicolumn{1}{c|}{-758.1109601} & \multicolumn{1}{c|}{-758.15714}   & -758.118457  \\ \hline
\multicolumn{1}{|c|}{Conformer 6}            & \multicolumn{1}{c|}{-755.9671993} & \multicolumn{1}{c|}{-758.0970568} & \multicolumn{1}{c|}{-758.1413555} & -758.1042623 \\ \hline
\multicolumn{1}{|c|}{Conformer 7}            & \multicolumn{1}{c|}{-755.9859496} & \multicolumn{1}{c|}{-758.1075671} & \multicolumn{1}{c|}{-758.1543593} & -758.1138361 \\ \hline
\multicolumn{1}{|c|}{Conformer 8}            & \multicolumn{1}{c|}{-756.0080427} & \multicolumn{1}{c|}{-758.1258409} & \multicolumn{1}{c|}{-758.1732388} & -758.134931  \\ \hline
\multicolumn{1}{|c|}{Conformer 9}            & \multicolumn{1}{c|}{-755.9765094} & \multicolumn{1}{c|}{-758.101256}  & \multicolumn{1}{c|}{-758.1468017} & -758.1088486 \\ \hline
\multicolumn{1}{|c|}{Conformer 10}           & \multicolumn{1}{c|}{-756.0035965} & \multicolumn{1}{c|}{-758.1233642} & \multicolumn{1}{c|}{-758.1702097} & -758.1323347 \\ \hline
\end{tabular}
\end{table}

\begin{table}[h!]
\caption{Aspirin Conformer Ranking with downfolding- Single Point energies computed from downfolding with 6-31g basis incorporating singles and doubles clusters starting from single reference state. In 6-31g basis the Cholesky is of dimension $N_{tf}=2200$ tensor factors, $N_{aux}=1094$ density fitting vectors, $N_{o}=47$ occupied orbitals. $N_{v}=86$ virtual orbitals.The rankings of conformers obtained are as follows- SCF:(1,4,8,3,10,2,6,9,5,7), MP2:(1,4,3,8,10,6,2,7,9,5), CCSD:(1,4,3,8,10,2,9,5,6,7), DOWNFOLDING:(1,4,8,10,3,6,2,7,9,5)}
\label{tab:aspirin-631g}
\begin{tabular}{c|cccc|}
\cline{2-5}
                                        & \multicolumn{4}{c|}{Level of Theory (basis: 6-31g)}                                                                      \\ \hline
\multicolumn{1}{|c|}{Aspirin Conformers} & \multicolumn{1}{c|}{SCF}          & \multicolumn{1}{c|}{RMP2}         & \multicolumn{1}{c|}{CCSD}         & DOWNFOLDING  \\ \hline
\multicolumn{1}{|c|}{Conformer 1}       & \multicolumn{1}{c|}{-644.6610982} & \multicolumn{1}{c|}{-645.9585356} & \multicolumn{1}{c|}{-646.0036644} & -645.9924406 \\ \hline
\multicolumn{1}{|c|}{Conformer 2}       & \multicolumn{1}{c|}{-644.6389322} & \multicolumn{1}{c|}{-645.9373075} & \multicolumn{1}{c|}{-645.9826633} & -645.9714558 \\ \hline
\multicolumn{1}{|c|}{Conformer 3}       & \multicolumn{1}{c|}{-644.6423715} & \multicolumn{1}{c|}{-645.9449257} & \multicolumn{1}{c|}{-645.9897701} & -645.9770809 \\ \hline
\multicolumn{1}{|c|}{Conformer 4}       & \multicolumn{1}{c|}{-644.6581017} & \multicolumn{1}{c|}{-645.9553979} & \multicolumn{1}{c|}{-646.0005532} & -645.9893473 \\ \hline
\multicolumn{1}{|c|}{Conformer 5}       & \multicolumn{1}{c|}{-644.631892}  & \multicolumn{1}{c|}{-645.9364658} & \multicolumn{1}{c|}{-645.9811905} & -645.966734  \\ \hline
\multicolumn{1}{|c|}{Conformer 6}       & \multicolumn{1}{c|}{-644.6331812} & \multicolumn{1}{c|}{-645.9373469} & \multicolumn{1}{c|}{-645.9808511} & -645.9719798 \\ \hline
\multicolumn{1}{|c|}{Conformer 7}       & \multicolumn{1}{c|}{-644.6300114} & \multicolumn{1}{c|}{-645.9366897} & \multicolumn{1}{c|}{-645.9805286} & -645.9686558 \\ \hline
\multicolumn{1}{|c|}{Conformer 8}       & \multicolumn{1}{c|}{-644.645868}  & \multicolumn{1}{c|}{-645.9439862} & \multicolumn{1}{c|}{-645.9895043} & -645.9782853 \\ \hline
\multicolumn{1}{|c|}{Conformer 9}       & \multicolumn{1}{c|}{-644.6319342} & \multicolumn{1}{c|}{-645.936523}  & \multicolumn{1}{c|}{-645.9812477} & -645.9667718 \\ \hline
\multicolumn{1}{|c|}{Conformer 10}      & \multicolumn{1}{c|}{-644.6418289} & \multicolumn{1}{c|}{-645.9415013} & \multicolumn{1}{c|}{-645.9862054} & -645.9771029 \\ \hline
\end{tabular}
\end{table}

\newpage
\begin{table}[h!]
\caption{Aspirin Conformer Ranking with downfolding- Single Point energies computed from downfolding with def2-svp basis incorporating singles and doubles clusters starting from single reference state. In def2-svp basis the Cholesky is of dimension $N_{tf}=2200$ tensor factors, $N_{aux}=1127$ density fitting vectors, $N_{o}=47$ occupied orbitals. $N_{v}=175$ virtual orbitals.The rankings of conformers obtained are as follows- SCF:(1,4,8,10,3,2,9,5,6,7), RMP2:(1,4,3,8,10,6,9,5,7,2), CCSD:(1,4,3,8,10,2,6,9,5,7), DOWNFOLDING:(1,4,3,8,10,6,7,2,9,5).}
\label{tab:asprin-def2svp}
\begin{tabular}{c|cccc|}
\cline{2-5}
                                         & \multicolumn{4}{c|}{Level of Theory (basis: def2-svp)}                                                                   \\ \hline
\multicolumn{1}{|c|}{Aspirin Conformers} & \multicolumn{1}{c|}{SCF}          & \multicolumn{1}{c|}{RMP2}         & \multicolumn{1}{c|}{CCSD}         & DOWNFOLDING  \\ \hline
\multicolumn{1}{|c|}{Conformer 1}        & \multicolumn{1}{c|}{-644.4563718} & \multicolumn{1}{c|}{-646.3873246} & \multicolumn{1}{c|}{-646.4372907} & -646.4186128 \\ \hline
\multicolumn{1}{|c|}{Conformer 2}        & \multicolumn{1}{c|}{-644.434291}  & \multicolumn{1}{c|}{-646.3662881} & \multicolumn{1}{c|}{-646.4163628} & -646.3961012 \\ \hline
\multicolumn{1}{|c|}{Conformer 3}        & \multicolumn{1}{c|}{-644.4373907} & \multicolumn{1}{c|}{-646.3732314} & \multicolumn{1}{c|}{-646.4225723} & -646.4036205 \\ \hline
\multicolumn{1}{|c|}{Conformer 4}        & \multicolumn{1}{c|}{-644.4528848} & \multicolumn{1}{c|}{-646.3840958} & \multicolumn{1}{c|}{-646.4339928} & -646.4134617 \\ \hline
\multicolumn{1}{|c|}{Conformer 5}        & \multicolumn{1}{c|}{-644.4279823} & \multicolumn{1}{c|}{-646.3672113} & \multicolumn{1}{c|}{-646.4160287} & -646.3949031 \\ \hline
\multicolumn{1}{|c|}{Conformer 6}        & \multicolumn{1}{c|}{-644.4271181} & \multicolumn{1}{c|}{-646.3691688} & \multicolumn{1}{c|}{-646.4163438} & -646.3975334 \\ \hline
\multicolumn{1}{|c|}{Conformer 7}        & \multicolumn{1}{c|}{-644.4264025} & \multicolumn{1}{c|}{-646.3671135} & \multicolumn{1}{c|}{-646.4151305} & -646.3969465 \\ \hline
\multicolumn{1}{|c|}{Conformer 8}        & \multicolumn{1}{c|}{-644.4406671} & \multicolumn{1}{c|}{-646.3722646} & \multicolumn{1}{c|}{-646.4225624} & -646.4029735 \\ \hline
\multicolumn{1}{|c|}{Conformer 9}        & \multicolumn{1}{c|}{-644.4280075} & \multicolumn{1}{c|}{-646.3672411} & \multicolumn{1}{c|}{-646.4160588} & -646.3949112 \\ \hline
\multicolumn{1}{|c|}{Conformer 10}       & \multicolumn{1}{c|}{-644.437574}  & \multicolumn{1}{c|}{-646.3715154} & \multicolumn{1}{c|}{-646.420843}  & -646.3999801 \\ \hline
\end{tabular}
\end{table}

\begin{table}[h!]
\caption{Aspirin Conformer Ranking with downfolding- Single Point energies computed from downfolding with ccpvdz basis incorporating singles and doubles clusters starting from single reference state. In ccpvdz basis the Cholesky is of dimension $N_{tf}=2200$ tensor factors, $N_{aux}=1094$ density fitting vectors, $N_{o}=47$ occupied orbitals. $N_{v}=175$ virtual orbitals.The rankings of conformers obtained are as follows- SCF:(1,4,8,10,3,2,9,5,6,7), RMP2:(1,4,3,8,10,6,9,7,5,2), CCSD:(1,4,8,3,10,6,2,9,5,7), DOWNFOLDING:(1,4,8,3,10,2,7,6,9,5).}
\label{tab:aspirin-ccpvdz}
\begin{tabular}{c|cccc|}
\cline{2-5}
                                         & \multicolumn{4}{c|}{Level of Theory (basis: ccpvdz)}                                                                     \\ \hline
\multicolumn{1}{|c|}{Aspirin Conformers} & \multicolumn{1}{c|}{SCF}          & \multicolumn{1}{c|}{RMP2}         & \multicolumn{1}{c|}{CCSD}         & DOWNFOLDING  \\ \hline
\multicolumn{1}{|c|}{Conformer 1}        & \multicolumn{1}{c|}{-644.9997392} & \multicolumn{1}{c|}{-646.9318783} & \multicolumn{1}{c|}{-646.9805528} & -646.9555228 \\ \hline
\multicolumn{1}{|c|}{Conformer 2}        & \multicolumn{1}{c|}{-644.9778063} & \multicolumn{1}{c|}{-646.9109991} & \multicolumn{1}{c|}{-646.9597428} & -646.936159  \\ \hline
\multicolumn{1}{|c|}{Conformer 3}        & \multicolumn{1}{c|}{-644.9806198} & \multicolumn{1}{c|}{-646.917197}  & \multicolumn{1}{c|}{-646.9652798} & -646.9410966 \\ \hline
\multicolumn{1}{|c|}{Conformer 4}        & \multicolumn{1}{c|}{-644.9963784} & \multicolumn{1}{c|}{-646.928845}  & \multicolumn{1}{c|}{-646.9774418} & -646.9527424 \\ \hline
\multicolumn{1}{|c|}{Conformer 5}        & \multicolumn{1}{c|}{-644.9713608} & \multicolumn{1}{c|}{-646.9110801} & \multicolumn{1}{c|}{-646.9586147} & -646.9339922 \\ \hline
\multicolumn{1}{|c|}{Conformer 6}        & \multicolumn{1}{c|}{-644.9706391} & \multicolumn{1}{c|}{-646.914223}  & \multicolumn{1}{c|}{-646.959752}  & -646.9343443 \\ \hline
\multicolumn{1}{|c|}{Conformer 7}        & \multicolumn{1}{c|}{-644.9698768} & \multicolumn{1}{c|}{-646.9110831} & \multicolumn{1}{c|}{-646.957796}  & -646.9350735 \\ \hline
\multicolumn{1}{|c|}{Conformer 8}        & \multicolumn{1}{c|}{-644.9839666} & \multicolumn{1}{c|}{-646.9168041} & \multicolumn{1}{c|}{-646.965791}  & -646.9417124 \\ \hline
\multicolumn{1}{|c|}{Conformer 9}        & \multicolumn{1}{c|}{-644.9713855} & \multicolumn{1}{c|}{-646.9111134} & \multicolumn{1}{c|}{-646.9586482} & -646.934041  \\ \hline
\multicolumn{1}{|c|}{Conformer 10}       & \multicolumn{1}{c|}{-644.98115}   & \multicolumn{1}{c|}{-646.9163835} & \multicolumn{1}{c|}{-646.9642987} & -646.9396566 \\ \hline
\end{tabular}
\end{table}

\newpage
\begin{figure}[h!]
\begin{subfigure}[b]{0.32\textwidth}
\centering
\includegraphics[width=\textwidth]{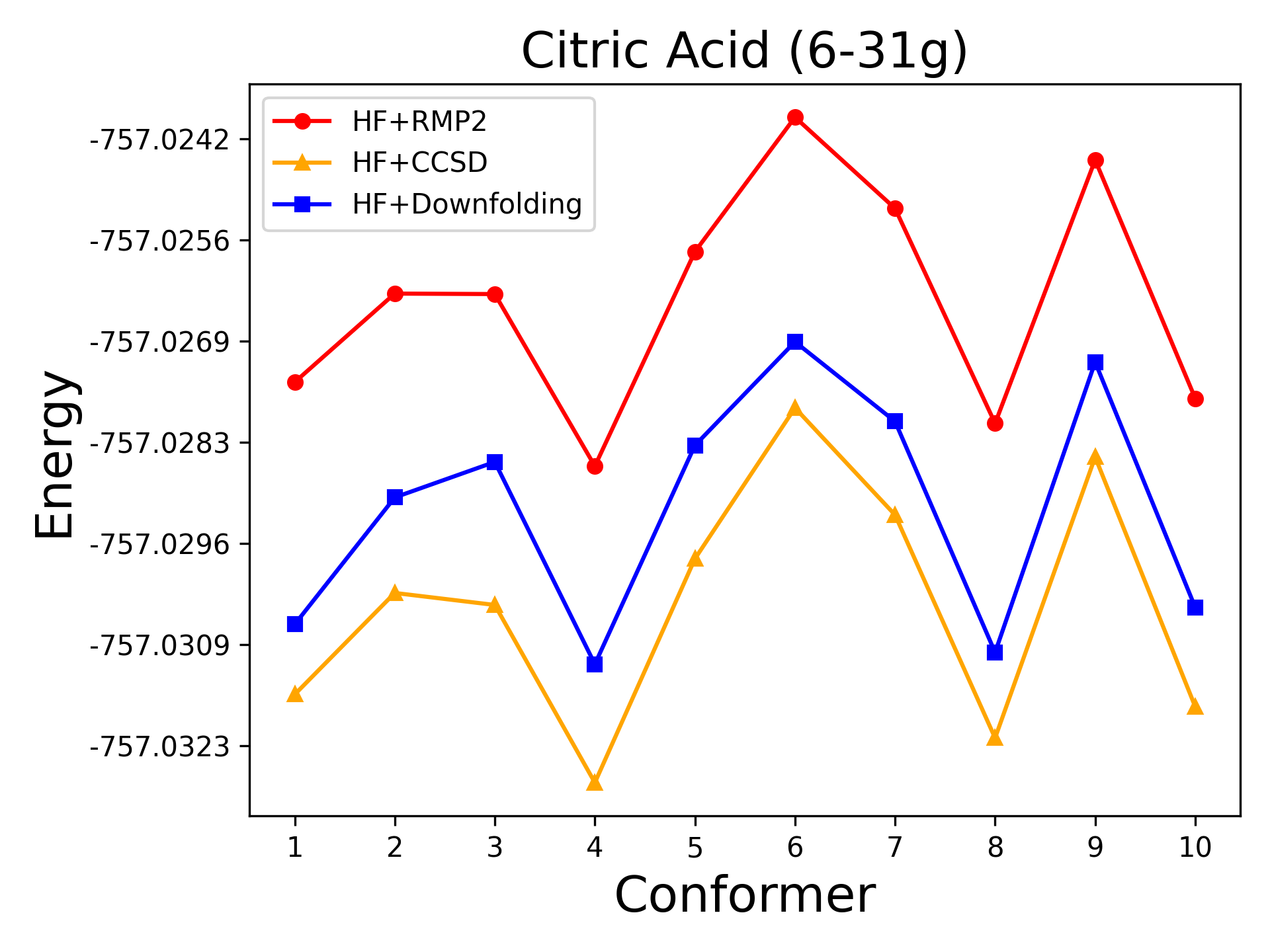}
\caption{}
\end{subfigure}
\hfill
\begin{subfigure}[b]{0.32\textwidth}
\centering
\includegraphics[width=\textwidth]{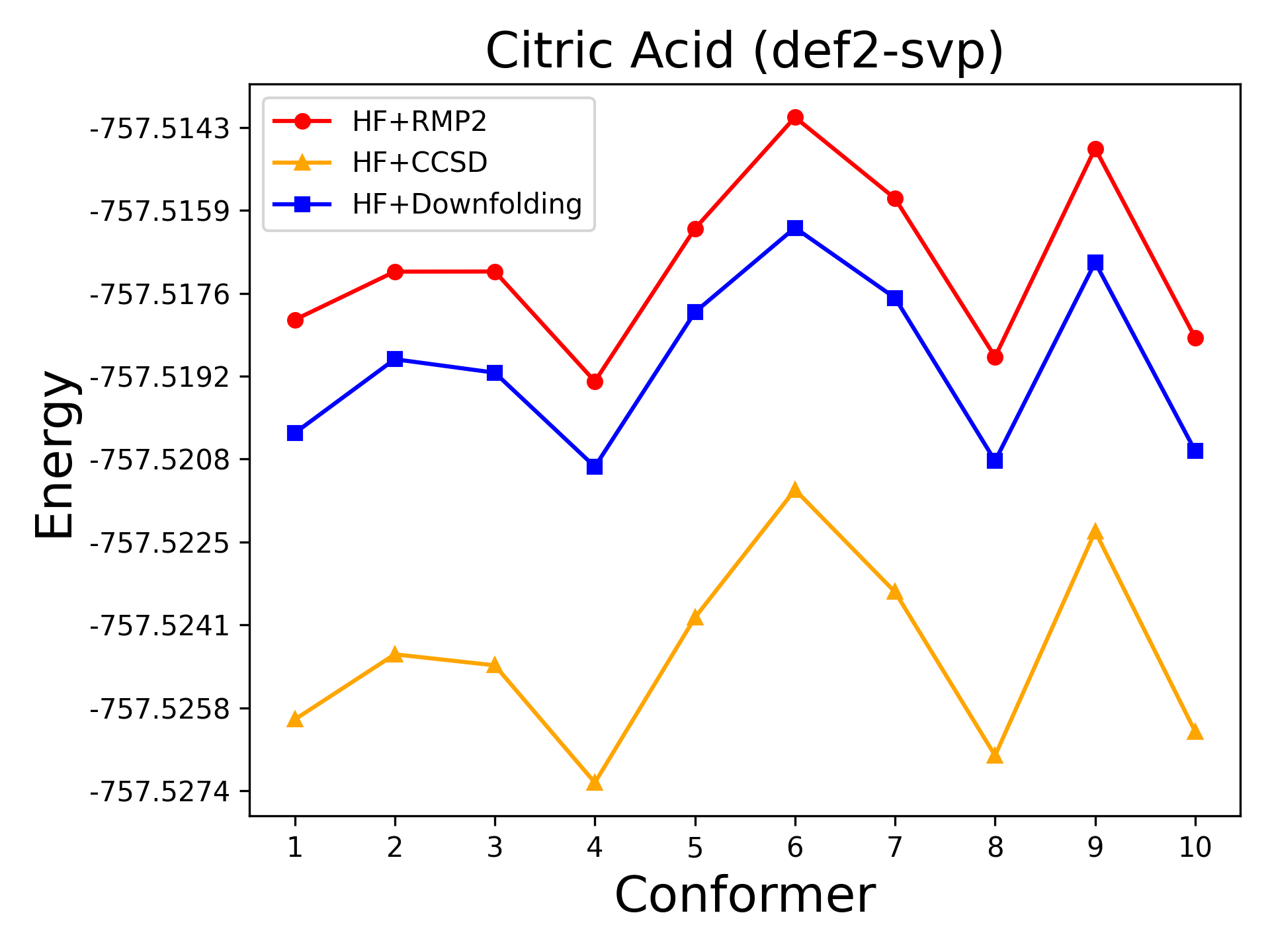}
\caption{}
\end{subfigure}
\hfill
\begin{subfigure}[b]{0.32\textwidth}
\centering
\includegraphics[width=\textwidth]{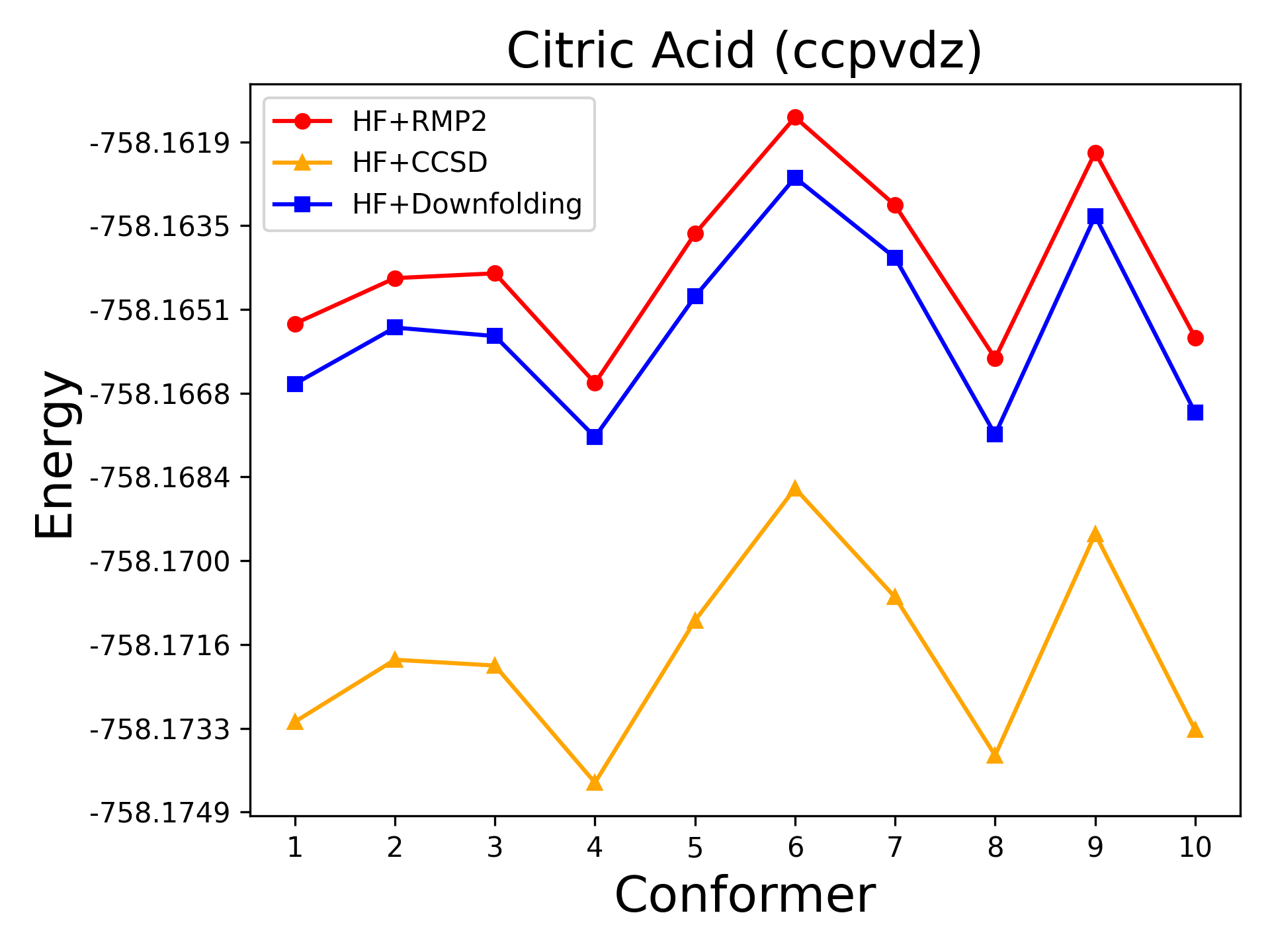}
\caption{}
\end{subfigure}
\caption{Energy Calculations in different basis sets for Citric Acid Conformers}
\label{fig:citric_benchmark}
\end{figure}

\begin{figure}[h!]
\begin{subfigure}[b]{0.32\textwidth}
\centering
\includegraphics[width=\textwidth]{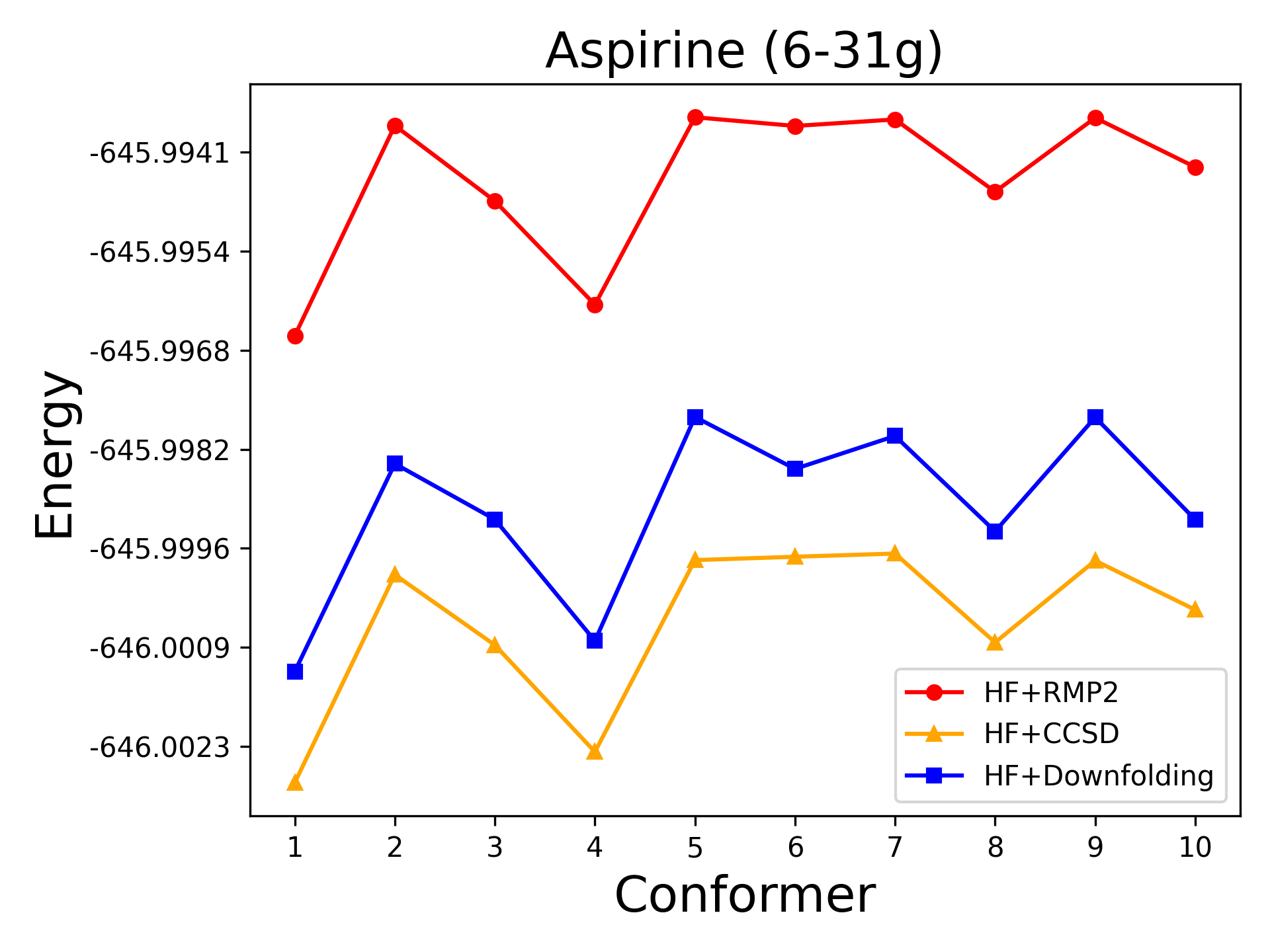}
\caption{}
\end{subfigure}
\hfill
\begin{subfigure}[b]{0.32\textwidth}
\centering
\includegraphics[width=\textwidth]{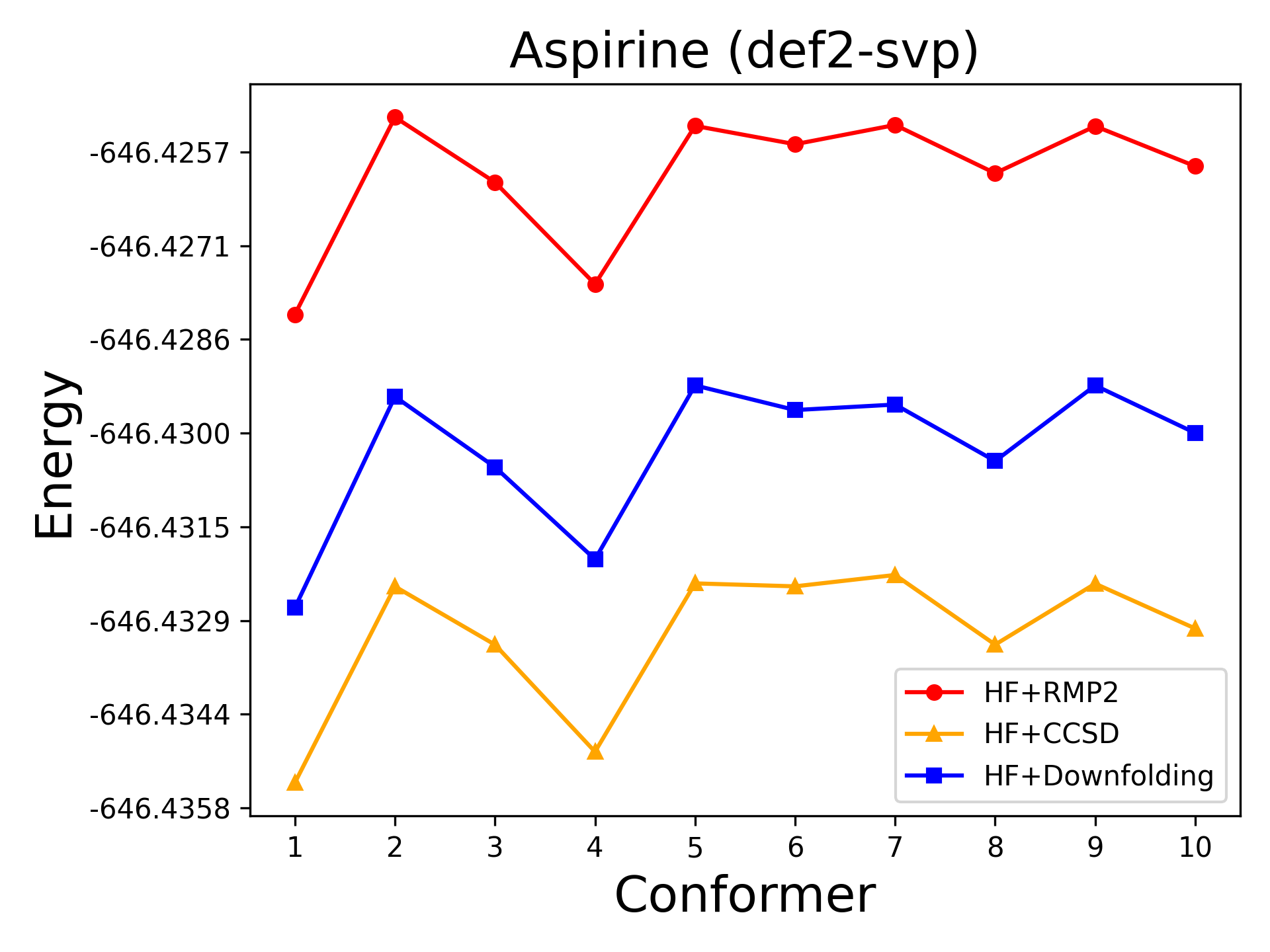}
\caption{}
\end{subfigure}
\hfill
\begin{subfigure}[b]{0.32\textwidth}
\centering
\includegraphics[width=\textwidth]{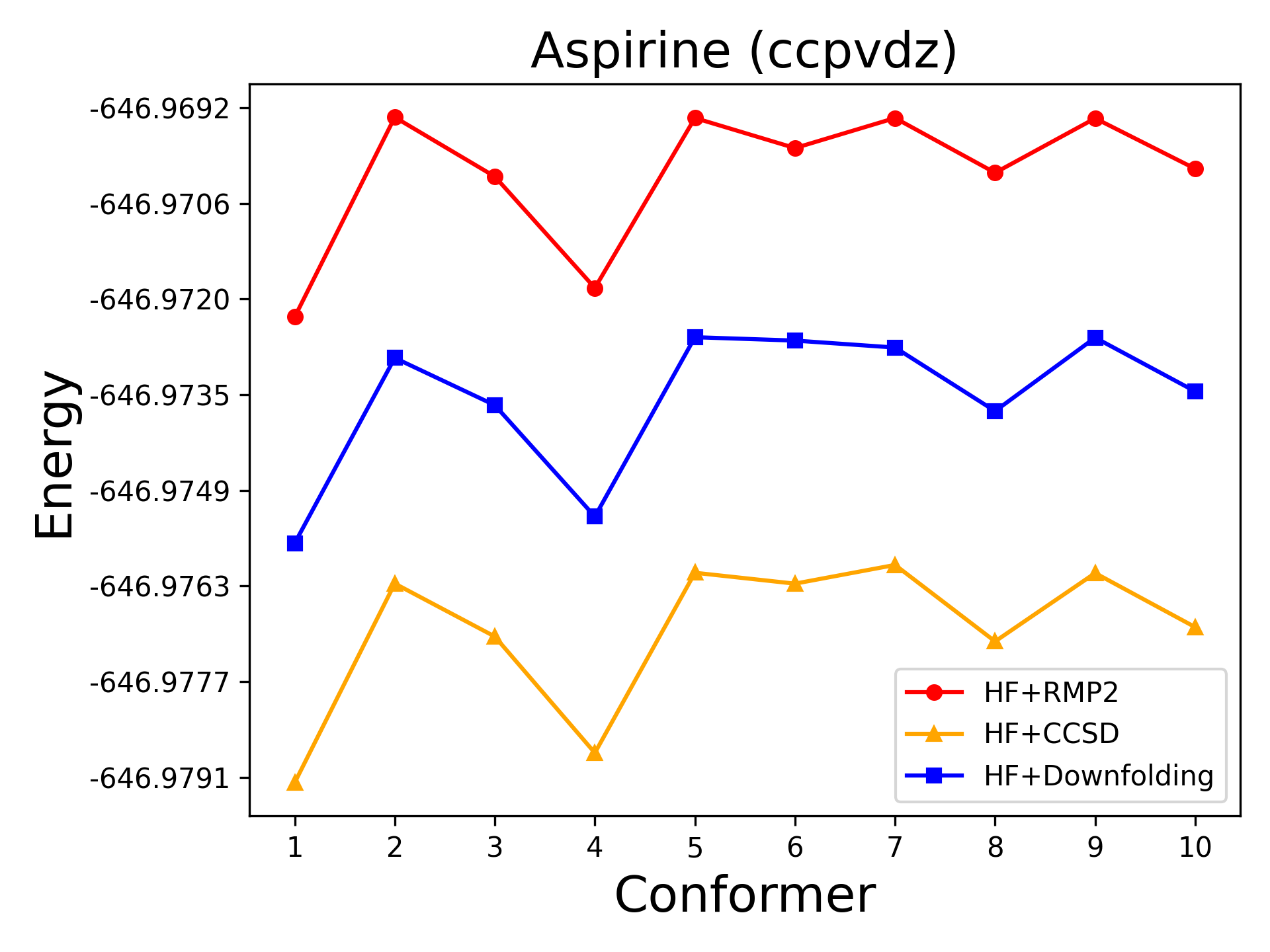}
\caption{}
\end{subfigure}
\caption{Energy Calculations in different basis sets for Aspirin Conformers}
\label{fig:aspirin-benchmark}
\end{figure}

\begin{table}[h!]
\hspace{-5em}
\caption{The table represents the calculated correlation energies of molecules $1 \to$ 1,3-hexadiene, $2 \to$ 1-heptyne, $3 \to$ benzene, $4 \to$ NaBr, $5 \to$ NaCl, $6 \to$ propyne in the basis: 6-31g, ccpvdz and def2-tzvp for the methods: MP2, RHD(Downfolding) and CCSD}
\label{tab:miscbenchmark}
\centering
\begin{tabular}{|c|ccc|ccc|ccc|}
\hline
\multirow{2}{*}{\#} & \multicolumn{3}{c|}{Correlation   Energy (6-31g)}                         & \multicolumn{3}{c|}{Correlation   Energy (ccpvdz)}                        & \multicolumn{3}{c|}{Correlation   Energy (def2-tzvp)}                     \\ \cline{2-10} 
                    & \multicolumn{1}{c|}{MP2}     & \multicolumn{1}{c|}{RHD(SD)} & CCSD    & \multicolumn{1}{c|}{MP2}     & \multicolumn{1}{c|}{RHD(SD)} & CCSD    & \multicolumn{1}{c|}{MP2}     & \multicolumn{1}{c|}{RHD(SD)} & CCSD    \\ \hline
1                   & \multicolumn{1}{c|}{-0.5429} & \multicolumn{1}{c|}{-0.5904}     & -0.6090 & \multicolumn{1}{c|}{-0.8374} & \multicolumn{1}{c|}{-0.8843}     & -0.9068 & \multicolumn{1}{c|}{-1.0795} & \multicolumn{1}{c|}{-1.1285}     & -1.1437 \\ \hline
2                   & \multicolumn{1}{c|}{-0.6437} & \multicolumn{1}{c|}{-0.6886}     & -0.7154 & \multicolumn{1}{c|}{-0.9868} & \multicolumn{1}{c|}{-1.0337}     & -1.0657 & \multicolumn{1}{c|}{-1.2705} & \multicolumn{1}{c|}{-1.3188}     & -1.3441 \\ \hline
3                   & \multicolumn{1}{c|}{-0.5236} & \multicolumn{1}{c|}{-0.5546}     & -0.5667 & \multicolumn{1}{c|}{-0.7986} & \multicolumn{1}{c|}{-0.8300}     & -0.8370 & \multicolumn{1}{c|}{-1.0416} & \multicolumn{1}{c|}{-1.0787}     & -1.0739 \\ \hline
4                   & \multicolumn{3}{c|}{N/A}                                                  & \multicolumn{1}{c|}{-0.1530} & \multicolumn{1}{c|}{-0.1612}     & -0.1632 & \multicolumn{1}{c|}{-0.5395} & \multicolumn{1}{c|}{-0.5427}     & -0.5380 \\ \hline
5                   & \multicolumn{1}{c|}{-0.0516} & \multicolumn{1}{c|}{-0.0592}     & -0.0616 & \multicolumn{1}{c|}{-0.1637} & \multicolumn{1}{c|}{-0.1775}     & -0.1785 & \multicolumn{1}{c|}{-0.4908} & \multicolumn{1}{c|}{-0.5088}     & -0.5065 \\ \hline
6                   & \multicolumn{1}{c|}{-0.2729} & \multicolumn{1}{c|}{-0.3033}     & -0.2988 & \multicolumn{1}{c|}{-0.4034} & \multicolumn{1}{c|}{-0.4361}     & -0.4323 & \multicolumn{1}{c|}{-0.5236} & \multicolumn{1}{c|}{-0.5595}     & -0.5499 \\ \hline
\end{tabular}
\end{table}

\newpage
\begin{figure}[h!]
\begin{minipage}[b]{0.45\textwidth}
\centering
\includegraphics[width=\textwidth]{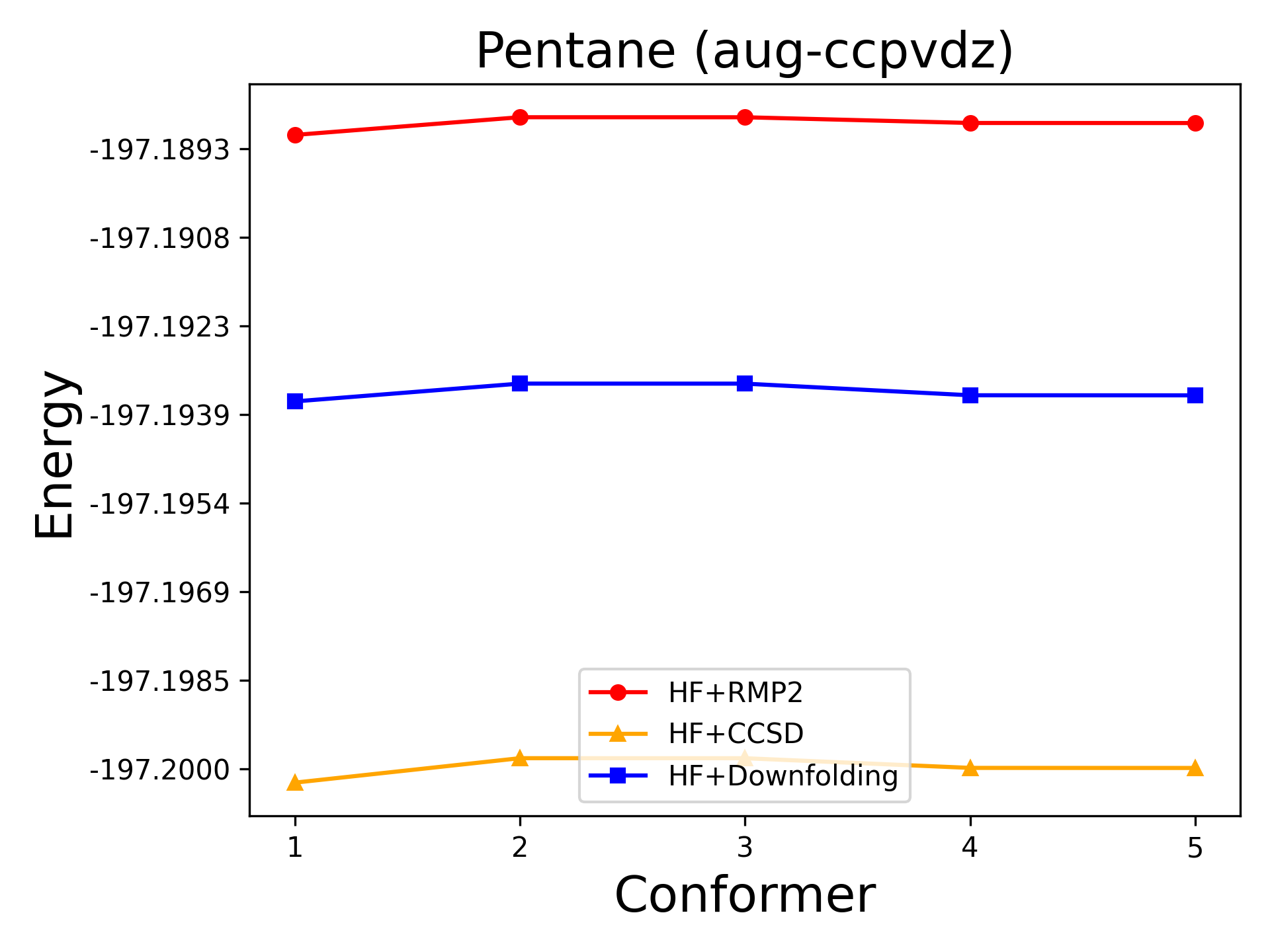}
\caption{Energy Calculations for Pentane Conformers in highly diffuse basis }
\label{fig:pentane-benchmark}
\end{minipage}
\hfill
\begin{minipage}[b]{0.45\textwidth}
\centering
\includegraphics[width=\textwidth]{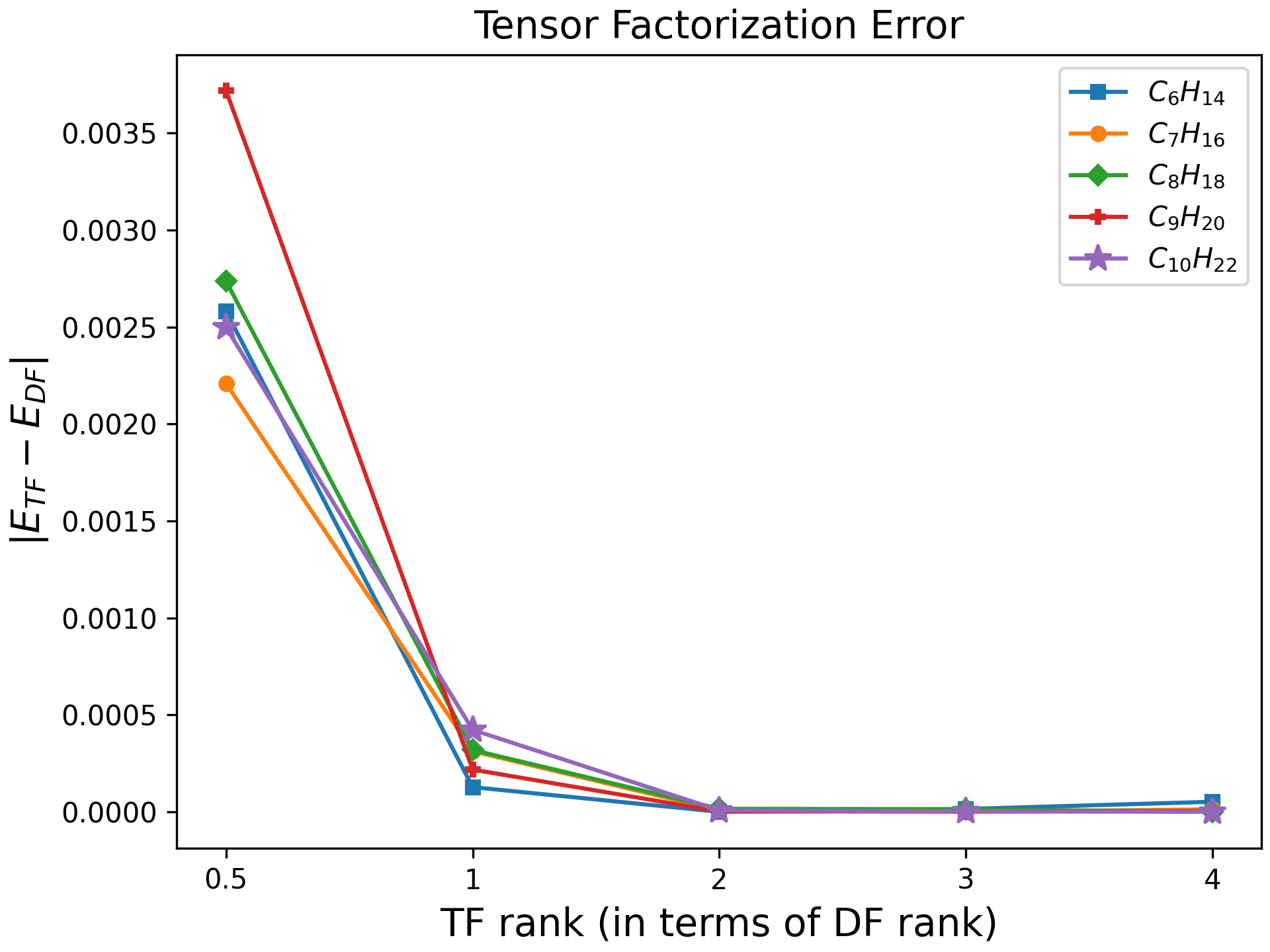}
\caption{Plot of Tensor factorization errors in Energy Calculations with tf-rank}
\label{fig:tf-error}
\end{minipage}
\end{figure}

\begin{figure}[h!]
\begin{subfigure}[b]{0.32\textwidth}
\centering
\includegraphics[width=\textwidth]{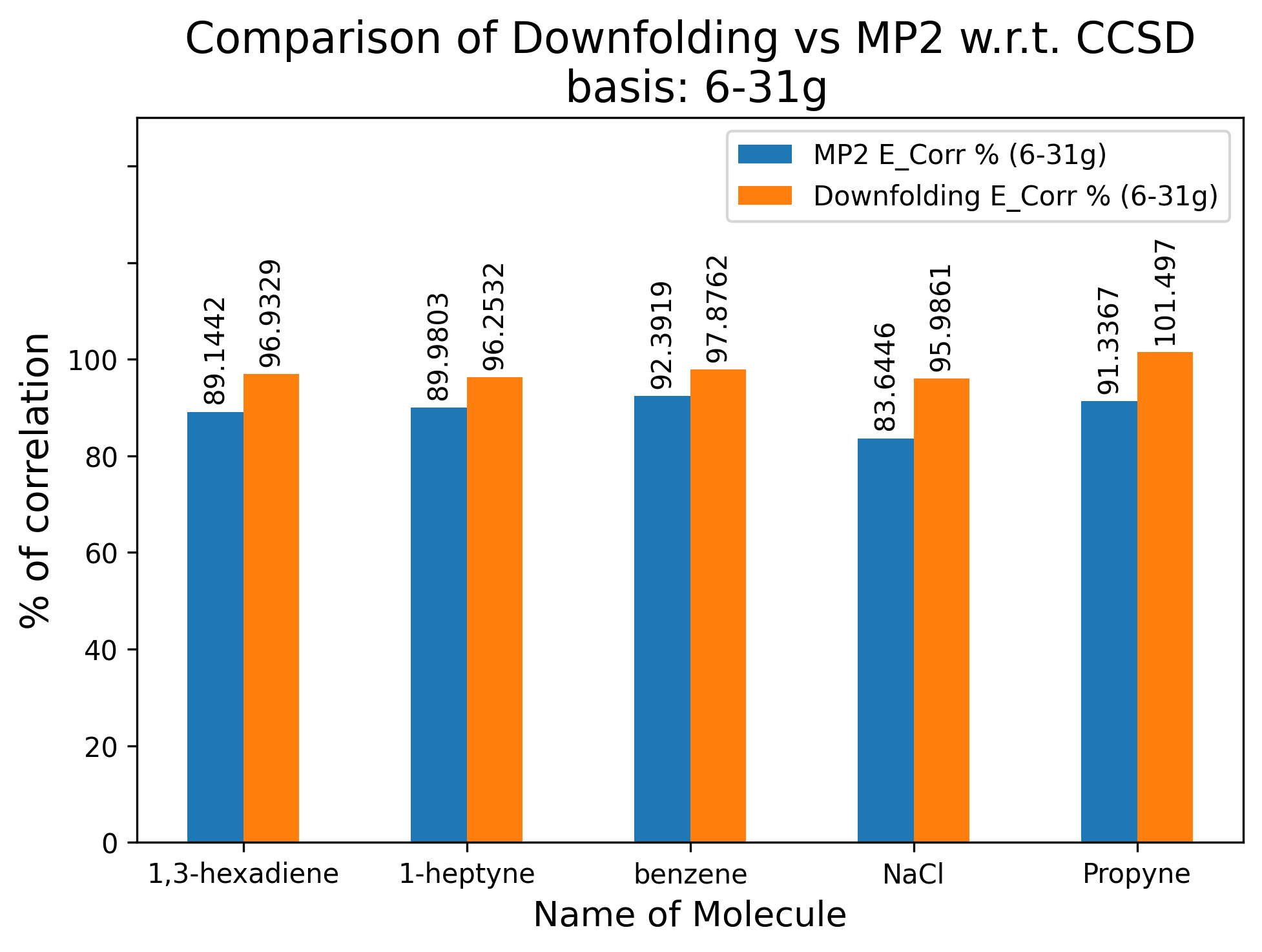}
\caption{}
\end{subfigure}
\hfill
\begin{subfigure}[b]{0.32\textwidth}
\centering
\includegraphics[width=\textwidth]{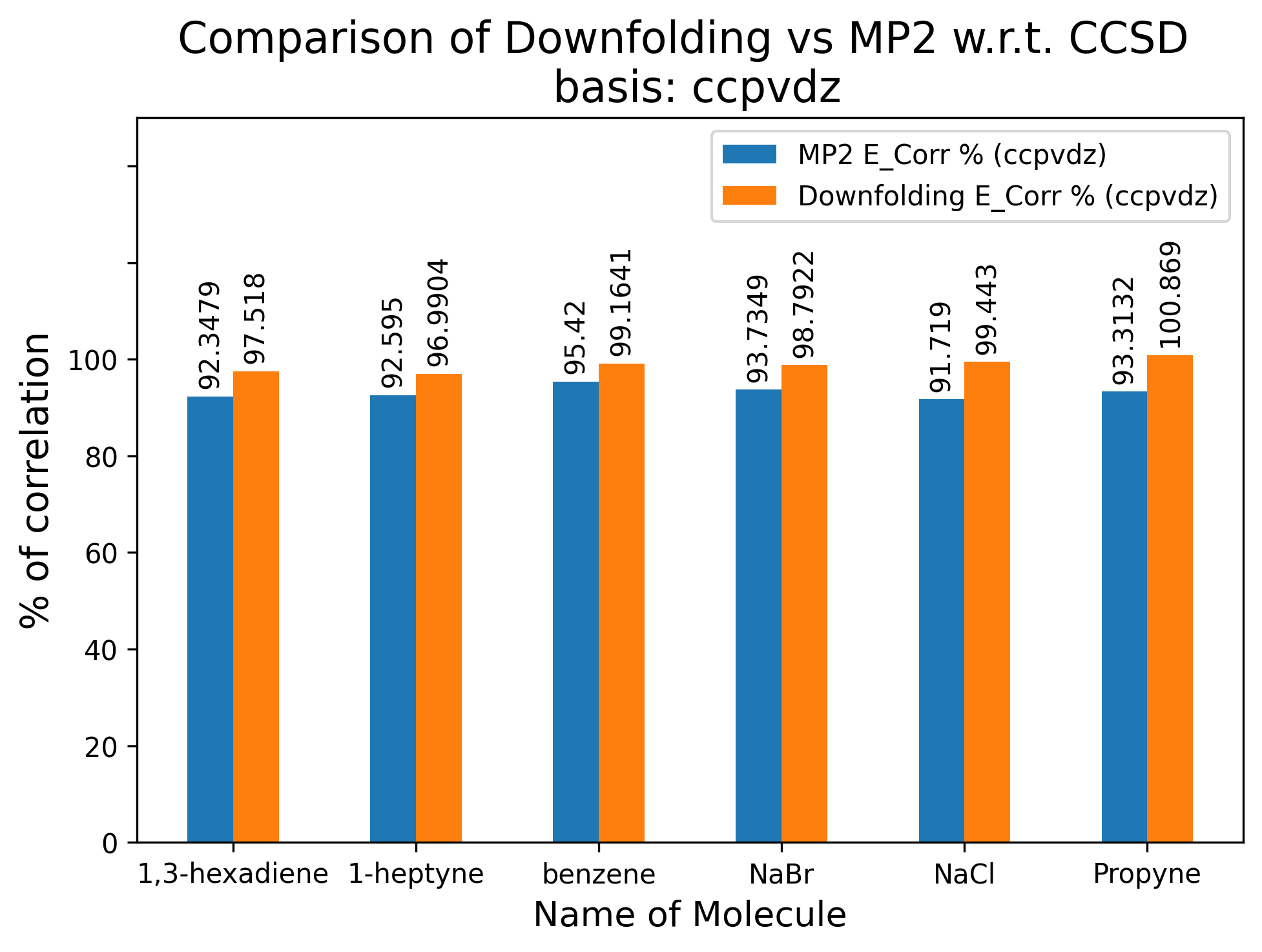}
\caption{}
\end{subfigure}
\hfill
\begin{subfigure}[b]{0.32\textwidth}
\centering
\includegraphics[width=\textwidth]{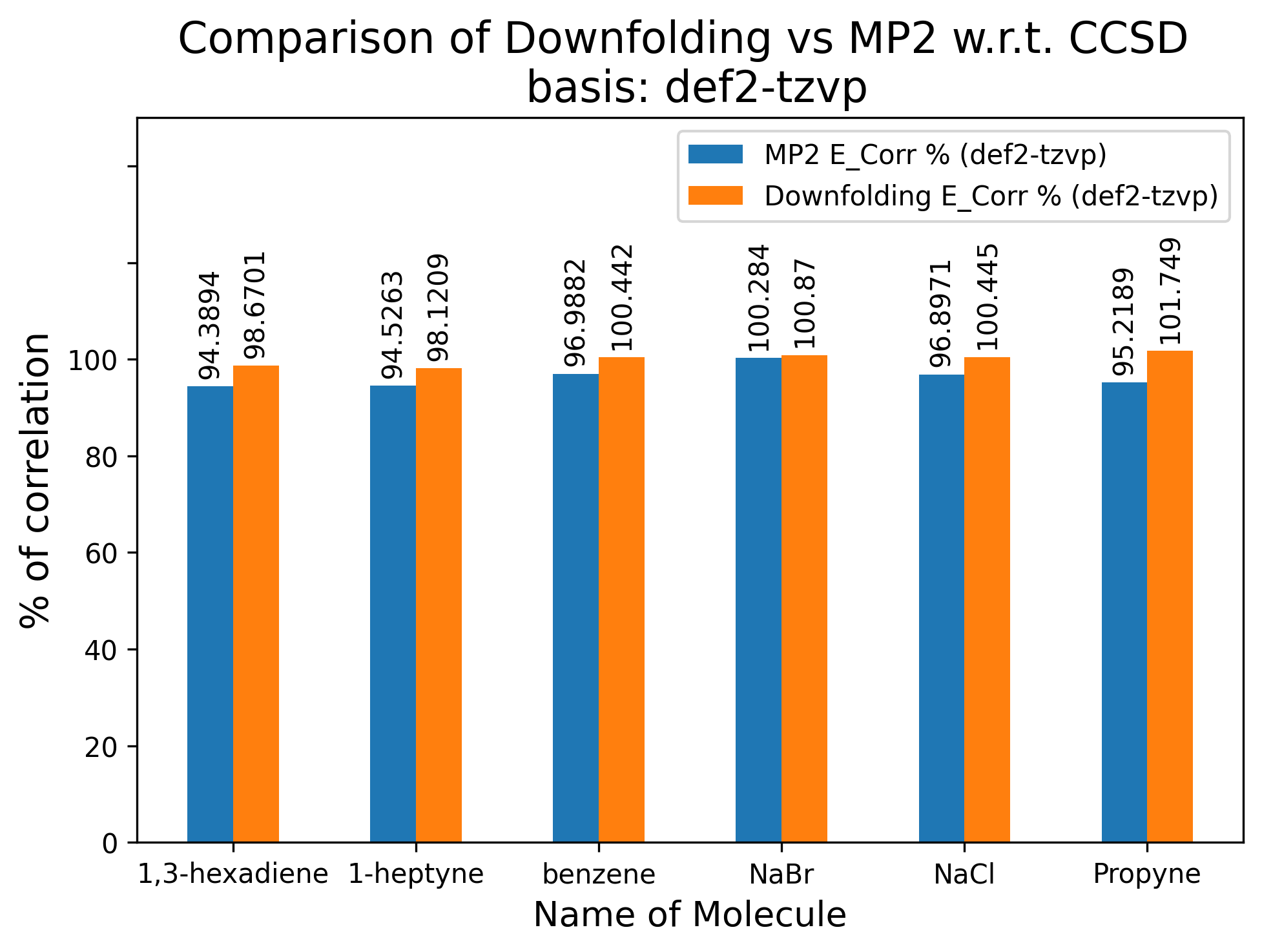}
\caption{}
\end{subfigure}
\caption{Energy Calculations in different basis sets for different molecules}
\label{fig:energy-benchmark-misc}
\end{figure}

\begin{figure}[h!]
\begin{subfigure}[b]{0.4\textwidth}
\centering
\includegraphics[width=\textwidth]{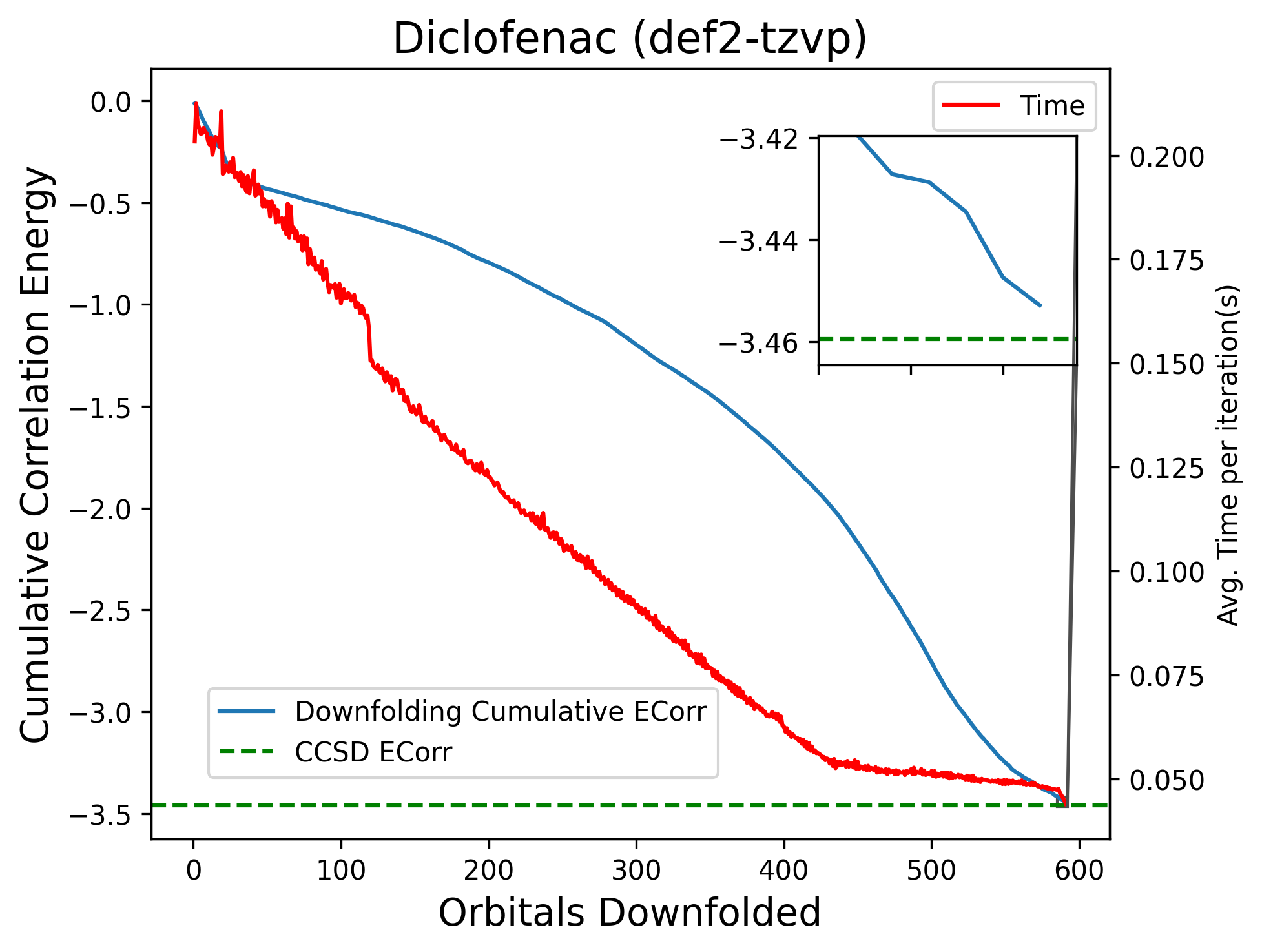}
\caption{}
\end{subfigure}
\hfill
\begin{subfigure}[b]{0.4\textwidth}
\centering
\includegraphics[width=\textwidth]{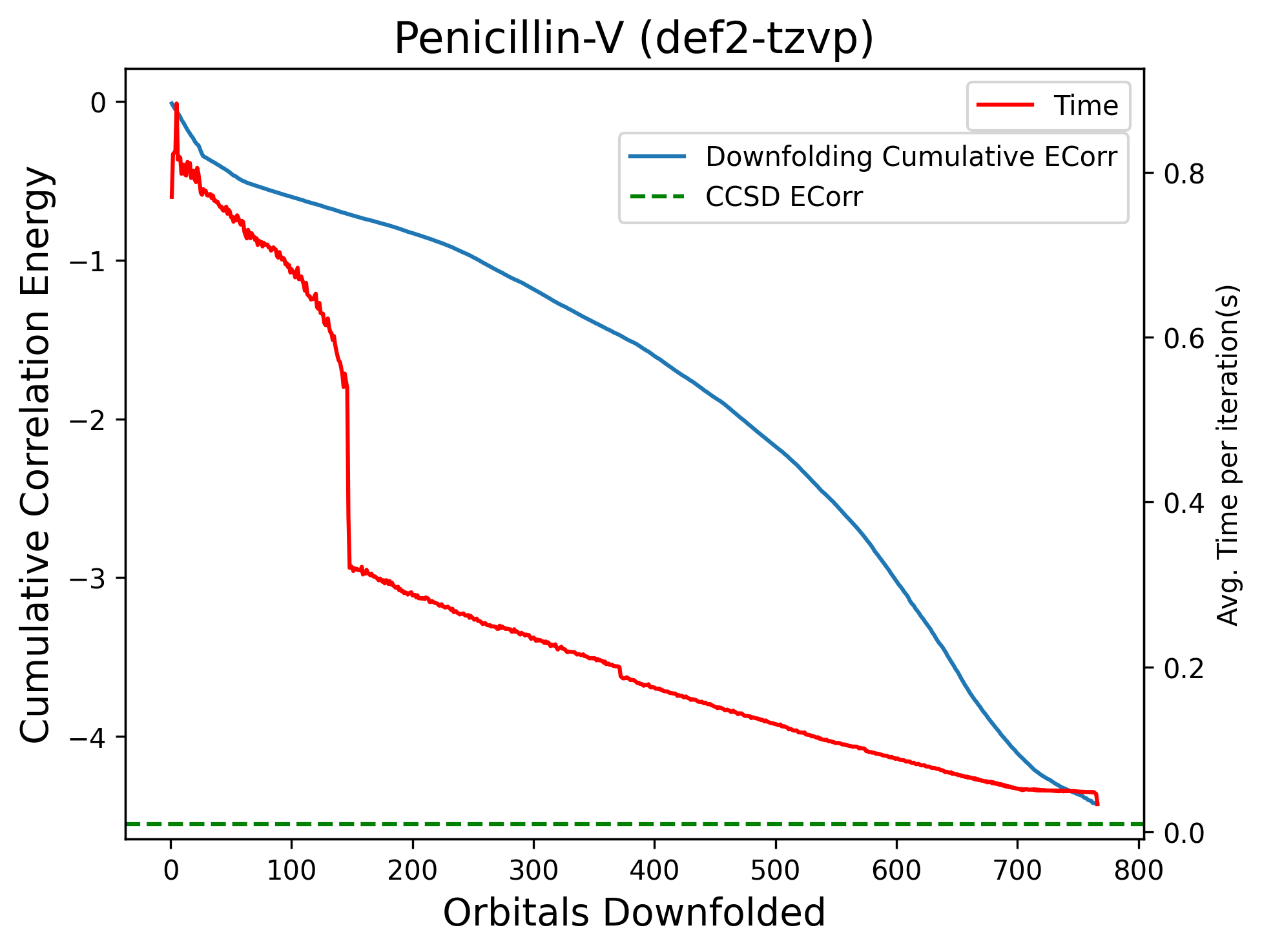}
\caption{}
\end{subfigure}
\caption{Downfolding Cumulative Correlation energy and iteration time with orbital decoupling steps for Diclofenac and Penicillin V}
\label{fig:iteration_time}
\end{figure}

\newpage
\begin{table}[h!]
\caption{Pentane Conformer Ranking with downfolding- Single Point energies computed from downfolding with ccpvdz basis incorporating singles and doubles clusters starting from single reference state. In ccpvdz basis the Cholesky is of dimension $N_{tf}=2200$ tensor factors, $N_{aux}=814$ density fitting vectors, $N_{occ}=21$ occupied orbitals. $N_{virt}=202$ virtual orbitals.The rankings of conformers obtained are as follows- SCF:(1,5,4,2,3), RMP2:(1,5,4,3,2), CCSD:(1,5,4,3,2), DOWNFOLDING:(1,5,4,3,2)}
\label{tab:pentane-augccpvdz}
\begin{tabular}{c|cccc|}
\cline{2-5}
                                         & \multicolumn{4}{c|}{Level of theory (basis: aug-ccpvdz)}                                                                 \\ \hline
\multicolumn{1}{|c|}{Pentane Conformers} & \multicolumn{1}{c|}{SCF}          & \multicolumn{1}{c|}{RMP2}         & \multicolumn{1}{c|}{CCSD}         & DOWNFOLDING  \\ \hline
\multicolumn{1}{|c|}{Conformer 1}        & \multicolumn{1}{c|}{-196.3530103} & \multicolumn{1}{c|}{-197.1284168} & \multicolumn{1}{c|}{-197.2015472} & -197.1585037 \\ \hline
\multicolumn{1}{|c|}{Conformer 2}        & \multicolumn{1}{c|}{-196.3475753} & \multicolumn{1}{c|}{-197.126418}  & \multicolumn{1}{c|}{-197.1987854} & -197.1564927 \\ \hline
\multicolumn{1}{|c|}{Conformer 3}        & \multicolumn{1}{c|}{-196.3475749} & \multicolumn{1}{c|}{-197.1264208} & \multicolumn{1}{c|}{-197.1987886} & -197.1564972 \\ \hline
\multicolumn{1}{|c|}{Conformer 4}        & \multicolumn{1}{c|}{-196.3501434} & \multicolumn{1}{c|}{-197.1270664} & \multicolumn{1}{c|}{-197.1998807} & -197.1578099 \\ \hline
\multicolumn{1}{|c|}{Conformer 5}        & \multicolumn{1}{c|}{-196.3501481} & \multicolumn{1}{c|}{-197.1270735} & \multicolumn{1}{c|}{-197.1998892} & -197.1578173 \\ \hline
\end{tabular}
\end{table}

\begin{table}[h!]
\caption{Single point energies are computed from downfolding at the level of 6-31g basis for different alkane molecules with 1 to 10 carbon atoms using tensor factorization approximation at varying  ranks: $N_{tf}=0.5N_{aux}$, $N_{tf}= N_{aux}$, $N_{tf}=2N_{aux}$, $N_{tf}=3N_{aux}$, $N_{tf}=4N_{aux}$ with $N_{aux}$ being the density fitting auxiliary basis rank. The energy values obtained using the density fitting Hamiltonian downfolding for the corresponding molecules is also presented for comparison}
\label{tab:tf-error}
\begin{tabular}{c|cccccc|}
\cline{2-7}
                              & \multicolumn{6}{c|}{Level of Approximation (basis: 6-31g)}                                                                                                                                         \\ \hline
\multicolumn{1}{|c|}{Alkane} & \multicolumn{1}{c|}{DF} & \multicolumn{1}{c|}{TF-0.5DF}    & \multicolumn{1}{c|}{TF-1DF}      & \multicolumn{1}{c|}{TF-2DF}      & \multicolumn{1}{c|}{TF-3DF}      & TF-4DF      \\ \hline
\multicolumn{1}{|c|}{1}       & \multicolumn{1}{c|}{-38.43259267}   & \multicolumn{1}{c|}{-38.43215632} & \multicolumn{1}{c|}{-38.43259267} & \multicolumn{1}{c|}{-38.43259267} & \multicolumn{1}{c|}{-38.43259267} & -38.43259267 \\ \hline
\multicolumn{1}{|c|}{2}       & \multicolumn{1}{c|}{-79.40927911}   & \multicolumn{1}{c|}{-79.40844159} & \multicolumn{1}{c|}{-79.40922995} & \multicolumn{1}{c|}{-79.40927911} & \multicolumn{1}{c|}{-79.40927911} & -79.40927911 \\ \hline
\multicolumn{1}{|c|}{3}       & \multicolumn{1}{c|}{-118.5268271}   & \multicolumn{1}{c|}{-118.5268112} & \multicolumn{1}{c|}{-118.526724}  & \multicolumn{1}{c|}{-118.5268378} & \multicolumn{1}{c|}{-118.5268271} & -118.5268271 \\ \hline
\multicolumn{1}{|c|}{4}       & \multicolumn{1}{c|}{-157.6403285}   & \multicolumn{1}{c|}{-157.6393513} & \multicolumn{1}{c|}{-157.6401943} & \multicolumn{1}{c|}{-157.6403458} & \multicolumn{1}{c|}{-157.6403307} & -157.6403285 \\ \hline
\multicolumn{1}{|c|}{5}       & \multicolumn{1}{c|}{-196.7550323}   & \multicolumn{1}{c|}{-196.7539008} & \multicolumn{1}{c|}{-196.7549078} & \multicolumn{1}{c|}{-196.7550365} & \multicolumn{1}{c|}{-196.7550666} & -196.7550323 \\ \hline
\multicolumn{1}{|c|}{6}       & \multicolumn{1}{c|}{-235.8671152}   & \multicolumn{1}{c|}{-235.8645352} & \multicolumn{1}{c|}{-235.8669872} & \multicolumn{1}{c|}{-235.8671161} & \multicolumn{1}{c|}{-235.8671307} & -235.8671678 \\ \hline
\multicolumn{1}{|c|}{7}       & \multicolumn{1}{c|}{-274.9818615}   & \multicolumn{1}{c|}{-274.9796533} & \multicolumn{1}{c|}{-274.9815484} & \multicolumn{1}{c|}{-274.9818699} & \multicolumn{1}{c|}{-274.9818612} & -274.9818751 \\ \hline
\multicolumn{1}{|c|}{8}       & \multicolumn{1}{c|}{-314.0939191}   & \multicolumn{1}{c|}{-314.0911834} & \multicolumn{1}{c|}{-314.0935996} & \multicolumn{1}{c|}{-314.0939026} & \multicolumn{1}{c|}{-314.0939334} & -314.0939195 \\ \hline
\multicolumn{1}{|c|}{9}       & \multicolumn{1}{c|}{-353.2063983}   & \multicolumn{1}{c|}{-353.2026801} & \multicolumn{1}{c|}{-353.2061796} & \multicolumn{1}{c|}{-353.206401}  & \multicolumn{1}{c|}{-353.2064015} & -353.2064029 \\ \hline
\multicolumn{1}{|c|}{10}      & \multicolumn{1}{c|}{-392.3185959}   & \multicolumn{1}{c|}{-392.3160985} & \multicolumn{1}{c|}{-392.3181725} & \multicolumn{1}{c|}{-392.3186047} & \multicolumn{1}{c|}{-392.3185993} & -392.318597  \\ \hline
\end{tabular}
\end{table}

\newpage
\begin{table}[h!]
\caption{A comparison of times taken for running calculations using CCSD and Downfolding techniques is listed below. Alkanes with 1 to 18 carbon atoms are considered. In a Nvidia V-100 GPU, the calculations were executed. GPU4PySCF implementation of CCSD couldn't perform calculations for alkanes with more than 13 carbon atoms due to storage constraints. For alkanes with more than 18 carbon atoms, the scf didn't converge under the same convergence criterion as for the rest of the molecules; hence downfolding data has not been provided for those molecules in this comparison}
\label{tab:alkane-speed}
\begin{tabular}{|c|cc|}
\hline
Basis: 6-31g & \multicolumn{2}{c|}{Time taken (s)}            \\ \hline
Alkanes      & \multicolumn{1}{c|}{CCSD}        & DOWNFOLDING \\ \hline
1            & \multicolumn{1}{c|}{10.49121623} & 28.33036164 \\ \hline
2            & \multicolumn{1}{c|}{1.83537627}  & 20.23007961 \\ \hline
3            & \multicolumn{1}{c|}{2.350443989} & 28.49350869 \\ \hline
4            & \multicolumn{1}{c|}{5.125217073} & 37.14023147 \\ \hline
5            & \multicolumn{1}{c|}{7.213134691} & 40.32105076 \\ \hline
6            & \multicolumn{1}{c|}{13.12267528} & 43.23324662 \\ \hline
7            & \multicolumn{1}{c|}{23.01328656} & 54.20392644 \\ \hline
8            & \multicolumn{1}{c|}{30.59317699} & 64.25860699 \\ \hline
9            & \multicolumn{1}{c|}{76.35492604} & 67.0198999  \\ \hline
10           & \multicolumn{1}{c|}{94.40636803} & 74.4054365  \\ \hline
11           & \multicolumn{1}{c|}{112.6176385} & 81.65810721 \\ \hline
12           & \multicolumn{1}{c|}{186.0039043} & 87.60531043 \\ \hline
13           & \multicolumn{1}{c|}{347.5013103} & 96.58040202 \\ \hline
14           & \multicolumn{1}{c|}{-}           & 100.6939708 \\ \hline
15           & \multicolumn{1}{c|}{-}           & 119.8199068 \\ \hline
16           & \multicolumn{1}{c|}{-}           & 121.2039495 \\ \hline
17           & \multicolumn{1}{c|}{-}           & 136.8925819 \\ \hline
18           & \multicolumn{1}{c|}{-}           & 165.9558665 \\ \hline
\end{tabular}
\end{table}

\begin{figure}[h!]
\begin{minipage}[b]{0.4\textwidth}
\centering
\includegraphics[width=\textwidth]{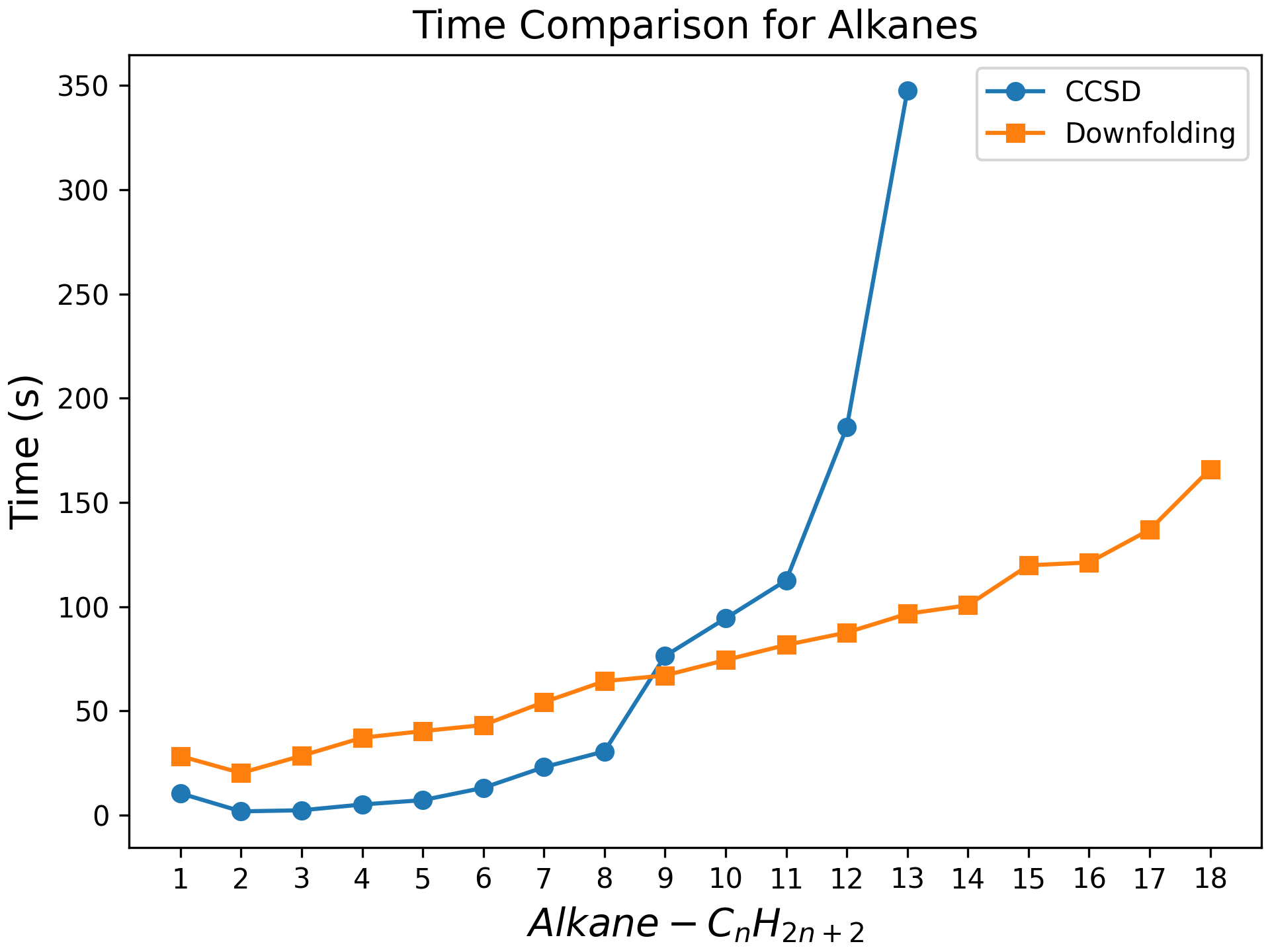}
\caption{Comparative runtimes of Downfolding and CCSD for different alkanes}
\label{fig:alkane-speed}
\end{minipage}
\hfill
\begin{minipage}[b]{0.4\textwidth}
\centering
\includegraphics[width=\textwidth]{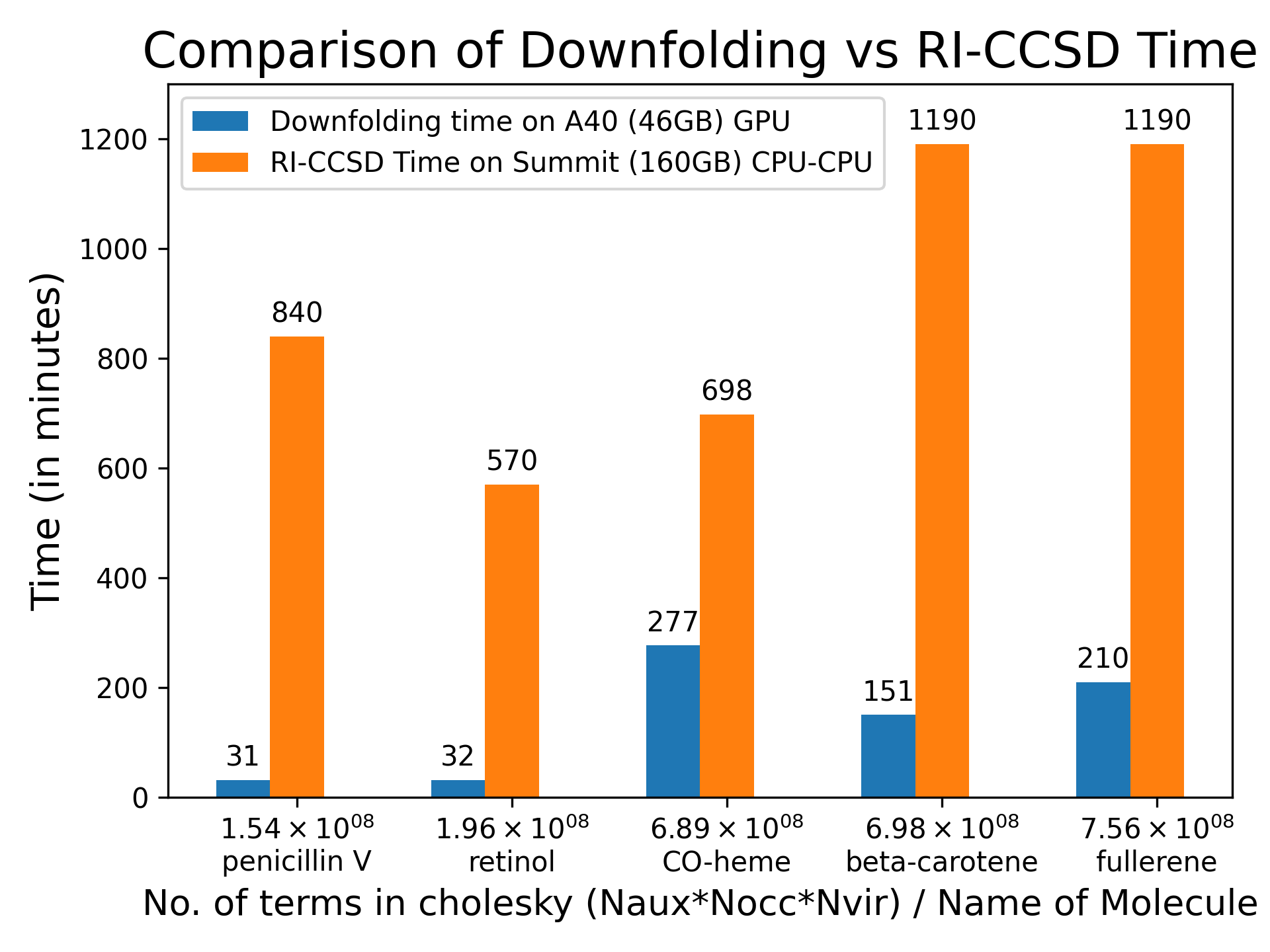}
\caption{Comparative runtimes of Downfolding and RI-CCSD for different molecules}
\label{fig:big-molecule-speed}
\end{minipage}
\end{figure}

\newpage
\begin{table}[h!]
\centering
\hspace{-5em}
\caption{Table for Downfolding energies runtimes and memory usage for molecules 1:$\beta-$Carotene, 2:Retinol, 3: $C_{60}$, 4: CO-Heme, 5: Penicillin-V, 6:Diclofenac on Intel Xeon , A40 Nvidia GPU 46 GB GPU, 75 GB RAM and comparison to state of the art. Electronic integrals for Molecules is represented in basis/aux-basis 1 in A\:(cc-pVDZ/aug-cc-pVTZ-RI),2 in B\:(def2-TZVPP/def2-TZVPP-RI), 3 in (cc-pVDZ/aug-cc-pvdz-RI),4 in C\:(Fe and five nitrogens around it is def2-tzvp and rest in def2-SVP and the auxilliary basis is def2-TZVP-RI), 5 and 6 in E,F\:(def2-tzvp/def2-tzvp-RI).The Time-1 and Mem-1 represents the state of the art times (in minutes) and memory requirements for RI-CCSD(T) implementation on Polaris supercomputer where two Nvidia A100 GPUs of total 80 GB GPU is being used\cite{datta2023accelerating}. The Time-1 is estimated assuming 100 iterations of CC, per CC iteration time is presented in ref\cite{datta2023accelerating}. The Time represents the downfolding time (in minutes)}
\label{tab:big-molecule-speed}
\begin{tabular}{|c|c|c|c|c|c|c|c|c|c|l} 
\cline{1-10}
\textbf{Mol} & \textbf{Rep.}  & $\mathbf{N_O}$ & $\mathbf{N_V}$ & $\mathbf{N_{aux}}$ & \textbf{Downfolding} & \textbf{Time} & \textbf{Mem} & \textbf{Time-1} & \textbf{Mem-1} &                       \\ 
\cline{1-10}
1        & A   & 148             & 692             & 6816            & -1552.42171499              & 151.28        &    (8/16)GB                   &    1190                &                  64 GB          &                       \\ 
\cline{1-10}
2           & B & 79              & 992             & 2496            & -853.61229138               & 31.8          & (4/10)GB                      &     570                &                       32 GB       &                      \\ 
\cline{1-10}
3            & C   & 180             & 660             & 6360            & -2278.79962753              & 210.16        & (8.44/16)GB                       &         1190           &  64GB                           & \multicolumn{1}{c}{}  \\ 
\cline{1-10}
4     & D                        & 185             & 840             & 4431            &                -3428.60238789                  & 277.35        & 10/40GB                  & 698                &        80 GB               &  \multicolumn{1}{c}{}  \\ 
\cline{1-10}
5      & E                & 92              & 766             & 2191            & -1501.94189954              & 30.71         & 6.1/24.2GB          & 141.9              & 68.1 GB                     &                       \\ 
\cline{1-10}
6        & F                & 76              & 591             & 1735            & -1663.48077496              & 10.05         & 2.9GB/15.3 GB         & 165.14             & 74.41 GB                    &                       \\
\cline{1-10}
\end{tabular}
\end{table}

\begin{table}[h!]
\hspace{-5em}
\caption{The table represents resource estimates i.e. Number of Qubits and Depth in the Clifford+T basis for emulating downfolding on Quantum Circuits for molecules: $1\to$ $\beta$-Carotene, $2\to$ Retinol, $3\to$ $C_{60}$, $4\to$ CO-Heme. The plots shows variations of resources for different tensor factors and different precision of representing the integrals and the cluster amplitudes on the Quantum Circuit.}
\label{tab:big-molecule-quantum-estimate}
\begin{tabular}{|c|c|c|c|c|cccc|}
\hline
\multirow{2}{*}{Mol}         & \multicolumn{1}{l|}{\multirow{2}{*}{norbs}} & \multirow{2}{*}{\# TF} & \multirow{2}{*}{Error} & \multirow{2}{*}{\# Qubits} & \multicolumn{4}{c|}{Depth(S,CNOT,H,T) for Precision}                                                                         \\ \cline{6-9} 
                                  & \multicolumn{1}{l|}{}                                    &                                        &                        &                                & \multicolumn{1}{c|}{1E-02}      & \multicolumn{1}{c|}{1E-03}      & \multicolumn{1}{c|}{1E-04}      & 1E-05      \\ \hline
\multirow{4}{*}{1} & \multirow{4}{*}{840}                                     & 6816                                   & $3.48\times 10^{-4}$           & 117                            & \multicolumn{1}{c|}{$6.71\times 10^8$}  &
\multicolumn{1}{c|}{$1.01\times 10^9$} & \multicolumn{1}{c|}{$1.34\times 10^9$} & $1.67\times 10^9$ \\ \cline{3-9} 
                                  &                                                          & 10224                                  & $3.5\times 10^{-4}$           & 121                            & \multicolumn{1}{c|}{$1.01 \times 10^9$} & \multicolumn{1}{c|}{$1.51\times 10^9$} & \multicolumn{1}{c|}{$2.01\times 10^9$} & $2.51\times 10^9$ \\ \cline{3-9} 
                                  &                                                          & 13632                                  & $3.5\times 10^{-4}$           & 121                            & \multicolumn{1}{c|}{$1.34\times 10^9$} & \multicolumn{1}{c|}{$2.01\times 10^9$} & \multicolumn{1}{c|}{$2.68\times 10^9$} & $3.35\times 10^9$ \\ \cline{3-9} 
                                  &                                                          & 17040                                  & -                      & 125                            & \multicolumn{1}{c|}{$1.67\times 10^9$} & \multicolumn{1}{c|}{$2.52\times 10^9$} & \multicolumn{1}{c|}{$3.35\times 10^9$} & $4.2\times 10^9$ \\ \hline
\multirow{4}{*}{2}          & \multirow{4}{*}{1071}                                    & 2496                                   & $7.43\times 10^{-4}$            & 108                            & \multicolumn{1}{c|}{$1.17\times 10^8$}  & \multicolumn{1}{c|}{$1.75\times 10^8$}  & \multicolumn{1}{c|}{$2.34\times 10^8$}  & $2.92\times 10^9$  \\ \cline{3-9} 
                                  &                                                          & 3744                                   & $7.1\times 10^{-4}$            & 108                            & \multicolumn{1}{c|}{$1.75\times 10^8$}  & \multicolumn{1}{c|}{$2.63\times 10^8$}  & \multicolumn{1}{c|}{$3.51\times 10^8$}  & $4.38\times 10^8$  \\ \cline{3-9} 
                                  &                                                          & 4992                                   & $7.58\times 10^{-4}$            & 112                            & \multicolumn{1}{c|}{$2.34\times 10^8$}  & \multicolumn{1}{c|}{$3.51\times 10^8$}  & \multicolumn{1}{c|}{$4.68\times 10^8$}  & $5.85\times 10^8$  \\ \cline{3-9} 
                                  &                                                          & 6240                                   & $7.51\times 10^{-4}$            & 112                            & \multicolumn{1}{c|}{$2.92\times 10^8$}  & \multicolumn{1}{c|}{$4.38\times 10^8$}  & \multicolumn{1}{c|}{$5.85\times 10^8$}  & $7.31\times 10^8$  \\ \hline
\multirow{4}{*}{3}        & \multirow{4}{*}{840}                                     & 6360                                   & $2.29\times 10^{-4}$           & 117                            & \multicolumn{1}{c|}{$5.89\times 10^8$}  & \multicolumn{1}{c|}{$8.84\times 10^8$}  & \multicolumn{1}{c|}{$1.18\times 10^9$} & $1.47\times 10^9$ \\ \cline{3-9} 
                                  &                                                          & 9540                                   & $2.52\times 10^{-4}$           & 121                            & \multicolumn{1}{c|}{$8.84\times 10^8$}  & \multicolumn{1}{c|}{$1.32\times 10^9$} & \multicolumn{1}{c|}{$1.76\times 10^9$} & $2.21\times 10^9$ \\ \cline{3-9} 
                                  &                                                          & 12720                                  & $2.55\times 10^{-4}$           & 121                            & \multicolumn{1}{c|}{$1.18\times 10^9$} & \multicolumn{1}{c|}{$1.77\times 10^9$} & \multicolumn{1}{c|}{$2.36\times 10^9$} & $2.95\times 10^9$ \\ \cline{3-9} 
                                  &                                                          & 15900                                  & $2.54\times 10^{-4}$           & 121                            & \multicolumn{1}{c|}{$1.47\times 10^9$} & \multicolumn{1}{c|}{$2.21\times 10^9$} & \multicolumn{1}{c|}{$2.94\times 10^9$} & $3.68\times 10^9$ \\ \hline
\multirow{4}{*}{4}    & \multirow{4}{*}{1025}                                    & 4431                                   & -                      & 117                            & \multicolumn{1}{c|}{$3.1\times 10^8$}  & \multicolumn{1}{c|}{$4.65\times 10^8$}  & \multicolumn{1}{c|}{$6.2\times 10^8$}  & $7.75\times 10^8$  \\ \cline{3-9} 
                                  &                                                          & 6646                                   & -                      & 117                            & \multicolumn{1}{c|}{$4.65\times 10^8$}  & \multicolumn{1}{c|}{$6.97\times 10^8$}  & \multicolumn{1}{c|}{$9.3\times 10^8$}  & $1.16\times 10^9$ \\ \cline{3-9} 
                                  &                                                          & 8862                                   & -                      & 121                            & \multicolumn{1}{c|}{$6.2\times 10^8$}  & \multicolumn{1}{c|}{$9.3\times 10^8$}  & \multicolumn{1}{c|}{$1.24\times 10^9$} & $1.55\times 10^9$ \\ \cline{3-9} 
                                  &                                                          & 11077                                  & -                      & 121                            & \multicolumn{1}{c|}{$7.75\times 10^8$}  & \multicolumn{1}{c|}{$1.16\times 10^9$} & \multicolumn{1}{c|}{$1.55\times 10^9$} & $1.93\times 10^9$ \\ \hline
\end{tabular}
\end{table}

\newpage
\begin{table}[h!]
\caption{The table represents the Number of Toffoli's and Number of Qubits for the quantum phase estimation circuits for different molecules for different sizes of tensor factors}
\label{tab:big-molecule-qpe}
\begin{tabular}{|c|c|c|c|}
\hline
\textbf{Molecule}                   & \textbf{\# TF's} & \textbf{\# Toffoli's in QPE} & \textbf{\# Qubits in QPE} \\ \hline
\multirow{4}{*}{retinol}            & 2496             & 5.79566E+13           & 23756                     \\ \cline{2-4} 
                                    & 3744             & 6.42007E+13           & 23757                     \\ \cline{2-4} 
                                    & 4992             & 6.84462E+13           & 23762                     \\ \cline{2-4} 
                                    & 6240             & 7.40703E+13           & 23763                     \\ \hline
\multirow{2}{*}{$\beta$-carotene}   & 6816             & 2.3614E+13            & 40748                     \\ \cline{2-4} 
                                    & 10224            & 2.77809E+13           & 42800                     \\ \hline
\multirow{4}{*}{$C_{60}$ Fullerene} & 6360             & 3.25465E+13           & 40750                     \\ \cline{2-4} 
                                    & 9540             & 3.8797E+13            & 42802                     \\ \cline{2-4} 
                                    & 12720            & 4.57649E+13           & 42804                     \\ \cline{2-4} 
                                    & 15900            & 5.15462E+13           & 83763                     \\ \hline
\multirow{4}{*}{CO-bound Heme}      & 4431             & 4.67249E+13           & 22748                     \\ \cline{2-4} 
                                    & 6646             & 5.84054E+13           & 41120                     \\ \cline{2-4} 
                                    & 8862             & 5.90529E+13           & 43172                     \\ \cline{2-4} 
                                    & 11077            & 6.51382E+13           & 43172                     \\ \hline
\end{tabular}
\end{table}


\begin{figure}[h!]
\begin{minipage}[b]{0.45\textwidth}
\centering
\includegraphics[width=\textwidth]{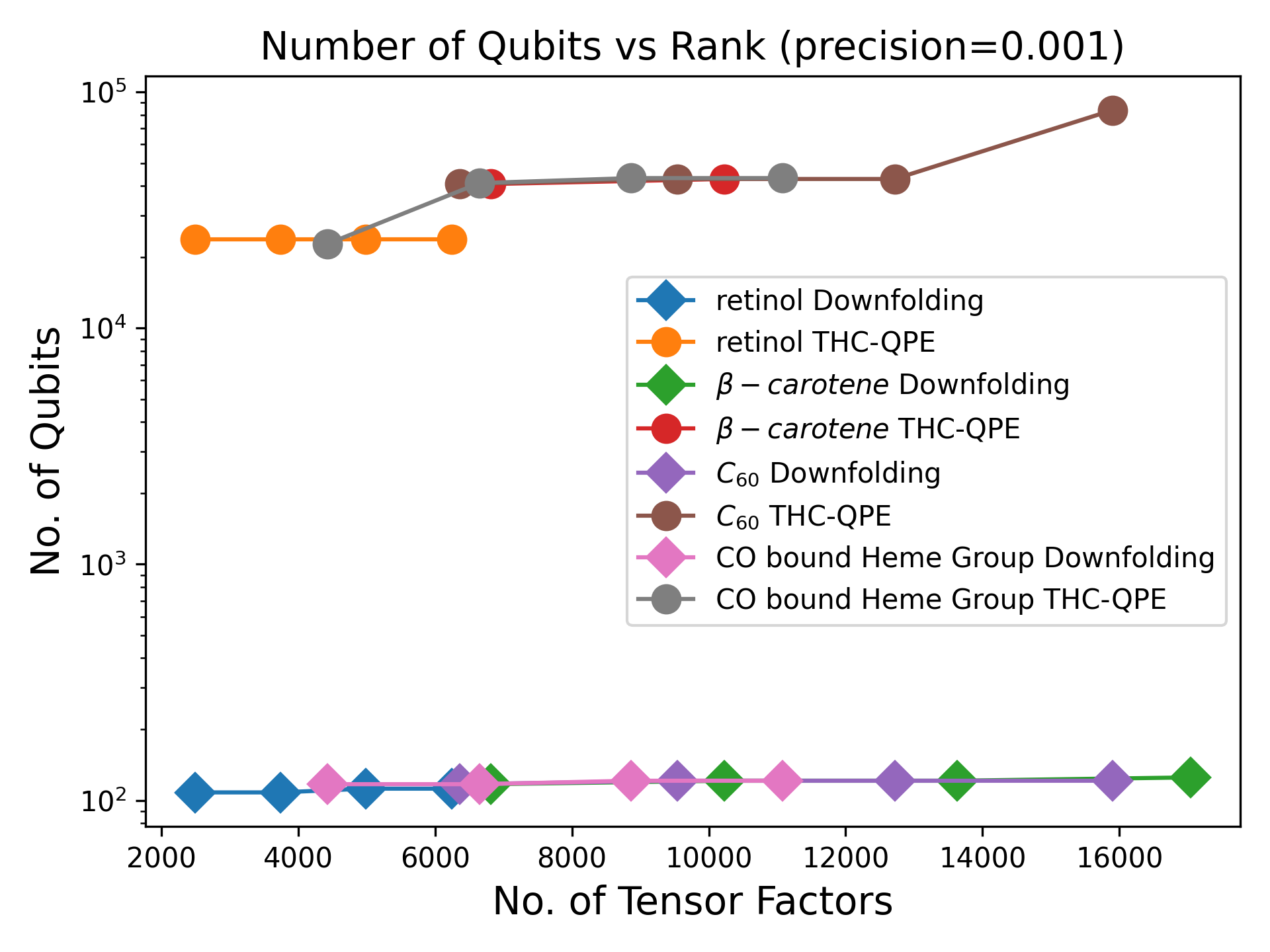}
\caption{Plot for Number of Qubits vs Rank}
\label{fig:quvsrank}
\end{minipage}
\hfill
\begin{minipage}[b]{0.45\textwidth}
\centering
\includegraphics[width=\textwidth]{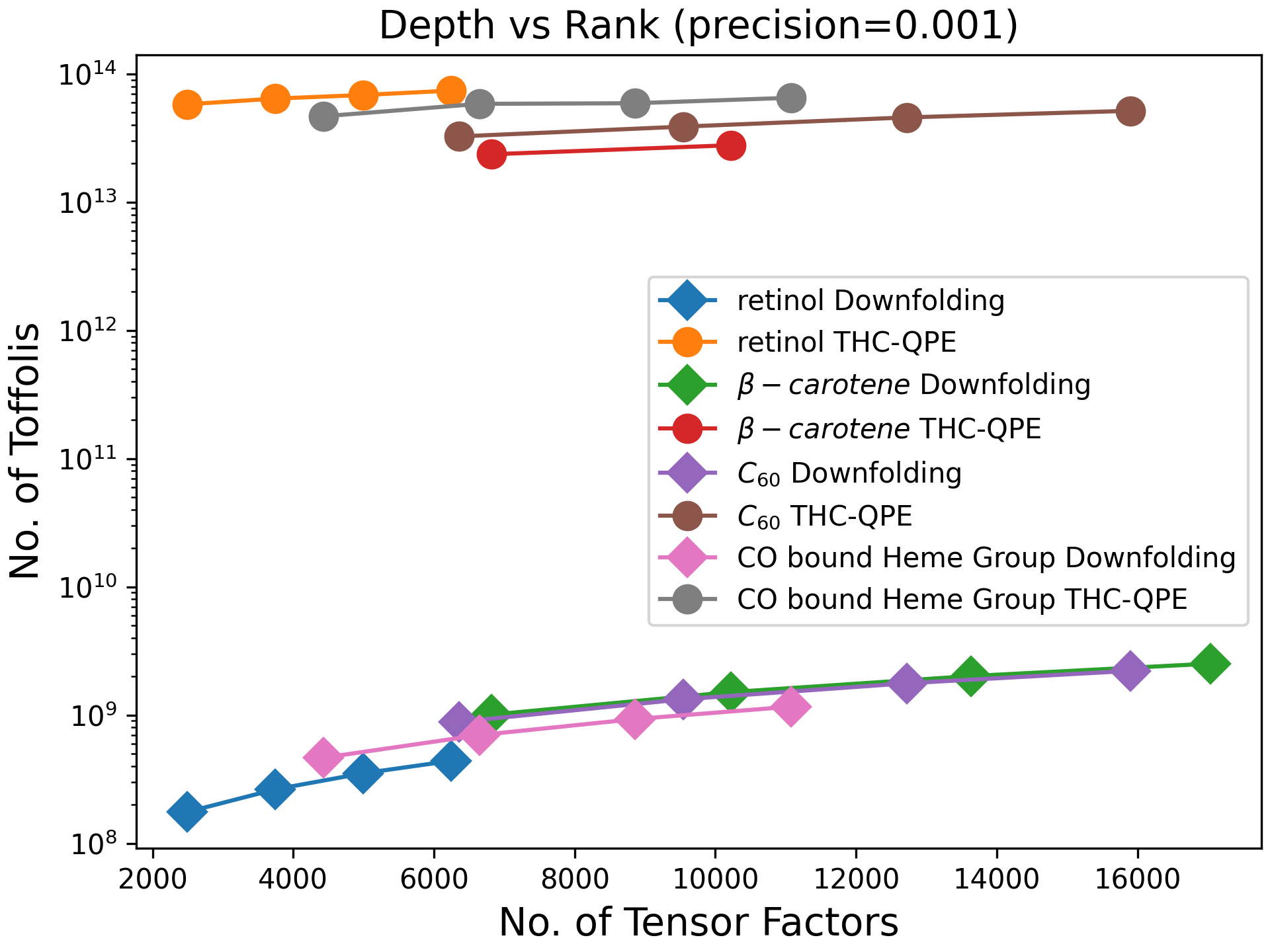}
\caption{Plot for Depth vs Rank}
\label{fig:depthvsrank}
\end{minipage}
\end{figure}

\newpage
\begin{figure}[h!]
\begin{subfigure}[b]{0.45\textwidth}
\centering
\includegraphics[width=\textwidth]{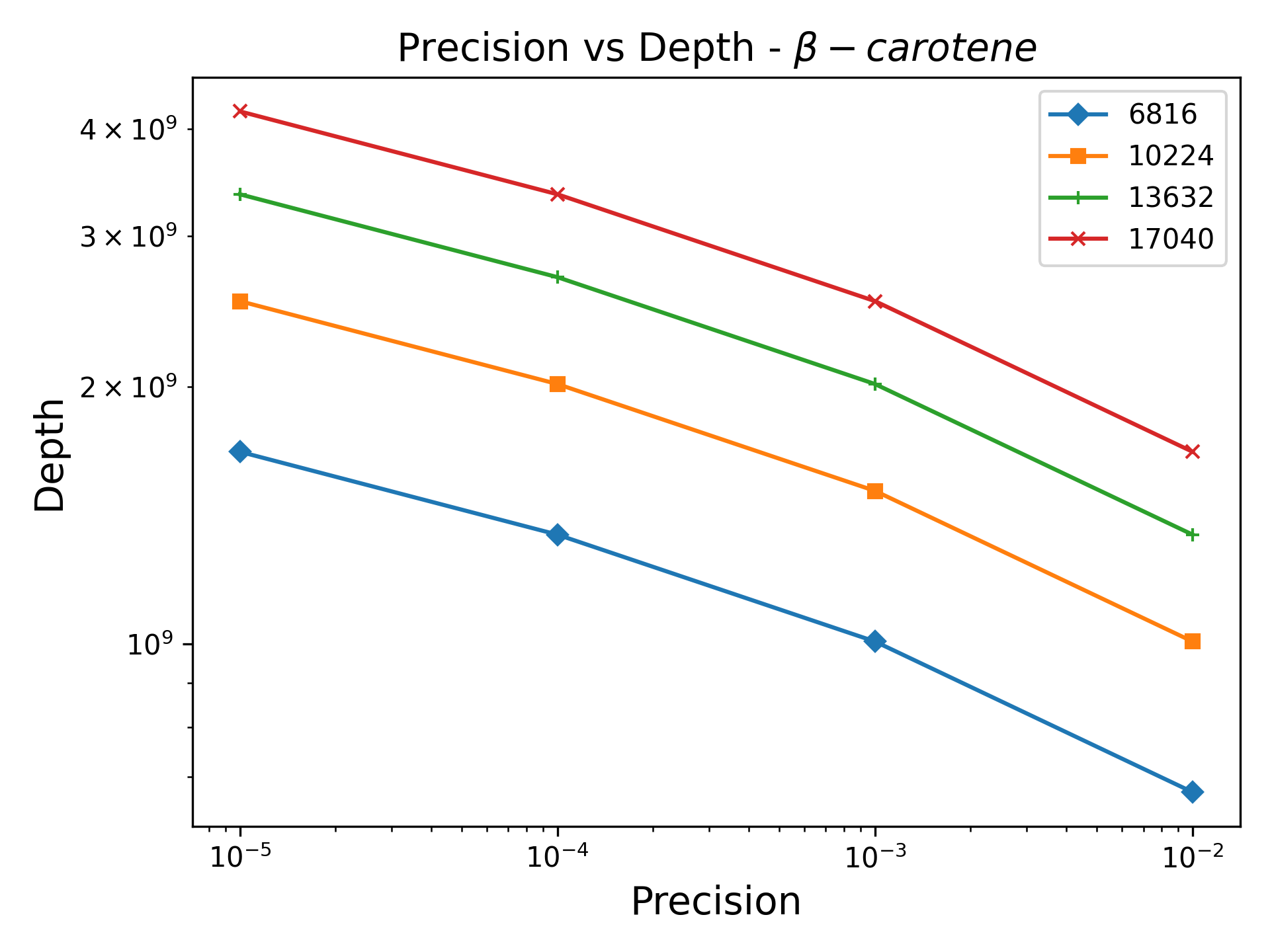}
\caption{}
\end{subfigure}
\hfill
\begin{subfigure}[b]{0.45\textwidth}
\centering
\includegraphics[width=\textwidth]{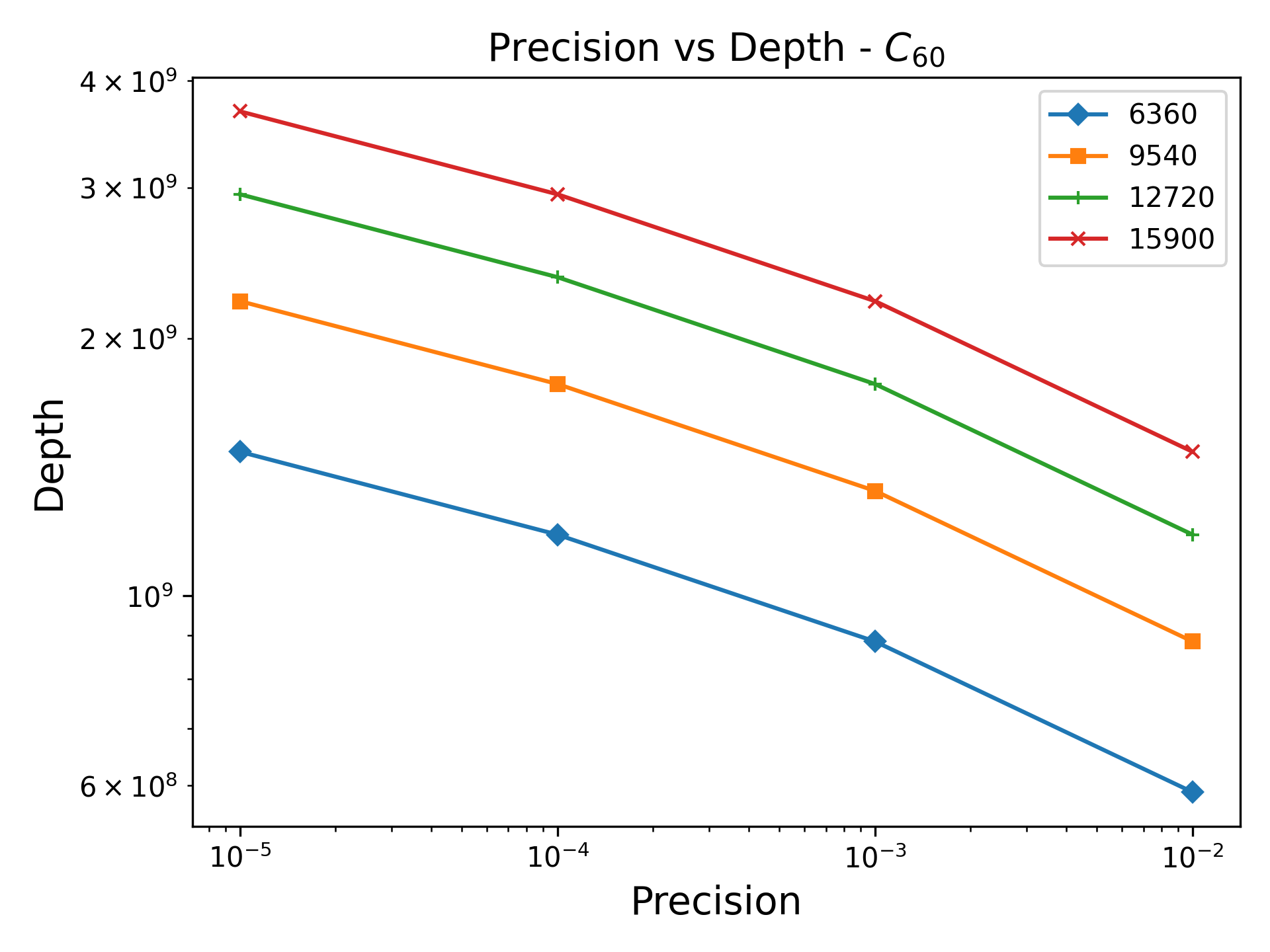}
\caption{}
\end{subfigure}
\hfill
\begin{subfigure}[b]{0.45\textwidth}
\centering
\includegraphics[width=\textwidth]{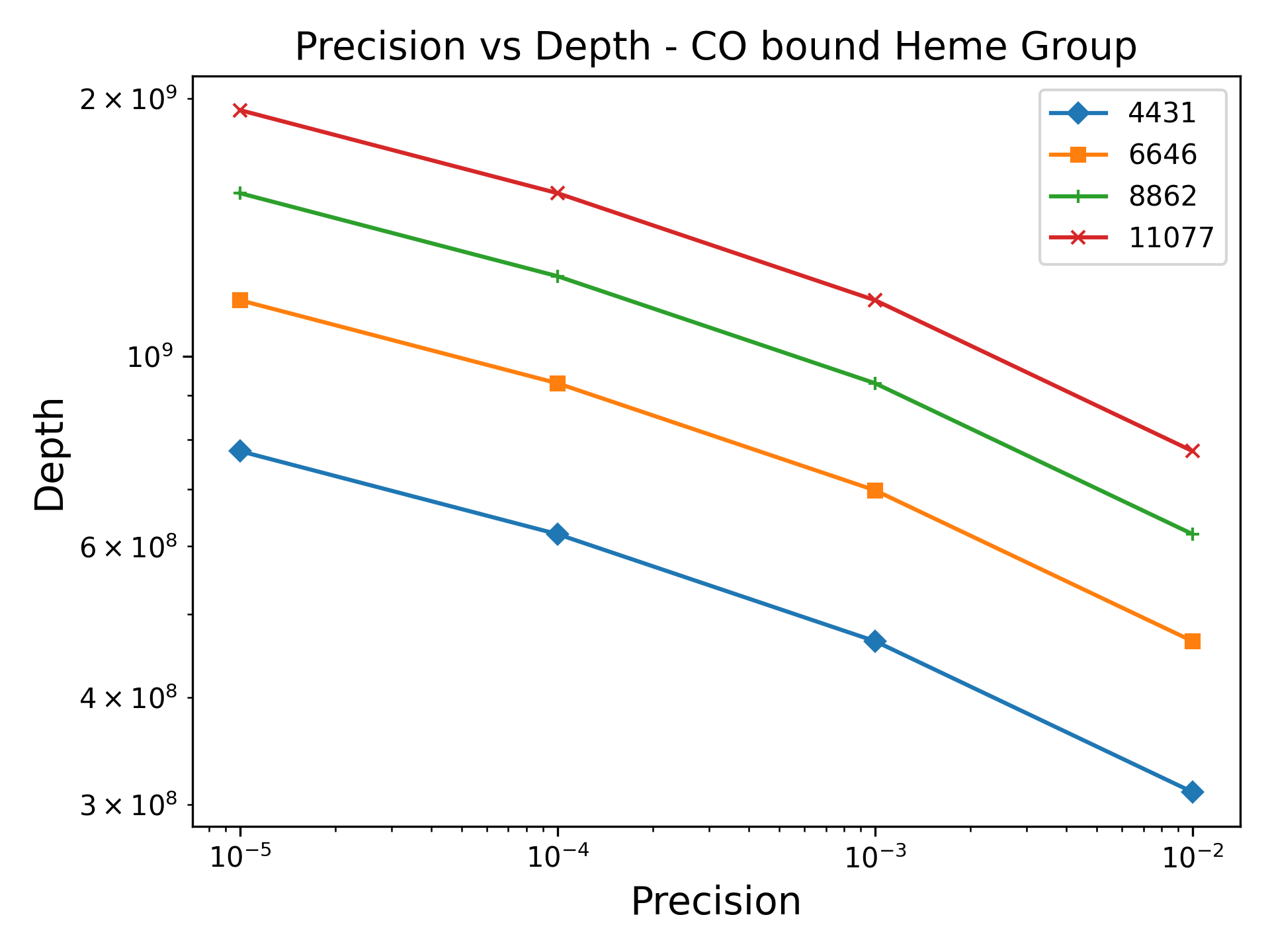}
\caption{}
\end{subfigure}
\hfill
\begin{subfigure}[b]{0.45\textwidth}
\centering
\includegraphics[width=\textwidth]{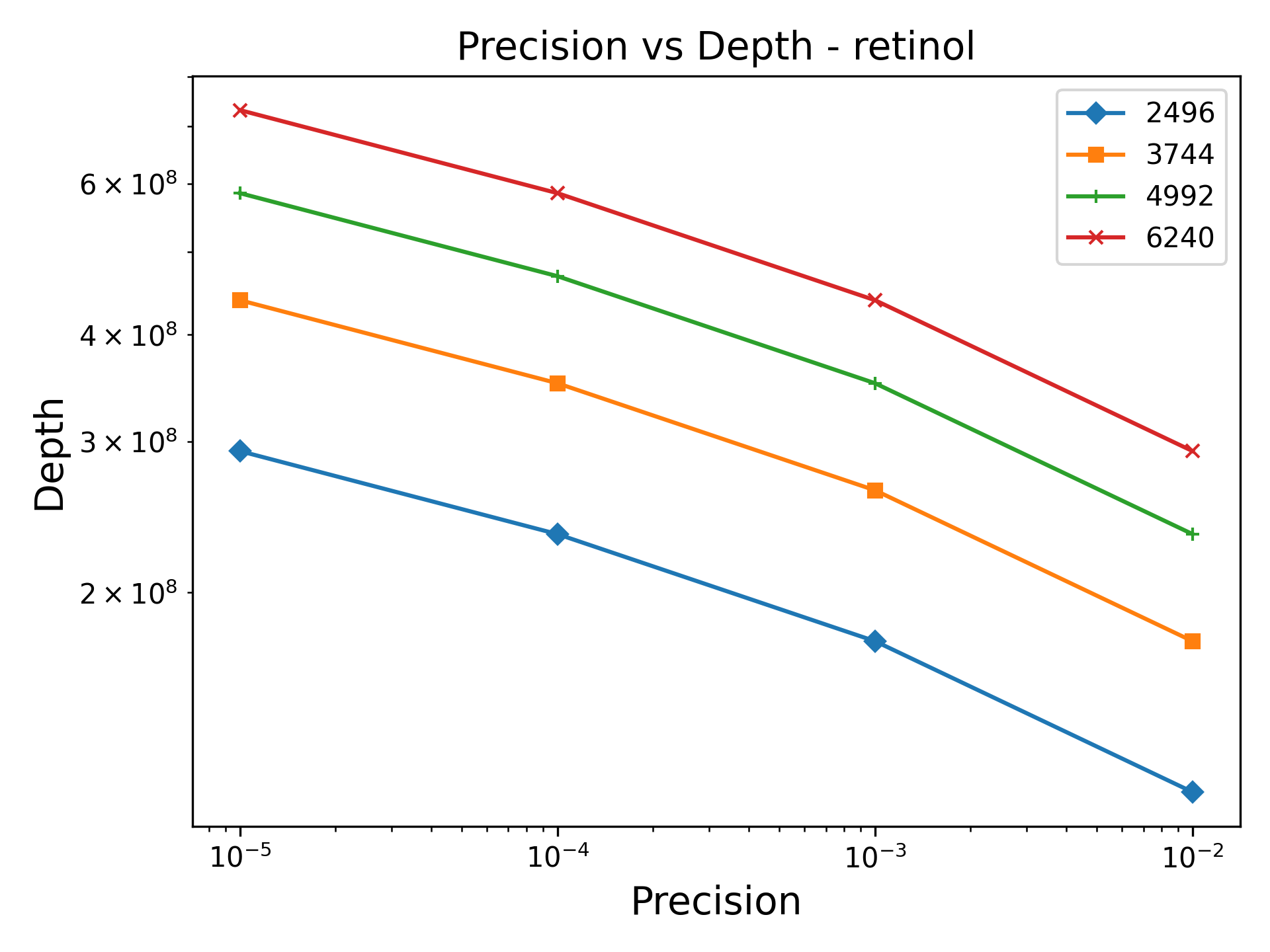}
\caption{}
\end{subfigure}
\caption{Number of Qubits vs Rank of Tensor factors for different molecules}
\label{fig:qubitsvsrank}
\end{figure}
\subsection{Full–Orbital FeMoCo Benchmark}
\label{sec:femoco}

\begin{figure}
    \includegraphics[scale=0.4]{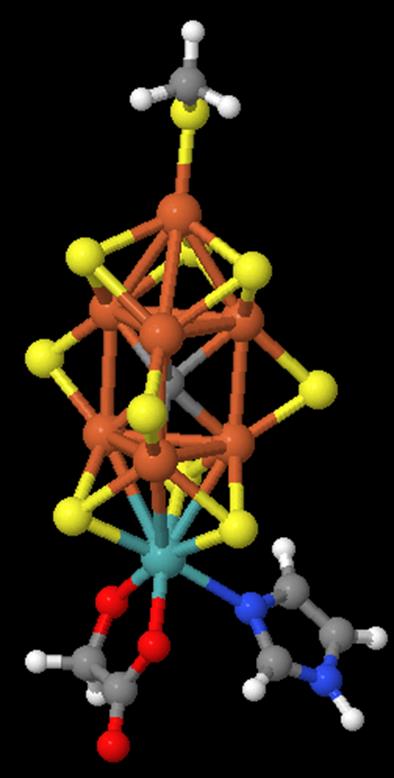}
    \caption{ Optimised “Structure I’’ model of the iron–molybdenum co-factor (FeMoCo) of nitrogenase, adopted from Ref. \cite{reiher2017elucidating}. The cluster consists of a distorted 
$[MoFe_7S_9C]^{3+}$ core with a $\mu_{6}$ -carbide (black) at the centre, three $\mu_{2}$-bridging sulfides (yellow), seven Fe atoms (orange) coordinated by sulfur and histidine/arginine ligands, and a terminal Mo atom (cyan) bound to a homocitrate ligand (grey/light red).}
    \label{fig:FeMoCo}
\end{figure}

In this study we consider structure 1 Fig\ref{fig:FeMoCo} of the FeMoCo proposed in
Reiher .et.al\cite{reiher2017elucidating}. This structure is a key intermediates along the catalytic cycle of the Molybdenum-Iron protein. \emph{Structure I} corresponds to the crystallographic resting state of the cofactor—an \(\mathrm{[MoFe_7S_9C]^{3+}}\) core with a central carbide and three $\mu$\(_2\)-bridging sulfides—embedded in a truncated ligand environment. After protonation-state geometry optimisation with B3LYP density-functional theory the model carries a formal charge of \(+3\) and an equal number of \(\alpha\)- and \(\beta\)-spin electrons, an overall singlet ground state (\(S=0\)). In Reiher.et.al the  “FeMoCo Hamiltonian’’ was obtained by projecting this \emph{Structure I} wave function onto a 54-orbital complete-active-space (CAS), yielding a strongly correlated, multireference problem that is an important benchmark for quantum-resource estimates in transition-metal catalysis\,\cite{reiher2017elucidating,von2021quantum,kim2022fault}. 
\par\noindent
In the present work we dispense with any CAS truncation and treat the
\emph{entire} $N_{\text{occ}}=235$, $N_{\text{vir}}=916$ spin‐orbital
manifold in the same \texttt{def2‐TZVP} basis.  
Our orbital-wise tensor-factorized downfolding compresses the full
valence–semi-core Hamiltonian to rank-2 form \emph{on the fly}, and the
block-encoded qubitized downfolding circuit achieves the same target
accuracy with only $67$–$121$ logical qubits and
$\sim\!10^{9}$ non-Clifford gates (Table~\ref{tab:femoco_resources}),
surpassing the Reiher baseline by roughly two orders of magnitude in
qubits and five in depth while eliminating the need for an empirically
selected active space.
The iron–molybdenum co-factor (FeMoCo) of nitrogenase \cite{lee2021even,lee2021even} is among the most challenging transition-metal systems considered in quantum chemistry.  
Unlike previous studies that restrict the problem to a $\sim\!50$–$100$-orbital active space, the present benchmark treats the \emph{entire} all-electron Hilbert space in a \texttt{def2-TZVP} basis.  
The resulting mean-field reference has
$\text{charge}=+3$, 
$N_{\text{ao}}=1151$, 
$N_{\text{occ}}=235$, 
$N_{\text{vir}}=916$, and an SCF energy  
$E_{\text{SCF}}=-14,547.734$~Ha.

\paragraph*{Classical TFHD performance.}
Orbital-wise tensor-factorized Hamiltonian downfolding (TFHD) was executed on eight NVIDIA A100 GPUs.  
The 916 virtual orbitals were eliminated sequentially; the cumulative wall-time was 

\[
t_{\text{TFHD}} = 3.53~\text{h},
\qquad
t_{\text{factorisation}} = 1.66~\text{h},
\]

corresponding to an average of $13.9$s per orbital—a runtime consistent with the $O(N^{3})$ prediction of Sec.~\ref{SR-TFHD}.  
The total correlation energy recovered is  

\[
E_{\text{corr}}^{\text{TFHD}} = -9.7885~\text{Ha},
\quad
E_{\text{tot}} = E_{\text{SCF}} + E_{\text{corr}} = -14\,557.523~\text{Ha}.
\]

The calculation therefore delivers a chemically significant correction for \emph{every} valence and semi-core orbital at cubic classical cost.

\paragraph*{Fault-tolerant resource comparison.}
Table~\ref{tab:femoco_resources} compares tensor-hypercontraction QPE (baseline) with block-encoded qubitized downfolding (QD) for four target accuracies.  
QD resources are quoted for the largest of the three oracle fragments (\texttt{PHQ}); the two smaller fragments are listed in parentheses.

\begin{table}[h]
  \centering
  \caption{Fault-tolerant resources for the full-orbital FeMoCo Hamiltonian
           ($N=1151$).  Depth is reported in non-Clifford ($T$ or Toffoli)
           gates.  Qubit counts for QD correspond to the largest oracle
           fragment (\texttt{PHQ}); the two smaller fragments require
           67 and 97 logical qubits, respectively.}
  \label{tab:femoco_resources}
  \renewcommand{\arraystretch}{1.05}
  \begin{tabular}{|c|cc|cc|}
    \hline
    $\epsilon$ (Ha) &
    \multicolumn{2}{c|}{QPE (THC)} &
    \multicolumn{2}{c|}{QD (block-encoded)} \\ \cline{2-5}
       & Qubits & Depth & Qubits & Depth \\
    \hline
    $10^{-2}$ & $8.44\times 10^{4}$  & $5.15\times 10^{13}$ & $121$ & $9.20\times 10^{8}$  \\
    $10^{-3}$ & $8.44\times 10^{4}$  & $5.15\times 10^{13}$ & $121$ & $1.38\times 10^{9}$ \\
    $10^{-4}$ & $8.44\times 10^{4}$  & $5.15\times 10^{13}$ & $121$ & $1.84\times 10^{9}$ \\
    $10^{-5}$ & $8.44\times 10^{4}$  & $5.15\times 10^{13}$ & $121$ & $2.30\times 10^{9}$ \\
    \hline
  \end{tabular}
\end{table}

Across all accuracy targets, QD reduces the logical-qubit requirement
from \(\sim\!8.4\times10^{4}\) to fewer than \(1.3\times10^{2}\) and
cuts non-Clifford depth by five orders of magnitude.  The depth growth
with \(1/\epsilon\) follows the predicted
\(D_{\text{QD}} = \mathcal{O}\bigl(N^{2}\log(1/\epsilon)\bigr)\)
behaviour, while the qubit count remains logarithmic in \(N\), placing
the full-orbital FeMoCo problem squarely within the projected budgets
of next-generation neutral-atom and superconducting platforms.

\paragraph*{Significance.}
This end-to-end demonstration confirms that tensor-factorized,
orbital-wise downfolding can (i) recover dynamical correlation for a
$>1100$-orbital, charge-three transition-metal cluster at cubic
classical cost and (ii) map the problem to a fault-tolerant circuit
with logarithmic qubit scaling and quadratic depth.  The results
therefore substantiate the claim that block-encoded qubitized
downfolding opens a realistic pathway to quantum advantage for catalysis
and other chemically complex, strongly correlated systems.
\section{Results and Discussion}
We have distributed the results into three components the first component of the result deals with determining the computational complexity of the optimized quantum chemistry calculations resulting from downfolding the single reference CC, MRCC and the general full configuration calculations. The downfolding calculations are carried out in a tensor factorized representation. We also prove a bunch of theorems that enable porting this tensor factorized computations onto quantum circuits using the block-encoding formalism. The second component of our results deal with constructing the tensor networks Fig\ref{TensorOps1},Fig\ref{TensorOps2} to showcase pictorially how the complexity is curbed on the classical computers.  We also build the Quantum circuits Fig\ref{Circuits} for implementing the tensor network operation arising from downfolding on Quantum computers. The third component of our results deals with getting bench-marking data that demonstrate the accuracy towards ranking conformers, accuracy of energy values with respect to CCSD and MP2, time and memory requirements for determining post-HF correlation energy from downfolding for medium to large molecules and or chemical complexes. Comparison of energies to state-of-the-art. We then also present a variety of data on the Quantum circuit resources for large molecules, the variation of the number of qubits and variation of the depth of the quantum circuit in the Clifford+T basis with the number of tensor factors. The accuracy towards approximating the electronic integrals in the tensor factorized rentation. We provide comparisons to the state-of-the-art Qubitized Phase estimation in the tensor hyper-contraction representation of \cite{lee2021even}. We also demonstrate two cases of downfolding (i)closed form expression for the unitary variant of the downfolding similarity transformation for the coupled cluster singles and doubles case  Sec\ref{Unitary} and (ii) we demonstrate a family of downfolding similarity transformation for singles and paired doubles where the renormalized Hamiltonian within the two electron interaction regime remains self-similar Sec\ref{MR-Downfolding-Singles-Paired-Doubles}. Below we discuss the results for all the three components we just discussed.
\par\noindent
In the approach of Hamiltonian Downfolding at every step one orbital is decoupled, i.e. all the interactions terms coupling the one-orbital via one-particle, two-particle and higher order terms become zero. This is requirement of the Bloch equation with the form of generator of similarity transformation $\eta$ eq.\eqref{FormOfEta}. With a tensor factorized representation of the generator $\eta$ and the electronic integrals eq. \eqref{tensor_factors_electronic} and the general expression for ordering fermionic strings(strings of creation and annhiliation operators) into irreducible representations the Bloch equation can be reduced to hierarchy of families of cluster equations that is given by eq.\eqref{FullClusterDownfoldingEquations}. From the computation of the order of complexity eq.\eqref{CubicComplexity}, generic tensor operation diagram Fig.\ref{fig:GeneralTensorOps} and the tensor factorized form of the Hamiltonian we conclude that all the Multiconfigurational Quantum Chemistry calculations have a cubic complexity.
\par\noindent 
For single configuration downfolding with singles and doubles for small molecules we present the benchmarking results below. All the calculations performed below are done on Nvidia V100 32 GB GPU. The results showcase the overall correlation capture efficacy of downfolding compared to MP2 and also efficacy towards conformer ranking. In tables \ref{tab:aspirin-631g},\ref{tab:asprin-def2svp}, \ref{tab:aspirin-ccpvdz} and figure \ref{fig:aspirin-benchmark} we show that for different Aspirin conformers downfolding captures more than $98\%$ correlation energy of CCSD for all three basis sets, 6-31g, def2-svp and ccpvdz. On the other-hand  for both the molecules- Citric acid and Aspirin MP2 reproduces the electronic correlation $>97\%$ of CCSD correlations). So we conclude that for small organic molecules with H, C, O downfolding captures more correlations than MP2. 
\par\noindent
Importantly also for both Citric acid and Aspirin conformers, for all basis sets studied as seen from figures Fig.\ref{fig:aspirin-benchmark},Fig.\ref{fig:citric_benchmark} downfolding reproduces the energy rankings accurately with small deviations. However, the shape of the energy curves agree with CCSD showcasing the robustness of the downfolding protocol. The deviations between the aforementioned curves can be attributed to the differential capture of electronic correlations via tensor factorized downfolding compared to one the one-shot similarity transformation in CCSD that decouples the HF state from its excitations.
\par\noindent
The result below demonstrates the efficiency towards conformer ranking with increasing basis set size for MP2 and downfolding compared to CCSD. We show that for pentane with a large basis set, aug-ccpvdz, (table \ref{tab:pentane-augccpvdz} and figure \ref{fig:pentane-benchmark}) the energy rankings is same for all three methods: MP2, CCSD and downfolding. These captures the accuracy of the downfolding technique for the cases where also both MP2 and CCSD accurately captures the conformer rankings. This shows that for highly correlated basis sets downfolding, CCSD and MP2 captures differential rankings between n-alkane chain conformers in the same way. 
\par\noindent \emph{Differential capture of electronic correlations:}Below we demonstrate cases where downfolding captures significantly more correlations than MP2 compared to CCSD. In table \ref{tab:miscbenchmark} and figure \ref{fig:energy-benchmark-misc} we study a variety of molecules where MP2 cannot accurately calculate the electronic correlations due to presence of one or more of the following: strong electrov alent bonds, delocalization of electrons, triple bonds, etc. For 6-31g basis sets, downfolding reproduces more than $95.9\%$ of CCSD correlations for all molecules studied, whereas MP2 could capture $83.7\%-89.1\%$ correlations. For larger basis sets(ccpvdz, def2-tzvp) downfolding consistently captures more correlations than MP2. In some cases downfolding shows higher correlation energies(magnitudes) than CCSD. This is again owing to the way downfolding captures correlation differently from CCSD. 
\par\noindent \emph{Accuracy of TF:} In table \ref{tab:tf-error} and figure \ref{fig:tf-error} we provide a detailed description of how the size of auxiliary basis sets for tensor factorization affect downfolding energies. For different n-alkanes with carbon atoms, 1 to 10, we show a drastic reduction of error in the total energy by increasing the size of the the auxiliary basis set. For basis set sizes more than $N_{aux}$, the auxiliary basis set size for Cholesky Decomposition, Tensor factorization gives errors of less than 1 mHa in the energy values. For auxiliary basis sizes of $2N_{aux}$ and more, the tensor factorization error is negligible.
\par\noindent\emph{Comparative Timings:} Downfolding is inherently faster than CCSD due to its reduced operational scaling complexity. Also orders of magnitude reduction in storage complexities allow for much greater parallelization which in turn, increases the speed of this formalism drastically than CCSD. In figure \ref{fig:iteration_time}, we study two molecules, Penicillin V and Diclofenac and show how the downfolding correlation energy accumulates to give the total correlation energy. Here we can see how the average run time for each iteration step reduces significantly as we downfold orbitals one by one, as a result of a reduction in dimension of the effective Hamiltonian. From table \ref{tab:alkane-speed} and figure \ref{fig:alkane-speed}, it is evident how downfolding provides a significant advantage in run-time speeds of high accuracy quantum chemistry calculations than CCSD. From table \ref{tab:big-molecule-speed} we can see the comparison of downfolding with RI-CCSD(T) with respect to speedup and memory usage. For this calculations of the big molecules we used the A40 46 GB GPU. We can also see that CCSD goes out of memory where as downfolding can work on much smaller storage requirements.
\par\noindent
\emph{Quantum Resources-}The Quantum Circuit results in table\ref{tab:big-molecule-quantum-estimate}, table\ref{tab:big-molecule-qpe}  and figures Fig.\ref{fig:quvsrank}, Fig.\ref{fig:depthvsrank}, Fig.\ref{fig:qubitsvsrank} demonstrate that our qubitized-block encoding based Hamiltonian downfolding approach  offers a substantial improvement over the quantum phase estimation\cite{lee2021even}. In QPE, achieving high precision requires repeated applications of controlled-unitary operations that scale exponentially as $O(1/\epsilon)$, where $\epsilon$ is the desired precision in phase estimation. Moreover, QPE requires a register of $O(log(1/\epsilon)$ ancilla qubits to store phase information.
\par\noindent
Our results show that for a target precision of $10^-5$ the circuit depth in QPE is of the magnitude of $10^{13}$ while our approach achieves similar precision with depths of $10^8-10^9$ even for  molecules, representing a nearly $10000x$ reduction in circuit depth. A key factor accounting for this drastic reduction is due to the block-encoding scheme in our approach, where the scaling with precision is only $O(log(1/\epsilon)$ \cite{ross2016optimalancillafreecliffordtapproximation,Kliuchnikov_2013}. By directly encoding the Hamiltonian onto the circuit we eliminate the need for repeated controlled-unitary operations, a major bottleneck in QPE\cite{MartynChuang2021}.
\par\noindent
Furthermore, our approach requires significantly fewer qubits. For example the number of qubits needed for complex systems like the Heme bound CO complex is in the range of 100-150 qubits, while QPE requires upwards of 40,000 qubits Fig\ref{fig:qubitsvsrank} for the same system. The reduction in logical qubit count further emphasizes the feasibility of our approach on near-term fault tolerant quantum devices.
\section{Pseudocode for Tensor Factorized Hamiltonian Downfolding}
\subsection*{Canonical Polyadic ALS Tensor Factorization}
\begin{algorithm}
\caption{\textsc{cp3 gpu} (in \texttt{factorization module})}
\begin{algorithmic}[1]
\Require 3‑tensor list ${L_{pq}^{(\chi)}}$ distributed over GPUs, target rank $R$, optional initial factors $(A,B,C)$
\If{saved factors exist on disk}
\State Load $(A,B,C)$ and \Return
\EndIf
\State Randomly initialize factor matrices if not provided
\For{$iter \gets 1$ \textbf{to} $\mathsf{max_iter}$}
\For{$mode \in {0,1,2}$}
\State $X \gets$ unfold tensor(s) along \emph{mode}
\State $lhs \gets V_{\neg mode}^\top V_{\neg mode}$ (Gram)
\State $rhs \gets$ \Call{update factor new}{$X,A,B,C,mode$}
\State Solve $lhs,\Theta = rhs$ for updated factor in shared memory
\EndFor
\If{residual $<\mathsf{tol}$} \textbf{break}
\EndIf
\EndFor
\State Persist $(A,B,C)$ to disk (\texttt{TFPath})
\Return $(A,B,C)$
\end{algorithmic}
\end{algorithm}
\subsection*{Driver: \texttt{downfolding}}
\begin{algorithm}
\caption{\textsc{DownfoldingWorkflow}}
\begin{algorithmic}[1]
\Require CLI arguments $\langle\mathsf{geom file},;\mathsf{bit},;\mathsf{basis},;p {div list},;\mathsf{charge},;\mathsf{aux},;\mathsf{scf cycle},;r,;\mathsf{start},;\mathsf{stop}\rangle$
\State Read molecular geometry from $\mathsf{geom file}$ and instantiate a \textsc{PySCF} \texttt{Mole} object $\mathcal{M}$
\State Perform (or load) RHF calculation; save checkpoint to disk
\Statex\Comment Results stored in $\mathsf{SCFPath}$ and exposed as $mfd$ (molecular Fock dictionary)
\State Build or load Cholesky vectors $L_{pq}$ via \Call{create cholesky distributed} {$mfd,\mathcal{M},\mathsf{auxbasis}$}
\State Compute total energy $E_{\mathrm{DF}}$ via \Call{converge ccsd}{$mfd,\mathsf{tfFound},r,n_{\mathrm{aux}},L_{pq},\mathsf{start},\mathsf{stop}$}
\State Output $E_{\mathrm{DF}}$ and timings
\end{algorithmic}
\end{algorithm}

\subsection*{Tensor‑Factorized CCSD Downfolding}
\begin{algorithm}[H]
\caption{$\mathsf{converge ccsd}$ (in $\mathsf{tf downfolding module}$)}
\begin{algorithmic}[1]
\Require RHF checkpoint $mfd$, tensor‑factor flag $tfFound$, rank factor $r$, $n_{\mathrm{aux}}$, Cholesky slices $L_{pq}$
\Statex Optional: orbital window $[\mathsf{stop},,\mathsf{start}]$
\If{$tfFound = \textsc{False}$}
\State $R \gets r,n_{\mathrm{aux}}$
\State $(X,Y,Z) \gets$ \Call{cp3 gpu}{$L_{pq},R$} \Comment factorize $L_{pq}$ on multi‑GPU
\Else
\State Load pre‑computed $(X,Y,Z)$ from disk
\EndIf
\State Partition factor matrices: $oY,oZ$ (occupied) and $vY,vZ$ (virtual)
\State Precompute intermediates ${\mathcal{O}V}$ using batched \textsc{einsum} 
\State Initialize cluster amplitudes $(t_1,t_2,t_{21})$ via \Call{init amps}{\dots}
\For{$n\gets \mathsf{start}$ \textbf{downto} $\mathsf{stop}$}
\Repeat
\State $(t_1,t_2,t_{21},p_{div}) \gets$ \Call{update amps}{$t_1,t_2,t_{21}$, intermediates, $p_{div}$}
\State $\Delta E \gets E_{\mathrm{corr}}^{\text{new}}-E_{\mathrm{corr}}^{\text{old}}$
\Until{$|\Delta E|<\varepsilon_E$ \textbf{and} $|\Delta t|<\varepsilon_t$ \textbf{or} $k=\mathsf{max_cycle}$}
\State Freeze the last virtual orbital, shrink Fock and factor matrices, regenerate intermediates
\EndFor
\Return $E_{\mathrm{HF}}+\sum E_{\mathrm{corr}},; \sum E_{\mathrm{corr}},;\textsc{Converged?}$
\end{algorithmic}
\end{algorithm}

\subsection*{Amplitude Update Kernel (\texttt{update amps}) —  rules}
\subsection*{Symbol Legend}
\begin{itemize}
\item $i,j,k,\ell$ – occupied indices; $a,b,c,d$ – active virtual; $N$ – last (downfolded) virtual.
\item $t_1^{i}$, $t_2^{ija}$, $t_{21}^{ijb}$ – cluster amplitudes.
\item $F^{ov}$, $F^{oo}$, $F^{vv}$ – Fock‑like intermediates.
\item $\mathcal{O}V_{\dots}$ – three‑index Coulomb vertices obtained from CP‑ALS factors $X,oY,oZ,vY,vZ$.
\item $W^{\ast}$ – composite “$W$‑tensors’’ (e.g.\ $W^{oooo},,W^{voov},\dots$).
\end{itemize}

\subsection*{Grouped $t_1$ Equation}
\noindent The updated amplitude is a sum of 11 groups ($G1\dots G11$):

\begin{enumerate}
\item \textbf{Linear Fock},: $\mathcal{T}^{(1)} = F^{ov}{iN}$
\item \textbf{Quadratic Fock},: $\mathcal{T}^{(2)} = -2 F^{ov}{kN} t_1^{k} t_1^{i}$
\item \textbf{Virtual shift},: $\mathcal{T}^{(3)} = (F^{vv}{NN} + F^{v}{N}) t_1^{i}$
\item \textbf{Occupied shift},: $\mathcal{T}^{(4)} = -F^{oo}{ki} t_1^{k}$
\item \textbf{Mixed Fock--CCD},: $\mathcal{T}^{(5)} = (2-P{ij}) F^{ov}{kc} t_2^{ikc}$
\item \textbf{Cubic Fock},: $\mathcal{T}^{(6)} = F^{ov}{kN} t_1^{k} t_1^{i}$
\item \textbf{Direct Coulomb},: $\mathcal{T}^{(7)} = (2-P_{ij}),\mathcal{O}V_{kiN},t_1^{k}$
\item \textbf{Screened Coulomb},: $\mathcal{T}^{(8)} = (2-P_{ij})\bigl[\mathcal{O}V_{kcN} t_2^{ikc} + \mathcal{O}V_{kdN} t_{21}^{ikd}\bigr]$
\item \textbf{Triple vertex},: $\mathcal{T}^{(9)} = \mathcal{O}V_{NNN} t_1^{k} t_1^{i}$
\item \textbf{3‑body exchange},: $\mathcal{T}^{(10)} = (2-P_{ij}) W^{oooV}{k\ell c} t_2^{kl}$ (GPU‑tiled)
\item \textbf{Triple exchange},: $\mathcal{T}^{(11)} = (2-P{ij}) W^{oooV}{k\ell c} t_1^{\ell} t_1^{k}$
\end{enumerate}
\vspace{0.3em}
\noindent Here $P{ij}$ is the antisymmetrizer: $P_{ij} X_{ij}=X_{ij}-X_{ji}$.

\subsection*{Residual Construction for $t_2$ and $t_{21}$}
\begin{itemize}
\item Build residuals $R_{2}^{ija}$ and $R_{21}^{ijb}$ as the sum of 12 $W$‑tensor groups $\mathcal{W}^{(g)}$.
\item Each group is evaluated by a two‑level tiling loop over CP‑ALS factor indices $(q,p)$.
\item GPU memory feedback updates the partition vector $p_{div}$ on each call.
\item Final amplitudes: $t_2^{new} = t_2 + R_2/\varepsilon_{ij,aN}$, $; t_{21}^{new} = t_{21} + R_{21}/\varepsilon_{ij,Nb}$.
\end{itemize}

\begin{algorithm}[H]
\caption{\textsc{update amps} (pseudo-code)}
\begin{algorithmic}[1]
\Require $(t_1,t_2,t_{21}),F,\mathcal{O}V,W,X,oY,oZ,vY,vZ,p_{div}$
\State Compute intermediates $F^{ov},F^{oo},F^{vv},L^{oo},L^{vv}$
\State $t_1^{new}\leftarrow \sum_{g=1}^{11} c_g,\mathcal{T}^{(g)}$ \Comment grouped rules above
\State $(R_2,R_{21})\leftarrow$ \Call{Build Residuals}{$t_1,t_2,t_{21},W,\mathcal{O}V,X,oY,oZ,vY,vZ,p_{div}$}
\State $t_2^{new}\leftarrow t_2 + R_2/\varepsilon_{ij,aN}$
\State $t_{21}^{new}\leftarrow t_{21} + R_{21}/\varepsilon_{ij,Nb}$
\State \Return $t_1^{new},t_2^{new},t_{21}^{new},p_{div}$
\end{algorithmic}
\end{algorithm}

\section{Future Directions}
In this work we have presented the orbital-wise Hamiltonian Downfolding that is efficiently implemented in tensor factorized representation on GPUs and can be implemented on future Quantum Computers. Future work will be pursued to in one hand develop computational chemistry packages that incorporate tensor factorization and multi-configuration effects where downfolding is used to speedup those calculations and alongside further benchmarks for d-orbital block systems will be provided. 
Further integrating Hamiltonian downfolding with tree tensor networks (TTNs) \cite{nakatani2013efficient,murg2015tree} and density matrices presents a fresh avenue to scaling Quantum Chemistry computations for molecules at cheaper computational cost.
\appendix
\section{Lowdin Decomposition of The Hamiltonian H For Primary Space (P) and Secondary Space (Q)}\label{LowdinDecomposition}
For decoupling the Nth Molecular Orbital where $N\in \mathcal{V}$ the projection operator for primary space $P$ and secondary space $Q$ are given by,
\begin{eqnarray}
 P_{(N)}&=&(1-\hat{n}_{N\uparrow})(1-\hat{n}_{N\downarrow}),Q_{(N)}=\hat{n}_{N\uparrow}(1-\hat{n}_{N\downarrow})+\hat{n}_{N\downarrow}(1-\hat{n}_{N\uparrow})+\hat{n}_{N\uparrow}\hat{n}_{N\downarrow}~~~~
\end{eqnarray}
With this definition we can compute the Lowdin decomposition of the Hamiltonian as,
\begin{small}
\begin{eqnarray}
Q_{(N)}H_{(N)}P_{(N)}=\sum_{j\sigma}h^{1,\sigma}_{Nj}(1-\hat{n}_{N-\sigma})f^{\dagger}_{N\sigma}f_{j\sigma}&+&\sum_{jkl,\sigma\sigma'}h^{2,\sigma\sigma'}_{Nklj}(1-\hat{n}_{N-\sigma})f^{\dagger}_{N\sigma}f_{j\sigma}f^{\dagger}_{k\sigma'}f_{l\sigma'}\nonumber\\
&+&\sum_{kl}h^{2}_{NNkl}f^{\dagger}_{N\uparrow}f^{\dagger}_{N\downarrow}f_{k\downarrow}f_{l\uparrow}
\end{eqnarray}
\begin{eqnarray}
Q_{(N)}H_{(N)}Q_{(N)}=H^{Q}_{(N)}&+&h^{2}_{N}\hat{n}_{N\uparrow}\hat{n}_{N\downarrow}+\sum_{j\sigma}\hat{n}_{N-\sigma}h^{1,\sigma}_{N}(f^{\dagger}_{j\sigma}f_{N\sigma}+h.c.)\nonumber\\ &+&\sum_{\sigma}h^{1,\sigma}_{NN}\hat{n}_{N\sigma}+\sum_{jkl\sigma\sigma'}\hat{n}_{N-\sigma}h^{2,\sigma\sigma'}_{Njkl}(f^{\dagger}_{N\sigma}f^{\dagger}_{j\sigma'}f_{k\sigma'}f_{l\sigma}+h.c.).
\end{eqnarray}
\end{small}
To derive the analytical expression from the operator Bloch equation eq.\eqref{BlochEqn} we normal order the fermionic operators and obtain the criteria for every normal ordered fermionic term to be zero. And $P_{(N)}H_{(N)}P_{(N)}$ has the same form as $H$ with orbital indices running from $1 \to N$. 
\subsection{Normal Ordering \texorpdfstring{$\eta_{(N)}P_{(N)}HQ_{(N)}\eta_{(N)}$}{Contribution-1}}
The Bloch equation eq.\eqref{BlochEqn} can be equivalently written as,
\begin{align}
Q_{(N)}S_{(N)}^{-1}H_{(N)}S_{(N)}P_{(N)}=Q_{(N)}&H_{(N)}P_{(N)}-\eta_{(N)}P_{(N)}H_{(N)}P_{(N)}\nonumber\\
+ &Q_{(N)}H_{(N)}Q_{(N)}\eta_{(N)}-\eta_{(N)}P_{(N)}H_{(N)}Q_{(N)}\eta_{(N)}~~~~~~\label{expandedBlochEqn}
\end{align}
For the electronic Hamiltonian eq.\eqref{Hamiltonian} and the transformation generator eq.\eqref{FormOfEta} we first list down all the different Fermionic operator products comprising,
\begin{eqnarray}
\eta_{(N)}P_{(N)}HQ_{(N)}\eta_{(N)}=\sum_{i=1}^{12}T_{i},
\end{eqnarray}
where the $T_{i}$'s are given as,
\begin{align}
T_{1}&=\sum_{\substack{ijk,\\\sigma\mu\nu}}t^{1,\sigma}_{i}h^{1,\mu}_{jN}t^{1,\nu}_{k}(1-\hat{n}_{N-\sigma})f^{\dagger}_{N\sigma}f_{i\sigma}(1-\hat{n}_{N-\mu})f^{\dagger}_{j\mu}f_{N\mu}(1-\hat{n}_{N-\nu})f^{\dagger}_{N\nu}f_{k\nu}\nonumber\\~~~~
T_{2}&=\sum_{\substack{ijklm\\\sigma\mu\nu\rho}}t^{1,\sigma}_{i}h^{2,\mu\nu}_{jklN}t^{1,\rho}_{m}(1-\hat{n}_{N-\sigma})f^{\dagger}_{N\sigma}f_{i\sigma}(1-\hat{n}_{N-\mu})f^{\dagger}_{j\mu}f^{\dagger}_{k\nu}f_{l\nu}f_{N\mu}(1-\hat{n}_{N-\rho})f^{\dagger}_{N\rho}f_{m\rho}\nonumber\\
T_{3}&=\sum_{\substack{ijkl \\ \sigma\nu}} t^{1,\sigma}_{i}h^{2,\mu\nu}_{jk}t^{1,\nu}_{l}(1-\hat{n}_{N-\sigma})f^{\dagger}_{N\sigma}f_{i\sigma}f^{\dagger}_{j\uparrow}f^{\dagger}_{k\downarrow}f_{N\downarrow}f_{N\uparrow}(1-\hat{n}_{N-\nu})f^{\dagger}_{N\nu}f_{l\nu}\nonumber\\
T_{4}&=\sum_{\substack{ijkl,\\\sigma\mu}}t^{1,\sigma}_{i}h^{1,\mu}_{jN}t^{2}_{kl}(1-\hat{n}_{N-\sigma})f^{\dagger}_{N\sigma}f_{i\sigma}(1-\hat{n}_{N-\mu})f^{\dagger}_{j\mu}f_{N\mu}f^{\dagger}_{N\uparrow}f^{\dagger}_{N\downarrow}f_{k\downarrow}f_{l\uparrow}\nonumber\\
T_{5}&=\sum_{\substack{ijklmn,\\ \sigma\mu\nu}}t^{1,\sigma}_{i}h^{2,\mu\nu}_{jklN}t^{2}_{mn}(1-\hat{n}_{N-\sigma})f^{\dagger}_{N\sigma}f_{i\sigma}(1-\hat{n}_{N-\mu})f^{\dagger}_{j\mu}f^{\dagger}_{k\nu}f_{l\nu}f_{N\mu}f^{\dagger}_{N\uparrow}f^{\dagger}_{N\downarrow}f_{m\downarrow}f_{n\uparrow}\nonumber\\
T_{6}&=\sum_{\substack{ijklm,\\ \sigma}}t^{1,\sigma}_{i}h^{2}_{jkNN}t^{2}_{lm}(1-\hat{n}_{N-\sigma})f^{\dagger}_{N\sigma}f_{i\sigma}f^{\dagger}_{j\uparrow}f^{\dagger}_{k\downarrow}f_{N\downarrow}f_{N\uparrow}f^{\dagger}_{N\uparrow}f^{\dagger}_{N\downarrow}f_{l\downarrow}f_{m\uparrow}\nonumber\\
T_{7}&=\sum_{\substack{ijkl,\\\sigma\mu}}t^{2}_{ij}h^{1,\sigma}_{kN}t^{1,\mu}_{l}f^{\dagger}_{N\uparrow}f^{\dagger}_{N\downarrow}f_{i\downarrow}f_{j\uparrow}(1-\hat{n}_{N-\sigma})f^{\dagger}_{k\sigma}f_{N\sigma}(1-\hat{n}_{N-\mu})f^{\dagger}_{N\mu}f_{l\mu}\nonumber\\
T_{8}&=\sum_{\substack{ijklmn,\\\sigma\mu\nu}}t^{2}_{ij}h^{2,\sigma\mu}_{klmN}t^{1,\nu}_{n}f^{\dagger}_{N\uparrow}f^{\dagger}_{N\downarrow}f_{i\downarrow}f_{j\uparrow}(1-\hat{n}_{N-\sigma})f^{\dagger}_{k\sigma}f^{\dagger}_{l\mu}f_{m\mu}f_{N\sigma}(1-\hat{n}_{N-\nu})f^{\dagger}_{N\nu}f_{n\nu}\nonumber\\
T_{9}&=\sum_{\substack{ijklm,\\\mu}}t^{2}_{ij}h^{2}_{klNN}t^{1,\mu}_{m}f^{\dagger}_{N\uparrow}f^{\dagger}_{N\downarrow}f_{i\downarrow}f_{j\uparrow}f^{\dagger}_{k\uparrow}f^{\dagger}_{l\downarrow}f_{N\downarrow}f_{N\uparrow}(1-\hat{n}_{N-\mu})f^{\dagger}_{N\mu}f_{m\mu}\nonumber\\
T_{10}&=\sum_{\substack{ijklm,\\\sigma}}t^{2}_{ij}h^{1,\sigma}_{kN}t^{2}_{lm}f^{\dagger}_{N\uparrow}f^{\dagger}_{N\downarrow}f_{i\downarrow}f_{j\uparrow}f^{\dagger}_{k\sigma}f_{N\sigma}f^{\dagger}_{N\uparrow}f^{\dagger}_{N\downarrow}f_{l\downarrow}f_{m\uparrow}\nonumber\\
T_{11}&=\sum_{\substack{ijklmno,\\\sigma\mu}}t^{2}_{ij}h^{2,\sigma\mu}_{klmN}t^{2}_{no}f^{\dagger}_{N\uparrow}f^{\dagger}_{N\downarrow}f_{i\downarrow}f_{j\uparrow}f^{\dagger}_{k\sigma}f^{\dagger}_{l\mu}f_{m\mu}f_{N\sigma}f^{\dagger}_{N\uparrow}f^{\dagger}_{N\downarrow}f_{n\downarrow}f_{o\uparrow}\nonumber\\
T_{12}&=\sum_{ijklmn}t^{2}_{ij}h^{2}_{klNN}t^{2}_{mn}f^{\dagger}_{N\uparrow}f^{\dagger}_{N\downarrow}f_{i\downarrow}f_{j\uparrow}f^{\dagger}_{k\uparrow}f^{\dagger}_{l\downarrow}f_{N\downarrow}f_{N\uparrow}f^{\dagger}_{N\uparrow}f^{\dagger}_{N\downarrow}f_{m\downarrow}f_{n\uparrow}\label{Term-Set-1}
\end{align}
From the Pauli blockade conditions i.e., $(f^{\dagger}_{N\sigma})^{2}=0,(f_{N\sigma})^{2}=0$ the terms $T_{3}$,$T_{4}$,$T_{5}$ and $T_{9}$, $T_{10}$,$T_{11}$ have zero contribution and are eliminated. As a next step we normal order(N.O.) the fermionic operators (denoted as $::$) within the remaining six terms in eqs.\eqref{Term-Set-1}, in the process of doing we use the Pauli-blockade conditions to balance the expressions,
\begin{small}
\begin{eqnarray}
:T_{1}:&=&\sum_{\substack{ijk,\\\sigma\mu\nu}}t^{1,\sigma}_{i}h^{1,\mu}_{jN}t^{1,\nu}_{k}\delta_{\mu\nu}\bigg(\delta_{ij}\delta_{\sigma\mu}(1-\hat{n}_{N-\sigma})f^{\dagger}_{N\sigma}(1-\hat{n}_{N-\mu})(1-\hat{n}_{N-\nu})f_{k\nu}\nonumber\\
&+&(1-\hat{n}_{N-\sigma})f^{\dagger}_{N\sigma}(1-\hat{n}_{N-\mu})f^{\dagger}_{j\mu}f_{k\nu}f_{i\sigma}\bigg)\nonumber\\
:T_{2}:&=&\sum_{\substack{ijklm\\\sigma\mu\nu\rho}}t^{1,\sigma}_{i}h^{2,\mu\nu}_{jklN}\bigg((1-\hat{n}_{N-\sigma})(1-\hat{n}_{N-\mu})(1-\hat{n}_{N-\rho})\delta_{ij}\delta_{\sigma\mu}\delta_{\mu\rho}f^{\dagger}_{N\sigma}f^{\dagger}_{k\nu}f_{l\nu}f_{m\rho}\nonumber\\&-&(1-\hat{n}_{N-\sigma})\delta_{ik}\delta_{\sigma\nu}\delta_{\mu\rho}f^{\dagger}_{N\sigma}f^{\dagger}_{j\mu}(1-\hat{n}_{N-\mu})f_{l\sigma}(1-\hat{n}_{N-\rho})f_{m\mu}+(1-\hat{n}_{N-\sigma})\delta_{\mu\rho}f^{\dagger}_{N\sigma}f^{\dagger}_{j\mu}f^{\dagger}_{k\nu}f_{l\nu}f_{m\rho}f_{i\sigma}\bigg)\nonumber\\
:T_{6}:&=&\sum_{\substack{ijklm,\\ \sigma}}t^{1,\sigma}_{i}h^{2}_{jkNN}t^{2}_{lm}\bigg(\delta_{ij}\delta_{\sigma\uparrow}(1-\hat{n}_{N-\sigma})f^{\dagger}_{N\sigma}f^{\dagger}_{k\downarrow}f_{l\downarrow}f_{m\uparrow}+\delta_{ik}\delta_{\sigma\downarrow}(1-\hat{n}_{N-\sigma})f^{\dagger}_{N\sigma}f^{\dagger}_{j\uparrow}f_{m\uparrow}f_{l\downarrow}\nonumber\\
&+&(1-\hat{n}_{N-\sigma})f^{\dagger}_{N\sigma}f^{\dagger}_{j\uparrow}f^{\dagger}_{k\downarrow}f_{l\downarrow}f_{m\uparrow}f_{i\sigma}\bigg)\nonumber\\
:T_{7}:&=&\sum_{\substack{ijkl,\\\sigma\mu}}t^{2}_{ij}h^{1,\sigma}_{kN}t^{1,\mu}_{l}\bigg(\delta_{jk}\delta_{\sigma\uparrow}\delta_{\mu\sigma}f^{\dagger}_{N\uparrow}f^{\dagger}_{N\downarrow}f_{i\downarrow}f_{l\mu}+\delta_{ik}\delta_{\sigma\downarrow}\delta_{\mu\sigma}f^{\dagger}_{N\uparrow}f^{\dagger}_{N\downarrow}f_{l\mu}f_{j\uparrow}+\delta_{\mu\sigma}f^{\dagger}_{k\sigma}f^{\dagger}_{N\uparrow}f^{\dagger}_{N\downarrow}f_{i\downarrow}f_{j\uparrow}f_{l\sigma}\bigg)\nonumber\\
:T_{8}:&=&\sum_{\substack{ijklmn,\\\sigma\mu\nu}}t^{2}_{ij}h^{2,\sigma\mu}_{klmN}t^{1,\nu}_{n}\bigg(\delta_{\sigma\nu}\delta_{jk}\delta_{\sigma\uparrow}\delta_{il}\delta_{\mu\downarrow}f^{\dagger}_{N\uparrow}f^{\dagger}_{N\downarrow}f_{m\downarrow}f_{n\uparrow}+\delta_{\sigma\nu}\delta_{jk}\delta_{\sigma\uparrow}f^{\dagger}_{N\uparrow}f^{\dagger}_{N\downarrow}f^{\dagger}_{l\mu}f_{m\mu}f_{i\downarrow}f_{n\uparrow}\nonumber\\
&+&\delta_{\sigma\nu}\delta_{ik}\delta_{\sigma\downarrow}\delta_{jl}\delta_{\mu\uparrow}f^{\dagger}_{N\uparrow}f^{\dagger}_{N\downarrow}f_{n\downarrow}f_{m\uparrow}+\delta_{\sigma\nu}\delta_{ik}\delta_{\sigma\downarrow}f^{\dagger}_{N\uparrow}f^{\dagger}_{N\downarrow}f^{\dagger}_{l\mu}f_{m\mu}f_{n\downarrow}f_{j\uparrow}+\delta_{jl}\delta_{\mu\uparrow}\delta_{\sigma\nu}f^{\dagger}_{N\uparrow}f^{\dagger}_{N\downarrow}f^{\dagger}_{k\sigma}f_{n\sigma}f_{i\downarrow}f_{m\uparrow}\nonumber\\
&+&\delta_{il}\delta_{\mu\downarrow}\delta_{\sigma\nu}f^{\dagger}_{N\uparrow}f^{\dagger}_{N\downarrow}f^{\dagger}_{k\sigma}f_{n\sigma}f_{m\downarrow}f_{j\uparrow}+\delta_{\sigma\nu}f^{\dagger}_{N\uparrow}f^{\dagger}_{N\downarrow}f^{\dagger}_{k\sigma}f^{\dagger}_{l\mu}f_{m\mu}f_{n\sigma}f_{i\downarrow}f_{j\uparrow}\bigg)\nonumber\\
:T_{12}:&=&\sum_{ijklmn}t^{2}_{ij}h^{2}_{klNN}t^{2}_{mn}\bigg(\delta_{il}\delta_{jk}f^{\dagger}_{N\uparrow}f^{\dagger}_{N\downarrow}f_{m\downarrow}f_{n\uparrow}+\delta_{jk}f^{\dagger}_{N\uparrow}f^{\dagger}_{N\downarrow}f^{\dagger}_{l\downarrow}f_{m\downarrow}f_{i\downarrow}f_{n\uparrow}-\delta_{il}f^{\dagger}_{N\uparrow}f^{\dagger}_{N\downarrow}f^{\dagger}_{k\uparrow}f_{j\uparrow}f_{m\downarrow}f_{n\uparrow}\nonumber\\
&+&f^{\dagger}_{N\uparrow}f^{\dagger}_{N\downarrow}f^{\dagger}_{k\uparrow}f^{\dagger}_{l\downarrow}f_{i\downarrow}f_{j\uparrow}f_{m\downarrow}f_{n\uparrow}\bigg)\label{NormalOrdered1}
\end{eqnarray}
\end{small}
\section{Normal ordering \texorpdfstring{$\eta_{(N)}P_{(N)}HP_{(N)}$}{Contribution-2}}
We carry out the term multiplications within $\eta_{(N)}P_{(N)}HP_{(N)}=V_{1}+V_{2}+V_{3}+V_{4}$ and write down the fermionic terms,
\begin{small}
\begin{eqnarray}
V_{1}&=&\sum_{ijk\sigma\nu}t^{1,\sigma}_{i}h^{1,\nu}_{jk}(1-\hat{n}_{N-\sigma})f^{\dagger}_{N\sigma}f_{i\sigma}f^{\dagger}_{j\nu}f_{k\nu}\nonumber\\
V_{2}&=&\sum_{\substack{ijkl\\ \sigma}}t^{2}_{ij}h^{1,\sigma}_{kl}f^{\dagger}_{N\uparrow}f^{\dagger}_{N\downarrow}f_{i\downarrow}f_{j\uparrow}f^
{\dagger}_{k\sigma}f_{l\sigma}\nonumber\\
V_{3}&=&\sum_{\substack{ijklm\\ \sigma\nu\rho}}t^{1,\sigma}_{i}h^{2,\nu\rho}_{jklm}(1-\hat{n}_{N-\sigma})f^{\dagger}_{N\sigma}f_{i\sigma}f^{\dagger}_{j\nu}f^
{\dagger}_{k\rho}f_{l\rho}f_{m\nu}\nonumber\\ V_{4}&=&\sum_{\substack{ijklmn,\\\sigma\nu}}t^{2}_{ij}h^{2,\sigma\nu}_{klmn}f^{\dagger}_{N\uparrow}f^{\dagger}_{N\downarrow}f_{i\downarrow}f_{j\uparrow}f^{\dagger}_{k\sigma}f^{\dagger}_{l\nu}f_{m\nu}f_{n\sigma}
\label{Term-Set-2}
\end{eqnarray}
Upon normal ordering the eqs.\eqref{Term-Set-2} we get,
\begin{eqnarray}
:V_{1}:&=&\sum_{ijk\sigma\nu}\sum_{ijk\sigma\nu}t^{1,\sigma}_{i}h^{1,\nu}_{jk}\bigg((1-\hat{n}_{N-\sigma})\delta_{ij}\delta_{\sigma\nu}f^{\dagger}_{N\sigma}f_{k\sigma}+(1-\hat{n}_{N-\sigma})f^{\dagger}_{N\sigma}f^{\dagger}_{j\nu}f_{k\nu}f_{i\sigma}\bigg)\nonumber\\
:V_{2}:&=&\sum_{\substack{ijkl\\ \sigma}}t^{2}_{ij}h^{1,\sigma}_{kl}\bigg(\delta_{jk}\delta_{\sigma\uparrow}f^{\dagger}_{N\uparrow}f^{\dagger}_{N\downarrow}f_{i\downarrow}f_{l\uparrow}+\delta_{ik}\delta_{\sigma\downarrow}f^{\dagger}_{N\uparrow}f^{\dagger}_{N\downarrow}f_{l\downarrow}f_{j\uparrow}+f^{\dagger}_{N\uparrow}f^{\dagger}_{N\downarrow}f^
{\dagger}_{k\sigma}f_{l\sigma}f_{i\downarrow}f_{j\uparrow}\bigg)\nonumber\\
:V_{3}:&=&\sum_{\substack{ijklm\\ \sigma\nu\rho}}t^{1,\sigma}_{i}h^{2,\nu\rho}_{jklm}\bigg(\delta_{ij}\delta_{\sigma\nu}(1-\hat{n}_{N-\sigma})f^{\dagger}_{N\sigma}f^
{\dagger}_{k\rho}f_{l\rho}f_{m\nu}+\delta_{ik}\delta_{\sigma\rho}(1-\hat{n}_{N-\sigma})f^{\dagger}_{N\sigma}f^{\dagger}_{j\nu}f_{m\nu}f_{l\sigma}\nonumber\\
&+&(1-\hat{n}_{N-\sigma})f^{\dagger}_{N\sigma}f^{\dagger}_{j\nu}f^
{\dagger}_{k\rho}f_{l\rho}f_{m\nu}f_{i\sigma}\bigg)\nonumber\\
:V_{4}:&=&\sum_{\substack{ijklmn,\\\sigma\nu}}t^{2}_{ij}h^{2,\sigma\nu}_{klmn}\bigg(\delta_{il}\delta_{\nu\downarrow}\delta_{jk}\delta_{\sigma\uparrow}f^{\dagger}_{N\uparrow}f^{\dagger}_{N\downarrow}f_{m\nu}f_{n\sigma}+\delta_{jk}\delta_{\sigma\uparrow}f^{\dagger}_{N\uparrow}f^{\dagger}_{N\downarrow}f^{\dagger}_{l\nu}f_{m\nu}f_{i\downarrow}f_{n\uparrow}\nonumber\\
&+&\delta_{ik}\delta_{\sigma\downarrow}\delta_{jl}\delta_{\nu\uparrow}f^{\dagger}_{N\uparrow}f^{\dagger}_{N\downarrow}f_{n\downarrow}f_{m\uparrow}+\delta_{ik}\delta_{\sigma\downarrow}f^{\dagger}_{N\uparrow}f^{\dagger}_{N\downarrow}f^{\dagger}_{l\nu}f_{m\nu}f_{n\downarrow}f_{j\uparrow}+\delta_{jl}\delta_{\nu\uparrow}f^{\dagger}_{N\uparrow}f^{\dagger}_{N\downarrow}f^{\dagger}_{k\sigma}f_{n\sigma}f_{i\downarrow}f_{m\uparrow}\nonumber\\
&+&\delta_{il}\delta_{\nu\downarrow}f^{\dagger}_{N\uparrow}f^{\dagger}_{N\downarrow}f^{\dagger}_{k\sigma}f_{n\sigma}f_{m\downarrow}f_{j\uparrow}+f^{\dagger}_{N\uparrow}f^{\dagger}_{N\downarrow}f^{\dagger}_{k\sigma}f^{\dagger}_{l\nu}f_{m\nu}f_{n\sigma}f_{i\downarrow}f_{j\uparrow}\bigg)\label{NormalOrdered2}
\end{eqnarray}
\end{small}
\section{Normal ordering \texorpdfstring{$Q_{(N)}HQ_{(N)}\eta_{(N)}$}{Contribution-4}}
We carry out the term multiplications and write down the fermionic terms comprising $Q_{(N)}HQ_{(N)}\eta_{(N)}=\sum_{i=1}^{15}W_{i}$,\\
\begin{align}
W_{1}&=\sum_{ij\sigma\nu}\hat{n}_{N-\sigma}h^{1,\sigma}_{iN}t^{1,\nu}_{j}f^{\dagger}_{i\sigma}f_{N\sigma}(1-\hat{n}_{N-\nu})f^{\dagger}_{N\nu}f_{j\nu}\nonumber\\
W_{2}&=\sum_{ij\sigma\nu}\hat{n}_{N-\sigma}h^{1,\sigma}_{iN}t^{1,\nu}_{j}f^{\dagger}_{N\sigma}f_{i\sigma}(1-\hat{n}_{N-\nu})f^{\dagger}_{N\nu}f_{j\nu}\nonumber\\
W_{3}&=\sum_{ijk\sigma}\hat{n}_{N-\sigma}h^{1,\sigma}_{iN}t^{2}_{jk}f^{\dagger}_{i\sigma}f_{N\sigma}f^{\dagger}_{N\uparrow}f^{\dagger}_{N\downarrow}f_{j\downarrow}f_{k\uparrow}\nonumber\\
W_{4}&=\sum_{ijk\sigma}\hat{n}_{N-\sigma}h^{1,\sigma}_{iN}t^{2}_{jk}f^{\dagger}_{N\sigma}f_{i\sigma}f^{\dagger}_{N\uparrow}f^{\dagger}_{N\downarrow}f_{j\downarrow}f_{k\uparrow}\nonumber\\
W_{5}&=\sum_{i\sigma}h^{2}_{NNNN}t^{1,\sigma}_{i}\hat{n}_{N\uparrow}\hat{n}_{N\downarrow}(1-\hat{n}_{N-\sigma})f^{\dagger}_{N\sigma}f_{i\sigma}\nonumber\\ W_{6}&=\sum_{ij}h^{2}_{NNNN}t^{2}_{ij}\hat{n}_{N\uparrow}\hat{n}_{N\downarrow}f^{\dagger}_{N\uparrow}f^{\dagger}_{N\downarrow}f_{i\downarrow}f_{j\uparrow}\nonumber\\ W_{7}&=\sum_{\substack{ijkl,\\\sigma\nu\rho}}\hat{n}_{N-\sigma}h^{2,\sigma\nu}_{Nijk}t^{1,\rho}_{l}f^{\dagger}_{N\sigma}f^{\dagger}_{i\nu}f_{j\nu}f_{k\sigma}(1-\hat{n}_{N-\rho})f^{\dagger}_{N\rho}f_{l\rho}\nonumber\\
W_{8}&=\sum_{\substack{ijkl,\\\sigma\nu\rho}}\hat{n}_{N-\sigma}h^{2,\sigma\nu}_{Nkji}t^{1,\rho}_{l}f^{\dagger}_{i\sigma}f^{\dagger}_{j\nu}f_{k\nu}f_{N\sigma}(1-\hat{n}_{N-\rho})f^{\dagger}_{N\rho}f_{l\rho}\nonumber\\
W_{9}&=\sum_{\substack{ijklm,\\\sigma\nu}}\hat{n}_{N-\sigma}h^{2,\sigma\nu}_{Nijk}t^{2}_{lm}f^{\dagger}_{N\sigma}f^{\dagger}_{i\nu}f_{j\nu}f_{k\sigma}f^{\dagger}_{N\uparrow}f^{\dagger}_{N\downarrow}f_{l\downarrow}f_{m\uparrow}\nonumber\\ W_{10}&=\sum_{\substack{ijklm,\\\sigma\nu}}\hat{n}_{N-\sigma}h^{2,\sigma\nu}_{Nijk}t^{2}_{lm}f^{\dagger}_{k\sigma}f^{\dagger}_{j\nu}f_{i\nu}f_{N\sigma}f^{\dagger}_{N\uparrow}f^{\dagger}_{N\downarrow}f_{l\downarrow}f_{m\uparrow}\nonumber\\
W_{11}&=\sum_{i\sigma}h^{1,\sigma}_{NN}t^{1,\sigma}_{i}f^{\dagger}_{N\sigma}f_{i\sigma}+\sum_{ij\sigma}h^{1,\sigma}_{NN}t^{2}_{ij}f^{\dagger}_{N\uparrow}f^{\dagger}_{N\downarrow}f_{i\downarrow}f_{j\uparrow}\nonumber\\
W_{12}&=\sum_{ijk\sigma\nu}h^{1,\nu}_{ij}t^{1,\sigma}_{k}(1-\hat{n}_{N-\sigma})f^{\dagger}_{N\sigma}f^{\dagger}_{i\nu}f_{j\nu}f_{k\sigma}\nonumber\\
W_{13}&=\sum_{\substack{ijkl,\\ \sigma}}h^{1,\sigma}_{ij}t^{2}_{kl}f^{\dagger}_{N\uparrow}f^{\dagger}_{N\downarrow}f^{\dagger}_{i\sigma}f_{j\sigma}f_{k\downarrow}f_{l\uparrow}\nonumber\\
W_{14}&=\sum_{\substack{ijklm,\\ \sigma\nu\rho}}h^{2,\nu\rho}_{ijkl}t^{1,\sigma}_{m}(1-\hat{n}_{N-\rho})f^{\dagger}_{N\sigma}f^{\dagger}_{i\nu}f^{\dagger}_{j\rho}f_{k\rho}f_{l\nu}f_{m\sigma}\nonumber\\
W_{15}&=\sum_{\substack{ijklmn,\\ \sigma\nu}}h^{2,\sigma\nu}_{ijkl}t^{2}_{mn}f^{\dagger}_{N\uparrow}f^{\dagger}_{N\downarrow}f^{\dagger}_{i\sigma}f^{\dagger}_{j\nu}f_{k\nu}f_{l\sigma}f_{m\downarrow}f_{n\uparrow}~~~~~\label{Term-Set-3}
\end{align}
From the Pauli blockade conditions the terms $W_{1}$,$W_{4}$,$W_{5}$ and $W_{8}$, $W_{9}$ have zero contribution and are eliminated. Also note that expressions
$W_{6}$, $W_{11}$, $W_{12}$, $W_{13}$, $W_{14}$, $W_{15}$ are already normal ordered. Next we normal order(N.O.) eqs.\eqref{Term-Set-3} the fermionic operators within the remaining four terms, 
\begin{small}
\begin{eqnarray}
:W_{2}:&=&\sum_{ij\sigma\nu}h^{1,\sigma}_{iN}t^{1,\nu}_{j}\bigg(\delta_{\sigma\uparrow}\delta_{\nu\downarrow}f^{\dagger}_{N\uparrow}f^{\dagger}_{N\downarrow}f_{j\downarrow}f_{i\uparrow}+\delta_{\sigma\downarrow}\delta_{\nu\uparrow}f^{\dagger}_{N\uparrow}f^{\dagger}_{N\downarrow}f_{i\downarrow}f_{j\uparrow}\bigg)\nonumber\\
:W_{3}:&=&\sum_{ijk\sigma}h^{1,\sigma}_{iN}t^{2}_{jk}\bigg(\delta_{\sigma\uparrow}f^{\dagger}_{N\downarrow}f^{\dagger}_{i\uparrow}f_{k\uparrow}f_{j\downarrow}+\delta_{\sigma\uparrow}f^{\dagger}_{N\uparrow}f^{\dagger}_{i\downarrow}f_{j\downarrow}f_{k\uparrow}\bigg)\nonumber\\
:W_{7}:&=&\sum_{\substack{ijkl,\\\sigma\nu\rho}}h^{2,\sigma\nu}_{Nijk}t^{1,\rho}_{l}\bigg(\delta_{\sigma\uparrow}\delta_{\rho\downarrow}f^{\dagger}_{N\uparrow}f^{\dagger}_{N\downarrow}f^{\dagger}_{i\nu}f_{j\nu}f_{l\downarrow}f_{k\uparrow}+\delta_{\sigma\downarrow}\delta_{\rho\uparrow}f^{\dagger}_{N\uparrow}f^{\dagger}_{N\downarrow}f^{\dagger}_{i\nu}f_{j\nu}f_{k\downarrow}f_{l\uparrow}\bigg),\nonumber\\
:W_{10}:&=&\sum_{\substack{ijklm,\\\sigma\nu}}h^{2,\sigma\nu}_{Nijk}t^{2}_{lm}\bigg(\delta_{\sigma\uparrow}f^{\dagger}_{N\downarrow}f^{\dagger}_{k\uparrow}f^{\dagger}_{j\nu}f_{i\nu}f_{m\uparrow}f_{l\downarrow}+\delta_{\sigma\downarrow}f^{\dagger}_{N\uparrow}f^{\dagger}_{k\downarrow}f^{\dagger}_{j\nu}f_{i\nu}f_{l\downarrow}f_{m\uparrow}\nonumber\\
&+&f^{\dagger}_{N\uparrow}f^{\dagger}_{N\downarrow}f^{\dagger}_{k\sigma}f^{\dagger}_{j\nu}f_{i\nu}f_{N\sigma}f_{l\downarrow}f_{m\uparrow}\bigg)\label{NormalOrdered3}
\end{eqnarray}
\end{small}
\subsection{Algebraic Downfolding equations
Deduced From Bloch Equation}\label{Contribution-5}
Starting from the Bloch equation eq.\eqref{expandedBlochEqn} we used the normal ordered expressions for, $:\eta_{(N)}P_{(N)}H_{(N)}Q_{(N)}\eta_{(N)}:$ eqs.\eqref{NormalOrdered1}, $:\eta_{(N)}P_{(N)}H_{(N)}P_{(N)}:$ eqs.\eqref{NormalOrdered2}, $:Q_{(N)}H_{(N)}Q_{(N)}\eta_{(N)}:$ eqs.\eqref{NormalOrdered3} to obtain the N.O. Bloch equation,
\begin{eqnarray}
:Q_{(N)}S^{-1}_{(N)}H_{(N)}S_{(N)}P_{(N)}:&=&\sum_{i}A^{(N),\sigma}_{i}f^{\dagger}_{N\sigma}f_{i\sigma}+\sum_{ijk}B^{(N),\sigma\nu}_{ijk}f^{\dagger}_{N\sigma}f^{\dagger}_{i\nu}f_{j\nu}f_{k\sigma}+\sum_{ij}C^{(N)}_{ij}f^{\dagger}_{N\uparrow}f^{\dagger}_{N\downarrow}f_{i\downarrow}f_{j\uparrow}\nonumber\\
&+&\sum_{ijkl\sigma}D^{(N),\sigma}_{ijkl}f^{\dagger}_{N\uparrow}f^{\dagger}_{N\downarrow}f^{\dagger}_{i\sigma}f_{j\sigma}f_{k\downarrow}f_{l\uparrow}+\sum_{\substack{ijklm,\\ \sigma\mu\nu}}E^{(N),\sigma\nu\rho}_{ijklm}f^{\dagger}_{N\sigma}f^{\dagger}_{i\nu}f^{\dagger}_{j\rho}f_{k\rho}f_{l\nu}f_{m\sigma}\nonumber\\
&+&\sum_{\substack{ijklmn,\\ \sigma\nu}}F^{(N),\sigma\nu\rho}_{ijklm}f^{\dagger}_{N\uparrow}f^{\dagger}_{N\downarrow}f^{\dagger}_{i\sigma}f^{\dagger}_{j\nu}f_{k\nu}f_{l\sigma}f_{m\downarrow}f_{n\uparrow}
\end{eqnarray}
where $\mathbf{A^{(N),\sigma}}$ constitute the N.O. single-particle excitations contribution to the downfolding Bloch equation. Here $\mathbf{B^{(N),\sigma\nu}}$ represents the N.O. contribution of doubles excitation involving one of the downfolding spin-orbital ($N\sigma$),   $\mathbf{B^{(N),\sigma\nu}}$. The coefficient $\mathbf{C^{(N)}}$ represents the contribution of paired doubles excitation  corresponding to the downfolding orbital. Coefficient $\mathbf{D}^{\sigma}$ constitutes the contribution from triples excitations containing paired doubles.$\mathbf{E^{(N),\sigma\nu\rho}}$ and $\mathbf{F^{(N),\sigma\nu\rho}}$ are the triples and quadruples contribution to the Bloch equation, 
\begin{small}
\begin{eqnarray}
A_{i}^{(N),\sigma}&=&\sum_{k}t^{1,\sigma}_{k}\left(h^{1,\sigma}_{kN}t^{1\sigma}_{i}+h^{1,\sigma}_{ki}\right)-h_{NN}^{1,\sigma}t^{1,\sigma}_{i}-h^{1,\sigma}_{iN}=0\label{1-particle}\\
B_{ijk}^{(N),\sigma\nu}&=&t^{1,\sigma}_{k}h^{1,\nu}_{iN}t^{1,\nu}_{j}+\sum_{n}t^{1,\sigma}_{n}\left(h^{2,\sigma\nu}_{nijN}t^{1,\sigma}_{k}+h^{2,\nu\sigma}_{inkN}t^{1,\nu}_{j}+\delta_{\nu,-\sigma}h^{2}_{niNN}t^{2}_{jk}+h^{2,\sigma\nu}_{nijk}+h^{2,\nu\sigma}_{inkj}\right)\nonumber\\
&&~~~~~-\delta_{\nu,-\sigma}(\delta_{\sigma\downarrow}h^{1,\sigma}_{iN}t^{2}_{kj}
+\delta_{\sigma\uparrow}h^{1,\sigma}_{iN}t^{2}_{jk})-h^{2,\sigma\nu}_{Nijk}=0\label{2-particle}\\
C_{ij}^{(N)}&=&\sum_{n}\left(t^{2}_{in}\left(h^{1,\uparrow}_{nN}t^{1,\uparrow}_{j}+h^{1,\uparrow}_{nj}\right)+t^{2}_{nj}\left(h^{1,\downarrow}_{nN}t^{1,\downarrow}_{i}+h^{1,\downarrow}_{ni}\right)\right)-\left(h^{1,\uparrow}_{jN}t^{1,\downarrow}_{i}+h^{1,\downarrow}_{iN}t^{1,\uparrow}_{j}\right)-(h^{1,\uparrow}_{NN}+h^{1,\downarrow}_{NN})t^{2}_{ij}\nonumber\\
&+&\sum_{mn}t^{2}_{mn}\left(h^{2,\uparrow\downarrow}_{nmiN}t^{1,\uparrow}_{j}+h^{2,\downarrow\uparrow}_{mnjN}t^{1,\downarrow}_{i}+h^{2,\uparrow\downarrow}_{nmij}+h^{2,\downarrow\uparrow}_{mnji}+h^{2}_{mnNN}t^{2}_{ij}\right)-h^{2}_{NNNN}t^{2}_{ij}-h^{2}_{NNij}\label{paired-excitations}\\
D^{\sigma}_{ijkl}&=&t^{2}_{kl}\left(h^{1,\sigma}_{iN}t^{1,\sigma}_{j}+h^{1,\sigma}_{ij}\right)+
\sum_{n}\left(t^{2}_{kn}h^{2,\uparrow\sigma}_{nijN}t^{1,\uparrow}_{l}+t^{2}_{nl}h^{2,\downarrow\sigma}_{nijN}t^{1,\downarrow}_{k}+t^{2}_{nl}h^{2,\sigma\downarrow}_{inkN}t^{1,\sigma}_{j}+t^{2}_{kn}h^{2,\sigma\uparrow}_{inlN}t^{1,\sigma}_{j}\right)\nonumber\\
&+&\sum_{n}\left(\delta_{\sigma\downarrow}t^{2}_{kn}h^{2}_{niNN}t^{2}_{jl}-\delta_{\sigma\uparrow}t^{2}_{nk}h^{2}_{niNN}t^{2}_{jl}\right)+\sum_{n}\left(t^{2}_{kn}\left(h^{2,\uparrow\sigma}_{nijl}+h^{2,\sigma\uparrow}_{inlj}\right)+t^{2}_{nl}\left(h^{2,\downarrow\sigma}_{nijN}t^{1,\downarrow}_{k}+h^{2,\sigma\downarrow}_{inkN}t^{1,\sigma}_{j}\right)\right)\nonumber\\
&-&\left(h^{2,\uparrow\sigma}_{Nijl}t^{1,\downarrow}_{k}+h^{2,\downarrow\sigma}_{Nijk}t^{1,\uparrow}_{l}\right)-h^{1,\sigma}_{ij}t^{2}_{kl}\label{3-particle-exc}\\
\mathbf{E}^{(N),\sigma\nu}&=&\mathbf{t^{1,\sigma}}\otimes\mathbf{h^{2,\mu\nu}_{N}}\otimes \mathbf{t^{1,\mu}}+\delta_{\sigma,-\nu}\mathbf{h^{2}_{NN}}\otimes\mathbf{t^{2}}\otimes\mathbf{t^{1,\sigma}}-\delta_{\sigma\downarrow}
(\mathbf{h^{2,\uparrow\nu}_{N}}\otimes\mathbf{ t^{2}})_{32154}-\delta_{\sigma\downarrow}
(\mathbf{h^{2,\downarrow\nu}_{N}}\otimes\mathbf{ t^{2}})_{32154}\\
\mathbf{F}^{(N),\sigma\nu\rho}&=&\mathbf{h^{2,\sigma\mu}_{N}}\otimes\mathbf{t^{1,\sigma}}\otimes\mathbf{t^{2}}+\mathbf{h^{2}_{NN}}\otimes\mathbf{t^{2}}\otimes\mathbf{t^{2}}
\end{eqnarray}
\end{small}
In the above expressions $\mathbf{h^{2,\sigma\nu}}_{abcd}$ represents a permutation of indexes of the tensor for e.g. $((\mathbf{h^{2,\sigma\nu}})_{3124})_{ijkl}=(\mathbf{h^{2,\sigma\nu}})_{kijl}$. Here ($\otimes$) represents tensor product and ($\cdot$) represents tensor contraction.  In order to satisfy the Bloch equation we need the contributions $\mathbf{A}$ to $\mathbf{D}$ to become zero. This corresponds to a quadratic polynomial system. In the next section we will evaluate its Jacobian. 
\subsection{Tensor Factorization of three rank tensors-canonical polyadic decomposition of three rank tensors}
 A detailed step-wise description is presented below.
\\ \par
\begin{itemize}
    \item[1.]  We want to find a decomposition of the three rank tensor A in terms of two rank tensor factors X,Y,Z. The number of indices in a tensor is the rank. Each index can run over the sequence of integers starting from 1 to N, this running index is to be referred as direction in the later steps.
    \begin{eqnarray}
    A_{ijk}=\sum_{a}X_{ia}Y_{ja}Z_{ka}\label{TF_A}
    \end{eqnarray}
    \item[2.] Randomly initialize tensors X and Y and multiply the transposition of X (call it XT), along the first direction of the tensor A. This leads to a matrix B.  This matrix B also has three directions.
    \begin{eqnarray}
    B_{bjk}=\sum_{i}X_{bi}A_{ijk}
    \end{eqnarray}
    \item[3.] This matrix B is now multiplied with the transposition of Y(call it YT) along the second direction. This leads to matrix C. The matrix C has two directions now.
    \begin{eqnarray}
    C_{bk}=\sum_{i}Y_{bj}B_{bjk}
    \end{eqnarray}
    \item[4.] Now we do the matrix multiplications XT with X call V  and YT with Y call it W. 
    \begin{eqnarray}
    V=X^{T}X\\
    W=Y^{T}Y
    \end{eqnarray}
    \item[5.]
    And then we perform Hadamard product of the matrices V and W leading to P. 
    \begin{eqnarray}
    P=VW
    \end{eqnarray}
    \item[6.] Finally we invert P and multiply it to  C leading to solution for Z
    \begin{eqnarray}
    Z=P^{-1}C
    \end{eqnarray}
    \item[7.] We repeat steps 3 to step 6 by now randomly initializing Y and taking the Z computed in step 6 to compute X
    \item[8.] We repeat steps 3 to step 6 by  taking the Z and X computed in step 7 to compute Y.
    \item[9.] We start with the three factors X,Y,Z obtained from steps 1 to steps 8 and compute the error between the factorized representation and the original tensor 
    \begin{eqnarray}
    E=\sum_{ijk}|A_{ijk}-\sum_{a}X_{ia}Y_{ja}Z_{ka}|^2
    \end{eqnarray}
    \item[10.] If error is above threshold we start with the X,Y,Z computed from last step and then repeat steps 1 to 8.
\end{itemize}

\subsection{Qubitization circuit for Matrix-Matrix multiplication with isometries}
\subsubsection*{Theorem}
If $A$ and $B$ are general rectangular matrices of dimensions $dim(A)=(N,P)$ and $dim(B)=(P,M)$ with $N,M\ge 2$ then there is a unitary operation $U(A,B)$ of dimension $2^{n_{q}}\times 2^{n_{q}}$ that operates on a system of $n_{q}=p+\max(m,n)+2$ qubit registers : $|\cdot\rangle_{p}|\cdot\rangle_{\max(m,n)}|\cdot\rangle_{a_{1}}|\cdot\rangle_{a_{2}}$ (where $n=\lceil\log_{2} N\rceil$,$m=\lceil\log_{2} M\rceil$,$p=\lceil\log_{2} P\rceil$ )and block encodes the matrix multiplication of $A$ and $B$ s.t. 
\begin{equation*}
    \langle 0|_{p}\langle i|_{\max(m,n)}\langle 0|_{a_{1}}\langle 0|_{a_{2}}U(A,B)|0\rangle_{p}|j\rangle_{\max(m,n)}|1\rangle_{a_{1}}|0\rangle_{a_{2}}=\frac{1}{P^{2}}\frac{\sum_{k}A_{ik}B_{kj}}{||A||||B||}.
\end{equation*}
\textit{Proof-}
Lets define an isometry $T(A,B)$,
\begin{eqnarray}
T(A,B)&=&\frac{1}{\sqrt{2\max(N,M)}}\sum_{r,c}|c\rangle\langle c|\otimes|r\rangle\otimes \bigg[\frac{A_{rc}}{||A||}|0,0\rangle+\sqrt{1-\left(\frac{A_{rc}}{||A||}\right)^{2}}|0,1\rangle\nonumber\\
&+&\frac{B_{cr}}{||B||}|r,1,0\rangle+\sqrt{1-\left(\frac{B_{cr}}{||B||}\right)^{2}}|1,1\rangle\bigg] 
\end{eqnarray}
The isometry $T(A,B)$ has the property $T^{\dagger}(A,B)T(A,B)=I_{2}^{\otimes p}$ this can be checked as follows, 
\begin{eqnarray}
T^{\dagger}(A,B)T(A,B) &=& \frac{1}{2\max(N,M)}\sum_{c}|c\rangle\langle c|\otimes\left[\sum_{r}(|A_{rc}|^{2}+1-A^{2}_{rc}+B^{2}_{cr}+1-B^{2}_{cr})\right]\nonumber\\
&=&\sum_{c}|c\rangle\langle c|=I_{2}^{\otimes p}
\end{eqnarray}
Utilizing the above property we can define a unitary operator $W:=W(A,B)$,
\begin{eqnarray}
W(A,B)=2T(A,B)T^{\dagger}(A,B)-1
\end{eqnarray}
The unitarity  of W can be checked as follows,
\begin{eqnarray}
W^{\dagger}W&=&WW^{\dagger}=I\nonumber\\
    &=&(2T(A,B)T^{\dagger}(A,B)-1)(2T(A,B)T^{\dagger}(A,B)-1)\nonumber\\
    &=&4T(A,B)T^{\dagger}(A,B)-4T(A,B)T^{\dagger}(A,B)+1=1
\end{eqnarray}
To proceed further we normalizing the matrices $A':=A/(\sqrt{2}||A||)$ and $B':=B/(\sqrt{2}||B||)$. 
The form of the $W$ matrix in terms of registers is as follows,
\begin{eqnarray}
W&=&\sum_{c,r,r'}|c,r\rangle\langle c,r'|\otimes \bigg[(4A'_{rc}A'_{r'c}-\delta_{rr'})|0,0\rangle\langle 0, 0|+4A'_{rc}\sqrt{1-A^{'2}_{r'c}}|0,0\rangle\langle 0, 1|\nonumber\\
&+&4\sqrt{1-A^{'2}_{rc}}A'_{r'c}|0,1\rangle\langle 0, 0|
    +(4\sqrt{(1-A^{'2}_{rc})(1-A^{'2}_{r'c})}-\delta_{rr'})|0,1\rangle\langle 0, 1|\nonumber\\
    &+&(4B'_{cr}B'_{cr'}-\delta_{rr'})|1,0\rangle\langle 1, 0|
    +4B'_{cr}\sqrt{1-B^{'2}_{cr'}}|1,0\rangle\langle 1, 1|\nonumber\\
    &+&4\sqrt{1-B^{'2}_{cr}}B'_{cr'}|1,1\rangle\langle 1, 0|+(4\sqrt{(1-B^{'2}_{cr})(1-B^{'2}_{cr'})}-\delta_{rr'})|1,1\rangle\langle 1, 1|\nonumber\\
&+&4A'_{rc}B'_{cr'}|0,0\rangle\langle1,0|+4B'_{cr}A'_{r'c}|1,0\rangle\langle0,0|+4B'_{cr}\sqrt{1-A^{'2}_{r'c}}|1,0\rangle\langle 0, 1|+4\sqrt{1-A^{'2}_{rc}}B'_{cr'}|0,1\rangle\langle 1, 0|\nonumber\\
&+&4A'_{rc}\sqrt{1-B^{'2}_{cr'}}|0,0\rangle\langle 1,1|+4\sqrt{1-B^{'2}_{cr}}A'_{r'c}|1,1\rangle\langle 0,0|+4\sqrt{(1-B^{'2}_{cr})(1-A^{'2}_{cr'})}|1,1\rangle\langle 0,1|\nonumber\\
&+&4\sqrt{(1-B^{'2}_{cr'})(1-A^{'2}_{cr})}|0,1\rangle\langle 1,1|\bigg]
\end{eqnarray}
Starting from an initial state with Hadamard on the column qubit registers we obtain,
\begin{eqnarray}
H^{\otimes p}|0\rangle|j\rangle|1\rangle|0\rangle=\frac{1}{P}\sum_{c}|c\rangle|j\rangle|1\rangle|0\rangle.
\end{eqnarray}
Upon acting $W$,
\begin{eqnarray}
    WH^{\otimes p}|0\rangle|j\rangle|1\rangle|0\rangle&=&\frac{1}{2P}\sum_{c,r}|c,r\rangle\bigg[(2B'_{cr}B'_{cj}-\delta_{rj})|1,0\rangle+2\sqrt{1-A^{'2}_{rc}}B'_{cj}|0,1\rangle\nonumber\\
    &+&2\sqrt{1-B^{'2}_{rc}}B'_{cj}|1,1\rangle+2A'_{rc}B'_{cj}|0,0\rangle\bigg]
\end{eqnarray}
Taking overlap of $WH^{\otimes p}|0\rangle|j\rangle|1\rangle|0\rangle$ with the state $H^{\otimes p}|0\rangle|i\rangle|0\rangle|0\rangle$ we get,
\begin{eqnarray}
    \langle 0|\langle 0|\langle i|\langle 0|H^{\otimes p}WH^{\otimes p}|0\rangle|j\rangle|1\rangle|0\rangle=\frac{4}{4P^{2}}\sum_{k}A'_{ik}B'_{kj}=\frac{4}{P^{2}}(A'B')_{ij}=\frac{(AB)_{ij}}{P^{2}||A||||B||}
\end{eqnarray}
By construction we have proved the existence of $U(A,B)$,
\begin{eqnarray}
    U(A,B)=H^{\otimes p}(2T^{\dagger}(A,B)T(A,B)-1)H^{\otimes p}.
\end{eqnarray}
\subsection{Matrix-multiplication with Quantum circuits only with Unitary operators}
\subsubsection*{Theorem}
(Isometry free proof)If $A$ and $B$ are general rectangular matrices of dimensions $dim(A)=(N,P)$ and $dim(B)=(P,M)$ then there is a unitary operation $U(A,B)$ of dimension $2^{n_{q}}\times 2^{n_{q}}$ that operates on a system of $n_{q}=p+\max(m,n)+2$ qubit registers $|\cdot\rangle_{p}|\cdot\rangle_{\max(m,n)}|\cdot\rangle_{a_{1}}|\cdot\rangle_{a_{2}}$ (where $n=\lceil\log_{2} N\rceil$,$m=\lceil\log_{2} M\rceil$,$p=\lceil\log_{2} P\rceil$ )and block encodes the matrix multiplication of $A$ and $B$ s.t. 
\begin{equation*}
    \langle 0|_{p}\langle i|_{\max(m,n)}\langle 0|_{a_{1}}\langle 0|_{a_{2}}U(A,B)|0\rangle_{p}|j\rangle_{\max(m,n)}|1\rangle_{a_{1}}|0\rangle_{a_{2}}=\frac{1}{max(M,N)P}|\frac{\sum_{k}A_{ik}B_{kj}}{||A||||B||}.
\end{equation*}
\textit{Proof-}
Let us take the normalized matrices $A'=A/(\sqrt{2}||A||)$, $B'=B/(\sqrt{2}||B||)$. For these we define two unitary operators $V(A)$,$V(B)$,
\begin{eqnarray}
V_A&=&\sum_{c=0,r=0}^{2^{p},2^{\max(m,n)}}\left[|c,r,0\rangle\langle c,r,0|\otimes \left(A'_{rc}I+i\sqrt{1-A_{rc}^{'2}}Y\right)+ |c,r,1\rangle\langle c,r,1|\otimes I_{2}\right]\nonumber\\
V_B&=&\sum_{c=0,r=0}^{2^{p},2^{\max(m,n)}}\left[|c,r,0\rangle\langle c,r,0|\otimes I_{2}+ |c,r,1\rangle\langle c,r,1|\otimes\left(B_{cr}'I+i\sqrt{1-B_{cr}^{'2}}Y\right)\right]
\end{eqnarray}
We load the classical data of the B matrix using the state preparation oracle $V_BH^{\otimes p}$ on the initial state $|0\rangle|j\rangle|1\rangle|0\rangle$,
\begin{eqnarray}
   |\Phi_{B}\rangle= V_BH^{\otimes p}|0\rangle|j\rangle|1\rangle|0\rangle  = \frac{1}{\sqrt{P}}\sum_{c}\left[B'_{cj}|c,j,1,0\rangle+\sqrt{1-B_{cj}^{'2}}|c,j,1,1\rangle\right].
\end{eqnarray}
We load the classical data of the A matrix using the state preparation oracle $V_AH^{\otimes p}$,
\begin{eqnarray}
|\Phi_{A}\rangle&=& V_AH^{\otimes p}|0\rangle|i\rangle|0\rangle|0\rangle =\frac{1}{\sqrt{P}}\sum_{c}\left[A^{'}_{ic}|c,i,0,0\rangle+\sqrt{1-A^{'2}_{ic}}|c,i,0,1\rangle\right]
\end{eqnarray}
Note that the states $|\Phi_{A}\rangle$ and $|\Phi_{B}\rangle$ are orthogonal,
\begin{eqnarray}
   \langle\Phi_{A}|\Phi_{B}\rangle=0
\end{eqnarray}
Next we define diffusion operator $R$ acting on the row registers and the ancillas $a_{1}$, $a_{2}$,
\begin{eqnarray}
    R&=&I_{2}^{\otimes p}\otimes\left[ \left(H^{\otimes \max(m,n)}\otimes H\otimes I_{2}\right)\left(2|0,0,0\rangle\langle 0,0,0|-1\right)\left(H^{\otimes \max(m,n)}\otimes H\otimes I_{2}\right)\right]\nonumber\\
    &=&I_{2}^{\otimes p}\otimes\left[\sum_{k,l}2\frac{|k,+,0\rangle\langle l,+,0|}{\max(M,N)}-I\right]
\end{eqnarray}
Then the overlap between these two states $|\Phi_{A}\rangle$ and $R|\Phi_{B}\rangle$ is given by,
\begin{eqnarray}
\langle\Phi_{A}|R|\Phi_{B}\rangle&=&  \langle 0,i,0,0|H_{c}^{\otimes p}V^{\dagger}_ARV_BH^{\otimes p}_{c}|0,j,1,0\rangle\nonumber\\
&=&\frac{2\sum_{c}A'_{ic}B'_{cj}}{\max(M,N)P}=\frac{\sum_{c}A_{ic}B_{cj}}{\max(M,N)P||A||||B||}
\end{eqnarray}
By construction we have proved the existence of $U(A,B)$ that can be defined without any isometry,
\begin{eqnarray}
    U(A,B)=H_{c}^{\otimes p}V^{\dagger}_ARV_BH^{\otimes p}_{c}.
\end{eqnarray}

\appendix

\begin{acknowledgement}
The authors would like to thank his collegeaus Anil Sharma, Manoj Nambiar, Geetha Thiagarajan, Sriram Goverpet Srinivasan from TCS for their constructive feedback and support. 
\end{acknowledgement}

\section*{Declaration}
This work is based on two patents US patent No-20240202561 and Indian patent No. 202421039027.  
\bibliographystyle{plain}
\input{main.bbl}

\end{document}

%% file: main.bbl
\providecommand{\latin}[1]{#1}
\makeatletter
\providecommand{\doi}
  {\begingroup\let\do\@makeother\dospecials
  \catcode`\{=1 \catcode`\}=2 \doi@aux}
\providecommand{\doi@aux}[1]{\endgroup\texttt{#1}}
\makeatother
\providecommand*\mcitethebibliography{\thebibliography}
\csname @ifundefined\endcsname{endmcitethebibliography}
  {\let\endmcitethebibliography\endthebibliography}{}